%% file: credibility-analysis.tex
\pdfoutput=1
\input{head/settings_epfl_template.tex}

\input{head/settings_custom.tex}  % place your custom packages, etc... in this file!

\newtheorem{question}{\color{blue}{RQ:}}

\begin{document}

\frontmatter
\input{head/titlepage.tex}

\include{head/reviewer}
\include{head/dedication}

\setcounter{page}{0}
\setcounter{tocdepth}{5}
\input{head/acknowledgements}
%\include{head/preface}
\include{head/abstracts}

%\include{main/introduction/abstract}

\tableofcontents
\cleardoublepage
\phantomsection
\addcontentsline{toc}{chapter}{List of figures} % adds an entry to the table of contents
\listoffigures
\cleardoublepage
\phantomsection
\addcontentsline{toc}{chapter}{List of tables} % adds an entry to the table of contents
\listoftables
% your list of symbols here, if needed.

% space before each new paragraph according to the template guidelines.
%(needs to be after titlepage and frontmatter to keep the table of contents lists short)
\setlength{\parskip}{1em}

%%%%%%%%%%%%%%%%%%%%%%%%%%%%%%%%%%%%%%%%%%%%%%
%%%%% MAIN: The chapters of the thesis
%%%%%%%%%%%%%%%%%%%%%%%%%%%%%%%%%%%%%%%%%%%%%%
\mainmatter

%\chapter{Abstract}
%\chapter{Summary}
%\chapter{Introduction}
%\chapter{Background and Related Work}

%\include{main/introduction}
\include{main/chapter-introduction/main}
\include{main/chapter-related-work/main}

\include{main/chapter-credibility-analysis/main}

\include{main/chapter-temporal-evolution/main}

\include{main/chapter-credible-applications/main}

\include{main/chapter-conclusions/main}

%\section{Finding Useful Product Reviews}

%\section{Consisten Analysis of Credible Reviews with Limited Information}

%\include{main/chapter--1-summary}
%\include{main/chapter-0-introduction}
%\include{main/chapter-1-peopleondrugs/kdd14}
%\include{main/chapter-2-peopleonmedia/cikm15}
%\include{main/chapter-3-discrete-experience/icdm15}
%\include{main/chapter-4-helpful-prediction/icdm16}
%\include{main/chapter-5-continuous-experience/kdd16}
%\include{main/chapter-6-limited-information/ecml16}
%%%%%%%%%%%%%%%%%%%%%%%%%%%%%%%%%%%%%%%%%%%%%%
%%%%% TAIL: Bibliography, Appendix, CV
%%%%%%%%%%%%%%%%%%%%%%%%%%%%%%%%%%%%%%%%%%%%%%
%\include{tail/appendix}

%\chapter{Conclusions}

\backmatter
\include{tail/biblio}

% Add your glossary here
% Add your index here
% Photographic credits (list of pictures&images that have been used with names of the person holding the copyright for them)
%\include{tail/cv}

\end{document}

%% file: head/settings_epfl_template.tex
\pdfoutput=1
\documentclass[a4paper,11pt,fleqn]{book}
\pdfoutput=1
\usepackage[T1]{fontenc}
\usepackage[utf8]{inputenc}
\usepackage[french,german,english]{babel}

\usepackage{fourier} % Utopia font-typesetting including mathematical formula compatible with newer TeX-Distributions (>2010)
\usepackage{setspace} % increase interline spacing slightly

\usepackage{graphicx,xcolor}
\graphicspath{{images/}}

\usepackage{subcaption}
\usepackage{booktabs}
\usepackage{lipsum}
\usepackage{microtype}
\usepackage{url}
\usepackage[final]{pdfpages}
\usepackage{bbm}
\usepackage{caption}
\usepackage{xspace}
\usepackage[lined, ruled, boxed, commentsnumbered]{algorithm2e}
\usepackage{amssymb,amsfonts,amsmath}
\usepackage{verbatim}
\usepackage{wrapfig}
\usepackage{multirow}

\usepackage{fancyhdr}

\fancypagestyle{plain}{
	\fancyhf{}

	\fancyfoot[EL,OR]{\thepage}}
\fancypagestyle{addpagenumbersforpdfimports}{
	\fancyhead{}
	
	\fancyfoot{}
	\fancyfoot[RO,LE]{\thepage}
}

\usepackage{listings}
\usepackage{hyperref}
\hypersetup{pdfborder={0 0 0},
	colorlinks=true,
	linkcolor=blue,
	citecolor=violet,
	urlcolor=pinegreen}
\makeatletter
\def\cleardoublepage{\clearpage\if@twoside \ifodd\c@page\else
    \hbox{}
    \thispagestyle{empty}
    \newpage
    \if@twocolumn\hbox{}\newpage\fi\fi\fi}
\makeatother \clearpage{\pagestyle{plain}\cleardoublepage}

\usepackage{color}
\usepackage{tikz}
\usepackage[explicit]{titlesec}
\newcommand*\chapterlabel{}

\titleformat{\chapter}[display]  % type (section,chapter,etc...) to vary,  shape (eg display-type)
	{\normalfont\bfseries\Huge} % format of the chapter
	{\gdef\chapterlabel{\thechapter\ }}     % the label 
 	{0pt} % separation between label and chapter-title
 	  {\begin{tikzpicture}[remember picture,overlay]
    \node[yshift=-8cm] at (current page.north west)
      {\begin{tikzpicture}[remember picture, overlay]
        \draw[fill=black] (0,0) rectangle(35.5mm,15mm);
        \node[anchor=north east,yshift=-7.2cm,xshift=34mm,minimum height=30mm,inner sep=0mm] at (current page.north west)
        {\parbox[top][30mm][t]{15mm}{\raggedleft $\phantom{\textrm{l}}$\color{white}\chapterlabel}};  %the black l is just to get better base-line alingement
        \node[anchor=north west,yshift=-7.2cm,xshift=37mm,text width=\textwidth,minimum height=30mm,inner sep=0mm] at (current page.north west)
              {\parbox[top][30mm][t]{\textwidth}{\color{black}#1}};
       \end{tikzpicture}
      };
   \end{tikzpicture}
   \gdef\chapterlabel{}
  } % code before the title body

\titlespacing*{\chapter}{0pt}{50pt}{30pt}
\titlespacing*{\section}{0pt}{13.2pt}{*0}  % 13.2pt is line spacing for a text with 11pt font size
\titlespacing*{\subsection}{0pt}{13.2pt}{*0}
\titlespacing*{\subsubsection}{0pt}{13.2pt}{*0}

\newcounter{myparts}
\newcommand*\partlabel{}
\titleformat{\part}[display]  % type (section,chapter,etc...) to vary,  shape (eg display-type)
	{\normalfont\bfseries\Huge} % format of the part
	{\gdef\partlabel{\thepart\ }}     % the label 
 	{0pt} % separation between label and part-title
 	  {\setlength{\unitlength}{20mm}
	  \addtocounter{myparts}{1}
	  \begin{tikzpicture}[remember picture,overlay]
    \node[anchor=north west,xshift=-65mm,yshift=-6.9cm-\value{myparts}*20mm] at (current page.north east) % for unknown reasons: 3mm missing -> 65 instead of 62
      {\begin{tikzpicture}[remember picture, overlay]
        \draw[fill=black] (0,0) rectangle(62mm,20mm);   % -\value{myparts}\unitlength
        \node[anchor=north west,yshift=-6.1cm-\value{myparts}*20mm,xshift=-60.5mm,minimum height=30mm,inner sep=0mm] at (current page.north east)
        {\parbox[top][30mm][t]{55mm}{\raggedright \color{white}Part \partlabel $\phantom{\textrm{l}}$}};  %the phantom l is just to get better base-line alingement
        \node[anchor=north east,yshift=-6.1cm-\value{myparts}*20mm,xshift=-63.5mm,text width=\textwidth,minimum height=30mm,inner sep=0mm] at (current page.north east)
              {\parbox[top][30mm][t]{\textwidth}{\raggedleft \color{black}#1}};
       \end{tikzpicture}
      };
   \end{tikzpicture}
   \gdef\partlabel{}
  } % code before the title body

\usepackage{amsmath}
\makeatletter
\def\resetMathstrut@{%
  \setbox\z@\hbox{%
    \mathchardef\@tempa\mathcode`\(\relax
      \def\@tempb##1"##2##3{\the\textfont"##3\char"}%
      \expandafter\@tempb\meaning\@tempa \relax
  }%
  \ht\Mathstrutbox@1.2\ht\z@ \dp\Mathstrutbox@1.2\dp\z@
}
\makeatother

%% file: head/settings_custom.tex
\newcommand{\xhdrNoPeriod}[1]{\vspace{1mm}\noindent{{\bf #1}}}

\newcommand{\sectionrule}{\addlinespace[0.5ex]}

\definecolor{pinegreen}{rgb}{0.0, 0.47, 0.44}

\newcommand{\squishlist}{
   \begin{list}{$\bullet$}
    { \setlength{\itemsep}{0pt}      \setlength{\parsep}{3pt}
      \setlength{\topsep}{3pt}       \setlength{\partopsep}{0pt}
      \setlength{\leftmargin}{1.5em} \setlength{\labelwidth}{1em}
      \setlength{\labelsep}{0.5em} } }
\newcommand{\squishlisttwo}{
   \begin{list}{$\bullet$}
    { \setlength{\itemsep}{0pt}    \setlength{\parsep}{0pt}
      \setlength{\topsep}{0pt}     \setlength{\partopsep}{0pt}
      \setlength{\leftmargin}{0.5em} \setlength{\labelwidth}{0.5em}
      \setlength{\labelsep}{0.5em} } }

      \newcommand{\squishlistthree}{
 \begin{enumerate}
  { \setlength{\itemsep}{0pt}
     \setlength{\parsep}{3pt}
     \setlength{\topsep}{3pt}
     \setlength{\partopsep}{0pt}
     \setlength{\leftmargin}{1.5em}
     \setlength{\labelwidth}{1em}
     \setlength{\labelsep}{0.5em} } }
     
     \newcommand{\squishendthree}{\end{enumerate}}
\newcommand{\squishend}{
    \end{list} 
}

\newcommand*{\myindent}{\hspace*{0.4cm}}%

\newtheorem{exm}{Example}[section]

\newcommand{\example}[1]{\vspace*{-1em}\begin{exm}{#1}\end{exm}\vspace*{-1em}}
\newcommand{\comm}[1]{}
\renewcommand{\comment}[1]{}

\newcommand{\todo}[1]{
\textcolor{red}{\textbf{(TODO: #1)}}
}

%% file: head/titlepage.tex
\begin{titlepage}

\begin{center}
  % Title
  \textsf{\textbf{{\Huge Probabilistic Graphical Models for}}}\\ \vspace{1mm}
  \textsf{\textbf{{\Huge Credibility Analysis in}}}\\    \vspace{2mm}
  \textsf{\textbf{{\Huge Evolving Online Communities}}}  \vspace{1mm}

  \vspace{30mm}

  % Author
  \textbf{\Large{Subhabrata Mukherjee}}\\
  \Large{Max-Planck-Institut f\"{u}r Informatik}

  \vspace{30mm}

  \textbf{Dissertation}\\
  zur Erlangung des Grades\\
  {\em des Doktors der Ingenieurwissenschaften (Dr.-Ing.)} \\
  %der Naturwissenschaftlich-Technischen Fakult\"{a}ten \\
  der Fakult\"{a}t f\"{u}r Mathematik und Informatik \\
  der Universit\"{a}t des Saarlandes

  \vspace{35mm}

  Saarbr\"{u}cken\\
  March 2017

\end{center}

\end{titlepage}

%% file: head/reviewer.tex
\thispagestyle{empty}

\vspace*{3cm}

\begin{center}{
\large
\begin{tabular}{ p{3cm} p{10cm} }
  Dean & Prof. Dr. Frank-Olaf Schreyer \\
       & Faculty of Mathematics and Computer Sciences\\
       & Saarland University\\
       & Saarbr\"ucken, Germany\\\\
            
  Colloquium & July 6, 2017 \\
	    & Saarbr\"ucken, Germany \\\\ 

  {\bf Examination} & {\bf Board}\\\\
  Advisor and & Prof. Dr. Gerhard Weikum \\
  First Reviewer & Department of Databases and Information Systems \\ 
		 & Max Planck Institute for Informatics\\
		 & Saarbr\"ucken, Germany \\\\
		 		 
  Second Reviewer & Prof. Dr. Jiawei Han \\
		  & Department of Computer Science \\
		  & University of Illinois at Urbana-Champaign\\
		  & Urbana, USA\\\\
		  
  Third Reviewer & Prof. Dr. Stephan G\"unnemann \\
		 & Department of Informatics\\
		 & Technical University of Munich\\
		 & Munich, Germany\\\\
  Chairman & Prof. Dr. Dietrich Klakow \\
	   & Department of Computer Science\\
	   & Saarland University\\
	   & Saarbr\"ucken, Germany \\\\
	   
  Research & Dr. Rishiraj Saha Roy \\
  Assistant & Department of Databases and Information Systems \\ 
	   & Max Planck Institute for Informatics\\
	   & Saarbr\"ucken, Germany \\\\

\end{tabular}
}
\end{center}

%% file: head/dedication.tex
\cleardoublepage
\thispagestyle{empty}

\vspace*{3cm}

\begin{raggedleft}
    	``Note to self: every time you were convinced you couldn't go on, \\
	you did.''\\
     --- Unknown\\
\end{raggedleft}

\vspace{4cm}

\begin{center}
    To my (latent) support system --- my loving parents and brother, and my beautiful wife \\ Sarah \dots
\end{center}

%% file: head/acknowledgements.tex
\chapter*{Acknowledgements}
\markboth{Acknowledgements}{Acknowledgements}
\addcontentsline{toc}{chapter}{Acknowledgements}
% put your text here
\vskip0.5cm

First and foremost, I would like to express my deepest gratitude to my supervisor and mentor Gerhard Weikum for giving me the opportunity to pursue research under his guidance. His constant motivation, excellent scientific advice, wisdom, and vision have been of quintessential importance to make this work possible. I will always cherish our interactions that have helped me mature not only as a researcher, but also as a person.\\ 
%\lipsum[1-2]

I would like to thank the additional reviewers and examiners of my dissertation,
Jiawei Han, and Dietrich Klakow for their valuable feedback. I am extremely grateful to all my collaborators and co-authors --- Cristian Danescu-Niculescu-Mizil, Stephan G\"unnemann, Hemank Lamba, Kashyap Popat, Sourav Dutta, and Jannik Str\"otgen --- for actively contributing to, and shaping my dissertation. I am thankful to all my colleagues at the Max Planck Institute for Informatics for participating in discussions, and providing insightful ideas and valuable feedback during the course of my doctoral studies. I am thankful to all my friends here to have made my journey an enjoyable one, especially Arunav Mishra, Sarvesh Nikumbh, Tomasz Tylenda, Dilafruz Amanova, Sourav Dutta and Nikita Dutta. I would also like to thank all the administrative staff at the Max Planck Institute for being supportive and providing assistance whenever necessary, so I could freely indulge in my research. I owe many thanks to the International Max 
Planck Research School and the Max Planck Society for the financial support that allowed 
me to pursue my research, and present my work at conferences around the 
world.\\

Last but not least, I would like to thank my parents Sushama and Subrata Mukherjee, and my brother Subhojyoti Mukherjee for their continued support and encouragement. Most importantly, I thank my wife Sarah John for being by my side since the beginning of time.

\bigskip
 
\noindent\textit{Saarbr\"ucken, March 2017}
\hfill S.~M.

%% file: head/abstracts.tex
%\begingroup
%\let\cleardoublepage\clearpage

% English abstract
\cleardoublepage
\chapter*{Abstract}
%\markboth{Abstract}{Abstract}
\addcontentsline{toc}{chapter}{Abstract (English/Deutsch)} % adds an entry to the table of contents
% put your text here
\vskip0.5cm
%Key words: 
%put your text here
One of the major hurdles preventing the full exploitation of information from online communities is the widespread concern regarding the quality and credibility of user-contributed content. Prior works in this domain operate on a static snapshot of the community, making strong assumptions about the structure of the data (e.g., relational tables), or consider only shallow features for text classification.\\

To address the above limitations, we propose probabilistic graphical models that can leverage the joint interplay between multiple factors in online communities --- like user interactions, community dynamics, and textual content --- to automatically assess the credibility of user-contributed online content, and the expertise of users and their evolution with {\em user-interpretable explanation}. To this end, we devise new models based on Conditional Random Fields for different settings like incorporating partial expert knowledge for semi-supervised learning, and handling discrete labels as well as numeric ratings for fine-grained analysis. This enables applications such as extracting reliable side-effects of drugs from user-contributed posts in healthforums, and identifying credible content in news communities.\\

Online communities are dynamic, as users join and leave, adapt to evolving trends, and mature over time. To capture this dynamics, we propose generative models based on Hidden Markov Model, Latent Dirichlet Allocation, and Brownian Motion to trace the continuous evolution of user expertise and their language model over time. This allows us to identify expert users and credible content jointly over time, improving state-of-the-art recommender systems by explicitly considering the maturity of users. This also enables applications such as identifying helpful product reviews, and detecting fake and anomalous reviews with {limited} information.

%Interpretability forms an important component of these models whereby we show distributional word clusters, representative snippets, evolution traces etc. to explain the models' verdict to the end-user.

% German abstract
\begin{otherlanguage}{german}
\cleardoublepage
\chapter*{Kurzfassung}
%\markboth{Zusammenfassung}{Zusammenfassung}
% put your text here
%\lipsum[1-2]
\vskip0.5cm
%Stichwörter: 
%put your text here
Eine der gr\"o\ss ten H\"urden, die die vollst\"andige Nutzung von Informationen aus sogenannten Online-Communities verhindert, sind weitverbreitete Bedenken bez\"uglich der Qualit\"at und Glaubw\"urdigkeit von Nutzer-generierten Inhalten. Fr\"uhere Arbeiten in diesem Bereich gehen von einer statischen Version einer Community aus, machen starke Annahmen bez\"uglich der Struktur der Daten (z.B. relationale Tabellen) oder ber\"ucksichtigen nur oberfl\"achliche Merkmale zur Klassifikation von Texten.\\

Um die oben genannten Einschr\"ankungen zu adressieren, schlagen wir eine Reihe von probabilistischen graphischen Modellen vor, die das Zusammenspiel mehrerer Faktoren in Online-Communities ber\"ucksichtigen: Interaktionen zwischen Nutzern, die Dynamik in Communities und der textuell Inhalt. Dadurch k\"onnen die Glaubw\"urdigkeit von Nutzer-generierten Online Inhalten sowie die Expertise von Nutzern und ihrer Entwicklung mit {\em interpretierbaren Erkl\"arungen} bewertet werden. Hierf\"ur konstruieren wir neue, auf Conditional Random Fields basierende Modelle f\"ur verschiedene Szenarien, um beispielsweise partielles Expertenwissen mittels semi-\"uberwachtem Lernen zu ber\"ucksichtigen. Genauso k\"onnen diskrete Labels sowie numerische Ratings f\"ur pr\"azise Analysen genutzt werden. Somit werden Anwendungen erm\"oglicht wie etwa das automatische Extrahieren von Nebenwirkungen von Medikamenten aus Nutzer-erstellten Inhalten in Gesundheitsforen und das Identifizieren von vertrauensw\"urdigen Inhalten aus 
Nachrichten-Communities.\\

Online-Communities sind dynamisch, da Nutzer zu Communities hinzusto\ss en oder diese verlassen. Sie passen sich entstehenden Trends an und entwickeln sich \"uber die Zeit. Um diese Dynamik abzudecken, schlagen wir generative Modelle vor, die auf Hidden Markov Modellen, Latent Dirichlet Allocation und Brownian Motion basieren. Diese k\"onnen die kontinuierliche Entwicklung von Nutzer-Erfahrung sowie ihrer Sprachentwicklung \"uber die Zeit nachzeichnen. Dies erm\"oglicht uns, Expertennutzer und glaubw\"urdigen Inhalt \"uber die Zeit gemeinsam zu identifizieren, sodass die aktuell besten Recommender- Systeme durch das explizite Ber\"ucksichtigen der Entwicklung und der Expertise von Nutzern verbessert werden k\"onnen. Dadurch wiederum k\"onnen Anwendungen entwickelt werden, die n\"utzliche Produktbewertungen erkennen sowie fingierte und anomale Bewertungen mit geringem Informationsgehalt identifizieren.

\end{otherlanguage}

%\endgroup			
%\vfill

%% file: main/chapter-introduction/main.tex
\chapter{Introduction}

\section{Motivation}

In recent years, the explosion of social networking sites (e.g., Facebook, Twitter),
blogs (e.g., Mashable, Techcrunch), and online review portals (e.g., Amazon, TripAdvisor,
IMDB, Healthboards) provide overwhelming amount of information on various topics like health, politics, movies, music, travel, and more. However, the usability of such massive data is largely restricted due to concerns about the quality and credibility of user-contributed content.

Online communities are massive repositories of knowledge that are accessed by regular everyday users as well as expert professionals. For instance, $59\%$ of the adult U.S. population and nearly half of U.S. physicians consult online resources (e.g., Youtube and Wikipedia)~\cite{Fox:PewInternetAndAmericanLifeProject:2013,IMS2014} for health-related information. In the product domain, $40\%$ of online consumers would not buy electronics without consulting online reviews first~\cite{nielsen}. However, this user-contributed content is highly noisy, unreliable, and subjective with rampant amount of spams, rumors, and misinformation injected by users in their postings.  This has greatly eroded public trust and confidence on social media information. Some statistics show that $66\%$ of web-using U.S. adults do not trust social media information~\cite{Fox:PewInternetAndAmericanLifeProject:2016}. To counter these, stakeholders in the industry (e.g., {\tt \href{https://www.yelp.com}{Yelp}})   
have been developing their own defense mechanism\footnote{\url{https://www.yelpblog.com/2013/09/fake-reviews-on-yelp-dont-worry-weve-got-your-back}\\Yelp filter rejects $25\%$ of user-contributed reviews as non-reliable.
}. %For instance,  developed a filter that rejects $25\%$ of user-contributed reviews as non-reliable\footnote{https://www.yelpblog.com/2013/09/fake-reviews-on-yelp-dont-worry-weve-got-your-back}. 
In certain domains like healthforums, misinformation can have hazardous consequences --- as these  
are frequently accessed by users to find potential side-effects of drugs, symptoms of diseases, or getting advice from health professionals. To give an example, consider the following user-post from the online healthforum {\tt \href{http://www.healthboards.com/}{Healthboards}}.

\example{

I took a cocktail of meds. Xanax gave me hallucinations and a demonic feel. I can feel my skin peeling off.
}

The above post suggests that {``peeling-of-skin''} is a probable side-effect of the drug Xanax, although the {\em style} in which it is written renders its credibility doubtful. 

In this case, the user seems to be suffering from hallucinations; and the side-effect can also be attributed to the ``cocktail of meds'', and not Xanax alone.

Prior works in Natural Language Processing dealing with fake reviews and opinion spam \cite{Mihalcea2009,Ott2011,recasens2013,Li2014} would only analyze the linguistic cues and writing style of this post (e.g., distribution of unigrams and bigrams, affective emotions, part-of-speech tags, etc.) to find if it is subjective, biased, or fake.
However, it is difficult to arrive at a conclusion by analyzing the post in isolation. In general, online communities provide many other signals that can help us in this task. 
For instance, the above post may be refuted (or downvoted) by an {\em experienced} health professional in the community. Similarly, credible postings or statements may be {\em corroborated} (or upvoted) by other experienced users in the community. A significant challenge is that {\em a priori} we do not know which users are experienced or trustworthy --- that need to be inferred as a part of
the task. 
{\em These kinds of implicit or explicit feedback from other users, and their identities, prove to be helpful for credibility analysis in a community-specific setting.} 

Prior works in Data Fusion and Truth Discovery (cf.~\cite{YaliangLi:SIGKDD2015} for a survey) leverage such interactions between sources and queries in a general setting. Some typical queries are ``the height of Mount Everest'' that fetch different answers (e.g., ``29,035 feet'', ``29,002 feet'', ``29,029 feet'') from various sources, or ``the birthplace of Obama'' that includes answers as ``Hawaii'', ``USA'', and ``Africa''. These methods aim to resolve conflicts among these multi-source data by obtaining reliability estimates of the sources providing the information (e.g., Wikipedia being a trustworthy source provides an accurate answer to the above queries), and aggregating their responses to obtain the truth. However, these approaches operate over structured data (e.g., relational tables, structured query templates like ``Obama\_BornIn\_Kenya'' represented as a subject-predicate-object triple), and factual claims --- whereby they ignore the content and context of information. These approaches are not 
geared for online communities with more fine-grained interactions, subjective, and unstructured data. Context helps us in understanding the attitude and emotional state of the user writing the posts, the topics of the postings and users' topic-specific expertise, objectivity and rationality of the postings, etc. Similar principles hold true for any online community like music, travel, politics, and news. %For instance, a user experienced on anxiety disorders may be oblivious to the risks of nuclear radiation. Similar principles hold true for other communities like music, travel, politics, and news as well. This example demonstrates the complex interplay between several factors in online communities --- like writing style, cross-talk between users and interactions, user experience, and topics --- that influence the credibility of statements therein. 

The above discussion demonstrates the complex interplay between several factors in online communities --- like writing style, cross-talk between users and interactions, user experience, and topics --- that influences the credibility of statements therein. A natural way to represent these interactions and dependencies between various factors is provided by Probabilistic Graphical Models (PGM) (like, Markov Random Fields, Bayesian Networks, and Factor Graphs)~\cite{KollerFriedman2009}, where each of the above aspects can be envisioned as random variables with edges depicting interactions between them. 

PGMs provide a natural framework to compactly represent high-dimensional distributions over many random variables as a product of local factors over subsets of the variables, i.e., by factoring the joint probability distribution into marginal distributions over subsets of the variables. The conditional independence assumptions, and factorization help us to make the problem tractable. It is also effective in practice as any random 
variable interacts with only a subset of all the variables. During inference and learning, we estimate the joint probability distribution, the marginals, and other queries of interest. In terms of {\em interpretability}, output of probabilistic models (labels, probabilities of queries and factors) can be better explained to the end-user. For instance, a PGM may label two sources as ``trustworthy'' with corresponding probabilities as $0.9$ and $0.7$ --- which is easier to envision than obtaining corresponding raw estimates as $12.7$ and $9.6$. 

{\bf The key contribution of this work is in bringing all of these different aspects together in a computational model, namely, a probabilistic graphical model, for credibility analysis in online communities, and providing efficient inference techniques for the same.}%{\bf The key prIn this work, we develop principles and models to automatically assess the credibility of user-generated statements in online communities.

\comment{
\section{Challenges} 

\begin{itemize}
 \item Unstructured, subjective, and noisy data
 \item Complex interactions in online communities
 \item Discrete and continuous output space
 \item Supervised learning
 \item Limited data
 \item Temporal evolution
 \item Interpretability
\end{itemize}
}

\section{Challenges}

Analyzing the credibility of user-contributed content in online communities is a difficult task with the following challenges:
\squishlist
\item User-contributed postings in online forums are {\em unstructured, biased, and subjective} in nature. This is in contrast to the classical setting in prior works in Truth Discovery and Data Fusion that deal with structured and factual data.
\item Although reliable sources and users contribute credible information, {\em a priori} we do not know which of these sources and users are trustworthy (or experts).
\item Online communities are {\em complex} in nature with rich user-user and user-item interactions (like, upvote, downvote, share, comment, etc.) that are difficult to model computationally.
\item Online communities are {\em dynamic} in nature as users' interactions, maturity, and content evolve over time. 
\item Scarcity of labeled training data and rich statistics (e.g., activity history, meta-data) about users and items lead to data sparsity and difficulty in learning.
\item It is difficult to generate user-interpretable explanations of the models' verdict. 
\squishend

\section{Prior Work and its Limitations}

%\noindent {\bf Extracting reliable knowledge from noisy user-contributed content} 
%\noindent {\bf Dealing with subjectivity and noise in user-contributed content} 
Information extraction methods~\cite{Sarawagi2008,KollerFriedman2009} previously used for extracting information from user-contributed content do not account for the inherent bias, subjectivity, and noise in the data. Additionally, they also do not consider the role of language (e.g., stylistic features, emotional state and attitude of the writer, etc.) in assessing the reliability of the extracted statements. 

%\noindent {\bf Leveraging structure, context, and interactions} 
Prior works in Natural Language Processing~\cite{Mihalcea2009,Ott2011,recasens2013,Li2014}, dealing with opinion spam and fake reviews in online communities, consider postings in isolation, and analyze their writing style to capture bias and subjectivity. They typically {\em ignore} the identity of the users writing the postings, and interactions between them. Typically these works use bag-of-words features, and resources like WordNet~\cite{WordNet}, and SentiWordNet~\cite{Sentiwordnet} to create feature vectors that are fed into supervised machine learning models (e.g., Support Vector Machines) to classify the postings as credible, or otherwise. 

On the other hand, works in Data Fusion and Truth Discovery (cf.~\cite{YaliangLi:SIGKDD2015} for a survey) make strong assumptions about the nature and structure of the data (e.g., relational tables, factual claims, static data, subject-predicate-object triples, etc.) whereby they model the interactions between sources and queries as edges in a network, but {\em ignore} the textual content and context altogether. Typically, these works use approaches like belief propagation and label propagation (e.g., Markov random walks) to propagate reliability estimates in the network. Availability of ground-truth data is a typical problem faced by the works in this domain. Therefore, most of these prior works operate in an unsupervised fashion. However, some prior works show that the performance of these methods can be improved by using a small set of labeled data for training.

In the absence of proper ground-truth data, prior works~\cite{Liu2007,Liu2008,Liu2010,Liu2012,Liu2013,Liu2014,Rahman2015} make strong assumptions, e.g., duplicates and near-duplicates are 
fake, and harness rich information about users and items in the form of activity, posting history, and meta-data. Such profile history may not be readily available in several domains, especially for ``long-tail'' users and items in the community (e.g., newcomers and recently launched products). Also, such a policy tends to over emphasize long-term contributors and suppress outlier opinions off the mainstream.

%data for training. that do not require much labeled data for training their models. even though supervised models have been shown to outperform unsupervised ones in literature.

%\noindent{\bf Temporal evolution of online communities, and finding experienced users} 
Prior works in collaborative filtering \cite{korenKDD2008, koren2011advances,Liu2008, Tang:2013:CRH:2507157.2507183,Ma:2015:FFG:2783258.2783314} consider a static snapshot of the data whereby they {\em ignore} the temporal evolution of users and their interactions. These use activity history (e.g., frequency of postings, number of upvotes / downvotes, rating history) as a proxy to find experienced members in the community. Online communities are dynamic in nature as users join and leave, adopt new vocabulary, and adapt to evolving trends. Therefore, a user who was not experienced a decade before could have evolved into a matured user now with refined preferences, writing style, and trustworthiness. This dimension of user evolution is ignored in the static analysis.

%\noindent {\bf User-interpretable explanation and output} 
Most of the works involving classifiers and machine learning models generate discrete (e.g., binary) decision labels as output. These models have limited interpretability as they rarely explain why the model arrived at a particular verdict. Most of these are not geared for fine-grained analysis involving continuous data types. Additionally, most of the prior works output only raw scores, as estimates of reliability, that are difficult to explain to the end-user. %In principle, it is possible to rank the features by their weights in the learned classifier --- 

%\noindent {\bf Harvesting expert knowledge from large-scale non-expert data} 

\section{Contributions}

This work addresses the challenges outlined above developing principles and models to advance the state-of-the-art. In summary, it addresses the following research questions.

{\question How can we develop models that {jointly} leverage the context and interactions in online communities for analyzing the credibility of user-contributed content? How can we complement expert knowledge with large-scale non-expert data from online communities?}

We develop novel forms of probabilistic graphical models that capture the complex interplay between several factors: the writing style, user-user and user-item interactions, latent semantic factors like the topics of the postings and experience of the users, etc. %The intuition behind the proposed models is that trustworthy and experienced users contribute objective and credible postings that are corroborated by other trustworthy users in the community. 
Specifically, we develop Conditional Random Field (CRF) based models, where these factors (e.g., users, postings, statements) are modeled as random variables with edges between them depicting interactions. Furthermore, these variables have observable features that capture the context (e.g., stylistic features, subjectivity, topics, etc.) of the postings and relevant background information (e.g., user demographics and activity history). %There are several latent factors in the model (e.g., experience and trustworthiness of users, topics of postings) that are inferred during inference. 
We develop efficient joint probabilistic inference techniques for these models for classification and regression settings. Specifically, we develop:
%\vspace{-1em}
\squishlist
 \item A {\em semi-supervised} version of the CRF for credibility classification (presented at SIGKDD 2014 \cite{mukherjee2014}) that learns from {\em partial} expert supervision using Expectation - Maximization principle. We use this model in a healthforum {\tt \href{http://www.healthboards.com/}{Healthboards}} to identify rare or uncommon side-effects of drugs from user-contributed posts. This is one of the problems where large-scale non-expert data has the potential to complement expert medical knowledge. Our model leverages partial expert knowledge of drugs and their side-effects to jointly identify credible statements (or, drug side-effects), reliable postings, and trustworthy users in the community. 
 \item A {\em continuous} version of the CRF for more fine-grained credibility regression (presented at CIKM 2015 \cite{SubhoCIKM2015}) to deal with user-assigned {\em numeric ratings} in online communities. As an  application use-case, we consider news communities (e.g., {\tt \href{www.newstrust.net}{NewsTrust}}) that are plagued by misinformation, bias, and polarization induced by the style of reporting and political viewpoint of media sources and users. We show that the joint probability distribution function for the continuous CRF %--- with a judiciously chosen energy function capturing joint interactions between several factors (modeled as cliques) --- 
 is Multivariate Gaussian, and propose a constrained Gradient Ascent based algorithm for scalable inference.
\squishend

We released two large-scale datasets used in these works: 
%\vspace{-1em}
\squishlist
 \item The healthforum dataset\footnote{http://resources.mpi-inf.mpg.de/impact/peopleondrugs/data.tar.gz} contains $2.8$ million posts from $15,000$ {\em anonymized} users in the community {\tt \href{http://www.healthboards.com/}{Healthboards}}, along with their demographic information. Additionally, we also provide side-effects of $2,172$ drugs from $837$ drug families contributed by expert health professionals in {\tt \href{http://www.mayoclinic.org/}{MayoClinic}}. The drug side-effects --- categorized as most common, less common, rare, and unobserved --- are used as ground-truth in our evaluation.
 \item The news community dataset\footnote{http://resources.mpi-inf.mpg.de/impact/credibilityanalysis/data.tar.gz} consists of $84,704$ stories from {\tt \href{www.newstrust.net}{NewsTrust}} on $47,565$ news articles crawled from $5,658$ media sources (like BBC, WashingtonPost, New York Times). The dataset contains $134,407$ NewsTrust-member reviews on the articles, corresponding ratings on various qualitative aspects like objectivity, correctness of information, bias and credibility; as well as interactions (e.g., comments, upvotes/downvotes) between members, and their demographic information.
\squishend

%\pagebreak

%\vspace{-2em}
{\question {How can we quantify changes in users' maturity and experience in online communities?}
{How can we model users' evolution or progression in maturity?}
{How can we improve recommendation by considering a user's evolved maturity or experience at the (current) timepoint of consuming items?}
}

Online communities are dynamic as users mature over time with evolved preferences, writing style, experience, and interactions. We study the temporal evolution of users' experience with respect to item recommendation in a collaborative filtering framework in review communities (like, movies, beer,
and electronics). We propose two approaches to model this evolving user experience, and her writing style:
\squishlist
 \item The first approach (presented at ICDM 2015 \cite{Subho:ICDM2015}) considers a user's experience to progress in a {\em discrete} manner employing a Hidden Markov Model (HMM) -- Latent Dirichlet Allocation (LDA) model: where HMM traces her (latent) experience progression, and LDA models her facets of interest at any timepoint as a function of her (latent) experience. This framework (presented at SDM 2017 \cite{Subho:SDM2017}) is used to identify useful product reviews --- in terms of being {\em helpful} to the end-consumers --- in communities like Amazon, where useful reviews are buried deep within a heap of non-informative ones.
 \item The second approach (presented at SIGKDD 2016 \cite{Subho:KDD2016}) addresses several drawbacks of this discrete evolution, and develops a natural and {\em continuous} mode of temporal evolution of a user's experience, and her language model (LM) using Geometric Brownian Motion (GBM), and Brownian Motion (BM), respectively. %GBM has previously been used in financial domains for modeling stock price behavior. 
 We develop efficient inference techniques to combine discrete multinomial distributions for LDA (generating words per review) with the
continuous Brownian Motion processes (GBM and BM) for experience and LM evolution. To this end, we use a combination of Metropolis Hastings, Kalman Filter, and Gibbs sampling that are shown to work coherently to increase the data log-likelihood smoothly and continuously over time.
\squishend

%Analysis of the GBM trajectory of users offer interesting insights; for instance, users who reach a high level of experience progress faster than those who do not, and also exhibit a comparatively higher variance. Also, the number of reviews written by a user does not have a strong influence, unless they are written over a long period of time. (Latent) word clusters show interesting insights like experienced users in Beer communities use more ``fruity'' words to depict beer taste and smell; whereas, in News Communities experienced users talk about policies and regulations in contrast to amateurs who are more interested in polarizing topics.

%\vspace{-2em}
{\question How can we perform credibility analysis with limited information and ground-truth?}% data?}

We utilize latent topic models leveraging review texts, item ratings, and timestamps to derive {\em consistency features} without relying on extensive item/user histories, typically unavailable for ``long-tail'' items/users. These are used to learn inconsistencies such as discrepancy between the contents of a review and its rating, temporal ``bursts'', facet descriptions etc. We also propose an approach to transfer a model learned on the ground-truth data in one domain (e.g., Yelp) to another domain (e.g., Amazon) with missing ground-truth information. These results were presented at ECML-PKDD 2016~\cite{Subho:ECML2016}. 

All the above models for product review communities use
only the information of {\em a user reviewing an item at an
explicit timepoint}. This makes our approach fairly generalizable across all communities and domains with limited meta-data requirements.

{\question How can we generate user-interpretable explanations for the models' credibility verdict?}

For each of the above tasks, we provide {\em user-interpretable explanations} in the form of interpretable word clusters, representative snippets, evolution traces, etc. This way we can explain to the end-user why the model arrived at a particular verdict. 
Our model shows user-interpretable word clusters depicting user maturity that give interesting insights. For example, experienced users in Beer communities use more ``fruity'' words to depict beer taste and smell; in News Communities experienced users talk about policies
and regulations in contrast to amateurs who are more interested in polarizing topics. Similarly, evolution traces show that experienced users progress faster than amateurs in acquiring maturity, and also exhibit a higher variance.

\section{Organization}

This dissertation is organized as follows. Chapter~\ref{chap:relatedwork} discusses the state-of-the-art in this domain and related prior work. Chapter~\ref{chap:framework} lays the foundation of our credibility analysis framework. It develops probabilistic graphical models and methods for joint inference in online communities for credibility classification, and credibility regression. It also presents large-scale experimental studies on one of the largest health community and a sophisticated news community. Chapter~\ref{chap:temporal} develops approaches for modeling temporal evolution of users in online communities. It presents stochastic models for discrete and continuous modes of experience evolution of users in a collaborative filtering framework. It also presents large-scale experimental studies on five real world datasets like movies, beer, food, and news. Chapter~\ref{chap:applications} uses the principles and methods developed in 
earlier 
chapters for credibility 
analysis in product review communities for two tasks, namely: (i) finding useful product reviews that are helpful to the end-consumers in communities like Amazon, and (ii) detecting non-credible reviews with limited information about users and items in communities like Yelp, TripAdvisor, and Amazon. Chapter~\ref{chap:conclusions} presents conclusions and future research directions.

%% file: main/chapter-related-work/main.tex
\chapter{Related Work}
\label{chap:relatedwork}

This chapter presents an overview of the related work in several overlapping domains like truth discovery, sentiment analysis and opinion mining, information extraction, and collaborative filtering in online communities. It discusses the state-of-the-art in these domains, and their limitations.%, and our contributions over prior works. %Further references and technical differences of our work over related works are presented at the beginning of each chapter.

\section{Probabilistic Graphical Models}

In each of the following sections, we give a brief overview of the usage of Probabilistic Graphical Models (PGM) for related tasks. Since a full primer on PGMs is beyond the scope of this work, we refer the readers to \cite{KollerFriedman2009} for a general overview on PGMs.

Probabilistic graphical models use a graph-based representation to encode complex high-dimensional distributions involving many random variables. It provides a natural framework to model {\em probabilistic} interactions between them, represented as edges in the graph with random variables as the nodes. The objective is to probabilistically reason about the values of subsets of random variables, possibly given observations about some others. In order to do so, we need to construct a joint probability distribution function over the space of all possible value assignments to the random variables. This is often intractable. In practice, any random variable interacts with only a subset of the others. This allows us to represent the joint distribution as a product of {\em factors} composed of a smaller set of random variables, representing the marginals. This has several advantages. The factorization or decomposition can lead to a tractable solution, even though the complete specification over all possible value 
assignments can be asymptotically large. Secondly, it is easy to interpret the semantics of the model and output to users; highlight interactions between factors, and answer queries of interest with probabilistic interpretations. Thirdly, it also is easy to encode expert knowledge in the framework for specifying the structure of the graph in terms of (in)dependencies, and priors for the parameters.

\pagebreak

{\noindent \bf Markov Random Fields}

There are typically two families of PGMs: {\em Bayesian} networks that use a {\em directed} representation, and {\em Markov} networks (or, Markov Random Fields (MRFs)) that use an {\em undirected} representation. MRFs model the joint probability distribution over $X$ and $Y$ as $P(X,Y)$: $X$ representing multi-dimensional input (or, features), and $Y$ representing multi-dimensional output (or, labels/values). Since they are fully generative, they can be used to model arbitrary prediction problems. In our work, we mostly use Conditional Random Fields (CRFs), which are a specific type of MRF. They are discriminative in nature, and model the conditional distribution $P(Y|X=x)$. Since they directly model the conditional distribution that are of primary interest for standard prediction problems, they are more accurate for these settings. They can also be viewed as a structured extension over logistic regression, where the output (labels) can have dependencies between them. Please refer to \cite{Sutton2012} for an 
introduction to CRFs.

{\noindent \bf Topic Models}

Probabilistic topic models extend the principles of PGMs to discover {\em thematic} information in unstructured collection of documents. Latent Dirichlet Allocation (LDA) is the simplest type of topic model. These assume that documents have a distribution over topics (or, themes), and topics have a distribution over words. For example, a news article can talk about sports and politics, and use specific words to describe these topics. The topics are not known {\em a priori}, and are treated as hidden random variables, that need to be inferred from data. It uses a generative process to model these principles and assumptions. Refer to \cite{Blei:2012:PTM:2133806.2133826} for an overview on probabilistic topic models.

{\noindent \bf Inference}

A crucial component of PGMs involve {\em inference} algorithms for computing marginals, conditionals, and maximum a posteriori (MAP) probabilities efficiently for answering queries of interest. There are several variants of message passing or belief propagation algorithms (e.g., junction tree) for {\em exact inference}. However, the computational complexity is often exponential due to large size of cliques (subsets of nodes that are completely connected), and long loops for arbitrary graph structures. Therefore, we have to often resort to {\em approximate probabilistic inference}. 
There are two large classes of such inference techniques: Monte Carlo and Variational algorithms. 

{\noindent \em Monte Carlo methods:} These algorithms are based on the fact that although computing expectation of the original distribution $P(X)$ may be difficult, we can obtain {\em samples} from it or some closely related distribution to compute sample-based averages. In our work, we mostly use Gibbs sampling, and Metropolis Hastings. Gibbs sampling is a type of Markov Chain Monte Carlo (MCMC) algorithm, where samples are obtained from a Markov chain whose stationary distribution is the desired $P(X)$. We use Collapsed Gibbs Sampling \cite{Griffiths02gibbssampling} for inference in probabilistic topic models. Metropolis Hastings is also a type of MCMC algorithm. Instead of sampling from the true distribution --- that can be often quite complex --- it uses a proposal distribution that is proportional in density to the true distribution for sampling the
random variables. This is followed by an acceptance or rejection of the newly sampled value. That
is, at each iteration, the algorithm samples a value of a random variable --- where the current
estimate depends only on the previous estimate, thereby, forming a Markov chain. The principle advantage of Monte Carlo algorithms is that they are easy to implement, and quite general. However, it is difficult to guarantee their convergence, and the time taken to converge can be quite long. In our work, we empirically demonstrate fast convergence, under certain settings. 

{\noindent \em Variational mthods:} The other class of approximate inference involving Variational methods use a family of approximate distributions with their own variational parameters. The objective is to find a setting of these parameters to make the approximate distribution to be as close to the posterior of interest. Thereafter, these approximate distributions with the fitted parameters are used as a proxy for the true posterior.

Refer to \cite{jordan2002probabilistic} for an overview of the probabilistic inference methods for graphical models.% --- that have significantly contributed to the development of this dissertation. 

%%------------------------------------------------------------------ People On Drugs

\section{Truth Discovery}

In approaches to truth discovery, the goal is to resolve conflicts in multi-source data~\cite{Yin:2008:TDM:1399100.1399392,Dong:2009:ICD:1687627.1687690,Galland:2010:CID:1718487.1718504,Pasternack:2010:KB:1873781.1873880,Zhao:2012:BAD:2168651.2168656,DBLP:journals/pvldb/LiDLMS12,Pasternack:2013:LCA:2488388.2488476,Dong:2013:CED:2488388.2488422,Li:2014:RCH:2588555.2610509,Li:2015:DET:2783258.2783277,Ma:2015:FFG:2783258.2783314,Zhi:2015:MTE:2783258.2783339}. Input data is assumed to have a structured representation: an entity of interest (e.g., a person) along with its potential values provided by different sources (e.g., the person's birthplace). 

Truth discovery methods of this kind (see~\cite{YaliangLi:SIGKDD2015} for a survey), starting with the seminal work of
\cite{Yin:2008:TDM:1399100.1399392}, assume that claims follow a structured template
with clear identification of the questionable values \cite{DBLP:journals/pvldb/LiDLMS12,Li:2011:TVT:2004686.2005589} or correspond to subject-predicate-object triples obtained by information extraction~\cite{DBLP:conf/acl/NakasholeM14}. 
%or require manual annotation of the input claims. 
A classic example is \textit{``Obama is born in Kenya''} viewed as a triple
$\langle$\textit{Obama, born in, Kenya}$\rangle$ where \textit{``Kenya''} is the critical value.
The assumption of such a structure is crucial in order to identify alternative values
for the questionable slot (e.g., \textit{``Hawaii''}, \textit{``USA''}, \textit{``Africa''}), and is appropriate
when checking facts for tasks like knowledge-base curation. Such alternative values are provided by many other sources. The objective is to resolve the conflict between these multi-source data for a given query to obtain the truth.  
It is assumed that the conflicting values are already available. To resolve conflicts for a particular entity, these approaches exploit that reliable or trustworthy sources often provide correct information. To exploit this principle, these works propagate and aggregate scores (or, reliability estimates) over networks of objects, and sources that provide information about the objects. A significant challenge is that {\em a priori} we do not know which sources are reliable or trustworthy that need to be inferred during the task.

\pagebreak

\cite{Li:2011:TVT:2004686.2005589} uses information-retrieval techniques to
systematically generate alternative hypotheses for the given
statement, and assess the
evidence for each alternative. 
However, it relies on the {\em user} providing the {\em doubtful} portion of the input statement (e.g., the birthplace of ``Obama'' in the above example).
 %expects the portion of the input statement which is \textit{doubtful}. 
Making use of the doubtful unit, alternative statements (e.g., alternative birthplaces) are generated via web search and ranked to identify the correct statement. 
Work in \cite{DBLP:conf/acl/NakasholeM14} goes a step further by proposing a method to generate conflicting {\em values} or {\em fact candidates} from Web contents. They make use of linguistic features to detect the 
objectivity of the source reporting the fact. %However, the work still depends on structured input in the form of Subject-Predicate-Object (SPO) triples, obtained by applying Open Information Extraction.
Note that both of these approaches can handle only input statements for which alternative facts or values are given or can be retrieved a priori. 

\cite{Yin:2008:TDM:1399100.1399392,Pasternack:2010:KB:1873781.1873880,DBLP:conf/ijcai/PasternackR11} develop methods for statistical reasoning on the cues
for the statement being true vs. false.
\cite{DBLP:journals/pvldb/LiDLMS12} has developed
%similar 
approaches for 
structured data such as flight times or stock quotes,
where different Web sources often yield contradictory values.
\cite{Vydiswaran:2011:GID:2023582.2023589} addressed  truth assessment for
medical claims about diseases and their treatments (including drugs and general phrases such as ``surgery''), by an IR-style evidence-aggregation and
ranking method
over \textit{curated} health portals.

{\noindent \bf Probabilistic graphical models:} Recently, \cite{Pasternack:2013:LCA:2488388.2488476} presented an LDA-style
latent-topic model for discriminating true from false claims,
with various ways of generating incorrect statements
(guesses, mistakes, lies). \cite{Ma:2015:FFG:2783258.2783314} proposed an LDA-style model to capture expertise of users for different topics. They use it to model question content, and answer quality to find the best candidate answer. \cite{DBLP:journals/pvldb/ZhaoRGH12} proposed a Latent Truth Model based on a generative process of two types of errors (false positive and false negative) by modeling two different aspects of source quality. They also propose a sampling based algorithm for scalable inference. \cite{DBLP:journals/qdb/ZhaoRGH12} proposed a Gaussian Truth Model to deal with numerical data based on a generative process.

Most of the above approaches are limited to resolving conflicts amongst multi-source data --- where, input data is in a structured format and conflicting facts are always available. Although these are elaborate models, they do not take into account the language in which statements are reported in user postings, and trustworthiness of the users making the statements. None of these prior works have considered
online discussion forums where credibility of statements is intertwined with all of the above factors. Moreover, due to limited availability of ground-truth data in this problem setting, most of these models work in an unsupervised fashion.

In our work, we propose 
general approaches that do not require any %manual input in the form of 
alternative claims. Our approaches are geared for online communities with rich interactions between users, (language of) postings, and statements. Also,  our models can be partially or weakly supervised, as well as fully supervised depending on the availability of labeled data. Moreover, we provide user-interpretable explanations for our models' verdict, unlike many of the previous works.

\section{Trust and Reputation Management}

This area has received much attention, mostly motivated by
analyzing customer reviews for product recommendations, but also
in the context of social networks.
\cite{DBLP:conf/www/KamvarSG03,DBLP:conf/www/GuhaKRT04} are seminal works that modeled
the propagation of trust within a network of users. 
TrustRank \cite{DBLP:conf/www/KamvarSG03} has become a popular measure
of trustworthiness, based on random walks on (or spectral
decomposition of) the user graph.
Reputation management has also been studied in the context
of peer-to-peer systems, the blogosphere, and online interactions
\cite{Adler2007,DBLP:reference/db/AgarwalL09,DBLP:reference/db/Despotovic09a,DBLP:journals/cacm/AlfaroKPA11,Hang2013}.

All these works focused on explicit relationships
between users to infer authority and trust levels.
The only content-aware model for trust propagation 
is \cite{DBLP:conf/kdd/VydiswaranZR11}.
This work develops a HITS-style algorithm for propagating
trust scores in a heterogeneous network of claims, sources,
and documents. 
Evidence for a claim is collected from related documents using generic IR-style word-level measures. It also requires weak supervision at the evidence level in the form of human judgment on the trustworthiness of articles. However, it ignores the fine-grained interaction between users making the statements, their postings, and how these evolve over time. % in online communities.
%Our work considers online users, interactions, and rich language features for their postings taking into account their evolution over time. %This more demanding setting
%requires sophisticated model, like our CRF.
We show that all of these factors can be jointly captured using sophisticated probabilistic graphical models.

\section{Information Extraction (IE)}

There is ample work on extracting Subject-Predicate-Object (SPO) like statements
from natural-language text. 
The survey \cite{Sarawagi2008} gives an overview; 
\cite{Krishnamurthy2009,Bohannon2012,SuchanekWeikum2013} provide
additional references. 
State-of-the-art methods combine pattern matching with 
extraction rules and consistency reasoning.
This can be done either in a shallow manner, over sequences
of text tokens, or in combination with deep parsing and
other linguistic analysis.
The resulting SPO triples often have highly varying confidence,
as to whether they are really expressed in the text or picked
up spuriously.
Judging the credibility of statements is out-of-scope for IE itself. 
\cite{Sarawagi2008, KollerFriedman2009} give an overview of probabilistic graphical models used for Information Extraction.

{\noindent \bf IE on Biomedical Text}

For extracting facts about diseases, symptoms, and drugs, 
customized IE techniques have been developed to tap biomedical
publications like PubMed articles.
Emphasis has been on the molecular level, i.e. proteins, genes,
and regulatory pathways 
(e.g., \cite{DBLP:journals/bmcbi/BundschusDSTK08,Krallinger2008,DBLP:journals/bioinformatics/BjorneGPTS10}), and
to a lesser extent on biological or medical events from scientific
articles and from clinical narratives~\cite{DBLP:conf/ijcai/JindalR13,DBLP:journals/jamia/XuHTC12}.
\cite{DBLP:conf/naacl/PaulD13} has used LDA-style models for summarization
of drug-experience reports.
\cite{Ernst2014} has employed such techniques to build
a large knowledge base for life science and health. 
Recently, \cite{White2014} demonstrated how to derive insight on drug effects from query logs
of search engines. 
Social media has played a minor role in this prior
IE work.

\comment{
\noindent {\bf Learning to Rank}

Supervised models have also been developed to rank items from constructed item feature vectors~\cite{liu2009}. Such techniques optimize measures like Discounted Cumulative Gain, Kendall-Tau, and Reciprocal Rank to generate item ranking similar to the training data based on the feature vectors. We use one such technique, and show its performance can be improved by removing non-credible item reviews.
}

\section{Language Analysis for Social Media}

{\noindent \bf Sentiment Analysis} 

Work on sentiment analysis~\cite{pang2002, Turney2002, Dave2003, Yu2003, Pang2004,PangLee2007,Liu2012,Mukherjee2012} has looked into language features ---  based on phrasal and dependency relations, narratives, perspectives, modalities, discourse relations, lexical resources etc. --- in customer reviews to classify their sentiment as positive, negative, or objective.
Going beyond this special class of texts, \cite{greene2009,recasens2013} have studied the use of biased language in Wikipedia and similar collaborative communities. Even more broadly, the task of characterizing subjective language has been addressed, among others, in \cite{Wiebe2005,Lin2011}.
The work by \cite{Wiebe2011} has explored benefits between subjectivity analysis and information extraction. %None of this prior work has addressed the specifics of discussions in online healthforums.

Opinion mining methods for recognizing a speaker's stance in online debates are proposed in~\cite{DBLP:conf/acl/SomasundaranW09,DBLP:conf/naacl/WalkerAAG12}. Structural and linguistic features of users' posts are harnessed to infer their stance towards discussion topics in~\cite{sridhar:aclws14}. Temporal and textual information are exploited for stance classification over sequence of tweets in \cite{DBLP:conf/acl/LukasikSVBZC16}. 

{\noindent \bf Opinion Spam} 

Several existing works~\cite{Mihalcea2009,Ott2011,Ott2013} consider the textual content of user reviews for tackling fake reviews (or, opinion spam) by using word-level unigrams or bigrams as features, along with specific lexicons (e.g., LIWC~\cite{liwc} psycholinguistic lexicon, WordNet Affect~\cite{WNAffect}), to learn latent topic models and classifiers (e.g., \cite{Ott2013a}).
Some of these works learn linguistic features from 
artificially created fake review dataset, leading to biased features that are not dominant in 
real-world data. This was confirmed by a study on Yelp filtered reviews~\cite{Liu2013a}, where the $n$-gram features used in prior works performed poorly despite their outstanding performance 
on the artificial datasets. Additionally, linguistic features such as {\em text sentiment}~\cite{Yoo2009}, {\em readability score} 
(e.g., Automated readability index (ARI), Flesch reading ease, etc.)~\cite{Hu2012}, {\em textual coherence}~\cite{Mihalcea2009}, and rules based on 
{\em Probabilistic Context Free Grammar} (PCFG)~\cite{Feng2012} have been studied.

{\noindent \bf Aspect Rating Prediction from Review Text}

Aspect rating prediction has received vigorous interest in recent times. A shallow dependency parser is used to learn product aspects and aspect-specific opinions in~\cite{DBLP:conf/acl/YuZWC11} by jointly considering the aspect frequency and the consumers' opinions about each aspect. \cite{DBLP:conf/www/MukherjeeBJ13} presents an approach to capture user-specific aspect preferences, but requires manual specification of a fixed set of aspects to learn from. \cite{DBLP:conf/naacl/SnyderB07} jointly learns ranking models for individual aspects by modeling dependencies between assigned ranks by analyzing meta-relations between opinions, such as agreement and contrast.

{\noindent \bf Probabilistic graphical models:} 
Latent Aspect Rating Analysis Model (LARAM)~\cite{DBLP:conf/kdd/WangLZ10, wang2011} jointly identifies latent aspects, aspect ratings, and weights placed on the aspects in a review. However, the model ignores user identity and writing style, and learns parameters \textit{per review}. % is described in. The work considers seed facets like `food', `ambience', `service' \textit{etc.} and uses dependency parsing with a lexicon to find the sentiment about each facet. 
% in contrast to our model which learns the latent parameters \textit{per author basis}.
A rated aspect summary of short comments is done in~\cite{DBLP:conf/www/LuZS09}. Similar to LARAM, the statistics are aggregated at the comment-level. % in this work and not at the author-level.
A topic model is used in~\cite{DBLP:conf/acl/TitovM08} to assign words to a set of induced topics. The model is extended through a set of maximum entropy classifiers, one per each rated aspect, that are used to predict aspect specific ratings. 

A joint sentiment topic model (JST) is described in~\cite{linCIKM2009} which detects sentiment and topic simultaneously from text. In JST, each document has a sentiment label distribution. Topics are associated to sentiment labels, and words are associated to both topics and sentiment labels. In contrast to~\cite{DBLP:conf/acl/TitovM08} and some other similar works~\cite{DBLP:conf/kdd/WangLZ10, wang2011, DBLP:conf/www/LuZS09} which require some kind of supervised setting like ratings for the aspects or overall rating~\cite{DBLP:conf/www/MukherjeeBJ13}, JST is fully unsupervised. The CFACTS model~\cite{lakkarajuSDM2011} extends the JST model to capture facet coherence in a review using Hidden Markov Model. This is further extended by \cite{mukherjee2014JAST} to capture author preferences, and writing style, while being completely unsupervised.

All these generative models have their root in Latent Dirichlet Allocation Model~\cite{Blei2003LDA}. LDA assumes a document to have a probability distribution over a mixture of topics and topics to have a probability distribution over words. In the Topic-Syntax Model~\cite{Griffiths02gibbssampling}, each document has a distribution over topics; and each topic has a distribution over words being drawn from classes, whose transition follows a distribution having a Markov dependency. In the Author-Topic Model~\cite{DBLP:conf/uai/Rosen-ZviGSS04}, each author is associated with a multinomial distribution over topics. Each topic is assumed to have a multinomial distribution over words.

However, these models --- with the exception of \cite{DBLP:conf/uai/Rosen-ZviGSS04,mukherjee2014JAST} that are not geared for credibility analysis ---  do not consider the users writing the reviews, their preferences for different topics, experience, or writing style. Our models capture all of these user-centric factors, as well interactions between them to capture credibility of user-contributed content in online communities.
%incorporate any authorship information to incorporate \textit{author preference} for the facets or \textit{author style} information for maintaining coherence in reviews.

%A WordNet similarity metric is used to assign each facet to a seed facet. Thereafter, they use linear regression to learn author preference for the seed facets from review ratings. The work is restrictive as it considers only manually given seed facets, topic-ratings are subjected to the lexicon coverage, and it does not incorporate review coherence.

\section{Information Credibility in Social Media}

Prior research for credibility assessment of social media posts exploits {\em community-specific} features for detecting rumors, fake, and deceptive content~\cite{DBLP:conf/www/CastilloMP11,DBLP:conf/ecai/LavergneUY08,DBLP:conf/emnlp/QazvinianRRM11,DBLP:conf/coling/XuZ12,Yang:2012:ADR:2350190.2350203}. Temporal, structural, and linguistic features were used to detect rumors on Twitter in \cite{DBLP:conf/icdm/KwonCJCW13}. \cite{DBLP:conf/www/0003LKJ13} addresses the problem of detecting fake images in Twitter based on influence patterns and social reputation. A study on Wikipedia hoaxes is done in~\cite{DBLP:conf/www/Kumar0L16}. They propose a model which can determine whether a Wikipedia article is a hoax or not --- by measuring how long they survive before being debunked, how many page-views they receive, and how heavily they are referred to by documents on the web compared to legitimate articles. %Also, the nature of successful hoaxes was characterized by 
%comparing them to legitimate articles and to failed hoaxes that were discovered shortly. 
\cite{castillo2011} analyzes micro-blog postings in Twitter related to trending topics, and classifies them as credible or not, based on features from user posting and re-posting behavior.
%They state that several factors influence the credibility of information in social media like ``the reactions that certain topics generate and the emotion conveyed by users discussing the topic: e.g. if they use opinion expressions representing positive or negative sentiments about the topic" and the level of certainty of the users and their characteristics in propagating that information.
%
\cite{kang2012} focuses on credibility of users, harnessing the dynamics of
information flow in the underlying social graph and tweet content.
\cite{canini2011} analyzes both topical content of information sources and
social network structure 
to find credible information sources in
social networks. Information credibility in tweets 
%corresponding to fourteen high impact news events of 2011 
has been studied in \cite{agupta2012}.  \cite{VZRPCIKM12} conducts a {\em user study} to analyze various factors like contrasting viewpoints and expertise affecting the truthfulness of controversial claims.

All these approaches are geared for specific forums, making use of several {community-specific} characteristics (e.g., Wikipedia edit history, Twitter follow graph, etc.) that cannot be generalized across domains, or other communities. Moreover, none of these prior works analyze the joint interplay between \textit{sources, language, topics,} and \textit{users} 
that influence the credibility of information in online communities.

{\noindent \bf Rating and Activity Analysis for Spam Detection} 

The influence of different kinds of bias in online user ratings has been studied in~\cite{fang2014,sloanreview}. \cite{fang2014} proposes an approach to handle users who might be subjectively different or strategically dishonest.

In the absence of proper ground-truth data, prior works make strong assumptions, e.g., duplicates and near-duplicates are 
fake, and make use of {\em extensive} background information like brand name, item description, user history, IP addresses and location, etc.~\cite{Liu2007,Liu2008,Liu2010,Liu2011,Liu2012,Liu2013,Liu2013a,Liu2014,Rahman2015}. 
Thereafter, regression models trained on all these features are used to classify reviews as credible or deceptive. Some of these works also use crude or ad-hoc language 
features like content similarity, presence of literals, numerals, and capitalization.

In contrast to these works, our approach uses limited information about users and items --- that may not be available for ``long-tail'' users and items in the community --- catering to a wide range of applications. We harvest several semantic and consistency features --- only from the information of a user reviewing an item at an explicit timepoint --- that also give user-interpretable explanation as to why a user posting should be deemed non-credible.

{\noindent \bf Citizen journalism}

\cite{wemedia} defines citizen journalism as ``the act of a citizen or group of citizens playing an active role in the process of collecting, reporting, analyzing and dissemination of news and information to provide independent, reliable, accurate, wide-ranging and relevant information that a democracy requires.''
\cite{allan2007} focuses on user activities like blogging in community news websites.
%\cite{lewis2010} define citizen journalists as those who comment on stories, respond to polls, and submit video, audio and text to traditional media companies. \cite{garbett2014} present a set of design implications for building systems that support interaction between citizen and professional journalists.
Although the potential of citizen journalism is greatly highlighted in the recent Arab Spring~\cite{howard2011}, misinformation can be quite dangerous when relying on users as news sources (e.g., the reporting of the Boston
Bombings in 2013~\cite{boston}).

Our proposed approaches automatically identify the trustworthy and experts users in the community, and extract credible statements from their postings.

\comment{

Sunita Sarawagi: Information Extraction. Foundations and Trends in Databases 1(3): 261-377 (2008)

Fabian M. Suchanek, Gerhard Weikum: Knowledge harvesting from text and Web sources. ICDE 2013: 1250-1253

Rajasekar Krishnamurthy, Yunyao Li, Sriram Raghavan, Frederick Reiss, Shivakumar Vaithyanathan, Huaiyu Zhu: Web Information Extraction. Encyclopedia of Database Systems 2009: 3473-3478

Philip Bohannon, Nilesh N. Dalvi, Yuval Filmus, Nori Jacoby, Sathiya Keerthi, Alok Kirpal: Automatic web-scale information extraction. SIGMOD Conference 2012: 609-612

Jin-Dong Kim, Ngan Nguyen, Yue Wang, Jun'ichi Tsujii, Toshihisa Takagi, Akinori Yonezawa: The Genia Event and Protein Coreference tasks of the BioNLP Shared Task 2011. BMC Bioinformatics 13(S-11): S1 (2012)

Yan Xu, Kai Hong, Junichi Tsujii, Eric I.-Chao Chang: Feature engineering combined with machine learning and rule-based methods for structured information extraction from narrative clinical discharge summaries. JAMIA 19(5): 824-832 (2012)

Jin-Dong Kim, Tomoko Ohta, Sampo Pyysalo, Yoshinobu Kano, Jun'ichi Tsujii: Extracting Bio-molecular Events from literature - the BioNLP'09 Shared Task. Computational Intelligence 27(4): 513-540 (2011)

Yoshinobu Kano, William A. Baumgartner Jr., Luke McCrohon, Sophia Ananiadou, K. Bretonnel Cohen, Lawrence Hunter, Jun'ichi Tsujii: U-Compare: share and compare text mining tools with UIMA. Bioinformatics 25(15): 1997-1998 (2009)

}

%%------------------------------------ People on Media

\section{Collaborative Filtering for Online Communities}

%\noindent \textbf{Rating prediction in online communities:}

State-of-the-art recommenders based on collaborative filtering \cite{korenKDD2008, koren2011advances} exploit user-user and item-item similarities by latent factors. The temporal aspects leading to bursts in item popularity, bias in ratings, or the evolution of the entire community as a whole is studied in~\cite{KorenKDD2010, xiongSDM2010, XiangKDD2010}. Other papers have studied temporal issues for anomaly detection~\cite{Gunnemann2014}, detecting changes in the social neighborhood~\cite{MaWSDM2011} and linguistic norms~\cite{DanescuWWW2013}. However, none of this prior work has considered the evolving experience and behavior of individual users.

\cite{mcauleyWWW2013} modeled and studied the influence of evolving user experience on rating behavior and for targeted recommendations. However, it disregards the vocabulary and writing style of users in their reviews. In contrast, our work considers the review texts for additional insight into facet preferences and experience progression. 
We address the limitations 
by means of language models that are specific to the experience level of an
individual user, and by modeling transitions between experience levels of users with a Hidden
Markov Model. Even then these models are limited to {\em discrete} experience levels leading to abrupt changes in both experience and language model of users. To address this, and other related drawbacks, we further propose continuous-time models for the smooth evolution of both
user experience, and their corresponding language models.

{\noindent \bf Probabilistic graphical models:} Sentiment analysis over reviews aimed to learn latent topics ~\cite{linCIKM2009}, latent aspects and their ratings~\cite{lakkarajuSDM2011, wang2011} using topic models, and user-user interactions~\cite{West-etal:2014} using Markov Random Fields. \cite{mcauleyrecsys2013} unified various approaches to generate user-specific ratings of reviews. \cite{mukherjee2014JAST} further leveraged the author writing style. However, all of these approaches operate in a static, snapshot-oriented manner, without considering time at all.

%%%GW: added one par on dynamic topic models
From the modeling perspective, some approaches learn a document-specific discrete rating~\cite{linCIKM2009, ramageKDD2011}, whereas others learn the facet weights outside the topic model \cite{lakkarajuSDM2011, mcauleyrecsys2013, mukherjee2014JAST}. In order to incorporate continuous ratings,~\cite{bleiNIPS2007} proposed a complex and computationally expensive Variational Inference algorithm, and~\cite{mimnoUAI2008} developed a simpler approach using Multinomial-Dirichlet Regression. The latter inspired our technique for incorporating supervision in our discrete-version of the experience model.

\cite{Wang2006} modeled topics over time. However, the topics
themselves were constant, and time was only
used to better discover them. Dynamic topic models have been introduced in  
\cite{BleiDTM,BleiCTM}. This prior work developed generic models
based on Brownian Motion, and applied them
to news corpora. 
\cite{BleiCTM} argues that the continuous model
avoids making choices for discretization and is also
more tractable compared to fine-grained discretization.
Our language model is motivated by the latter. We substantially extend it to capture evolving user behavior and experience
in review communities using Geometric Brownian Motion.
%\cite{BleiCTM} presents variational inference techniques, 
%whereas our work adopts Metropolis-Hastings
%sampling.

Our models therefore unify several 
dimensions to jointly study the role of language, users, and topics over time for collaborative filtering in online communities.

%GW: is the last sentence - "inspired ..." - adequate?

%%-------------------------------------------- Continuous Experience

%%%GW: re-worded the following paragraph, to make it look different from the ICDM paper

%%%GW: I believe the following can be dropped
%%% in the case it is revived, it needs to be re-worded a bit, to make the wording differ from the ICDM paper
%
%From the modeling perspective, some approaches learn a document-specific discrete %rating~\cite{linCIKM2009, ramageKDD2011}, whereas others learn the facet weights outside the %topic model~\cite{lakkarajuSDM2011, mcauleyRecSys2013, mukherjeeSDM2014}. In order to %incorporate continuous ratings,~\cite{bleiNIPS2007} proposed a complex and computationally %expensive Variational Inference algorithm, and~\cite{mimnoUAI2008} developed a simpler %approach using Multinomial-Dirichlet Regression. 

%% -------------------------------- Credible Review Detection

%% ---------------------------------------- Helpful Review Detection

{\noindent \bf Detecting Helpful Reviews}

Prior works on predicting review helpfulness~\cite{Kim:2006:AAR:1610075.1610135,Lu:2010:ESC:1772690.1772761} exploit shallow syntactic features to classify extremely opinionated reviews as not helpful. Similar features are also used in finding review spams~\cite{Liu2008, Liu2013}. Similarly, few other approaches utilize features like frequency of user posts, average ratings of users and items to distinguish between helpful and unhelpful reviews. Community-specific features with explicit user network are used in~\cite{Tang:2013:CRH:2507157.2507183,Lu:2010:ESC:1772690.1772761}. However, these shallow features do not analyze what the review is {\em about}, and, therefore, cannot {\em explain} why it should be helpful for a given product. 

Approaches proposed in~\cite{icdm2008,Kim:2006:AAR:1610075.1610135} also utilize item-specific meta-data like {\em explicit} item facets and product brands to decide the helpfulness of a review. However, these approaches heavily rely on a large number of meta-features which make them less generalizable. Some of the related approaches~\cite{O'Mahony:2009:LRH:1639714.1639774,icdm2008} also identify {\em expertise} of a review's author as an important feature. However, they do not explicitly model the user expertise. 

We use our own approach for finding expert users in a community using experience-aware collaborative filtering models, and leverage the distributional similarity in the semantics (e.g, writing style, facet descriptions) and consistency of expert-contributed reviews to identify useful product reviews.
%Recent work in \cite{mcauleyWWW2013} studies the influence of rating behavior on evolving user experience. However, it does not consider the vocabulary and writing style of users in reviews, and their temporal progression. 
%In contrast, our work gains additional insights by considering the review texts to capture the facet preferences and experience progression. 
%we use principles and techniques developed in our 

%Our model for capturing user expertise draws motivation from~\cite{mukherjee2015jertm},\cite{mukherjee2016KDD},\cite{mcauleyWWW2013} with significant differences. Unlike prior works, the facet preferences in our model is conditioned only on the user expertise and not the user. This significantly reduces the dimensionality of the model making it more efficient; as well as the distributions are learned directly from minimizing the mean squared error for predictions, allowing all parameters to be coupled tightly, unlike in~\cite{mukherjee2015jertm}. 

%% -------------------------------- WWW Snopes

%%--------------------------------------------- Joint Author Sentiment Topic Model

%% file: main/chapter-credibility-analysis/main.tex
\chapter{Credibility Analysis Framework}
\label{chap:framework}

\input{main/chapter-credibility-analysis/introduction}

\section{Problem Statement}

Given a set of {\em users} and {\em sources} generating {\em postings}, and other users (or sources) reviewing these postings with mutual interactions (e.g., likes, shares, upvotes/downvotes etc.) --- where each of these factors can have several features --- our objective is to {\em jointly} identify: (i) {\em trustworthy} sources, (ii) {\em credible} postings and statements (extracted from postings), and (iii) {\em expert} users for {\em classification} and {\em regression} tasks.

In this process, we want to analyze the influence of various factors like the writing style of a posting, its topic distribution, viewpoint and expertise of the users and sources for {\em credibility analysis}.

\input{main/chapter-credibility-analysis/overview}

\section{Model Components}
\label{sec:features}

In this section, we outline the different components, and features used in our probabilistic models for credibility analysis {\em with a focus on health and news communities}. These features are extracted from the postings of users in online communities, and their interactions with other users and sources. Since the features are fairly generic, and not community-specific --- they are easily applicable to other communities like travel, food, and electronics.

\input{main/chapter-credibility-analysis/language}
\input{main/chapter-credibility-analysis/users}
\input{main/chapter-credibility-analysis/topics}

\input{main/chapter-credibility-analysis/sources}

\section{Probabilistic Inference}
\label{sec:cred-inference}

\input{main/chapter-credibility-analysis/models}

\section{Experimental Evaluation: Health Communities}
\label{sec:experiments-health}

\input{main/chapter-credibility-analysis/experiments-health}

\section{Experimental Evaluation: News Communities}
\label{sec:evaluations}

\input{main/chapter-credibility-analysis/experiments-news}

\newpage
\section{Conclusions}

\input{main/chapter-credibility-analysis/conclusions}

%% file: main/chapter-credibility-analysis/introduction.tex
\section{Introduction and Motivation}

%social media in general and health forums in particular
Online social media includes a wealth of topic-specific communities and discussion forums about politics, music, health, and many other domains. User-contributed contents in such communities offer a great potential for distilling and analyzing facts and opinions. For instance, online health communities constitute an important source of information for patients and doctors alike,
with 59\% of the adult U.~S. population consulting online health resources \cite{Fox:PewInternetAndAmericanLifeProject:2013},
and nearly half of U.~S. physicians relying on online resources for professional use \cite{IMS2014}.

One of the major hurdles preventing the full exploitation of information from online communities is the widespread concern regarding the quality and credibility of user-contributed content \cite{Peterson:JMedInternetRes:2003,white2014health,nber,gallop}; as the information obtainable in the raw form is very noisy and subjective due to the personal bias and perspectives injected by the users in their postings. 

{\noindent \bf State-of-the-Art and Its Limitations:} Although information extraction methods using probabilistic graphical models \cite{Sarawagi2008,KollerFriedman2009} have been previously employed to extract statements from user generated content, they do not account for the the inherent bias, subjectivity and misinformation prevalent in online communities. 
%
%Furthermore, standard information extraction techniques \cite{Krishnamurthy2009,Bohannon2012,SuchanekWeikum2013} do not consider the role of language can have in assessing the credibility of the extracted statements. 
%
Unlike standard information extraction techniques \cite{Krishnamurthy2009,Bohannon2012,SuchanekWeikum2013}, our method considers the role language can have in assessing the credibility of the extracted statements. 
%In this work, we perform linguistic analyses to extract stylistic and affective features from user posts. 
For instance, stylistic features --- such as the use of modals and inferential conjunctions --- help identify accurate statements, while affective features help determine the emotional state of the user making those statements (e.g., anxiety, confidence).

Prior works in truth discovery and fact finding (see \cite{YaliangLi:SIGKDD2015} for a survey) make strong assumptions about the nature and structure of the data --- e.g., {\em factual claims and structured input} in the form of subject-predicate-object triples like {\tt Obama\_BornIn\_Kenya}, or relational tables \cite{Dong:2015:KTE:2777598.2777603,DBLP:journals/pvldb/LiDLMS12,Li:2011:TVT:2004686.2005589,Li:2015:DET:2783258.2783277}).
%and/or online communities with specific characteristics
%like user metadata, who-replied-to-whom, who-edited-what, etc. (e.g., \cite{kumar2016disinformation,Mukherjee:2014:PDC:2623330.2623714}). 
%Truth-finding methods of this kind, starting with the seminal work of
%\cite{Yin:2008:TDM:1399100.1399392}, assume that claims follow a structured template
%with clear identification of the questionable values \cite{DBLP:journals/pvldb/LiDLMS12,Li:2011:TVT:2004686.2005589} or correspond to subject-predicate-object triples obtained by information extraction~\cite{DBLP:conf/acl/NakasholeM14}. 
%or require manual annotation of the input claims. 
%A classic example is \textit{``Obama is born in Kenya''} viewed as a triple
%$\langle$\textit{Obama, born in, Kenya}$\rangle$ where \textit{``Kenya''} is the critical value.
%The assumption of such a structure is crucial in order to identify alternative values
%for the questionable slot (e.g., \textit{``Hawaii''}, \textit{``USA''}, \textit{``Africa''}), and is appropriate
%when checking facts for tasks like knowledge base curation.
These approaches, also, do not consider the role of language, writing style and trustworthiness of the users, and their interactions that limit their coverage and applicability in online communities.

To address these issues, we propose probabilistic graphical models that can automatically assess the credibility of statements made by users of online communities by analyzing the joint interplay between several factors like the community interactions (e.g., user-user, user-item links), language of postings, trustworthiness of the users etc. Our model settings, features, and inference are generic enough to be applicable to {\em any} online community; however, as use-case studies for validating our framework we focus on two disparate communities: namely {\em health}, and {\em news}. Unlike the healthforums focusing mostly on drugs and their side-effects, the latter community is highly heterogeneous covering topics ranging from sports, politics, environment, to current affairs --- thereby testing the generalizability of our framework.

\subsection{Use-case Study: Health Communities} 

As our first use-case, consider healthforums such as {\tt \href{http://www.healthboards.com/}{healthboards.com}} or {\tt \href{http://www.patient.co.uk/}{patient.co.uk}}, where patients engage in discussions about their experience with medical drugs and therapies, including negative side-effects of drugs or drug combinations. From such user-contributed postings, we focus on extracting rare or unknown side-effects of drugs --- this being one of the problems where large scale non-expert data has the potential to complement expert medical knowledge \cite{White2014}, but where misinformation can have hazardous consequences \cite{cline2001consumer}.

%factors: subjectivity of language, trustworthiness of users, credibility of statements
The main intuition behind the proposed model is that there is an important interaction between the {\em credibility} of a statement, the {\em trustworthiness} of the user making that statement, and the {\em language} used in the posting containing that statement. Therefore, we consider the mutual interaction between the following factors:
\begin{itemize}
\item {\em Users:} the overall {\em trustworthiness} (or authority) of a user, corresponding to her status and engagement in the community.
\item {\em Language:} the {\em objectivity}, rationality (as opposed to emotionality), and general quality of the language in the users' postings.
Objectivity is the quality of the posting to be free from preference, emotion, bias and prejudice of the author.
\item {\em Statements:} the {\em credibility} (or truthfulness) of medical statements contained within the postings. Identifying accurate drug side-effect statements is a goal of the model.
\end{itemize}

%cross-talk of the three aspects
These factors have a strong influence on each other. Intuitively, a statement is more credible if it is posted by a trustworthy user and expressed using confident and objective language. As an example, consider the following review about the drug Depo-Provera by a senior member of {\tt \href{http://www.healthboards.com/}{healthboards.com}}, one of the largest online health communities:\\

\example{\dots Depo is very dangerous as a birth control and has too many long term side-effects like reducing bone density \dots}

This posting contains a credible statement that a potential side-effect of Depo-Provera is to ``reduce bone density''. Conversely, highly subjective and emotional language suggests lower credibility of 
%%C: edit
%statements and lower believability in that user's contents
the user's statements. A negative example along these lines is:
%subho:camera-ready

\example{I have been on the same cocktail of meds (10 mgs. Elavil at bedtime/60-90 mgs. of Oxycodone during the day/1/1/2 mgs. Xanax a day....once in a while I have really bad hallucination type dreams. I can actually ``feel" someone pulling me of the bed and throwing me around. I know this sounds crazy but at the time it fels somewhat demonic.}\label{ex:cocktail}

Although this posting suggests that taking Xanax can lead to hallucination,
the style in which it is written renders the credibility of this statement doubtful.
%GW: we may want to pick a different example - hallucinations may be among the true side effects of Prozac
%
These examples support the intuition that to identify credible medical statements, we also need to assess the trustworthiness of users and the objectivity of their language. In this work we leverage this intuition through a {\em joint analysis of statements, users, and language}
in online health communities. %To the best of our knowledge, this problem has not been studied in prior work.

%%%%%%%%%%%%%%%%%%%%%%%%%%%%%%%%%%%%%%%%%%%%%%%%%%%

%\subsection{Approach and Contribution}

%our approach in a nutshell: feature model first
% We identify a rich set of features for each of the three types of factors.  On the language side, we perform linguistic analyses to extract {stylistic} and {affective} features from user postings. Discourse features help to identify authoritative user statements by examining the usage of modals, negation, inferential and violating conjunctions, hypothetical statements, definitives, density of nouns (names of drugs, diseases, symptoms), frequency of adjectives and adverbs etc. Affective features help to identify objective statements  by analyzing user emotions in the posting like anxiety, depression, esteem, confidence, sympathy, coolness, etc.  We establish a baseline for identifying potential side-effects of drugs
% from user postings by training an SVM over this comprehensive feature model.
%GW: the following sentence is potentially misleading, hence dropped
%Since the number of positive side-effects of a drug is typically less,
%we use the notion of \textit{distant supervision} to create an expanded training set.

%our approach in a nutshell: now the probabilistic model
\noindent{\bf Approach:} The first technical contribution of our work is a probabilistic graphical model for {\em classifying} a statement as credible or not --- which is tailored to the problem setting as to facilitate joint inference over users, language, and statements.
%Our model captures cross-talk between users and their statements,
%for example, a user gaining trust by making statements that agree with highly reputed users.
We devise a Markov Random Field (MRF) with individual users, postings, and statements as nodes, as summarized in Figure \ref{fig:triangle}.
The quality of these nodes---trustworthiness, objectivity, and credibility---is 
modeled as binary random variables.
The model is semi-supervised with a subset of training (side-effect) statements
derived from expert medical databases, labeled as true or false.
In addition, the model relies on linguistic and user features
that can be directly observed in online communities.
Inference and parameter estimation is done via an EM (Expectation-Maximization)
framework, where MCMC sampling is used
%for MAP inference on our MRF model
in the \textit{E-step} for estimating the label of unknown statements and the Trust Region Newton method~\cite{Lin2008} is used in the \textit{M-step} to compute feature weights.

\comment{
%results: experiments, findings
We apply our method to
%$2.7$
$2.8$ %see experiments section
million postings contributed by $15,000$ users of one of the largest online health community {\tt \href{http://www.healthboards.com/}{healthboards.com}}. Our model achieves an overall accuracy of $82\%$ in
identifying drug side-effects,
bringing an improvement of $13\%$ over an SVM baseline using the same features and an improvement of 4\% over a stronger SVM classifier which uses distant supervision to account for feature sparsity.  We further evaluate how the proposed model performs in two realistic use cases: discovering rare side-effects of drugs and identifying trustworthy users in a community.

%novel contributions
To summarize, this work brings the following main contributions:
\squishlist
\item {\em Model:} It proposes a model that captures the interactions between user trustworthiness, language objectivity, and statement credibility
in social media (Section~\ref{sec:model}), and devises a comprehensive feature set to this end (Section~\ref{sec:features});
\item {\em Method:} It introduces a method for joint inference over users, language, and statements (Section~\ref{sec:inference}) 
%%C: judiciously means something different
%by a judiciously designed
through a  
 probabilistic graphical model;
\item {\em Application:} It applies this methodology to the problem of extracting side-effects of medical drugs from online health forums (Section~\ref{sec:experiments});
\item {\em Use-cases:} It evaluates the performance of the model in the context of two realistic practical tasks (Section~\ref{sec:usecases}).
\squishend
}

\subsection{Use-case Study: News Communities}

As a second use-case, consider the role of media in the public dissemination of information about events. %However 70 percent of Americans believe that there is either a great deal or a fair %amount of media bias in news coverage~\cite{nber}. A June 2013 Gallup %poll~\cite{gallop,global-research} indicates that nearly 4 out of 5 Americans among %younger generations from age 21-64 do not trust the major news networks in the age %of super-mergers; when corporations like General Electric, Comcast and possibly Time %Warner own media in the likes of NBC and MSNBC.
%%%GW: need to make wording much crisper
Many people find online information and blogs as useful as TV or magazines.
%%%GW: no citation needed here
At the same time, however, people also believe that there is substantial media bias in
news coverage \cite{nber,gallop}, %cite{nber}
%%%GW: we should reconsider if we really need 3 citations here
especially in view of inter-dependencies and cross-ownerships
of media companies and other industries (like energy).

Several factors affect the coverage and presentation 
of news in media incorporating potentially biased information induced via the fairness and style of reporting.
%%%GW: isn't this trivial: Politics always comes with political orientation,
%%% so political news are naturally more susceptible to bias than say sports
%One such factor is the \emph{topical coverage} of news. 
%We found that $54\%$ of the news coverage on any topic (and $62\%$ of the news %articles) is related to \emph{Politics} {explicitly}, which, in turn, is influenced by the %political orientation of the media (e.g., left, center, or right). The political viewpoint %and perspective of the users and news web-sources (e.g., BBC, CNN) lead to polarization in %the community. The influence of Politics in media~\cite{chomsky1988} and vice %versa~\cite{dellavigna2007} has received vigorous interest in recent times.
%%%subho: influence of Politics on even sports (election) or environment (e.g., global warming)
%GW: keep the intro straight, don't add digression on secondary issues
%For instance, we found significant influence of {\em Politics} on almost all topics covered by the media.
News 
%on such topics 
are often presented in a polarized way 
depending on the political viewpoint of the media source (newspapers, TV stations, etc.). 
In addition, other  source-specific properties like {\em viewpoint, expertise, and format} of news may also be indicators of information credibility. 
%\cite{flanagin2000} found that people consider online information to be as credible as that obtained from television, radio, and magazines, %but not as credible as newspaper information. \cite{johnson2007} reported that politically interested Internet users find blogs to be %moderately credible web-sources for news and other information.

%\noindent{\bf News Community:} 
In this use-case, we embark on an in-depth study and formal modeling 
of these factors and
inter-dependencies within \textit{news communities} for {\em credibility analysis}. A news community is a news aggregator site (e.g., \href{http://www.reddit.com}{reddit.com}, \href{http://www.digg.com}{digg.com}, \href{http://www.newstrust.net}{newstrust.net}) where users can give explicit feedback (e.g., rate, review, share) on the quality of news and
can interact (e.g., comment, vote) with each other. Users can rate and review news, point out differences, bias in perspectives, unverified claims etc. However, 
this adds user subjectivity to the evaluation process, as users incorporate their own bias and perspectives in the framework. Controversial topics create polarization among users which influence their ratings. \cite{sloanreview,fang2014} state that online ratings are one of the most trusted sources of user feedback; however they are systematically {\em biased} and easily manipulated.

%\subsection{Research Questions}
%{\color{red}
%\noindent{\bf Research Questions:}
%\input{objectives}
%}

\noindent{\bf Approach:} 
%%%GW: claiming the study of this data as a contribution was adequate for ICSWM 
%%%but not for CIKM - here emphasis must be on methodology
%In this work we present an in-depth study of the different kinds of interactions at play %in a typical news community that influences the credibility of the information content %therein. One of our contribution is in harvesting a rich (and hitherto untapped) %dataset to analyze these interactions, consisting of $~62$K news articles from $~6$K %news web-sources that are rated and analyzed on different facets by $~6$K expert users %contributing $~134$K reviews in {\em newstrust.net}.
%
Unlike the healthforums focusing on a single topic, news communities are heterogeneous in nature, discussing on topics ranging from sports, politics, environment to food, movies, restaurants etc. Therefore, we propose a more {\em general} framework to analyze the factors and inter-dependencies in such a heterogeneous community; specifically, with additional factors for sources and topics, as well as allowing for inter user and inter source interactions. We develop a sophisticated probabilistic graphical model for {\em regression} to assign credibility {\em rating} to postings, as opposed to binary classification; specifically, we develop a Continuous Conditional Random Field (CCRF) model, 
which exploits several {\em moderate} signals of interaction {\em jointly} between the following factors to 
derive a {\em strong} signal for information credibility {(refer to Figures~\ref{subfig1:interaction} and \ref{subfig2:instance})}. In particular, the model captures the following factors.

\begin{itemize}
\item \textit{Language and credibility of a posting}: \textit{objectivity}, rationality, and general quality of language in the posting. Objectivity is the quality of the news to be free from emotion, bias and prejudice of the author. The \textit{credibility} of a posting refers to presenting an unbiased, informative and balanced narrative of an event.
\item \textit{Properties and trustworthiness of a source}: \textit{trustworthiness} of a source in the sense of generating credible postings based on source properties like viewpoint, expertise 
%type
%%%subho: type - newspaper, radio, blog; format - editorial, investigative, research
%%%GW: what is "type"? whay is it not subsumed under "format"?
and format of news.
\item \textit{Expertise of users and review ratings}: \textit{expertise} of a user,
in the community, in properly judging the credibility of postings. 
Expert users should provide objective evaluations --- in the form of reviews or ratings --- of postings, corroborating with the evaluations of other expert users. These can be used to identify potential ``citizen journalists'' \cite{lewis2010} in the community.
\end{itemize}

We show that the CCRF performs better than sophisticated collaborative filtering approaches based on latent factor models, and regression methods that do not consider these interactions. 

The proposed approach (CCRF) aggregates information (e.g., ratings) from various factors (e.g., users and sources), taking into account their interactions and topics of discussion, and presents a consolidate view (e.g., aggregated rating) about an item (e.g., posting). Therefore, this is similar to {\em ensemble learning}, and {\em learning to rank} based approaches, and can improve those methods by explicitly considering interaction between the participating factors.

\comment{Although this work is focused on news communities, the framework can also be used for instance, in health communities (e.g. {\tt healthboards.com}) where users write postings on drug usage --- the objective being to {\em jointly} rank postings, drug side-effects, and users based on their quality, credibility, and trustworthiness respectively.
}

In this work, the attributes {\em credibility} and {\em trustworthiness} are
always associated with a posting and a source, respectively. The joint
interaction between several factors also captures that a source garners
trustworthiness by generating credible postings, which are highly rated by
expert users. Similarly, the likelihood of a posting being credible
increases if it is generated by a trustworthy source.

%%%GW: important to state the following, as we claim that the Continuous CRF is a specific contribution
Some communities offer users {\em fine-grained scales for rating} different
aspects of postings and sources. For example, the {\tt \href{http://www.newstrust.net}{newstrust.net}} community analyzes a posting on $15$ aspects like insightful, fairness, style and factual. These are aggregated into an overall {\em real-valued} rating after weighing the aspects based on their importance, expertise of the user, feedback from the community, and more.
%Newstrust, for example, has aspects like {\tt GW: .......... ?????  (fill in)}, each of which can be rated in 0.1 steps between 0 and 5, and can further be aggregated into an overall rating.
This setting cannot be easily discretized without
blow-up or risking to lose information. Therefore, we model ratings as
real-valued variables in our CCRF.

\subsection{Contributions}

To summarize, this chapter introduces the following novel elements:
\begin{itemize}
\item {\em Model:} It proposes probabilistic graphical models that capture the mutual interactions and dependencies between trustworthiness of sources, credibility of postings and statements, objectivity of language, and expertise of users
in online communities (Section~\ref{sec:model}), and devises a comprehensive feature set to this end (Section~\ref{sec:features}).
\item {\em Method:} It introduces methods for joint inference over users, sources, language of postings, and statements (Section~\ref{sec:cred-inference}) 
%%C: judiciously means something different
%by a judiciously designed
through probabilistic graphical models for credibility classification (Section~\ref{sec:classification}) and credibility regression (Section~\ref{sec:regression}).
\item {\em Application:} 
\begin{itemize}
 \item A large-scale experimental study on one of the largest online health community {\tt \href{http://www.healthboards.com/}{healthboards.com}} --- where, we apply our method to $2.8$ million postings contributed by $15,000$ users for extracting side-effects of medical drugs from user-contributed posts (Section~\ref{sec:experiments-health}).
 \item A large-sale experimental study with data from {\tt \href{http://www.newstrust.net}{newstrust.net}}, one of
the most sophisticated news communities with a focus on quality
journalism (Section~\ref{sec:evaluations}).
\end{itemize}
\item {\em Use-cases:} It evaluates the performance of these models in the context of practical tasks like: (i) discovering rare side-effects of drugs (Section~\ref{subsec:usecaseRare}) and (ii) identifying trustworthy users (Section~\ref{subsec:usecaseTrust}) in a health community; (iii) finding trustworthy sources (Section~\ref{subsec:expert-sources}), and (iv) expert users (Section~\ref{subsec:expert-users}) in a news community who can play the role of {\em citizen journalists}.
\end{itemize}

\comment{
\noindent{\bf Contributions:}
%%%GW: now a few sentences on contributions, so far was more on approach
The paper introduces the following novel elements:

\squishlisttwo
\item A continuous CRF that captures the mutual dependencies between
credibility of postings, trustworthiness of sources, expertise of users,
and expresses real-valued ratings.
\item An inference method for the CCRF that allows us to {\em jointly} (a) predict ratings;
and (b) rank postings, sources, and users by their credibility, trustworthiness,
and expertise, respectively.
\item A large experimental study with data from {\em newstrust.net},
one of the most sophisticated news communities with a focus on quality journalism.
\squishend

The rest of the paper is organized as follows.
Section 2 presents how we model news communities, and which
factors we include in the model.
Section 3 develops the CCRF that captures the interaction between all the factors.
Section 4 introduces the dataset that we use for experimental evaluation
and further studies.
Section 5 presents our experimental results followed by discussion.
}

%% file: main/chapter-credibility-analysis/overview.tex
\section{Overview of the Model}
\label{sec:model}

\subsection{Credibility Classification}
\label{sec:cred-class-overview}

\begin{figure*}[htbp]
\centering
\includegraphics[scale=0.55]{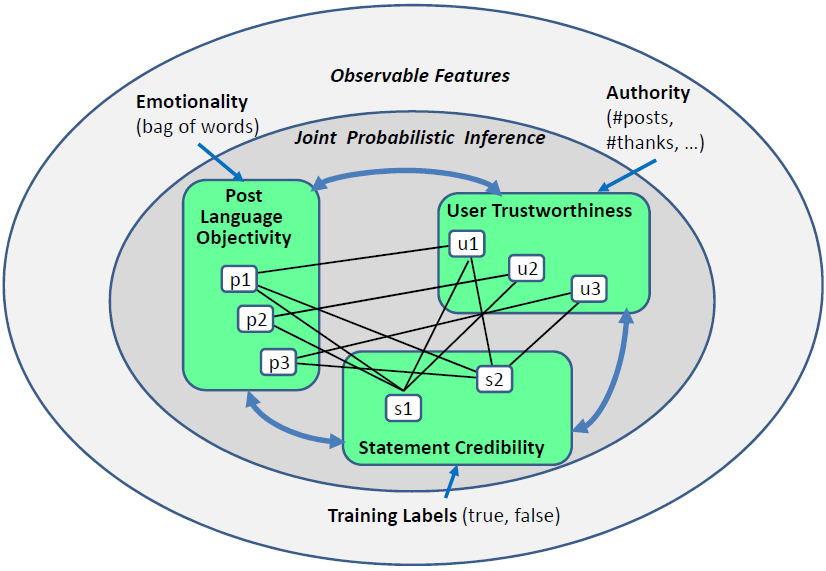}
\caption{Overview of the proposed model, which captures the interactions between statement credibility, posting objectivity, and user trustworthiness.}
\label{fig:triangle}
\end{figure*}

Our approach leverages the intuition that there is an important interaction between statement credibility,
linguistic objectivity, and user trustworthiness.  We therefore model these factors jointly through a probabilistic graphical model, more specifically a Markov Random Field (MRF), where
% Our task is to jointly assess the credibility of statements, 
% objectivity of posts' language, and trustworthiness of users.  We 
% model this joint interaction through a probabilistic graphical model, more specifically,
% a Markov Random Field (MRF).
each statement, posting and user is associated with a binary
random variable. Figure \ref{fig:triangle} provides an overview of our model.  For a given statement, the corresponding variable should have value $1$ if the
statement is credible, and $0$ otherwise. Likewise, the values of
posting and user variables reflect the objectivity and trustworthiness
of postings and users. 

%features and training data

\xhdrNoPeriod{Nodes, Features and Labels:} 
Nodes associated with users and postings have observable features, which can be extracted from the online community.
For users, we derive engagement features (number of questions and answers posted),
interaction features (e.g., replies, giving thanks),
and demographic information (e.g., age, gender).
For postings, we extract linguistic features in the form of discourse markers
and affective phrases.  Our features are presented in details in Section~\ref{sec:features}.
While for statements there are no observable features, we can derive
distant training labels for a subset of statements from expert databases, like the Mayo Clinic,\footnote{\tt \href{http://www.mayoclinic.org/drugs-supplements/}{mayoclinic.org/drugs-supplements/}}
which lists typical as well as rare side-effects of widely used drugs.

%using features and training labels
% \xhdrNoPeriod{Baseline System}
% The labels and features can be straightforwardly used to train standard SVM classifier
% for statement credibility. This will constitute the baseline for
% our experiments (Section~\ref{sec:experiments}).

%own method with RV couplings
\xhdrNoPeriod{Edges:} 
% The method developed in this paper aims at leveraging the intuition that there is an important interaction between statement credibility,
% linguistic objectivity, and user trustworthiness.  Therefore, we build a graphical model that exploits couplings between these variables.
The primary goal of the proposed system is to retrieve the credibility label of unobserved statements given \textit{some} expert labeled statements and the observed features by leveraging the mutual influence between the model's variables.
To this end, 
%%C: edit
%the corresponding (undirected) edges in the MRF are as follows:
the MRF's nodes are connected by the following (undirected) edges:
\begin{itemize}
\item each user  is connected to all her postings;
\item each statement is connected to all postings from
which it can be extracted (by state of the art information extraction methods);
\item each user is connected to statements
that appear in at least one of her postings.
\end{itemize}

%{\tt GW: is this statement that we are interested in MAP correct?}
%S:We are doing MAP and MLE both in the 2 steps of EM. I am not sure if writing only MAP is correct for CRF usage, where the more dominant term is MLE. For general MRF, MAP is fine. Since we do not refer to MAP in the inferencing section, I am bypassing it.
Configured this way, the model has the capacity to capture important interactions between statements, postings, and users --- for example, credible statements can boost a user's trustworthiness, whereas some false statements may bring it down.  Furthermore, since the inference (detailed in Section \ref{sec:classification}) is centered around the cliques in the graph (factors) and multiple cliques can share nodes, more complex ``cross-talk'' is also captured.  For instance, when several highly trustworthy users agree on a statement and one user disagrees, this reduces the trustworthiness of the disagreeing user.

In addition to classifying statements as credibility or not, the proposed system also computes individual likelihoods as a by-product of the inference process, and therefore can output rankings for all statements, users, and postings, in descending order of credibility, trustworthiness, and objectivity.

\subsection{Credibility Regression}
\label{sec:newscom}

%%%GW: make the figure 2-column wide (again) - right now it looks cramped
\begin{figure*}[!h]
\begin{subfigure}{\textwidth}
        \centering
	\includegraphics[width=0.6\textwidth]{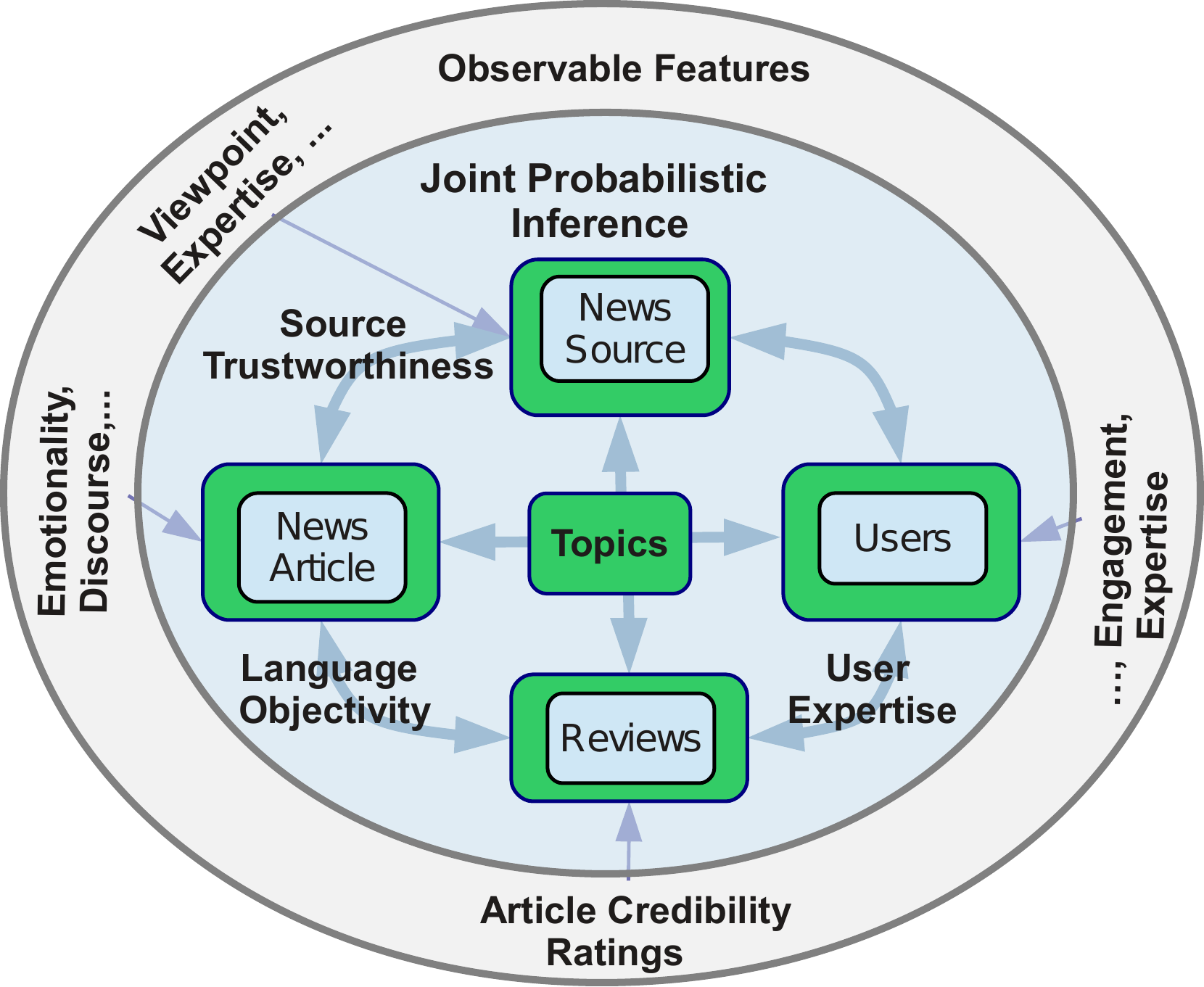}
	\caption{Interactions between source trustworthiness, posting (i.e. article) credibility, language objectivity, and user expertise.}
	\label{subfig1:interaction}
    \end{subfigure}
    \hfill
    \begin{subfigure}[b]{0.4\linewidth}
        \centering
	\includegraphics[width=\linewidth]{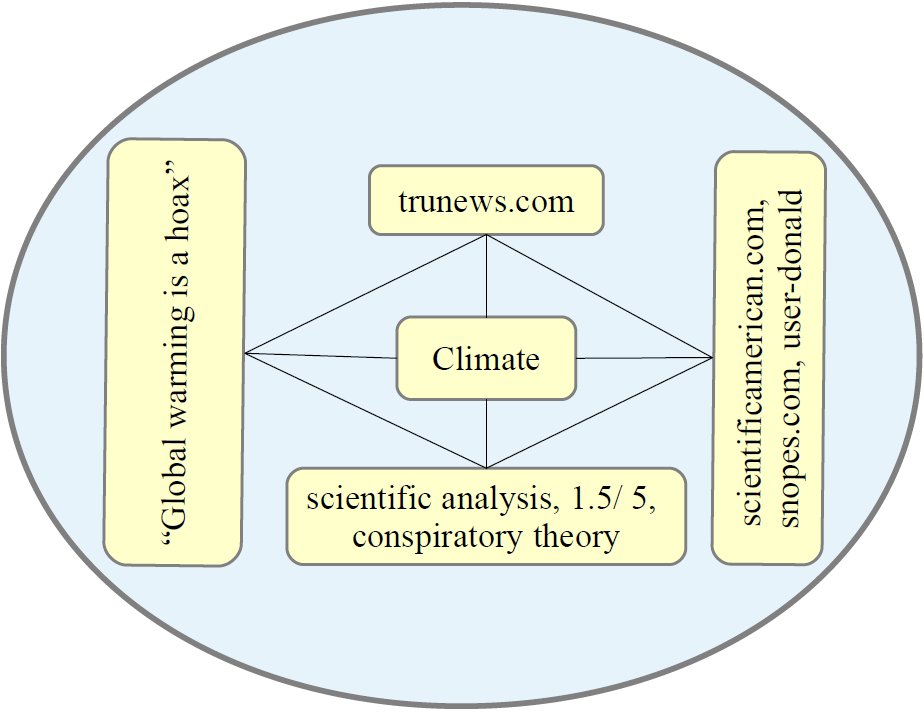}
	\caption{Sample instantiation.}
	\label{subfig2:instance}
    \end{subfigure}
        \begin{subfigure}[b]{0.6\linewidth}
        \centering
	\includegraphics[ height=70mm]{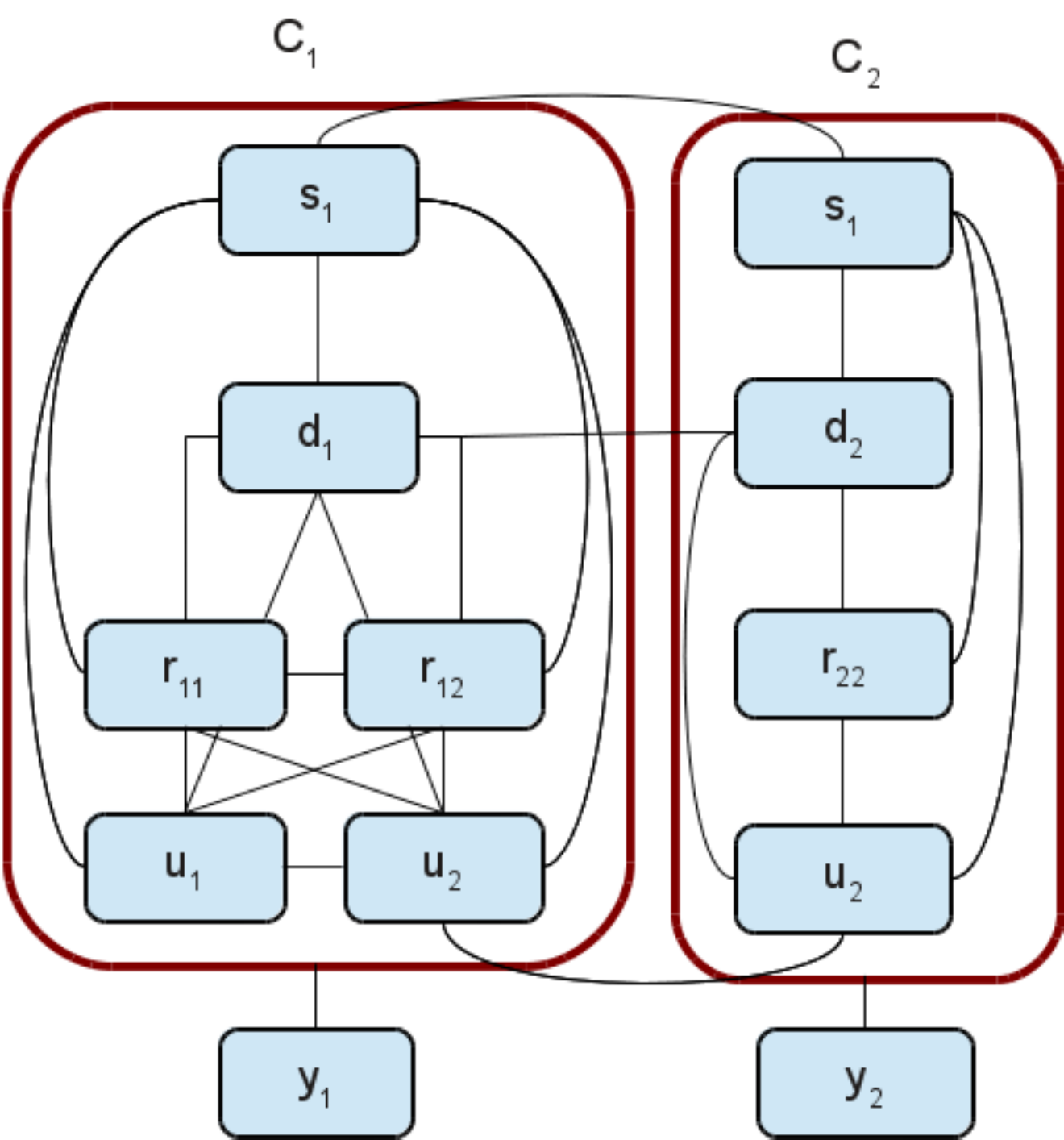}
	\caption{Clique representation.}
	\label{subfig3:clique}
    \end{subfigure}
    \caption{Graphical model representation.}
    \label{fig:0}
    \vspace{-1em}
\end{figure*}

%%%GW: included model overview in the modeling section
%{\tt GW: the whole text here needs more polishing - sometimes it is very compact and hard to follow, sometimes it is pretty verbose}\\

%\label{subfig1:interaction},\label{subfig1:clique}, \label{subfig1:instantiation} 

The earlier model is used for classifying statements as credible or not. However, in many scenarios for a more fine-grained credibility analysis, we want to assign a real-valued {\em credibility rating} to a posting. Additionally, we want to address several drawbacks of the earlier model, and propose a more general framework that models topics, users, sources, and explicit interactions between them --- as is prevalent in {\em any} online community.

%Our approach exploits the rich interaction taking place between the different factors in a news community. 
%We propose a {\em probabilistic graphical model} that leverages the interplay between news credibility, language objectivity, web-source trustworthiness, and user expertise. 

Refer to Figure~\ref{fig:0} for the following discussion. Consider a set of sources $\langle s \rangle$ (e.g., $s_1$ in Figure~\ref{subfig3:clique}) generating postings $\langle p \rangle$ which are reviewed and analyzed by users $\langle u \rangle$ for their credibility. Consider $r_{ij}$ to be the review by user $u_j$ on posting $p_i$. The overall credibility rating of the posting $p_i$ is given by $y_i$.

%In our model, each news web-source (e.g., $s_1$ in Figure~\ref{fig:0}), news article (e.g., $a_1$), user (e.g., $u_1$) and his rating or review (e.g., $r_11$) is associated with a continuous random variable $\in [1 - 5]$, that indicates its trustworthiness, objectivity, expertise and credibility respectively. $5$ denotes the maximum quality that can be achieved by any node.

In this model, each source, posting, user and her rating or review, and overall rating of the posting is associated with a continuous random variable %{\color{red}$y$} 
{$r.v.\in [1 \ldots 5]$}, that indicates its trustworthiness, objectivity, expertise, and credibility, respectively. $5$ indicates the best quality that an item can obtain, and $1$ is the worst. {Discrete ratings, being a special case of this setting, can be easily handled.}

Each node is associated with a set of observed features that are extracted from the community. For example, a source has properties like topic specific expertise, viewpoint and format of news; a posting has features like topics, and style of writing from the usage of discourse markers and subjective words in the posting. For users we extract their topical perspectives and expertise, engagement features (like the number of questions, replies, reviews posted) and various interactions with other users (like upvotes/downvotes) and sources in the community.

The objective of our  model is to 
%retrieve 
predict
credibility ratings $\langle y \rangle$ of postings $\langle d \rangle$ by exploiting the mutual interactions between different variables. The following edges between the variables capture their interplay:
\begin{itemize}
\item Each posting is connected to the source from where it is extracted (e.g., $s_1-p_1$, $s_1-p_2$)
\item Each posting is connected to its review or rating by a user (e.g., $p_1-r_{11}$, $p_1-r_{12}$, $p_2-r_{22}$)
\item Each user is connected to all her reviews (e.g., $u_1-r_{11}$, $u_2-r_{12}$, $u_2-r_{22}$)
\item Each user is connected to all postings rated by her (e.g., $u_1-p_1$, $u_2-p_1$, $u_2-p_2$)
\item Each source is connected to all the users who rated its postings (e.g., $s_1-u_1$, $s_1-u_2$)
\item Each source is connected to all the reviews of its postings (e.g., $s_1-r_{11}$, $s_1-r_{12}$, $s_1-r_{22}$)
\item For each posting, all the users and all their reviews on the posting are inter-connected (e.g., $u_1-r_{12}$, $u_2-r_{11}$, $u_1-u_2$). This captures user-user interactions (e.g., $u_1$ upvoting/downvoting $u_2$'s rating on $p_1$) influencing the overall post rating.
\end{itemize}

Therefore, a {\em clique} (e.g., $C_1$) is formed between a posting, its source, users and their reviews on the posting. Multiple such cliques (e.g., $C_1$ and $C_2$) share information via their common sources (e.g., $s_1$) and users (e.g., $u_2$).

\textit{Topics} play a significant role on information credibility. Individual users in community (and sources) have their own perspectives and expertise on various topics (e.g., environmental politics). Modeling user-specific topical perspectives explicitly captures credibility judgment better than a user-independent model.
However, many postings do not have explicit topic tags. 
%%%GW: correct? LDA is latent topics, not explicit - here, too?
%%%subho: yes
Hence we use Latent Dirichlet Allocation (LDA)~\cite{Blei2003LDA} in conjunction with Support Vector Regression (SVR)~\cite{drucker97} to learn words associated to each (latent) topic, and user (and source) perspectives for the topics. Documents are assumed to have a distribution over
%Latent Dirichlet Allocation assumes a document to have a distribution over topics, and topics to have a distribution over words.
topics as latent variables, with words as observables. Inference is by Gibbs sampling.
This LDA model is a component of the overall model, discussed next.
%and hyper-parameter optimization is done by Support Vector Regression to guide the sampling process by attributing more mass to the topics for which the users have a higher preference.

We use a probabilistic graphical model, specifically a Conditional Random Field (CRF), to model all factors jointly. 
The modeling approach is related to the model discussed in the previous Section~\ref{sec:cred-class-overview}.
%the prior work of~\cite{mukherjee2014}.
%%%GW: toned down the wording here: "related to" 
However, unlike that model and traditional CRF models, our problem setting requires a \textit{continuous} version of the CRF (CCRF) to deal with real-valued ratings instead of discrete labels. In this work, we follow an approach similar to~\cite{qinNIPS2008,radosavljevicECAI2010,tadas14} in learning the parameters of the CCRF. We use Support Vector Regression~\cite{drucker97} to learn the elements of the feature vector for the CCRF.

The inference is centered around cliques of the form
 $\langle$ source, posting, $\langle$ users $\rangle$, $\langle$ reviews $\rangle \rangle$.
An example is the two cliques $C_1: ~ s_1-p_1-\langle u_1, u_2 \rangle - \langle r_{11}, r_{12} \rangle$
and $C_2: ~ s_1-p_2-u_2-r_{22}$ in the instance graph of Figure \ref{subfig3:clique}.
This captures the ``cross-talk'' between different cliques sharing nodes. 
A source garners trustworthiness by generating multiple credible postings. Users attain expertise by correctly identifying credible postings that corroborate with other expert users. Inability to do so brings down their expertise. Similarly, a posting attains credibility if it is generated by a trustworthy source and highly rated by an expert user.
The inference algorithm for the CCRF is discussed in detail in Section \ref{sec:regression}.

%%%GW: added explicit structure
In the following section, 
we discuss the various feature groups that are considered in
our credibility model.

%% file: main/chapter-credibility-analysis/language.tex
\subsection{Postings and their Language}
\label{sec:language}

%Health Communities

The style in which a post is written %news is presented to the reader
plays a pivotal role in understanding its credibility. The
desired property for a posting is to be objective and unbiased. In our model 
we use \textit{stylistic}
%,
%\textit{semantic operators}, 
and \textit{affective} features to assess a posting's objectivity and quality.

%In this section, we examine the different {\em stylistic} indicators of credibility. All the lexicons used in this section are compiled from~\cite{recasens2013,mukherjee2014}.

%The linguistic characteristics of a post can convey the author's attitude towards her statements as well as her emotional state~\cite{Coates1987}.
%certainty, or lack thereof~\cite{Coates1987}.
%In our model 
%we use \textit{stylistic}
%,
%\textit{semantic operators}, 
%and \textit{affective} features to assess a post's objectivity and quality.

\subsubsection{Stylistic} 

Consider the following user posting in a {\em health} community:
\example{``I heard Xanax \underline{can} have pretty bad side-effects. You \underline{may} have peeling of skin, and apparently \underline{some} friend of mine told me you \underline{can} develop ulcers in the lips also. \underline{If} you take this medicine for a long time then you \underline{would probably} develop a lot of other physical problems. \underline{Which} of these did you experience ?''}
This posting 
%%C: edit
%presents a lot of possibilities and uncertainties
evokes a lot of uncertainty, 
and does not specifically point to the occurrence of any side effect from a first-hand experience. Note the usage of strong modals (depicting a high degree of uncertainty)
%\footnote{\small Strong modals express a higher degree of uncertainty in any situation, in contrast to the weak modals that express a lesser degree of uncertainty and more emphasis on certain events or situations.}
``can'', ``may'', ``would'', the 
indefinite determiner ``some'',  
%third person ``some friend'', 
 the conditional ``if'', the adverb of possibility ``probably'', and the question particle ``which''. Even the usage of too many named entities for drug and disease names can impact the credibility of a statement (refer the introductory Example~\ref{ex:cocktail}).

Contrast the above posting with the following one :
\example{``Depo is very dangerous as a birth control and has too many long term side-effects like reducing bone density. \underline{Hence}, I \underline{will} never recommend anyone using \underline{this} as a birth control. \underline{Some} women tolerate it well but \underline{those} are \underline{the} minority. \underline{Most} women have horrible long lasting side-effects from it.''}
This posting uses the inferential conjunction ``hence'' to draw conclusions from a previous argument, the definite determiners ``this'', ``those'', ``the'' and ``most'' to pinpoint entities and the 
 %%C: edit
 % highly certain
 highly certain weak
  modal ``will''. 

Table~\ref{tab:discourse} shows a set of linguistic features which we deem suitable
for discriminating between these two kinds of postings. Many of the features related to epistemic modality have been discussed in prior linguistic literature \cite{Coates1987, Westney1986} and features related to discourse coherence have also been employed in earlier computational work (e.g., \cite{Mukherjee2012,Wolf2004}).

\comment{For each stylistic feature type $f_i$ shown in Table \ref{tab:linguistic} and each posting $p_j$,
we compute the relative frequency of words of type $f_i$ occurring in $p_j$, thus
constructing a feature vector $F^L(p_j) =\langle freq_{ij} = \#(words~in~f_i) ~ / ~length(p_j) \rangle$.
We further aggregate these vectors over all postings $p_j$ by a user $u_k$ into
\begin{equation}
F^L(u_k) = \langle {\sum_{p_j~by~u_k} \#(words~in~f_i)}
~ / ~ {\sum_{p_j~by~u_k} length(p_j)} \rangle.
 \label{eq2}
\end{equation}
}

\comment{
\begin{table}
%\todo{Fix this table}
\centering
\small
\begin{tabular}{p{14.3cm}}
\toprule
{\bf Modals} depict the possibility of an event occurring under certain conditions. Strong modals depict a higher degree of uncertainty, whereas the weaker ones depict a lesser degree of uncertainty, and more emphasis on certain situations. \\\sectionrule
{\bf Conditionals} depict hypothetical situations.\\\sectionrule
{\bf Negation} contradicts or denies the possibility of an event. \\\sectionrule
{\bf Following, Inferential Conjunctions} assign more emphasis on the subsequent discourse segment.\\\sectionrule
{\bf Contrasting Conjunctions} oppose or refute the neighboring discourse segment.\\\sectionrule
{\bf Determiners} are used to mark nouns or identify entities for more details. Definite determiners refer to specific entities.\\\sectionrule
{\bf Person} : From the narrative point of view, `first person' is the narrator who talks about self-experience. `Second person' refers to the reader, whereas `third person' refers to any other character.\\\sectionrule
{\bf Question Particles} are grammatical elements used to form a question.\\\sectionrule
{\bf Parts-of-Speech} categories of words like nouns, verbs, adjectives and adverbs\\\sectionrule
\bottomrule
\end{tabular}
\caption{Brief description of some stylistic features.}
\label{tab:dis}
\end{table}
}

\begin{table}
\centering
\small
\begin{tabular}{p{2.5cm}p{4cm}|p{2.6cm}p{4cm}}
\toprule
{\bf Feature types} & {\bf Example values} & {\bf Feature types} & {\bf Example values}\\\midrule
Strong modals & might, could, can, would & First person & I, we, me, my, mine, us, our\\\sectionrule
Weak modals & should, ought, need, shall & Second person & you, your, yours\\\sectionrule
Conditionals & if & Third person & he, she, him, her, his, it, its\\\sectionrule
Negation & no, not, neither, nor, never & Question particles & why, what, when, which\\\sectionrule
Inferential conj. & therefore, thus, furthermore & Adjectives & correct, extreme, visible\\\sectionrule
Contrasting conj. &  until, despite, in spite & Adverbs &maybe, about, probably\\\sectionrule
Following conj. & but, however, otherwise, yet & Proper nouns & Xanax, Zoloft, Depo\\\sectionrule
Definite det. & the, this, that, those, these & &\\\sectionrule
\bottomrule
\end{tabular}
\caption{Stylistic features.}
\label{tab:discourse}
\end{table}

\subsubsection{Affective}

Each user has an \textit{affective state} that depicts her attitude and emotions that are reflected in 
her postings. Note that a user's affective state may change over time; so it is a property of
postings, not of users per se.
As an example, consider the following posting in a {\em health} community:
\example{``I've had chronic \underline{depression} off and on since adolescence. In the past I've taken Paxil (made me \underline{anxious}) and Zoloft (caused insomnia and stomach problems, but at least I was mellow ). I have been taking St. John's Wort for a few months now, and it helps, but not enough. I wake up almost every morning feeling very \underline{sad} and \underline{hopeless}. As afternoon approaches I start to feel better, but there's almost always at least a low level of \underline{depression} throughout the day.''}
The high level of depression and negativity in the posting makes one wonder if the statements on 
drug side-effects are really credible. Contrast this posting to the following one:
%from {\tt patient.co.uk}:\\
\example{``A diagnosis of GAD (Generalized Anxiety Disorder) is made if you suffer from excessive anxiety or worry and have at least three symptoms including...If the symptoms above, touch a chord with you, do speak to your GP. There are effective treatments for GAD, and Cognitive Behavioural Therapy in particular can help you ...''}
--- where the user objectivity and positivity in the posting make it much more credible.

We use the WordNet-Affect 
lexicon~\cite{WNAffect}, where each word sense (WordNet synset) is mapped to one of 285 attributes
of the affective feature space, like \textit{confusion, ambiguity, hope, anticipation, hate}. 
%The lexicon contains a mapping from WordNet synsets to an affective attribute. 
We do not perform word sense disambiguation (WSD), and instead simply take
the most common sense of a word (which is generally a good heuristics for WSD). 
%For each post, we create an affective feature vector $\langle F^E(p_j) \rangle$ using these features, analogous to the stylistic vectors $\langle F^L(p_j) \rangle$. 
Table~\ref{tab:aff} shows a sample of the affective features used in this work.

\comment{
\begin{table}
\centering
\small
\begin{tabular}{p{7cm}p{7cm}}
\toprule
{\bf Positive Features} & {\bf Negative Features}\\\midrule
levity, coolness, hysteria, eagerness, edginess, contentment, resignation, devotion, triumph, insecurity, antipathy, self esteem, fit, togetherness, fondness, sympathy, confidence%, encouragement, approval, surprise, downheartedness, affection, ingratitude 
& distress, weight, depression, wonder, misery, stupefaction, favor, guilt, creeps, jitteriness, malice, harassment, levity gaiety, embarrassment, anxiousness, compunction, humility\\
\bottomrule
\end{tabular}
\caption{Positive vs. negative affective features.}
\label{tab:aff}
\end{table}
}

\begin{table}
\centering
\small
\begin{tabular}{p{14cm}}
\toprule
{\bf Sample Affective Features}\\\midrule
affection, antipathy, anxiousness, approval, compunction, confidence, contentment, coolness, creeps, depression, devotion, distress, downheartedness, eagerness, edginess, embarrassment, encouragement, favor, fit, fondness, guilt, harassment, humility, hysteria, ingratitude, insecurity, jitteriness, levity, levitygaiety, malice, misery, resignation, selfesteem, stupefaction, surprise, sympathy, togetherness, triumph, weight, wonder\\
\bottomrule
\end{tabular}
\caption{Examples of affective features.}
\label{tab:aff}
\end{table}

\newcommand{\helpfulness}{helpfulness\xspace}
\newcommand{\helpful}{helpful\xspace}

\comment{
%subho:camera-ready: as per discussion with Cristian, creating a separate sub-section for Feature Exploration and moving the regression experiments here
\subsubsection{Preliminary Feature Exploration}

To test whether the linguistic features introduced so far are sufficiently informative of how \helpful a user is in the context of health forums, we conduct a preliminary experimental study.
In the \mbox{\small\tt \href{http://www.healthboards.com/}{healthboards.com}} forum,  community members have the option of
expressing their gratitude to a user if they find one of her postings helpful by giving ``thanks'' votes.
Solely for the purpose of this preliminary experiment,
we use the {\em total number of ``thanks'' votes} that a user received from all her postings
as a weak proxy measure for user \helpfulness.

We train a regression model on the per-user stylistic feature vectors $F^L(u_k)$ with
\#thanks normalized by \#postings for each user $u_k$ as response variable. We repeat the same experiment using only the per-user affective feature vectors $F^E(u_k)$ to identify the most important affective features.
%GW: usually, logistic regression is used only for 0-1 classifiers, not for numerical output ?

Figure~\ref{fig:langFeat} shows the relative weight of various stylistic and affective linguistic features in determining user \textit{helpfulness}, with positive weights being indicative of features contributing to a user 
being considered helpful by the community.
%users that receive a high number of thanks.
%The signs of features in the learned weight vector are indicators of objective and authoritative language vs. subjective and doubtful language.
%
Figure~\ref{fig:affFeat} shows that user confidence, pride, affection and positivity in statements are correlated with user \helpfulness, in contrast to misery, depression and negativity in attitude. Figure~\ref{fig:styFeat} shows that inferential statements about definite entities have a positive impact, as opposed to the use of hypothetical statements, contrasting sentences, doubts and queries.

\todo{polish this discussion}

From the above experiment, we can make the following observations: Function words (like \textit{conjunctions, articles, particles, modals etc.}) play a huge role in determining user authority in establishing discourse and ascertaining the speaker mood and modality. In fact, they are more important than content words like \textit{nouns}). The primary role played by them is in ascertaining the user confidence, tone and attitude (the latter is also captured in the affective space) in statements. Inferential conjunctions give credence to statements by drawing conclusions. Any conjunction that gives more weight to the subsequent discourse segment has a positive influence on ascertaining user credibility. Definite determiners that point to specific objects or events have a positive influence. Strong modals and conditionals that depict hypothetical situations or \textit{irrealis events} have a negative influence. Contrasting conjunctions that depict a contrast/violation between connecting discourse segments have a 
negative influence. Similarly asking questions, or having doubts (
as we will also see in the affective space) have a negative influence. First hand experience, and giving suggestions/advice (usage of second person) have a positive influence, than reporting incidents or events heard from others (usage of third person which has a negative weight). Content words like \textit{Adjectives} are important as they capture the user emotions strongly and help in finding the user attitude; whereas usage of \textit{Adverbs} in flowery languages (rhetorically ornate) decreases the credibility of the statements.

\textit{To summarize: Giving suggestions/advice to other people by making inferential statements about definite entities have a positive impact than the use of hypothetical statements, doubts/queries and flowery language.}

%%C: this uses CRF, too early to discuss (talked with Subho about it)
\comment{
Table~\ref{tab:useFeat} shows the relative importance of various user-specific features. The most significant feature is the ratio of the number of replies by a user to the questions posted by him in the community. The figure also depicts a gender-bias in the community, as $77.69\%$ active contributors in this health forum are female.
}
%%subho:camera-ready
%%moving these observations to discussion of results about feature informativeness

This experiment confirms that linguistic features can be informative 
%%C: edit
%of how \helpful a user is
in the context of online health communities.
Although we use ``thanks'' votes as a proxy for user \helpfulness, there is no guarantee that the information provided by \helpful users is actually correct.  A user can receive ``thanks'' for a multitude of reasons (e.g. being compassionate or supportive), and yet provide incorrect information. Hence, while the features described here are part of our final model, the feature weights learned in this preliminary experiment are not going to be used; instead, partially provided expert information is used to train our probabilistic model (refer to Section~\ref{sec:inference}).

\begin{figure*}
\begin{subfigure}{.6\textwidth}
\includegraphics[scale=0.42]{main/images/affFeat.jpg}
\caption{Weight of top 20 affective features.}
\label{fig:affFeat}
\end{subfigure}
\begin{subfigure}{.4\textwidth}
\includegraphics[scale=0.55]{main/images/styfeat.jpg}
\caption{Weight of stylistic features.}
\label{fig:styFeat}
\end{subfigure}
\caption{Relative importance of linguistic features for predicting user \helpfulness in a preliminary experiment.}
\label{fig:langFeat}
%\vspace*{-4mm}
\end{figure*}
}

\comment{
\begin{table}
\centering
\small
\begin{tabular}{lc}
\toprule
{\bf User Feature} & {\bf Scaled Feature Weight}\\\midrule
Replies / Queries & 305.72 \\
Gender (Female) & 159.43 \\
Postings & 137.09 \\
Thanks & 123.62 \\\bottomrule
\end{tabular}
\caption{User feature importance for predicting  \helpfulness.}
\label{tab:useFeat}
%\vspace*{-4mm}
\end{table}
}

\subsubsection{Bias and Subjectivity}

\comment{
The style in which news is presented to the reader
plays a pivotal role in understanding its credibility. The
desired property for news is to be objective and unbiased. In this section, we examine the different stylistic indicators of news credibility. All the lexicons used in this section are compiled from~\cite{recasens2013}.

\xhdrNoPeriod{Assertives}
``Assertive verbs are those whose complement clauses assert a proposition. The truth of the
proposition is not presupposed, but its level of certainty
depends on the asserting verb.''~\cite{recasens2013} For example, although verbs like \emph{say} and \emph{state} are neutral, \emph{point} and \emph{claim} induce an uncertainty in the proposition.
\example{ ``These folks while shrouding themselves in outward displays of patriotism and exaggerated
	\underline{claims} of support for the liberty and freedom of all decided to make their opposition personal"}
\example{
	``Deputy accused of prostitution, drug
	and theft crimes I-5 driver battled mental illness, \underline{say}
	two women who knew him"}

\xhdrNoPeriod{Factives}
These verbs presuppose the truth of their complement
clause. Note the usage of \emph{realize} and \emph{indicate} in
the following examples. The first pre-supposes a notion,
whereas the latter depicts an experimental observation.
\example{
	``I think that we \underline{realize} on the left
	that complete democratic control on every aspect of the
	economy is probably not a good idea ...''
	}
	%, but we need our
	%friends on the libertarian side to \underline{realize} that completely
	%unfettered free markets tend to spin wildly out of control ..."
\example{
	``Numerous studies on developing
	country livestock \underline{indicate} that annual mortality is
	high."
}

\xhdrNoPeriod{Hedges}
``It is a mitigating word used to reduce the impact of an
utterance or one's commitment to an assertion.''~\cite{recasens2013}. Note the contrast between the situations presented by \emph{will} and \emph{may} in the following examples.
\example{
	``Party \underline{will} continue to back Supreme
	Court candidate despite pending DWI case."
}
\example{
	``Party \underline{may} continue to back Supreme Court candidate ..."
}

\xhdrNoPeriod{Implicatives}
These words trigger pre-supposition in an utterance. For
example, usage of the word \emph{complicit} indicates participation in an activity in an unlawful way.
\example{
	``If one has evidence that the CEO of the
	company one works for is aware of and \underline{complicit} in
	doing something improper you're not going to expect ..."
}

\xhdrNoPeriod{Report Verbs}
These are used to indicate the attitude towards the
source, or report what someone said more accurately,
rather than using just \emph{say} and \emph{tell}. For example,
usage of the word \emph{argues} indicates the subject has evidence in support of the argument.
\example{
	``First, Lanier \underline{argues} that the digital
	architecture of computers causes things to end up in binary states at the macro level ..."
}

Table~\ref{tab:linguistic} shows a snapshot of the words belonging to
each of the linguistic categories as well as the number of
elements in each category.

News is supposed to be objective i.e. writers should not
convey their own opinions, feelings or prejudices in their
stories. For example, a news titled ``Why do conservatives hate your children?" is not considered credible or
objective journalism. We use a subjectivity lexicon, a list of positive and negative
opinionated words, and an affective lexicon to detect subjective clues in postings.

We furthermore use a lexicon of bias-inducing words
extracted from the Wikipedia edit history from~\cite{recasens2013}. Wikipedia has a Neutral Point of View policy to keep its posts ``fairly, proportionately, and as
far as possible without bias, all significant views that
have been published by reliable sources on a topic". Accordingly,
editors identify and rewrite biased passages that do
not conform to this, and mark them with NPOV
tags.
}

A posting is supposed to be objective: writers should not
convey their own opinions, feelings or prejudices in their
postings. For example, a posting titled ``Why do conservatives hate your children?'' is not considered %credible or
objective journalism in a {\em news} community. We use the following linguistic cues for detecting bias and subjectivity in user-written postings. A subset of these features has been earlier used in \cite{recasens2013,mukherjee2014}.
%All the lexicons used in this section are compiled from~\cite{recasens2013,mukherjee2014}.

%The style in which news is presented to the reader
%plays a pivotal role in understanding its credibility. The
%desired property for news is to be objective and unbiased. In this section, we examine the different stylistic indicators of news credibility. All the lexicons used in this section are compiled from~\cite{recasens2013,mukherjee2014}.

%%%GW: drastically trimmed this !!!

\noindent{\bf Assertives}: 
Assertive verbs (e.g., ``claim'') complement and modify a proposition in a sentence. 
They capture the degree of certainty to which a proposition holds. % \cite{recasens2013}. 

\noindent{\bf Factives}: Factive verbs (e.g., ``indicate'')
pre-suppose the truth of a proposition in a sentence.
%%%GW: "presuppose" is not clear to me - can we say this differently?
%%%subho: take for granted, presume

\noindent{\bf Hedges}: These are mitigating words (e.g., ``may'') to soften the degree of
commitment to a proposition.% \cite{recasens2013}.
%Note the contrast between the situations presented by \emph{will} and \emph{may} in the following examples.
%%%GW: why do we say assertion here and proposition previously? any diff?

\noindent{\bf Implicatives}:
These words trigger pre-supposition in an utterance. For
example, usage of the word \emph{complicit} indicates participation in an activity in an unlawful way.

\noindent{\bf Report verbs}: These verbs (e.g., ``argue'')
are used to indicate the attitude towards the
source, or report what someone said more accurately,
rather than using just \emph{say} and \emph{tell}. 

\pagebreak

\noindent{\bf Discourse markers}: These capture the degree of confidence, perspective, and certainty in the set of propositions made. For instance, strong modals (e.g., ``could''), 
probabilistic adverbs (e.g., ``maybe''), and conditionals (e.g., ``if'') depict a high degree of uncertainty and hypothetical situations, whereas weak modals (e.g., ``should'')
 and inferential conjunctions (e.g., ``therefore'') depict certainty.

\noindent{\bf Subjectivity}:
%News is supposed to be objective: writers should not
%convey their own opinions, feelings or prejudices in their
%stories. For example, a news titled ``Why do conservatives hate your children?'' is not considered %credible or
%objective journalism. 
We use a subjectivity lexicon\footnote{\small \href{http://mpqa.cs.pitt.edu/lexicons/subj\_lexicon/}{http://mpqa.cs.pitt.edu/lexicons/subj\_lexicon/}}, a list of positive and negative
opinionated words\footnote{\small \href{http://www.cs.uic.edu/~liub/FBS/opinion-lexicon-English.rar}{http://www.cs.uic.edu/~liub/FBS/opinion-lexicon-English.rar}}, and an affective lexicon\footnote{\small \href{http://wndomains.fbk.eu/wnaffect.html}{http://wndomains.fbk.eu/wnaffect.html}} to detect subjective clues in postings.
%The {\em affective features} capture the state of mind (like attitude and emotions) of the writer while writing an article or post (e.g., anxiousness, confidence, depression, favor, malice, sympathy etc.).

%{\tt GW: which lexcion(s)????? be explict!!!!!}

We additionally harness a lexicon of bias-inducing words
extracted from the Wikipedia edit history from~\cite{recasens2013} exploiting its Neutral Point of View Policy to keep its postings ``fairly, proportionately, and as far as possible without bias, all significant views that have been published by reliable sources on a {\em topic}''.

\begin{table}
\small
\centering
 \begin{tabular}{p{1.5cm}p{3.2cm}p{1cm}|p{1.5cm}p{3.2cm}p{1cm}}
 \toprule
\textbf{Category} & \textbf{Example Values} & \textbf{\#Count} & \textbf{Category} & \textbf{Example Values} & \textbf{\#Count}\\\midrule
{\bf Bias} & & &{\bf Subjectivity}&&\\\sectionrule
{Assertives} & think, believe, suppose, expect, imagine& 66 & Wiki Bias Lexicon & apologetic, summer, advance, cornerstone, & 354\\\sectionrule
{Factives} & know, realize, regret, forget, find out & 27 & Negative & hypocricy, swindle, unacceptable, worse & 4783\\\sectionrule
{Hedges} &  postulates, felt, likely, mainly, guess& 100 & Positive & steadiest, enjoyed, prominence, lucky & 2006\\\sectionrule
{Implicatives} & manage, remember, bother, get, dare& 32 & Subj. Clues & better, heckle, grisly, defeat, peevish & 8221\\\sectionrule
{Report} & claim, underscore, alert, express, expect & 181 & Affective & disgust, anxious, revolt, guilt, confident & 2978\\\sectionrule
\bottomrule
\end{tabular}
\caption{Subjectivity and bias features.}
\label{tab:linguistic}
\vspace{-1.5em}
\end{table}

\noindent {\bf Feature vector construction}: For each stylistic feature type $f_i$ 
%shown in Table \ref{tab:linguistic} 
and each posting $p_j$, we compute the relative frequency of words of type $f_i$ occurring in $p_j$, thus constructing a feature vector $F^L(p_j) =\langle freq_{ij} = \#(words~in~f_i) ~ / ~length(p_j) \rangle$. 

We further aggregate these vectors over all postings $p_j$ by a user $u_k$ into
\begin{equation}
F^L(u_k) = \langle {\sum_{p_j~by~u_k} \#(words~in~f_i)}
~ / ~ {\sum_{p_j~by~u_k} length(p_j)} \rangle.
 \label{eq2}
\end{equation}

Since our model allows users to interact with other users, and give feedback (reviews/comments) on their postings --- we also create feature vectors for the users' reviews to capture whether the feedbacks are credible or biased by the users' judgment.
Consider the review $r_{j,k}$ written by user $u_k$ on a posting $p_j$. For each such review, analogous to the per-posting stylistic feature vector $\langle F^L(p_j) \rangle$, we construct a {\em per-review} feature vector $\langle F^L(r_{j,k}) \rangle$.

%For each stylistic feature type $f_i$ shown in Table \ref{tab:linguistic} and each post $p_j$, we compute the relative frequency of words of type $f_i$ occurring in $p_j$, thus constructing a feature vector $F^L(p_j) =\langle freq_{ij} = \#(words~in~f_i) ~ / ~length(p_j) \rangle$.

%\todo{replace all articles and postings by postings and unify all symbols. d\_j with p\_j}

%%%%%%%%%%%%%%%%%%%%%%%%%%%%%%%%%%%%%%%%

%% file: main/chapter-credibility-analysis/users.tex
\subsection{User Expertise}

A user's expertise in judging credibility of other users' postings depends on many factors. \cite{einhorn1977} discusses the following traits for recognizing an expert.

\begin{itemize}
 \item An expert user needs to be recognized by other members. %in the community. This can be captured by how the community %members perceive the quality of the user's article ratings.
  \item Experience is an uncertain indicator of user expertise. %\textit{Community engagement} of the member is taken as an %indicator of his experience.
 \item Inter-expert agreement should be high.% and different %experts should converge on the same signs.
 \item Experts should be independent of bias.
\end{itemize}

%%%GW: this itemized list is merely the preview of the following paragraphs, hence commented out

%We validate all these hypothesis in Section~\ref{sec:hypotheses}. %Or proposed model captures the following aspects of user %expertise in identifying credible news.
 
\noindent{\bf Community Engagement:} of the user is an obvious measure for judging the user authority in the community. We capture this with different features: number of answers, 
ratings given, comments, ratings received, disagreement and number of raters. In case user demography information like age, gender, location, etc. are available, we also incorporate them as features.

%\noindent{\bf Intra-User Agreement}: 
\noindent{\bf Inter-User Agreement:}  
Expert users typically agree on what constitutes a credible posting. This is inherently captured in the proposed graphical model, where a user gains expertise by assigning credibility ratings to postings that corroborate with other expert users.

\noindent{\bf Topical Perspective and Expertise:} The potential for harvesting user preference and expertise in topics for rating prediction of reviews has been demonstrated in~\cite{mukherjee2014JAST,mcauleyrecsys2013}. For credibility analysis, the model needs to capture the user's {\em perspective} and {\em bias} towards certain topics based on their political inclination that bias their ratings, and their topic-specific {\em expertise} that allows them to evaluate postings on certain topics better as ``Subject Matter Experts''. These are captured as {\em per-user} feature weights for the stylistic indicators and topic words in the language of user-contributed reviews.

%In this work, we do not distinguish between the fine-grained %aspects of topic influence on users. Instead, we capture them %together as {\em per topic} feature weights for {\em each} user %for explicit and latent topic representations.
%%%GW: this is very vague: what features do we introduce here ?????

\noindent{\bf Interactions:}\label{sec:interactions} 
%There are different kinds of interactions at play between users and other facets in the community that can be harnessed to identify expert users and credible articles.
In a community, users can upvote ({\em digg, like, rate}) the ratings of users that they appreciate, and downvote the ones they do not agree with. High review ratings from expert users increase the value of a user; whereas low ratings bring down her expertise. Similar to this {\em user-user} interaction, there can be {\em user-posting}, {\em user-source} and {\em source-posting} interactions which are captured as edges in our graphical model (by construction).
Consider the following anecdotal example in the community showing an expert in nuclear energy \textit{downvoting} another user's rating on nuclear radiation:
\example{
``{\bf Non-expert}: Interesting opinion about health risks of nuclear radiation, from a physicist at Oxford University. He makes some reasonable points ...\\
Low rating by {\bf expert} to above review: Is it fair to assume that you have no background in biology or anything medical? While this story is definitely very important, it contains enough inaccurate and/or misleading statements...''
}

%. In the proposed probabilistic graphical model, each web-source, article or user is treated as a {\em random variable} with the {\em edges} between them modeling their interactions --- that capture expert corroboration, and the assigned ratings indicating the quality of a random variable.
%%%GW: which of the above has been covered already by other parts of the paper??? it sounds a bit generic and repetitive, but I may be wrong on this

\noindent {\bf Verbosity:} Users who write long postings tend to deviate from the topic, often
with highly emotional digression. On the other hand, short postings can be regarded as being crisp, objective and
on topic. Specifically, we compute the first three moments of each user's
posting-length distribution, in terms of sentences and in terms of words.

\noindent{\bf Feature vector construction} For each user $u_k$, we create an engagement feature vector $\langle F^E(u_k) \rangle$. In order to capture user {\em subjectivity}, in terms of different stylistic indicators of credibility, we consider the {\em per-review} language feature vector $\langle F^L(r_{j,k}) \rangle$ of user $u_k$ (refer to Section~\ref{sec:language}). To capture {\em user perspective and expertise} on different topics, we consider the {\em per-review} topic feature vector $\langle F^T(r_{j,k}) \rangle$ of each user $u_k$ (discussed in the next section).
% (refer to Section~\ref{sec:topics}).

\comment{
A user's expertise in judging news credibility depends on many factors. \cite{einhorn1977} discusses the following traits for recognizing an expert.

\begin{itemize}
 \item An expert user needs to be recognized by other members. %This can be captured by how the community perceive the quality of the user's postings.
\item  Experience is an uncertain indicator of user expertise. %\textit{Community engagement} of the member is taken as an %indicator of his experience.
 \item Inter-expert agreement should be high.% and different %experts should converge on the same signs.
\item Experts should be independent of bias.
\end{itemize}

%We validate all these hypothesis in Section~\ref{sec:hypotheses}. %Or proposed model captures the following aspects of user %expertise in identifying credible news.
 
%of the user is an obvious measure for judging the user authority in the community. We capture this with different features: number of answers, 
ratings given, comments, ratings received, disagreement and number of raters.

\noindent{\bf Community Engagement} of the user (e.g., reviews, answers, comments, number of raters, ratings given, ratings received) is an obvious measure for judging the user authority in the community. Table~\ref{tab:socialEngagement} shows the different user engagement features with their statistical significance.

\begin{table}
\begin{center}
\small
\centering
\begin{tabular}{p{14cm}}
\toprule
%Number of Raters***, Comments***, Answers***, Ratings Given**, Agreement***, Ratings Received***\\
Number of Raters, Comments, Answers, Agreement, Ratings Received, Ratings Given\\
\bottomrule
\end{tabular}
\end{center}
\caption{User feature significance in decreasing order of importance. Statistical Significance: $p<0.0001$}
\label{tab:socialEngagement}
\end{table}

%\noindent{\bf Intra-User Agreement}: 
\noindent{\bf Inter-User Agreement} 
Expert users typically agree on what constitutes a credible posting. This is inherently captured in the proposed graphical model, where a user gains expertise by assigning credibility ratings to postings that corroborate with other expert users.

We use Pearson's product-moment correlation test to find out the correlation between user disagreement and expertise in the community. There is an overall correlation of $-0.10$ (with \textit{p-value} $1.6e-07$) between user expertise, and his disagreement with other community members. A negative sign of correlation shows that there is a decrease in disagreement with increase in expertise.

%The overall correlation between the mean absolute disagreement of a user with other users across all the articles rated by him and the other community members, and his expertise level in the community is $-0.1$ with \textit{p-value} $1.6e-07$.
%The $95\%$ confidence interval is $[ -0.1,\ -0.071]$.
The effect is more pronounced for highly expert members (member level $\in [4.5, 5]$), with a correlation of $-0.34$ (with \textit{p-value} of $0.06$), who show more agreement on credibility rating of postings.

\noindent{\bf Topical Perspective and Expertise} The potential for harvesting user preference and expertise in topics for rating prediction of reviews has been demonstrated in~\cite{mukherjee2014JAST,mcauley2013}. For credibility analysis the model needs to capture the user's {\em perspective} and {\em bias} towards certain topics based on their political inclination that bias their ratings, and their topic-specific {\em expertise} that allows them to evaluate postings on certain topics better as ``Subject Matter Experts''. These are captured as {\em per-user} feature weights for the stylistic indicators and topic words in the language of user-contributed reviews.

We provide some example reviews to demonstrate the influence of user \textit{viewpoint} and \textit{bias} on their credibility rating of postings:

%\textit{Viewpoint}: Ratings of non-expert users in the community are often biased by their political viewpoint. E.g.:

\example{
{\bf Viewpoint}: Rating 2.8 by Preston Watts (Non Expert): Sounds like the right wing nut jobs are learning how to use traditional leftist pincko tricks. It sucks its messy but it works.
}

%\textit{Expertise}: Users have \textit{topic-specific expertise}, which allow them to evaluate articles on those topics better as Subject Matter Experts (SME). For example, the following review shows a user (Patricia) with expertise in healthcare and interest in nuclear energy \textit{commenting} on another expert user's rating on nuclear radiation.

%\textit{Preference}: Non expert users are often sensitive towards certain topics. For example, an article\footnote{\tiny http://www.telegraph.co.uk/news/worldnews/northamerica/usa/ barackobama/6062485/Racial-tension-simmers-on-Marthas-Vineyard-as-Barack-Obama-arrives.html} on racial tension simmering as Barack Obama visits a vineyard gets an overall rating of $2.7$ on a $5$ point scale. The first example shows an expert user objectively evaluating the article, whereas the second one shows the less expert user's rating being influenced by his sensitivity towards the topic.

Consider the following posting{http://www.telegraph.co.uk/news/worldnews/northamerica/usa/barackobama/6062485/Racial-tension-simmers-on-Marthas-Vineyard-as-Barack-Obama-arrives.html} on racial tension simmering as Barack Obama visits a vineyard with an overall rating of $2.7$ on a $5$ point scale. The following example shows an expert user objectively evaluating the posting, whereas the second one shows the less expert user's rating being influenced by his bias.

\example{
{\bf Bias}:
Rating 1.5 by Priscilla L. Davis (Moderate Expert): This is an attempt at race-baiting. Why is Obama connected with this incident at all. I am sure the Brazilian laborers were there when George Bush was in office, as well.
%This is just wrong. the caption states ``racial tensions simmers" not immigration tensions. This is complete hogwash!!

Rating 4.0 by Gordon Townsend (Non Expert): Great piece because it highlights the hypocrisy that the power elites in this country have lead us into. Obama worried about what clams to suck up while I am worried about my family getting fed.
%I no longer tell my kids that hard work and merit are what counts. My son can't get a job making a few bucks doing anything because illegals have all the entry positions. %
}

%In this work, we do not distinguish between the fine-grained %aspects of topic influence on users. Instead, we capture them %together as {\em per topic} feature weights for {\em each} user %for explicit and latent topic representations.
%%%GW: this is very vague: what features do we introduce here ?????

%. In the proposed probabilistic graphical model, each web-source, article or user is treated as a {\em random variable} with the {\em edges} between them modeling their interactions --- that capture expert corroboration, and the assigned ratings indicating the quality of a random variable.
%%%GW: which of the above has been covered already by other parts of the paper??? it sounds a bit generic and repetitive, but I may be wrong on this

\comment{
\begin{figure}
\centering
\begin{subfigure}{.27\textwidth}
  \centering
  \includegraphics[width=1\linewidth]{allExpert1.pdf}
  \caption{Expert users.}
  \label{fig:sub1}
\end{subfigure}%
\begin{subfigure}{.27\textwidth}
  \centering
  \includegraphics[width=1\linewidth]{allExpert2.pdf}
  \caption{Moderately expert users.}
  \label{fig:sub2}
\end{subfigure}
\caption{User agreement in community with expertise.}
\label{fig:userAgreementExpertise}
\end{figure}

\begin{figure*}
\centering
\begin{subfigure}{.3\textwidth}
  \centering
  \includegraphics[width=1\linewidth]{expert3.pdf}
  \caption{New user.}
  \label{fig:sub1}
\end{subfigure}%
\begin{subfigure}{.3\textwidth}
  \centering
  \includegraphics[width=1\linewidth]{expert2.pdf}
  \caption{Moderately experienced user.}
  \label{fig:sub2}
\end{subfigure}
\begin{subfigure}{.3\textwidth}
  \centering
  \includegraphics[width=1\linewidth]{expert1.pdf}
  \caption{Experienced user.}
  \label{fig:sub2}
\end{subfigure}
\caption{User agreement about source credibility in community with expertise.}
\label{fig:sampleUserAgreement}
\end{figure*}

\begin{figure}
\centering
\includegraphics[width=9cm]{allExpert.jpg}
\caption{User disagreement in community with expertise.}
\label{fig:expertUserAgreement}
%\vspace{-0.5em}
\end{figure}

\begin{figure}
\centering
\includegraphics[width=9cm]{expert.jpg}
\caption{Three typical kinds of user disagreement in community with expertise.}
\label{fig:sampleUserAgreement}
%\vspace{-1.5em}
\end{figure}
}

\noindent{\bf Demographics}
User demographics like age, gender and location, as well as engagement in the community 
reflected by the number of postings, questions, replies, or thanks received,
 are expected to correlate with user authority in social networks.

\noindent{\bf Verbosity} Users who write long postings tend to deviate from the topic, often
with highly emotional digression. On the other hand, short postings can be regarded as being crisp, objective and
on topic. For verbosity, we compute the first three moments of each user's
posting-length distribution (\#sentences and  \#words).

\noindent{\bf Feature vector construction}: For each user $u_k$, we create an engagement feature vector $\langle F^E(u_k) \rangle$. In order to capture user {\em subjectivity}, in terms of different stylistic indicators of credibility, we consider the {\em per-review} language feature vector $\langle F^L(r_{j,k}) \rangle$ of user $u_k$ (refer to Section~\ref{sec:language}). To capture {\em user perspective and expertise} on different topics, we consider the {\em per-review} topic feature vector $\langle F^T(r_{j,k}) \rangle$ of each user $u_k$.
% (refer to Section~\ref{sec:topics}).

%\textit{etc.} 
%GW: if we say "etc.", there should really be more, and we should be able to state this - otherwise: no "etc."

\section{Interactions}\label{sec:interactions}

\noindent{\bf User User Interaction}: In a community, users can upvote (or, digg, like, rate) the ratings of users that they appreciate, and downvote the ones they do not agree with. Good review ratings from expert users increase the value of a user; whereas low ratings bring down her expertise as judged by the community. 

%In a community, users can upvote ({\em digg, like, rate}) the ratings of users that they appreciate, and downvote the ones they do not agree with. High review ratings from expert users increase the value of a user; whereas low ratings bring down her expertise. Similar to this {\em user-user} interaction, there can be {\em user-article}, {\em user-source} and {\em source-article} interactions which are captured as edges in our graphical model (by construction).
Consider the following anecdotal example in the community showing an expert in nuclear energy \textit{downvoting} another user's rating on nuclear radiation:
\example{
Rating 3.9 by Fabrice Florin (Expert): Interesting opinion about the health risks of nuclear radiation, from a physicist at the University of Oxford. He makes some reasonable points based on factual evidence, ...

Comment by Patricia Blochowiak (Expert): Is it fair to assume that you have no background in biology or anything medical? While this story is definitely very important, it contains enough inaccurate and/or misleading statements that it should be interpreted with great caution.
}

We use Pearson's product-moment correlation test to find out the correlation between the ratings received by a user from other users (on his posting ratings), and his expertise in the community. There is a moderate correlation of $0.40$ (with \textit{p-value} $2.2e-16$) between user expertise, and the feedback from the community on his ratings.

\noindent {\bf Source Posting Interaction} We use Pearson's product-moment correlation test to find out the correlation between \textit{trustworthiness} of the sources and \textit{credibility} of the postings generated by them. Trustworthy sources generate credible postings having a moderate correlation of $0.47$ (with \textit{p-value} $2.2e-16$) between the two. %The $95\%$ confidence interval is $[ 0.44,\ 0.49]$.

\noindent{\bf User Posting Interaction} We use Pearson's product-moment correlation test to find out correlation between the error committed by a user in identifying credible postings, and his expertise in community. There is a negative correlation of $-0.29$ (with \textit{p-value} $2.2e-16$) indicating that the error in credibility judgment of a user decreases with increase in his expertise. As trustworthy sources generate credible postings and expert users can reliably identify them, the \textit{user source} interaction can be captured from the previous interactions.
}

%% file: main/chapter-credibility-analysis/topics.tex
\subsection{Postings and their Topics}
\label{sec:topics}

\comment{
According to the Prominence-Interpretation theory~\cite{fogg2003} on how people assess credibility, there are several factors that affect prominence 
%(noticeability) 
of news: involvement of the user, content of the
news, task and experience of the user, and individual differences. The underlying \textit{topic} of news, and correspondingly the viewpoint of users influences all of these factors in different ways.
}%\comment

%An earlier work in content driven trust propagation~\cite{vg2011} found that news coverage of topics such as ``Bush administration" and ``WikiLeaks" are fairly trustworthy; whereas the coverage on ``Republican policy" and ``Democratic policy" are not considered as trustworthy by users. They also found the articles on ``Obama administration" to be significantly less trustworthy than the coverage of ``Bush administration".

%We observe that significant difference in opinion between different users on the credibility
%rating of articles arises due to conflict between their political viewpoint and that of the %article. We perform some preliminary experiments to establish this hypothesis, which we %discuss in Section~\ref{sec:hypotheses}.

Topic tags for postings play an important role in user-perceived prominence, bias
and credibility, in accordance to the Prominence-Interpretation theory~\cite{fogg2003}. For example, the tag {\em Politics} is often viewed as an indicator
of potential bias and individual differences; whereas tags like {\em Energy} or {\em Environment} are perceived
as more neutral postings and therefore invoke higher agreement in the community on the associated postings' credibility. Obviously, this can be misleading as there is a significant influence of Politics on all topics.% in all format of news.%irrespective of genre.
%%%GW: what is "genre"???
%%%subho: topic is what the article is about, and genre is what category it falls under
%%%GW: I would avoid such forward pointers - you could give them in every subsection, and this would only distract the reader=reviewer; hence commented out the following
%We perform 
% experiments on the role of topics 
%%for new credibility
% in Section~\ref{subsec:discussions}.

\comment{
To show this we
%there is hardly any topic today that does not have a political flavor (incl. even sports, e.g.,  when the FIFA or the IOC make decisions about future world cups or olympic games).
%In order to find the influence of {\em Politics} in news content 
perform the following study on $39$K news articles extracted from
%$5.6$K 
ca. $5K$ sources including New York Times, BBC and CNN. 
%%%GW: really 5000 different web-web-sources??? that is, ca. 8 articles per source??? why this setup???
%Many of these news articles are tagged with explicit topics.
%a typical news community {\em www.newstrust.net}, where the community members tag many of the articles with explicit topic tags.
We count the number of times any explicit topic tag $t$ has \emph{Politics} as a co-occurring topic over all the news articles in the corpus. This gives us an average measure of relatedness ($\text{Rel}(t, \text{``Politics''}) = \frac{\text{\#Count}(t, \text{``Politics''})}{\text{\#Count}(t)}$) of $54\%$ of the topic to Politics.% with a variance of $30\%$.
Table~\ref{tab:polAss} shows the influence of {Politics} on some randomly chosen topics in the corpus.
{\tt GW: not sure if we should really keep this part - what is the bigger message here?????}

For any news article $a$ with a set of explicit topic tags $\langle t \rangle$, there is an average measure of association ($\text{Rel}(a, \text{``Politics''}) = $\\$\frac{\sum_t \text{Rel}(t, \text{``Politics''})}{\text{Cardinality}(\langle t \rangle)}$) of $62\%$ to \emph{Politics}.% with a variance of $19\%$.

\begin{table}
\scriptsize
\center
\begin{tabular}{lrlr}
\toprule
\textbf{Topic} & \multicolumn{1}{l}{\textbf{Association}} & \textbf{Topic} & \multicolumn{1}{l}{\textbf{Association}} \\\midrule
democrats & 1 & bad journalism & 0.64 \\
ron paul & 1 & islam & 0.62 \\
health care & 1 & iraq & 0.57 \\
government & 0.99 & european union & 0.55 \\
patriot act & 0.95 & latin america & 0.52 \\
cia & 0.84 & georgia & 0.5 \\
nicolas sarkozy & 0.78 & social networks & 0.47 \\
ethics in journalism & 0.74 & kenya & 0.39 \\
%west virginia & 0.71 & chile & 0.33 \\
%stanford university & 0.65 & israel & 0.28 \\
%california & 0.64 & high school & 0.18 \\
\bottomrule
\end{tabular}
\caption{Political association of randomly chosen topics.}
\label{tab:polAss}
\end{table}
}

%Political viewpoint of the users influence their credibility rating of articles.
%An earlier work~\cite{vg2011} found the news coverage on ``Bush administration" and ``WikiLeaks" to be more trustworthy than that on ``Republican policy" and ``Democratic policy".
Certain
users have topic-specific expertise that make them judge (or rate) postings on those topics better than others. Sources also have expertise on specific topics and provide a better coverage of postings on those topics than others. For example, {\tt National Geographic} provides a good coverage
of postings related to \textit{environment}, whereas {\tt The Wall Street
Journal} provides a good coverage on \textit{economic} policies.

However, 
%$33.12\%$ of the 
%%% such specific percentages need a reference, but can still not be quantitatively interpreted here, hence commented out
most
postings do not have any explicit topic
tag. In order to automatically identify the underlying theme of the posting, we use Latent
Dirichlet Allocation (LDA)~\cite{Blei2003LDA} to learn the latent topic distribution in the corpus. LDA assumes a document to have a distribution over a set of topics, and each topic to have a distribution over words. 
Table~\ref{tab:lattopics} shows an excerpt of the top topic words in each topic,
where we manually added illustrative labels for the topics. The latent topics also capture some subtle themes not detected by the explicit tags. For example, \emph{Amy Goodman} is an American broadcast journalist, syndicated columnist and investigative reporter who is considered highly credible in the community. Also, associated with that topic cluster is \emph{Amanda Blackhorse}, a Navajo
activist and plaintiff in the Washington Redskins case.

\noindent{\bf Feature vector construction}: For each posting $p_j$ and each of its review $r_{j,k}$, we create feature vectors $\langle F^T(p_j) \rangle$ and $\langle F^T(r_{j,k}) \rangle$ respectively, using the learned {\em latent} topic distributions, as well as the {\em explicit} topic tags. Section~\ref{subsec:latentTopics} discusses our method to learn the topic distributions.

%Another example latent topic shown there is \emph{Alternet} which is a progressive/liberal activist news service whose mission is to ``inspire citizen action and advocacy on the environment, human rights and civil liberties, social justice, media, and health care issues".

\begin{table}
\small
\centering
	\begin{tabular}{p{2.5cm}p{10cm}}
		\toprule
		\textbf{Latent Topics} & \textbf{Topic Words}\\\midrule
		Obama admin. & obama, republican, party, election, president, senate, gop, vote\\
		Citizen journ. & cjr, jouralism, writers, cjrs, marx, hutchins, reporting, liberty, guides\\
		US military & iraq, war, military, iran, china, nuclear, obama, russia, weapons\\
		AmyGoodman & democracy, military, civil, activist, protests, killing, navajo, amanda\\
		Alternet & media, politics, world news, activism, world, civil, visions, economy\\
		Climate & energy, climate, power, water, change, global, nuclear, fuel, warming\\
		\bottomrule
	\end{tabular}
	\caption{Latent topics (with illustrative labels) and their words.}
	\label{tab:lattopics}
\end{table}
%{\tt GW: trim the table: one row per topic!!!!!!}

%% file: main/chapter-credibility-analysis/sources.tex
\subsection{Sources}

A source is considered \emph{trustworthy} if it generates highly \emph{credible} postings. We examine the effect of different features of a source on its trustworthiness based on user assigned ratings in the community. 
%%%GW: the following repeats the intro
\comm{
Previous works~\cite{flanagin2000, johnson2007} 
found that people consider newspaper information to be more credible than online information, TV or magazines. However,  politically-interested internet users find online information to be more credible than traditional media counterparts. 
Besides the medium (print vs. broadcast vs. online), 
the level of experience of the users, and the format of news or type of information need plays a role in their perceived credibility of the news content. 
}%\comment
We consider the following source 
features (summarized in Table~\ref{tab:web-sources}) for a {\em news community}:
the type of \emph{media} (e.g., online, newspaper, tv, blog), \emph{format} of postings (e.g., news analysis, opinion, special report, news report, investigative report), (political) \emph{viewpoint} (e.g., left, center, right),
\emph{scope} (e.g., international, national, local), the top \emph{topics} covered by the source, and their topic-specific \emph{expertise}.
%The observations from our study are presented in %Section~\ref{subsec:qualanalysis}.r

%%%GW: counts are unnecessary and potentially distracting
\begin{table}
\small
\centering
 \begin{tabular}{p{2cm}p{8cm}}
 \toprule
\textbf{Category} & \textbf{Elements} %& \textbf{Count}
\\\midrule
{Media} & newspaper, blog, radio, magazine, online %& 9
\\\sectionrule
{Format} & editorial, investigative report, news, research %& %26
\\\sectionrule
{Scope} & local, state, regional, national, international %& %6
\\\sectionrule
{Viewpoint} & far left, left, center, right, neutral %& %6
\\\sectionrule
{Top Topics} & politics, weather, war, science,, U.S. military %& %282
\\\sectionrule
{Expertise on Topics}& U.S. congress, Middle East, crime, presidential election, Bush administration, global warming %& %28
\\
\bottomrule
\end{tabular}
\caption{Features for source trustworthiness.}
\label{tab:web-sources}
\end{table}

%creating feature vectors for regression
\noindent{\bf Feature vector construction}: For each source $s_l$, we create a feature vector $\langle F^S(s_l) \rangle$ using features in Table~\ref{tab:web-sources}. Each element $f_i^S(s_l)$ is $1$ or $0$ indicating presence or absence of a feature. Note that above features include the top (explicit) topics covered by any source, and its topic-specific expertise for a subset of those topics.

%% file: main/chapter-credibility-analysis/models.tex
\subsection{Semi-supervised Conditional Random Fields for Credibility Classification}
\label{sec:classification}

\comment{
Consider a set of $A$ authors in a community, denoted by $A = \{a_1,a_2, ... a_{|A|}\}$, who write about the various side-effects, symptoms and conditions related to a \textit{target} drug. Let each author $a_i$ write a set of $\{p_i\}$ postings in the community, where $p_{i,j}$ is the $j^{th}$ post by the author on the given drug. Let $D$ be the set of all documents written in the community. Let there be $S$ side-effects of the drug as observed in the community, denoted by $\{s_1, s_2 ... s_{|S|}\}$. Let $S^L$ be the set of side-effects labeled by an expert as $+1$ or $-1$, and $S^U$ be the set of unlabeled side-effects, where $S=S^U \cup S^L$. A positive label signifies that the side-effect is possible from the given drug and a negative label indicates it to be rare or unreported. The label of the side-effects depend on a number of \textit{language-specific} and \textit{author-specific} features weighed by a weight vector, say $W$, and the trustworthiness of the author.

%The objective is to learn the joint distribution of $P(A,S,D; W)$.
}%\comment

Given a set of users (or sources) contributing postings containing dubious statements --- in the first task, we want to classify the statements as {\em credible} or not. For instance, users in a health community can write postings about their experience with drugs and their side-effects, from where we want to extract the most credible side-effects of a given drug; sources can generate postings (i.e. articles) containing dubious claims, whereby we may be interested to find out if the claims are {\em authentic} or {\em hoaxes}.\\

We first propose a probabilistic model for classification with the following simplifications, which are addressed in Section~\ref{sec:regression}:

\begin{itemize}
 \item We do not model users and sources {\em separately} as factors.
 \item We do not take into account inter user or inter source interactions.
 \item We no not model topics implicitly or explicitly, assuming all discussions are on a homogeneous topic (e.g., health).
\end{itemize}

As outlined in Section~\ref{sec:model}, we model our learning task as a Markov Random Field (MRF), where the random variables are the users $U = \{u_1,u_2, ... u_{|U|}\}$, their postings {$P = \{p_1, p_2 ... p_{|P|}\}$},
and the distinct statements $S = \{s_1, s_2 ... s_{|S|}\}$ extracted from all postings --- whose credibility labels need to be inferred. For example, in a {\em health community} the statements are SPO (Subject-Predicate-Object) triples of the form {\tt ``X\_Causes\_Y''} ($X$: Drug, $Y$: Side-effect); in the open web the statements can be SPO claims like {\tt ``Obama\_BornIn\_Kenya''}.

%about drug side-effects extracted from all postings.

Our model is semi-supervised in that we harness ground-truth labels for a subset of statements, derived from the expert knowledge-bases.
%knowledge available at {\small\tt \href{http://www.mayoclinic.org/drugs-supplements/}{mayoclinic.org/drugs-supplements/}}. 
Let $S^L$ be the set of statements labeled by an expert as true or false, and let $S^U$ be the set of unlabeled statements. Our goal is to infer labels for the statements in $S^U.$  

The cliques in our MRF are 
%typically 
triangles consisting of a statement $s_i$, a posting $p_j$ that contains that statement, and a user $u_k$ who wrote this post.
As the same statement can be made in different postings by the same or other users, there are more cliques than statements. For convenient notation, let $S^*$ denote the set of statement instances that correspond to the set of cliques, with statements ``repeated'' when necessary.
%GW: I believe this is exactly the set of cliques, there are no larger cliques other than these triangles.

Let $\phi_i(S^*_i, p_j, u_k)$ be a potential function for clique $i$. 
Each clique has a set of associated feature functions $F_i$ with a weight vector $W$. We denote the individual features and their weights as $f_{il}$ and $w_{l}$. The features are constituted by the stylistic, affective, and user features explained in Section~\ref{sec:features}:
%
%\begin{equation} \label{eq3}
$  F_i = F^L(p_j) \cup F^E(p_j) \cup  F^U(u_k).$
%\end{equation}

Instead of computing the joint probability distribution $Pr(S,P,U;W)$ like in a standard MRF, we adopt the paradigm of Conditional Random Fields (CRF's)
%\cite{Lafferty2001} 
and settle for the simpler task of estimating the 
conditional distribution:

\begin{equation} \label{eq4.0}
 Pr( S | P, U; W) = \frac{1}{Z(P,U)} \prod_i\phi_i(S^*_i, p_j, u_k; W),
\end{equation}

\noindent with normalization constant $Z(P,U)$;

or with features and weights made explicit:

\begin{equation} \label{eq5}
 Pr( S | P, U; W) = \frac{1}{Z(P,U)} \prod_i \exp (\sum_l w_l \times f_{il}(S^*_i, p_j, u_k)).
\end{equation}

CRF parameter learning usually works on fully observed training data.
However, in our setting, only a subset of the $S$ variables have labels and we need to consider the partitioning of $S$ into $S^L$ and $S^U$:

\begin{equation} \label{eq6}
 Pr( S^U, S^L | P, U; W) = \frac{1}{Z(P,U)} \prod_i \exp (\sum_l w_l \times f_{il}(S^*_i, p_j, u_k)).
\end{equation}

For parameter estimation, we need to maximize the marginal log-likelihood:

\begin{equation} \label{eq7}
 LL(W) = \log Pr(S^L|Ps,U;W) = \log \sum_{S^U} Pr(S^L, S^U | P,U; W).
\end{equation}

We can clamp the values of $S^L$ to their observed values in the training data~\cite{Sutton2012, Zhu2005} and compute the distribution over $S^U$ as:

\begin{equation} \label{eq8}
 Pr(S^U|S^L,P,U; W) = 
 \frac{1}{Z(S^L,P,U)} \prod_i \exp (\sum_l w_l \times f_{il}(S^*_i, p_j, u_k)).
\end{equation}

%GW: this equation is not really needed at all
\comment{
From Equations~\ref{eq6} and ~\ref{eq8} we obtain:

\begin{equation} \label{eq9}
 Pr(S^L|P,U;W) = \frac{Z(S^L,P,U)}{Z(P,U)}
\end{equation}
}

There are different ways of addressing the optimization problem for finding the argmax of $LL(W)$. In this work, we choose the Expectation-Maximization (EM) approach~\cite{mccullum2005}.
We first estimate the labels of the variables $S^U$ from the posterior distribution using Gibbs sampling, and then maximize the log-likelihood to estimate the feature weights:

\label{eq10}
\begin{subequations}
\begin{flalign} \label{eq10.1}
 E-Step: q(S^U) &= Pr(S^U|S^L,P,U;W^{(\nu)}) &
\end{flalign}
\begin{flalign}\label{eq10.2}
 M-Step: W^{(\nu+1)} &= argmax_{W^{\prime}} \sum_{S^{U}} q(S^{U})\log Pr(S^L, S^{U}|P,U;W^{\prime}). &
\end{flalign}
\end{subequations}

The update step to sample the labels of $S^U$ variables by Gibbs sampling is given by:

\begin{equation} \label{eq11}
 Pr(S^U_i| P,U,S^L; W) \propto \prod_{\nu\in C} \phi_\nu (S^*_\nu, p_j, u_k; W),
\end{equation}

\noindent where 
$C$ denotes the set of cliques containing statement $S^U_i$.

%{\tt GW: in general, the notation with upper-case S seems to be overloaded: sets of variables vs. individual variables vs. cliques - we would ideally clean this up, but it is not that straightforward}

For the M-step in Equation~\ref{eq10.2}, we use an {$L_2$-regularized} Trust Region Newton Method~\cite{Lin2008}, suited for large-scale unconstrained optimization, where many feature values may be zero. For this we use an implementation of LibLinear~\cite{LibLinear}.

The above approach captures user trustworthiness implicitly via the weights of the feature vectors. However, we may want to model user trustworthiness in a way that explicitly aggregates over all the statements made by a user.
Let $t_k$ denote the trustworthiness of user $u_k$, measured as the fraction of her statements that were considered true
in the previous EM iteration:

\begin{equation} \label{eq12}
  t_k = \frac{\sum_i \mathbf{1}_{S_{i,k}= \mbox{True}}}{|S_k|},
\end{equation}
where $S_{i,k}$ is the label assigned to $u_k$'s statement $S_i$ in the previous EM iteration.
%, out of the statements $S_k$ made by him.
Equation~\ref{eq11} can then be modified into:

\begin{equation} \label{eq13}
 Pr(S^U_i| P,U,S^L; W) \propto \prod_{\nu\in C} t_k \times \phi_\nu (S^*_\nu, p_j, u_k; W)
\end{equation}

%subho:camera-ready
Therefore, the random variable for trustworthiness depends on the proportion of \textit{true} statements made by the user. The \textit{label} of a statement, in turn, is determined by the language objectivity of the postings and trustworthiness of all the users in the community that make the statement.

The inference is an iterative process consisting of the following $3$ main steps:

\begin{itemize}
 \item Estimate user trustworthiness $t_k$ using Equation~\ref{eq12}.
 \item Apply the {\emph E-Step} to estimate  $q(S^U;W^{(\nu)})$ \\
       For each $i$, sample $S_i^U$ from Equation~\ref{eq10.1} and~\ref{eq13}.
 \item Apply the {\emph M-Step} to estimate $W^{(\nu+1)}$ using Equation~\ref{eq10.2}.
%using Equation~\ref{eq10.2}
\end{itemize}

%GW: the following is just another variant that is likely to confuse the reader
%%%%%%%%%%%%%%%%%%%%%
\comment{
The given approach finds the label assignment of the unobserved side-effects, given the partial assignment of expert-given labels to some side-effects. Although, the learnt model weights can be used to infer the author trustworthiness based on the number of credible statements made by the author, there is no explicit random variable to directly capture it. To incorporate user trustworthiness directly in the model, we introduce another set of latent random variables $T$ in the model, where $t_i$ is the trustworthiness of user $a_i$. Let $S_{i,k}$ be the label assigned to $S_k$ by the $i^{th}$ user $a_i$, and $\{S_i\}$ be the set of all statements made the author. This introduces another set of cliques $\phi_m (a_i, t_i, S_i^m)$ in the above graphical model. We define the clique potential for capturing user trustworthiness as a ratio of the number of credible statements (positive side-effects of a drug) made by the author to her total number of statements. Note that by our model convention, a label 
of
$+1$ denotes a side-effect is possible from a given drug whereas $-1$ denotes it as rare or unreported.

\begin{equation} \label{eq12}
  \phi_m (a_i, t_i, S_i^m) = \frac{\sum_k \mathbbm{1}_{S_{i,k}= +1}}{|S_i|}
\end{equation}

Equation~\ref{eq11} can now be modified as:

\begin{equation} \label{eq13}
 Pr(S^U_k| A,D,S^L,S_{-k}^U; W) \propto \prod_{m\in C} \phi_m (a_i, t_i, S_{k}{^m}) \phi_m (S_k{^m}, a_i, p_{i,j}; W)
\end{equation}

//Explain intuitively what happens in each iteration
//Substitute q(Su), take derivative and show it to be similar to original one

}%\comment
%%%%%%%%%%%%%%%%

\subsection{Continuous Conditional Random Fields for Credibility Regression}
\label{sec:regression}

In the previous section, we discussed an approach for {\em classifying} statements as credible or not. However, in many scenarios we want to perform a more {\em fine-grained} analysis. Some communities (e.g., {\tt \href{http://www.newstrust.net}{newstrust.net}}) offer users fine-grained scales for rating different aspects of an item --- which are {\em aggregated} into an
overall real-valued rating after weighing the aspects based on their
importance, expertise of the user, feedback from the community,
and more. This setting cannot be easily discretized without blowup
or risking to lose information. Therefore, in this task we want to perform {\em regression} for fine-grained credibility analysis, whereby we want to assign a real-valued credibility rating (e.g. $2.5$ on a scale of $1$ to $5$) to a posting. 

We also address the earlier drawbacks of our model (discussed in Section~\ref{sec:classification}), whereby we {\em now} model users and sources as separate factors, taking into consideration the inter user and inter source interactions, as well as the influence of {\em topics} of discussions.

Consider a set of sources generating postings (i.e. articles), and a set of users providing feedback (i.e. writing reviews) on the postings with mutual interactions (i.e. a user can upvote/downvote, like, and share other users' reviews) --- our objective is to identify credible postings,
trustworthy sources, and expert users {\em jointly} in the community, incorporating the discussed features and insights (discussed in Section~\ref{sec:features}). 

Table~\ref{tab:symbol} summarizes the important notations used in this section.

%{\tt GW: this section is very heavy in notation. It would be great to add a reference table with all notation: Summary of variables and notation !!!!!}\\
%\vspace{-0.5em}
\begin{table}
\small
\centering
\begin{tabular}{p{2cm}p{3cm}p{7cm}}
\toprule
\textbf{Variables} & \textbf{Type} & \textbf{Description}\\\midrule
$p_j$ & Vector & Document with sequence of words $\langle w \rangle$\\
$s$ & Vector & Sources\\
$u$ & Vector & Users\\
$r_{j,k}$ & Vector & Review by user $u_k$ on document $p_j$\\
& & \myindent with sequence of words $\langle w \rangle$\\
$y_{j,k}$ & Real Number & Rating of $r_{j,k}$\\
$z$ & Vector & Sequence of topic assignments for $\langle w \rangle$\\
$\text{SVR}_{u_k}, \text{SVR}_{s_i}$ & Real Number & SVR prediction for users,
sources,\\
$\text{SVR}_L, \text{SVR}_T$ & \myindent $\in$ [1 \ldots 5] & \myindent
language, and topics\\
$\Psi = f(\langle \psi_j \rangle) $ & Real Number & Clique potential with
$\psi_j = \langle y_j, s_i, p_j,$ $ \langle u_k \rangle, \langle r_{j,k} \rangle \rangle
 $ for clique of $p_j$\\
\myindent ${\lambda=}$ $\langle \alpha_u, \beta_s, \gamma_1, \gamma_2 \rangle$ &
Vector & Combination weights for users $\langle u \rangle$, sources $\langle s
\rangle$, language and topic models\\
$y_{n \times 1}$ & Vector & Credibility rating of documents $\langle d \rangle$\\
$X_{n \times m}$ & Matrix & Feature matrix with $m = |U|+|S| + 2$\\
$Q_{n \times n}$ & Diagonal Matrix & $f(\lambda)$\\
$b_{n \times 1}$ & Vector & $f(\lambda, X)$\\
$\Sigma_{n \times n}$ & {\small CovarianceMatrix} & $f(\lambda)$\\
% & Matrix & \\
$\mu_{n \times 1}$ & Mean Vector & $f(\lambda, X)$\\
\bottomrule
\end{tabular}
%\vspace{-1em}
\caption{Symbol table.}
\label{tab:symbol}
%\vspace{-2em}
\end{table}

\subsubsection{Topic Model}
\label{subsec:latentTopics}

Consider a posting $d$ consisting of a sequence of $\{N_d\}$ words denoted by
${w_1,w_2,...w_{N_d}}$. Each word is drawn from a vocabulary $V$ having unique
words indexed by ${1,2,...V }$. Consider a 
%sequence 
set of topic assignments $z = \{z_1,z_2,...z_K\}$ 
for $d$, 
%%%GW: these are the topics for d - right?
%%%this is what the wording "topic assignment" suggests to me
%%%otherwise I would talk about "set of topics"
where
each topic $z_i$ can be from a set of $K$ possible topics. 

LDA~\cite{Blei2003LDA} assumes each document $d$ to be associated with a multinomial distribution $\theta_d$ over topics $Z$ with a symmetric dirichlet prior $\rho$. $\theta_d(z)$ denotes the probability of occurrence of topic $z$ in document $d$. Topics have a multinomial distribution $\phi_z$ over words drawn from a vocabulary $V$ with a symmetric dirichlet prior $\zeta$. $\phi_z(w)$ denotes the probability of the word $w$ belonging to the topic $z$.
Exact inference is not possible due to intractable coupling between $\Theta$ and $\Phi$. We use Gibbs sampling for approximate inference.

Let $n(d, z, w)$ denote the count of the word $w$ occurring in document $d$ belonging to the topic $z$. In the following equation, $(.)$ at any position in the above count indicates marginalization, i.e.,  summing up the counts over all values for the corresponding position in $n(d,z,w)$.
The conditional distribution for 
%the update of 
the latent variable $z$ (with components $z_1$ to $z_K$)
is given by:

%\vspace{-1em}
\begin{equation}
\label{eq.3}
\begin{aligned}
 P(z_i=k| &w_i=w, z_{-i}, w_{-i}) \propto\\
  \frac{n(d, k, .) + \rho}{\sum_{k}n(d, k, .) + K \rho} &\times \frac{n(., k, w) + \zeta}{\sum_{w}n(., k, w) + V \zeta}
 \end{aligned}
\end{equation}
%\vspace{-0.5em}
%%%GW: I moved propto to the first line, as it is easy to confuse propto with some coefficient rho and then believing that an equality sign is missing

Let $\langle T^E \rangle$ and $\langle T^L \rangle$ be the set of explicit topic tags and latent topic dimensions, respectively. The topic feature vector $\langle F^T \rangle$ for a posting or review
combines both explicit tags and latent topics and is constructed as follows:

%\vspace{-1.5em}
\begin{equation*}
\label{eq:topic}
F^T_t(d) =
\begin{cases}
\#freq(w,d),& \text{if} \ T^E_{t^{'}}=F^T_t \\
\#freq(w,d) \times \phi_{T^L_{{t'}}}(w),& \text{if} \ T^L_{t^{'}} = F^T_t \text{ and } \phi_{T^L_{t^{'}}}(w) > \delta\\
0 & \text{otherwise}
\end{cases}\hfill
\end{equation*}
%\vspace{-1em}

So for any word in the document matching an explicit topic tag, the corresponding element in the feature vector $\langle F^T \rangle$ 
is set to its occurrence count in the document. If the word belongs to any latent topic with probability greater than threshold $\delta$, the probability of the word belonging to that topic ($\phi_{t}(w)$) is added to the corresponding element in the feature vector, and set to $0$ otherwise.

\subsubsection{Support Vector Regression}
\label{subsec:SVR}

We use Support Vector Regression (SVR)~\cite{drucker97} to combine the different features discussed in Section~\ref{sec:features}. SVR is an extension of the max-margin framework for SVM classification to the regression problem. It solves the following optimization problem to learn weights $w$ for features $F$:

{
%\vspace{-2em}

\begin{multline}
\label{eq.5}
 \min_{w} \frac{1}{2}  {w} ^T{w} + C \times
 \sum_{d=1}^{N} (max(0, |y_d - w^TF| - \epsilon))^2 \hfill
%   \text{where $\langle .,. \rangle$ denotes a scalar product.}\hfill
 \end{multline}
 %\vspace{-1em}
}

%{\tt GW: why the special notation for scalar product -- w**T*w is also a scalar product ????? we should not use two different notations !!!!!}

\noindent \textbf{Posting Stylistic Model}: We learn a stylistic regression model $\text{SVR}_L$ using the {\em per-posting} stylistic feature vector $\langle F^L(p_j) \rangle$ for posting $p_j$ (or, $\langle F^L(r_{j,k}) \rangle$ for review $r_{j,k}$), with the overall credibility rating $y_j$ (or, $y_{j,k}$) of the posting as the response variable.

\noindent \textbf{Posting Topic Model}: Similarly, we learn a topic regression model $\text{SVR}_T$ using the {\em per-posting} topic feature vector $\langle F^T(p_j) \rangle$ for posting $p_j$ (or, $\langle F^T(r_{j,k}) \rangle$ for review $r_{j,k}$), with the overall credibility rating $y_j$ (or, $y_{j,k}$) of the posting as the response variable.

\noindent \textbf{Source Model}: We learn a source regression model $\text{SVR}_{s_i}$ using the {\em per-source} feature vector $\langle F^S(s_i) \rangle$ for source $s_i$, with the overall source rating as the response variable .

\noindent \textbf{User Model}: For each user $u_k$, we learn a user regression model $\text{SVR}_{u_k}$ with her {\em per-review} stylistic and topic feature vectors \\$\langle F^L(r_{j,k}) \cup F^T(r_{j,k}) \rangle$ for review $r_{j,k}$ for posting $p_j$, with her overall review rating $y_{j,k}$ as the response variable.

Note that we use {\em overall} credibility rating of the posting to train posting stylistic and topic models. For the user model, however, we take {\em user assigned} credibility ratings of the postings, and per-user features. This model captures user subjectivity and topic perspective. The source models are trained on source specific meta-data and its ground-truth ratings.

\subsubsection{Continuous Conditional Random Field}

%As outlined in Section~\ref{sec:newscom}, 
We model our learning task as a Conditional Random Field (CRF), where the random variables are the ratings of postings $\langle p_j \rangle$, sources $\langle s_i \rangle$, users $\langle u_k \rangle$, and reviews $\langle r_{j,k} \rangle$. The objective is to predict the credibility ratings $\langle y_j \rangle$ of the postings $\langle p_j \rangle$.

%Let $\langle u^j_k \rangle$ denote the set of users reviewing article $p_j$.

The cliques in the CRF consist of a posting $p_j$, its source $s_i$, set of users $\langle u_k \rangle$ reviewing it, and the corresponding user reviews $\langle r_{j,k} \rangle$ --- where $r_{j,k}$ denotes the review by user $u_k$ on posting $p_j$. Different cliques are connected via the common sources, and users. There are as many cliques as the number of postings.

Let $\psi_j(y_j, s_i, p_j, \langle u_k \rangle, \langle r_{j,k} \rangle)$ be a
potential function for clique $j$. Each clique has a set of associated {\em
vertex} feature functions.
%$F_j$ with a weight vector $W$.
In our problem setting, we associate features to each vertex. % and ignore the edge feature functions.
%We denote the individual features and their weights as $f_{m,n}$ and $w_{n}$.
The features constituted by the stylistic, topic, source and user features explained in Section~\ref{sec:newscom} are:
%
%\begin{equation} \label{eq3}
$ F^L(p_j) \cup F^T(p_j) \cup F^S(s_i) \cup_k (F^E(u_k) \cup F^L(r_{j,k}) \cup F^T(r_{j,k})).$
%\end{equation}

%In the following, $y, \mu, b$ are vectors, and $X, Q, \Sigma$ are matrices.

A traditional CRF model allows us to have a {\em binary} decision if a posting is {\em credible} ($y_j=1$) or not ($y_j=0$), by estimating the conditional distribution with the probability {\em mass} function of the discrete random variable $y$:

%\vspace{-1.5em}
\begin{equation} \label{eq4}
 Pr( y | D, S, U, R) = \frac{\prod_{j=1}^{n} exp(\psi_j(y_j, s_i, p_j, \langle u_k 
\rangle, \langle r_{j,k} \rangle))}{{\sum_{y} \prod_{j=1}^{n} exp(\psi_j(y_j, s_i,
p_j,                                                                        
\langle u_k \rangle, \langle r_{j,k} \rangle))}}
\end{equation}
%\vspace{-1.5em}

But in our problem setting, we want to estimate the credibility {\em rating} of a posting. Therefore, we need to estimate the conditional distribution with the probability {\em density} function of the continuous random variable $y$:\\
%{\tt GW: this sounds odd and confusing -- the equation for the continuous case is still a conditional distribution, not the density function ?????}\\
%{\tt GW: why are the bounds of the integral non-standard: from plus infinity to minus infinity ?????}\\

{
%\vspace{-3em}
\begin{equation}
\setlength{\mathindent}{0pt}
\label{eq.5}
 Pr( y | D, S, U, R) = \frac{\prod_{j=1}^{n} exp(\psi_j(y_j, s_i, p_j, \langle u_k
\rangle, \langle r_{j,k} \rangle))}{\int_{-\infty}^{\infty} \prod_{j=1}^{n}
exp(\psi_j(y_j, s_i, p_j, \langle u_k \rangle, \langle r_{j,k} \rangle))dy}\\
\end{equation}
%\vspace{-2em}
}

\comment{
where the potential function made explicit in terms of features and weights as:

\begin{equation}
\label{eq.6}
\begin{aligned}
\psi_j(s_i, p_j, \langle u_k \rangle, \langle r_{j,k} \rangle) &= exp (\sum_n
w_n \times f_{m,n}(s_i, p_j, \langle u_k \rangle, \langle r_{j,k} \rangle))
 \end{aligned}
\end{equation}
}

Given a posting $p_j$, its source id $s_i$, and a set of user ids $\langle u_k \rangle$ who reviewed the posting, the regression models $\text{SVR}_L(p_j)$, $\text{SVR}_T(p_j)$, $\text{SVR}_{s_i}$, $\langle \text{SVR}_{u_k}(p_j) \rangle$ (discussed in Section~\ref{subsec:SVR}) independently predict the rating of $p_j$. For notational brevity, hereafter, we drop the argument $p_j$ from the SVR function.
%%%GW: avoid repeating the features all the time
%based on its stylistic and topic features extracted from text, %source features extracted from the meta-data of its source, and %user models learned from the training data. 
These SVR predictors are for separate feature groups and
independent of each other. Now we combine the 
different SVR models to capture mutual interactions, such that the weight for each SVR model reflects our confidence on its quality. 
Errors by an SVR are penalized by the squared loss between the predicted credibility rating of the posting and the ground-truth rating. There is an additional constraint that for any clique {\em only} the regression models corresponding to the source and users present in it should be activated. This can be thought of as partitioning the input feature space into subsets, with the 
features inside a clique capturing {\em local} interactions, and the {\em global} weights capture the overall quality of the random variables via the shared information between the cliques (in terms of common sources, users, topics and language features) --- an ideal setting for using a CRF. Equation~\ref{eq.7} shows one such linear combination. 
Energy function of an individual clique is given by:
\vspace{-1em}

{
%\vspace{-1.6em}
\setlength{\belowdisplayskip}{0pt} \setlength{\belowdisplayshortskip}{0pt}
\setlength{\abovedisplayskip}{0pt} \setlength{\abovedisplayshortskip}{0pt}
\setlength{\mathindent}{0pt}
\centering
\begin{multline}
\label{eq.7}
\psi(y, s, d, \langle u \rangle, \langle r \rangle) =
-\sum_u \alpha_{u} \mathbb{I}_{u}(d) (y - \text{SVR}_{u})^2 \\ -
\sum_s \beta_s \mathbb{I}_{s}(d) (y - \text{SVR}_{s})^2
 -\gamma_1 (y - \text{SVR}_L)^2 -\gamma_2 (y - \text{SVR}_T)^2
\end{multline}
}
%\vspace{-2.5em}
 
{\noindent Indicator functions $\mathbb{I}_{u_k}(p_j)$ and $\mathbb{I}_{s_i}(p_j)$ are 1 if $u_k$ is a reviewer and $s_i$ is the source of posting $p_j$ respectively, and are $0$ otherwise.}

As the output of the SVR is used as an input to the CCRF in Equation~\ref{eq.7}, each element of the input feature vector is already predicting the output variable.  The learned parameters $\lambda = \langle \alpha, \beta, \gamma_1, \gamma_2 \rangle$ (with dimension$(\lambda)=|U|+|S|+2$) of the linear combination of the above features depict how much to trust individual predictors. Large $\lambda_k$ on a particular predictor places large penalty on the mistakes committed by it, and therefore depicts a higher quality for that predictor. $\alpha_u$ corresponding to user $u$ can be taken as a proxy for that user's \textit{expertise}, allowing us to obtain a ranked list of expert users.
Similarly, $\beta_s$ corresponding to source $s$ can be taken as a proxy for that source's \textit{trustworthiness}, allowing us to obtain a ranked list of trustworthy sources.

Overall energy function of all cliques is given by:
%{\tt GW: from here on it gets a bit messy -- regarding layout of equations and, especially, lack of guiding the reader through the math!!!!!}
{
%\vspace{-1em}
%\centering
\setlength{\mathindent}{0pt}
\setlength{\belowdisplayskip}{0pt} \setlength{\belowdisplayshortskip}{0pt}
\setlength{\abovedisplayskip}{0pt} \setlength{\abovedisplayshortskip}{0pt}
\begin{multline*}
\myindent \myindent \myindent \myindent
 \Psi = \sum_{j=1}^{n} \psi_j(y_j, s_i, p_j, \langle u_k \rangle, \langle r_{j,k}
\rangle)\\
%\vspace{-1em}
 \text{\normalsize (Substituting $\psi_j$ from Equation~\ref{eq.7} and
re-organizing terms)}\\
     \Psi = \sum_{j=1}^{n} (-\sum_{k=1}^{k=U} \alpha_{k} \mathbb{I}_{u_k}(p_j)
(y_j - \text{SVR}_{u_k})^2 \\ - \sum_{i=1}^{i=S} \beta_i \mathbb{I}_{s_i}(p_j)
(y_j - \text{SVR}_{s_i})^2 -\gamma_1 (y_j - \text{SVR}_L)^2 -\gamma_2 (y_j - \text{SVR}_T)^2)\\
      = -\sum_{j=1}^{n} y_j^2 [\sum_{k=1}^{k=U} \alpha_{k} \mathbb{I}_{u_k}(p_j)
+ \sum_{i=1}^{i=S} \beta_i \mathbb{I}_{s_i}(p_j) + \gamma_1 + \gamma_2] \\
      + \sum_{j=1}^{n} 2y_j[\sum_{k=1}^{k=U} \alpha_{k} \mathbb{I}_{u_k}(p_j)
\text{SVR}_{u_k} + \sum_{i=1}^{i=S} \beta_i \mathbb{I}_{s_i}(p_j)\text{SVR}_{s_i} + \gamma_1
\text{SVR}_L 
      + \gamma_2 \text{SVR}_T]\\ -\sum_{j=1}^{n} [\sum_{k=1}^{k=U} \alpha_{k}
\mathbb{I}_{u_k}(p_j) \text{SVR}_{u_k}^2 + \sum_{i=1}^{i=S} \beta_i
\mathbb{I}_{s_i}(p_j)\text{SVR}_{s_i}^2 + \gamma_1 \text{SVR}_L^2
      + \gamma_2 \text{SVR}_T^2]\\
\end{multline*}
\begin{multline*}
\text{\normalsize Organizing the bracketed terms into variables as follows:}\\
Q_{i,j} =
\begin{cases}
\sum_{k=1}^{k=U} \alpha_{k}\mathbb{I}_{u_k}(p_i) + \sum_{l=1}^{l=S} \beta_l \mathbb{I}_{s_l}(p_i) + \gamma_1 + \gamma_2 & i=j \\
0 & \ i \neq j
\end{cases}\\
b_i = 2 [\sum_{k=1}^{k=U} \alpha_{k} \mathbb{I}_{u_k}(p_i) \text{SVR}_{u_k} + \sum_{l=1}^{l=S} \beta_l \mathbb{I}_{s_l}(p_i) \text{SVR}_{s_l}+ \gamma_1 \text{SVR}_L + \gamma_2 \text{SVR}_T]\\
c = \sum_{j=1}^{n} [\sum_{k=1}^{k=U} \alpha_{k} \mathbb{I}_{u_k}(p_j)
\text{SVR}_{u_k}^2 + \sum_{i=1}^{i=S} \beta_i \mathbb{I}_{s_i}(p_j)\text{SVR}_{s_i}^2 +
\gamma_1 \text{SVR}_L^2 + \gamma_2 \text{SVR}_T^2]\\
\end{multline*}
}
We can derive:
{
%\vspace{-1em}
\setlength{\belowdisplayskip}{0pt} \setlength{\belowdisplayshortskip}{0pt}
\setlength{\abovedisplayskip}{0pt} \setlength{\abovedisplayshortskip}{0pt}
\begin{multline}
 \centering
 \Psi =  -y^TQy +y^Tb - c \hfill
\end{multline}
}

%\vspace{-0.5em}

\noindent Substituting $\Psi$ in Equation~\ref{eq.5}:

{
\setlength{\mathindent}{0pt}
\centering
%\vspace{-0.5em}
\setlength{\belowdisplayskip}{0pt} \setlength{\belowdisplayshortskip}{0pt}
\setlength{\abovedisplayskip}{0pt} \setlength{\abovedisplayshortskip}{0pt}
\begin{equation}
\label{eq.9}
\begin{split}
  P(y|X) &= \frac{\prod_{j=1}^{n} exp(\psi_j)}{\int_{-\infty}^{\infty}
\prod_{j=1}^{n} exp(\psi_j) dy}\\
 &= \frac{exp(\Psi)}{\int_{-\infty}^{\infty}  exp(\Psi) dy}\\
 &= \frac{exp(-y^TQy +y^Tb)}{\int_{-\infty}^{\infty} exp(-y^TQy +y^Tb) dy}\\
 &= \frac{exp(-\frac{1}{2}y^T\Sigma^{-1}y+y^T\Sigma^{-1}\mu)}{\int_{-\infty}^{\infty} exp(-\frac{1}{2}y^T\Sigma^{-1}y+y^T\Sigma^{-1}\mu)dy} \text{\myindent \myindent \myindent{\normalsize (Substituting $Q = \frac{1}{2}\Sigma^{-1}, b = \Sigma^{-1}\mu$)}}
 \end{split}
\end{equation}
}

%\vspace{0.5em}
%Making the substitutions:
%$Q = \frac{1}{2}\Sigma^{-1}, b = \Sigma^{-1}\mu,\\ \int_{-\infty}^{\infty}exp(-\frac{1}{2}y^T\Sigma^{-1}y+y^T\Sigma^{-1}\mu)dy = \frac{(2\pi)^{n/2}}{|\Sigma^{-1}|^{\frac{1}{2}}}exp(\frac{1}{2}\mu^T\Sigma^{-1}\mu)$, Eqn.~\ref{eq.9} can be transformed into a multivariate Gaussian distribution:
%{\tt GW: not clear what you are talking about here !!!!!}\\
\noindent Equation~\ref{eq.9} can be transformed into a multivariate Gaussian distribution after substituting $\int_{-\infty}^{\infty}exp(-\frac{1}{2}y^T\Sigma^{-1}y+y^T\Sigma^{-1}\mu)dy = \frac{(2\pi)^{n/2}}{|\Sigma^{-1}|^{\frac{1}{2}}}exp(\frac{1}{2}\mu^T\Sigma^{-1}\mu)$. Therefore obtaining,

{
%%\vspace{-1em}
\setlength{\belowdisplayskip}{0pt} \setlength{\belowdisplayshortskip}{0pt}
\setlength{\abovedisplayskip}{0pt} \setlength{\abovedisplayshortskip}{0pt}
 \begin{multline}
 \label{eq.8}
 P(y|X) = \frac{1}{{(2\pi)}^{\frac{n}{2}}{|\Sigma|}^\frac{1}{2}}exp(-\frac{1}{2}(y-\mu)^T\Sigma^{-1}(y-\mu)) \hfill
\end{multline}
}

$Q$ represents the contribution of $\lambda$ to the covariance matrix $\Sigma$. Each row of the vector $b$ and matrix $Q$ corresponds to one training instance, representing the {\em active} contribution of features present in it.
To ensure Equation~\ref{eq.8} represents a valid Gaussian distribution, the covariance matrix $\Sigma$ needs to be positive definite for its inverse to exist. For that the diagonal matrix $Q$ needs to be a positive semi-definite matrix. This can be ensured by making all the diagonal elements in $Q$ greater than $0$, by constraining $\lambda_k > 0$.

Since this is a constrained optimization problem, gradient ascent cannot be directly used. We follow the approach similar to~\cite{radosavljevicECAI2010} and maximize log-likelihood with respect to $log\ \lambda_k$, instead of $\lambda_k$ as in standard gradient ascent, making the optimization problem unconstrained as:

{
%\vspace{-0.5em}
\setlength{\mathindent}{0pt}
\setlength{\belowdisplayskip}{0pt} \setlength{\belowdisplayshortskip}{0pt}
\setlength{\abovedisplayskip}{0pt} \setlength{\abovedisplayshortskip}{0pt}
\begin{multline}
\label{eq.11}
 \frac{\partial logP(y|X)}{\partial log\lambda_k} = \alpha_k(\frac{\partial logP(y|X)}{\partial \lambda_k})\hfill
 \end{multline}

 %\vspace{0.5em}
 \text{\normalsize Taking partial derivative of the $log$ of Equation~\ref{eq.8} w.r.t $\lambda_k$}:
 
 \begin{multline}
\frac{\partial logP(y|X)}{\partial \lambda_k} = \frac{1}{2} \frac{\partial}{\partial \lambda_k} (-y^T\Sigma^{-1}y +2y^T\Sigma^{-1}\mu - \mu^T \Sigma^{-1} \mu + log|\Sigma^{-1}| + Constant)\hfill
\end{multline}
%\vspace{-0.5em}
}

Substituting the following in the above equation:

{
%%\vspace{-1em}
%\setlength{\mathindent}{0pt}
\setlength{\belowdisplayskip}{0pt} \setlength{\belowdisplayshortskip}{0pt}
\setlength{\abovedisplayskip}{0pt} \setlength{\abovedisplayshortskip}{0pt}
\centering
\begin{equation*}
\centering
 \begin{split}
  \frac{\partial \Sigma^{-1}}{\partial \lambda_k} &= 2\frac{\partial Q}{\partial \lambda_k}\\
&= 2I\\
\frac{\partial \Sigma^{-1}\mu}{\partial \lambda_k} &= \frac{\partial b}{\partial \lambda_k} \ \ [\because \mu = \Sigma b]\\
&= 2X_{(.),k} \text{\normalsize \hspace{3em} where, $X_{(.),k}$ indicates the $k^{th}$ column of the feature matrix $X$.}
\end{split}
\end{equation*}

\begin{equation*}
\begin{split}
\frac{\partial \Sigma}{\partial \lambda_k} &= -\Sigma \frac{\partial \Sigma^{-1}}{\partial \lambda_k}\Sigma \\
&= - 2\Sigma\Sigma\\
\frac{\partial}{\partial \lambda_k}(\mu^T \Sigma^{-1} \mu) &= \frac{\partial}{\partial \lambda_k}(b^T \Sigma b) \\
&= b^T\frac{\partial \Sigma b}{\partial \lambda_k} + \frac{\partial b^T}{\partial \lambda_k}\Sigma b\\
&= b^T(\Sigma \frac{\partial b}{\partial \lambda_k} + \frac{\partial \Sigma}{\partial \lambda_k}b) + \frac{\partial b^T}{\partial \lambda_k}\Sigma b\\
&= 4 X_{(.),k}\Sigma b -2b^T\Sigma \Sigma b \\
&= 4X_{(.),k} \mu - 2\mu^T\mu\\
\frac{\partial log|\Sigma^{-1}|}{\partial \lambda_k}&=\frac{1}{|\Sigma^{-1}|}\text{Trace}(|\Sigma^{-1}|\Sigma\frac{\partial
\Sigma^{-1}}{\partial \lambda_k})\\
&= 2\text{Trace}(\Sigma)\hfill
 \end{split}
\end{equation*}
}

We can derive the gradient vector:

{
\setlength{\belowdisplayskip}{0pt} \setlength{\belowdisplayshortskip}{0pt}
\setlength{\abovedisplayskip}{0pt} \setlength{\abovedisplayshortskip}{0pt}
\begin{equation}
\frac{\partial logP(y|X)}{\partial \lambda_k} = -y^Ty + 2y^TX_{(.),k} -
2X^T_{(.),k}\mu + \mu^T\mu + \text{Trace}(\Sigma)
\end{equation}
%\vspace{-0.5em}
}

Let $\eta$ denote the learning rate. The update equation is given by:
\vspace{-1em}

{
\setlength{\belowdisplayskip}{0pt} \setlength{\belowdisplayshortskip}{0pt}
\setlength{\abovedisplayskip}{0pt} \setlength{\abovedisplayshortskip}{0pt}
\begin{equation}
\centering
log\lambda_k^{new} = log\lambda_k^{old} + \eta \frac{\partial logP(y|X)}{\partial log\lambda_k} 
\end{equation}
}

Once the model parameters are learned using gradient ascent, the inference for the prediction $y$ of the credibility rating of the posting is straightforward. As we assume the distribution to be Gaussian, the prediction is the expected value of the function, given by the mean of the distribution:
$ y\prime = argmax_y\ P(y|X) = \mu = \Sigma  b$.\\
Note that $\Sigma$ and $b$ are both a function of $\lambda= \langle \alpha, \beta, \gamma_1, \gamma_2 \rangle$ which represents the combination weights of various factors to capture mutual interactions. The optimization problem determines the optimal $\lambda$ for reducing the error in prediction.
%{\tt GW: I am confused by this equation: if this is equal to Sigma*b, isn't this a constant solution given the input???? Sigma and b are input values, right???? even if I'm misinterpreting something here, this indicates that we need better reader guidance!!!!!}

%% file: main/chapter-credibility-analysis/experiments-health.tex
\newcommand{\MRF}{CRF\xspace}

In this section, we apply the predictive power of our probabilistic model for classification (refer to Section~\ref{sec:classification}) to the problem
of extracting credible side-effects of medical drugs from user-contributed postings in online
healthforums.

\subsection{Data}

We use data from the {\tt \href{http://www.healthboards.com}{healthboards.com}}, one of the largest online health communities,
with 
%over $10\ million$ monthly visitors, 
$850,000$ registered members
and over $4.5$ million posted messages. 
We sampled $15,000$ users based on their posting frequency and all of their postings, $2.8$\ million postings in total for experimentation.
%subho:camera-ready
Table~\ref{tab:author} shows the user categorization in terms of their community engagement. 
We employ an IE tool~\cite{Ernst2014} to extract side-effect statements from the postings. It generates tens of thousands of such SPO triple patters, although only a handful of them are credible ones. 
Details of the experimental setting are available on our website.\footnote{ \tt \href{http://www.mpi-inf.mpg.de/impact/peopleondrugs/}{http://www.mpi-inf.mpg.de/impact/peopleondrugs/}}

%GW: users chosen by which criterion? originally the text said "top users", but
%what exactly does this mean? also, didn't we sample users from the long tail as well?
%The postings were cleaned and part-of-speech tagged with Stanford POS-Tagger~\cite{}. 
%The POS tags are subsequently used during discourse feature extraction, 
%and synset-mapping during affective feature extraction. 
%Table~\ref{tab:dataset} shows the dataset statistics in terms of the number of users, postings %and documents written by them. 
%A document consists of multiple user postings on a given sub-topic. 

\comment{
\begin{table}
\centering
\begin{tabular}{p{1.5cm}p{2cm}p{2cm}p{1.5cm}}
\hline
{\bf Users}&{\bf Total Postings}&{\bf Total Documents}&{\bf Total Words}\\\hline
14,996 & 2,856,992 & 620,510 & 3,84,774,957 \\
\hline
\end{tabular}
\caption{Dataset Meta Statistics}
\label{tab:dataset}
\end{table}
}%\comment

\begin{table}
\centering
%
%\begin{tabular}{p{1.8cm}p{2cm}p{1.8cm}p{1.5cm}}
%\hline
%{\bf \% Female}&{\bf Total Postings}&{\bf Total Documents}&{\bf Total Words}\\\hline
%77.69 & 2,856,992 & 620,510 & 3,84,774,957 \\
%\hline
%\end{tabular}
%
{\small
\begin{tabular}{p{2.2cm}p{1.2cm}p{1.3cm}p{1.2cm}p{1.2cm}}
\toprule
{\bf Member Type}&{\bf Members}&{\bf Postings}&{\bf Average Qs.}&{\bf Average Replies}\\\midrule
Administrator & 1 & - & 363 & 934\\
Moderator & 4 & - & 76 & 1276\\
Facilitator & 16 & > 4700 & 83 & 2339\\
Senior veteran & 966 & > 500 & 68 & 571\\
Veteran & 916 & > 300 & 41 & 176\\
Senior member & 4321 & > 100 & 24 & 71\\
Member & 5846 & > 50 & 13 & 28\\
Junior member & 1423 & > 40 & 9 & 18\\
Inactive & 1433 & - & -& -\\
Registered user & 70 & - & - & -\\
\bottomrule
\end{tabular}
}%\small
\caption{User statistics.}
\label{tab:author}
\end{table}

%\subsection{Community Drug Statistics}

As ground truth for drug side-effects, we rely on data from the Mayo Clinic portal\footnote{\tt \href{http://www.mayoclinic.org/drugs-supplements/}{mayoclinic.org/drugs-supplements/}},
which contains curated expert information about drugs, with
side-effects being listed as \emph{more common, less common and rare}
for each drug.
%GW: we may want to comment on the "unobserved" and "overdose" categories, perhaps leave them out
We extracted $2,172$\ drugs which are categorized into $837$ drug\ families.
For our experiments, we select $6$ widely used drug families (based on {\tt \href{http://www.webmd.com}{webmd.com}}).
Table~\ref{tab:drug} provides information on this sample
and its coverage on {\tt \href{http://www.healthboards.com/}{healthboards.com}}. Table~\ref{tab:SE} shows the number of common, less common, and rare side-effects
for the six drug families as given by the Mayo Clinic portal.
\comment{The table also shows the number of community members who have reported
at least one side-effect for these drugs in {\tt \href{http://www.healthboards.com}{healthboards.com}}.}

\comment{To identify the postings that mention a drug and a side-effect,
we use exact matching for drug names (taking into consideration alias names
as given by the expert database) and approximate matching for
noun phrases that correspond to side-effects.
We use a high threshold for character-level Jaccard similarity
to accept a side-effect phrase.}
%GW: is this correct this way?

%%%%%%%%%%%%%%%%%%%%%%%%%%%%%%%%%%

\subsection{Baselines}

We compare our probabilistic model against the following baseline methods,
using the same set of features for all the models, and classifying the same set of side-effect candidates.

\begin{table}
\centering
{\small
\begin{tabular}{p{4cm}p{5cm}p{1cm}p{1cm}}
\toprule
{\bf Drugs}&{\bf Description}&{\bf Users}&{\bf Postings}\\\midrule
alprazolam, niravam, xanax & relieve symptoms of anxiety, depression, panic disorder & 2785 & 21,112 \\\midrule
ibuprofen, advil, genpril, motrin, midol, nuprin & relieve pain, symptoms of arthritis, such as inflammation, swelling, stiffness, joint pain & 5657 & 15,573\\\midrule
omeprazole, prilosec &treat acidity in stomach, gastric and duodenal ulcers, \dots & 1061 & 3884\\\midrule
metformin, glucophage, glumetza, sulfonylurea &treat high blood sugar levels, sugar diabetes & 779 & 3562\\\midrule
levothyroxine, tirosint &treat hypothyroidism: insufficient hormone production by thyroid gland & 432 & 2393 \\\midrule
metronidazole, flagyl &treat bacterial infections in different body parts & 492 & 1559\\
\bottomrule
\end{tabular}
}%\small
\caption{Information on sample drug families: number of postings and number of users reporting at least one side effect.}
\label{tab:drug}
\end{table}

\begin{table}[!h]
\centering
\small
\begin{tabular}{lccc}
\toprule
{\bf Drug family}&{\bf Common}&{\bf Less common}&{\bf Rare}\\\midrule
alprazolam & 35 & 91 & 45\\
ibuprofen & 30 & 1 & 94\\
omeprazole & - & 15 & 20\\
metformin & 24 & 37 & 5\\
levothyroxine & - & 51 & 7\\
metronidazole & 35 & 25 & 14 \\
\bottomrule
\end{tabular}
\caption{Number of common, less common, and rare side-effects listed by experts on Mayo Clinic.}
\label{tab:SE}
\end{table}

%%C: macro
%\noindent{\bf Frequency Baseline: }
\xhdrNoPeriod{Frequency Baseline:} 
%The first baseline, we consider, is the simple \textit{tf-idf} baseline. 
For each statement on a drug side-effect, we consider how frequently the statement
has been made in community. This gives us a ranking of side-effects.
%GW: why do you call this tf-idf? isn't it just frequency? what should idf be here?

%%C: macro
%\noindent{\bf SVM Baseline: }
\xhdrNoPeriod{SVM Baseline:} 
For each drug and possible side-effect we determine all postings where it
is mentioned and aggregate the features $F^L$, $F^E$, $F^U$,
described in Section~\ref{sec:features} over all these postings, thus creating a single feature vector for each side-effect.

We use the ground-truth labels from the Mayo Clinic portal to
train a Support Vector Machine (SVM) classifier with a linear kernel, $L_2$ loss,
and $L_1$ or $L_2$ regularization, for classifying unlabeled statements.

%GW: the user features F^U are not used here, are they?
%
\comment{
For every side-effect $s_k$, consider the set of all documents, $\{d_{i,j}^{s_k}\}$, containing a mention of the side-effect. Using the features, described in Section~\ref{sec:features}, we construct a single feature vector $F_{s_{k}}^L \cup F_{s_{k}}^E \cup F_{s_{k}}^A$ for each side-effect, by aggregating all the feature values element-wise across all the documents containing the mention, where
\begin{equation}
\begin{split}
F_{s_{k}}^L = \{f_{s_k,l}\} \\
f_{s_k,l} = \frac{\sum_i \sum_j f_{l,c}(d_{i,j}^{s_k}(w), c^{\prime})}{|d^{s_k}|}
\end{split} 
\end{equation}
}%\comment
%
%GW: no need to repeat or introduce heavy notation for a straightforward approach

%\noindent{\bf SVM Baseline with Distant Supervision: }
\xhdrNoPeriod{SVM Baseline with Distant Supervision:} 
As the number of common side-effects for any drug is typically small, the above approach to create a single feature vector for each side-effect results in a very small training set. Hence, we use the notion of \textit{distant supervision} to create a rich, expanded training set.

A feature vector is created for \textit{every mention} or instance of a side-effect in different user postings. The feature vector
%subho:camera-ready
$<S_i, p_j, u_k>$ has the label of the side-effect, and represents the set of cliques in Equation~\ref{eq4.0}. The semi-supervised CRF formulation in our approach further allows for information sharing between the cliques to estimate the labels of the unobserved statements from the expert-provided ones.  

This process creates a noisy training set, as a posting may contain multiple side-effects, positive and negative. This results in multiple similar feature vectors with different labels.
During testing, the same side-effect may get different labels from its different instances. We take a majority voting of the labels obtained by a side-effect, across predictions over its different instances, and assign a unique label to it.

%%%%%%%%%%%%%%%%%%%%%%%%%%%%%%%%%%%%

\subsection{Experiments and Quality Measures}

We conduct two lines of experiments, with different settings on what is
considered ground-truth.

%\noindent{\bf Experimental Setting I: }  
\xhdrNoPeriod{Experimental Setting I:}  
We consider
only {\em most common side-effects} listed by the Mayo Clinic portal
as positive ground-truth, whereas all other side-effects (less common, rare and unobserved) are considered to be negative instances (i.e., so unlikely that they should be
considered as false statements, if reported by a user).
The training set is constructed in the same way.
This setting aims to study the predictive power of our model in determining the common side-effects of a drug,
in comparison to the baselines.

%\noindent{\bf Experimental Setting II: }

\xhdrNoPeriod{Experimental Setting II:} 
Here we address our original motivation: discovering less common and rare side-effects.
Durring training, as positive ground-truth we consider common and less common side-effects (as stated
by the experts on the Mayo Clinic site), whereas all rare and unobserved
side-effects are considered negative instances.
%%C: Same as what? (It can not be same as setting 1, since there less common is considered un)
%%The model is trained with the same partitioning of true vs. false statements.
%
%We want the model to automatically label the rare side-effects as positive. From the side
%effects tagged positive by our model, we find out how many of those side-effects are %deemed as rare by experts from Mayo Clinic.
%subho:camera-ready
Our goal here is to test how well the model can identify \textit{less known} and \textit{rare}
side-effects as true statements.  We purposely do not consider rare side-effects as positive training examples, since we aim to evaluate the model's ability to retrieve such statements starting only from very reliable positive instances.  We measure performance on rare side-effects as the recall for such statements being labeled as true statements, in spite of considering \textit{only} common and less common
side-effects as positive instances durring training. 
%%C: edit
%(if indeed deemed credible according to postings and users)

%since users frequently talk about experiencing them in the community. Instead, the classifier is made to learn only from the most probable ones as positive instances.

%GW: discuss results more explicitly in the Results subsection, not here

\xhdrNoPeriod{Train-Test Data Split:}\label{lab:ds} 
For each drug family, we create multiple random splits of $80\%$ training data
and $20\%$ test data.  All results reported below
are averaged over $200$ such splits. 
All baselines and our CRF model use same test sets.
%As the number of negative instances is much greater than the number of positive instances, %the dataset is imbalanced.
%We undersample the negative instances to create a balanced training set.
%GW: either drop this sentence or explain what exactly is meant by "undersampling" here

%%C: macro, and connecting to the actual paragraph
%\noindent{\bf Evaluation Metrics: } 
\xhdrNoPeriod{Evaluation Metrics:} The standard measure for the quality of a binary classifier is {\em accuracy}:
$\frac{tp+tn}{tp+fn+tn+fp}$.
We also report the \textit{specificity} ($\frac{tn}{tn+fp}$) and \textit{sensitivity} ($\frac{tp}{tp+fn}$).
%GW: explain in terms of a binary classifier, not in terms of a clinical diagnosis test
Sensitivity measures the true positive rate or the model's ability to identify positive side-effects, whereas specificity measures true negative rate.
%%%%%%%%%%%%%%%%%%%%%%%%%%%%%%%%%%

\subsection{Results and Discussions}
\label{subsec:accuracy}

Table~\ref{tab:Acc} shows the accuracy comparison of our system (\MRF) with the baselines for different drug families in the first setting. The first naive baseline, which simply considers the frequency of postings containing the side-effect by different users, has an average accuracy of $57.65\%$ across different drug families.

Incorporating supervision in the classifier as the first SVM baseline (SVM w/o DS), along with a rich set of features for users, postings and language, achieves an average accuracy improvement of $11.4\%$. In the second SVM baseline (SVM DS), we represent each posting reporting a side-effect as a separate feature vector. This not only expands the training set leading to better parameter estimation, but also represents the set of cliques in Equation~\ref{eq4.0} (we therefore consider this to be a strong baseline). This brings an average accuracy improvement of $7\%$ when using $L_1$ regularization and $9\%$ when using $L_2$ regularization. Our model (\MRF), by further considering the coupling between users, postings and statements, allows information to flow between the cliques in a feedback loop bringing a further accuracy improvement of 
%$6\%$ over the strong { SVM DS $L_1$} baseline and 
$4\%$ over the strong {SVM DS $L_2$}  baseline.

\begin{table}
\centering
\small
\begin{tabular}{lccccc}
\toprule
\multicolumn{ 1}{l}{\bf Drugs} & \multicolumn{ 1}{c}{\bf \parbox{1cm}{Post\\Freq.}} & \multicolumn{ 3}{c}{\bf SVM} & \multicolumn{ 1}{c}{\bf CRF} \\ \cmidrule{ 3- 5}
\multicolumn{ 1}{l}{} & \multicolumn{ 1}{c}{} & \multicolumn{ 1}{l}{\bf w/o DS} & \multicolumn{ 2}{c}{\bf DS} & \multicolumn{ 1}{l}{} \\ \cmidrule{ 4- 5}
\multicolumn{ 1}{l}{} & \multicolumn{ 1}{c}{} & \multicolumn{ 1}{l}{} & $L_1$ & $L_2$ & \multicolumn{ 1}{l}{} \\ \midrule
Alprazolam & 57.82 & 70.24 & 73.32 & 73.05 & 79.44  \\
Metronidazole & 55.83 & 68.83 & 79.82 & 78.53 & 82.59  \\
Omeprazole & 60.62 & 71.10 & 76.75 & 79.15 & 83.23  \\
Levothyroxine & 57.54 & 76.76 & 68.98 & 76.31 & 80.49 \\
Metformin & 55.69 & 53.17 & 79.32 & 81.60 & 84.71 \\
Ibuprofen & 58.39 & 74.19 & 77.79 & 80.25 & 82.82 \\
\bottomrule
\end{tabular}
\caption{\hspace{0.3em}Accuracy comparison in setting I.}
\label{tab:Acc}
\end{table}

\begin{table}
\centering
\small
\begin{tabular}{p{2.5cm}p{2cm}p{2cm}p{2cm}p{1.2cm}}
\toprule
{\bf Drugs}&{\bf \parbox{2.5cm}{Sensitivity}}&{\bf \parbox{2.5cm}{Specificity}}&{\bf \parbox{2cm}{Rare SE\\Recall}}&{\bf \parbox{1.2cm}{Accuracy}} \\\midrule
Metformin & 79.82 & 91.17 & 99 & 86.08\\
Levothyroxine & 89.52 & 74.5 & 98.50 & 83.43\\
Omeprazole & 80.76 & 88.8 & 89.50 & 85.93\\
Metronidazole & 75.07 & 93.8 & 71 & 84.15\\
Ibuprofen & 76.55 & 83.10 & 69.89 & 80.86\\
Alprazolam & 94.28 & 68.75 & 61.33 & 74.69\\
\bottomrule
\end{tabular}
\caption{\hspace{0.3em}CRF performance in setting II.}
\label{tab:rare}
\vspace{-1.5em}
\end{table}

\begin{figure*}[!h]
\centering
\includegraphics[scale=0.4]{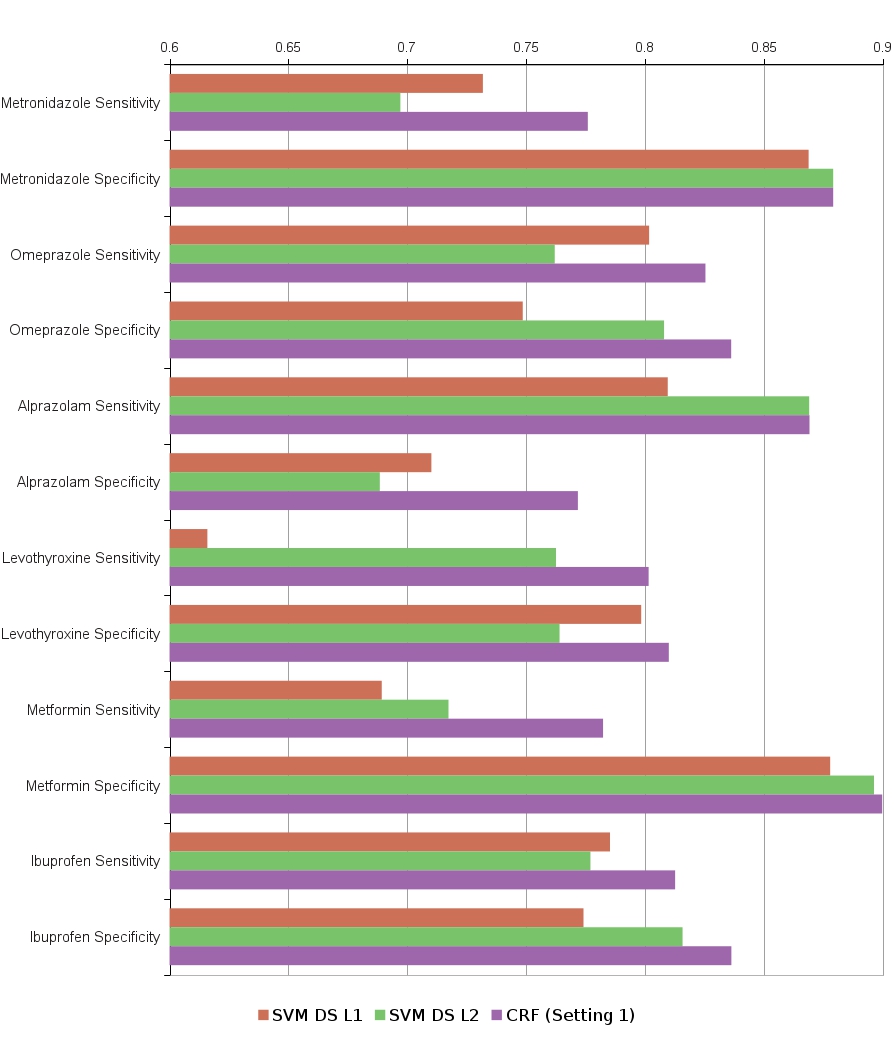}
\caption{Specificity and sensitivity comparison of models.}
\label{fig:modelComparison}
\vspace{-1.5em}
\end{figure*}

Figure~\ref{fig:modelComparison} shows the sensitivity and specificity comparison of the baselines with the CRF model. Our approach has an overall $5\%$ increase in sensitivity and $3\%$ increase in specificity over the { SVM $L_2$} baseline.

%\pagebreak

The specificity increase over the SVM $L_2$ baseline is maximum for the Alprazolam drug family at $8.33\%$ followed by Levothyroxine at $4.6\%$. The users taking anti-depressants like Alprazolam suffer from anxiety disorder, panic attacks, depression etc. and report a large number of side-effects of drugs. Hence, it is very difficult to negate certain side-effects, in which our model performs very well due to well-designed language features. Also, Alprazolam and Levothyroxine have a large number of expert-reported side-effects (refer Table~\ref{tab:SE}) and corresponding user-reported ones, and the model learns well for the negative class.

The drugs Metronidazole, Metformin and Omeprazole treat some serious physical conditions, have less number of expert and user-reported side-effects. Consequently, our model captures user statement corroboration well to attain a sensitivity improvement of $7.89\%, 6.5\%$ and $6.33\%$ respectively. Overall, classifier performs the best in these drug categories.

Table~\ref{tab:rare} shows the overall model performance, as well as the recall for identifying rare side-effects of each drug in the second setting.
The drugs Metformin, Levothyroxine and Omeprazole have much less number of side-effects, and the classifier does an almost perfect job in identifying all of them. Overall, the classifier has an accuracy improvement of $2-3\%$ over these drugs in Setting II. However, the classifier accuracy { significantly} drops for the anti-depressants (Alprazolam) after the introduction of ``less common'' side-effects as positive statements in Setting II. The performance drop is attributed to the loss of $8.42\%$ in specificity due to increase in the number of false-positives, as there is conflict between what the model learns from the language features (about negative side-effects) and that introduced as ground-truth.

\xhdrNoPeriod{Feature Informativeness:} In order to find the \textit{predictive power} of individual feature classes, tests are perfomed using $L_2$-loss and $L_2$-regularized Support Vector Machines over a split of the test data. Affective features are found to be the most informative, followed by document length statistics, which are more informative than user and stylistic features. Importance of document length distribution strengthens our observation that objective postings tend to be crisp, whereas longer ones often indulge in emotional digression.

Amongst the user features, the most significant one is the ratio of the number of replies by a user to the questions posted by her in the community, followed by the gender, number of postings by the user and finally the number of thanks received by her from fellow users. There is a gender-bias in the community, as $77.69\%$ active contributors in this health forum are female.

Individual F-scores of the above feature sets vary from $51\%$ to $55\%$ for Alprazolam; whereas the combination of all features yield $70\%$ F-score.

\comment{
\begin{tabular}{p{2cm}p{2cm}p{2cm}}
\hline
{\bf Classifier}&{\bf Sensitivity}&{\bf Specificity}\\\hline
SVM DS $L_1$ & 73.89 & 79.63 \\
SVM DS $L_2$ & 76.41 & 80.86 \\
CRF & 81.11 & 83.84
\hline
\end{tabular}
\caption{Average sensitive and specificity of classifiers.}
\label{tab:SE}
\end{table}
}%\small

\vspace{-0.5em}
\subsection{Discovering Rare Side Effects}
\label{subsec:usecaseRare}

Section~\ref{subsec:accuracy} has focused on evaluating the predictive power of our model and inference method. Now we shift the focus to two application-oriented use-cases: 1) discovering side-effects of drugs that are not covered by expert databases, and 2) identifying the most trustworthy users that one would want to follow for certain topics.

Members of an online community may report side-effects that are either flagged as very rare in an expert knowledge base (KB) or not listed at all. We call the latter {\em out-of-KB} statements. As before, we use the data from {\tt \href{http://www.mayoclinic.org}{mayoclinic.org}} as our KB, and focus on the following 2 drugs representing different kinds of medical conditions and patient-reporting styles: Alprazolam and Levothyroxine. For each of these drugs, we perform an experiment as follows.

For each drug $X$, we use our IE machinery to identify all side-effects $S$ that are reported for $X$, regardless of whether they are listed for $X$ in the KB or not. The IE method uses the set of all side-effects listed for {\em any} drug in the KB as potential result. For example, if ``hallucination'' is listed for some drug but not for the drug Xanax, we capture mentions of hallucination in postings about Xanax.
We use our probabilistic model to compute credibility scores for these out-of-KB side-effects, and compile a ranked list of 10 highest-scoring side-effects for each drug. This ranked list is further extended by 10 randomly chosen out-of-KB side-effects (if reported at least once for the given drug).
%GW: if the data is too sparse, we may want to back up to say top-5 plus 5 random ones

The ranked list of out-of-KB side-effects is shown to $2$ expert annotators who manually assess their credibility, by reading the complete discussion thread (e.g. expert replies to patient postings) and other threads that involve the users who reported the side-effect. The assessment is binary: true (1) or false (0); we choose the final label as majority of judges. This way, we can compute the quality of the ranked list in terms of the {\em NDCG (Normalized Discounted Cumulative Gain)}~\cite{Jarvelin:TOIS2002} measure
$NDCG_p = \frac{DCG_p}{IDCG_p}$, where
\begin{equation}
DCG_p = rel_1 + \sum_{i=2}^p \frac{rel_i}{\log_2 i}
\end{equation}

Here, $rel_i$ is the graded relevance of a result ($0$ or $1$ in our case) at position $i$. DCG penalizes relevant items appearing lower in the rank list, where the graded relevance score is reduced logarithmically proportional to the position of the result. As the length of lists may vary for different queries, DCG scores are normalized using the ideal score, IDCG
where the results of a rank list are sorted by relevance giving the maximum possible DCG score.
We also report the inter-annotator agreement using Cohen's Kappa measure.

Table~\ref{tab:useRare} shows the Kappa and NDCG score comparison between the baseline and our CRF model.
The baseline here is to rank side-effects by frequency \textit{i.e.}
how often are they reported in the postings of different users on the given drug.
The strength of Kappa is considered ``moderate'' (but significant), which depicts the difficulty in identifying the side-effects of a drug just by looking at user postings in a community. The baseline performs very poorly for the anti-depressant Alprazolam, as the users suffering from anxiety disorders report  a large number of side-effects most of which are not credible. On the other hand, for Levothyroxine (a drug for hypothyroidism), the baseline model performs quite well as the users report more serious symptoms and conditions associated with the drug, which also has much less expert-stated side-effects compared to Alprazolam (refer Table~\ref{tab:drug}). The CRF model performs perfectly for both drugs.

\comment{We compare our ranking against two baselines:
\squishlist
\item {\em Frequencies:} the rank of a side-effect is determined by how often it is observed in all postings about a given drug.
\item {\em Weighted Frequencies:} the frequencies are weighted by the average number of overall postings (across all drugs and topics) of the users who report a side-effect. Alternatively, we can use the total number of thanks received by a user, instead of counting postings.
%GW: we could also use the trust scores from our model, but then we are mixing a simple baseline with one of our advocated results, hence the simpler choise of #postings or #thanks
\squishend
}

\begin{table}
\centering
\small
\begin{tabular}{cccc}
\toprule
\multicolumn{ 1}{c}{Drug} & \multicolumn{ 1}{c}{Kappa} & \multicolumn{ 2}{c}{Model NDCG Scores} \\ \cmidrule{ 3- 4}
\multicolumn{ 1}{c}{} & \multicolumn{ 1}{c}{} & Frequency & CRF \\
Alprazolam, Xanax & \multicolumn{1}{c}{0.471} & \multicolumn{1}{c}{0.31} & \multicolumn{1}{c}{1} \\
Levothyroxine, Tirosint & \multicolumn{1}{c}{0.409} & \multicolumn{1}{c}{0.94} & \multicolumn{1}{c}{1} \\
\bottomrule
\end{tabular}
\caption{\hspace{0.3em}Experiment on finding rare drug side-effects.}
\label{tab:useRare}
\end{table}

\begin{table}
\centering
\small
\begin{tabular}{cccc}
\toprule
\multicolumn{ 1}{c}{Drug} & \multicolumn{ 1}{c}{Kappa} & \multicolumn{ 2}{c}{Model NDCG Scores} \\ \cmidrule{ 3- 4}
\multicolumn{ 1}{c}{} & \multicolumn{ 1}{c}{} & Frequency & CRF \\
Alprazolam, Xanax & \multicolumn{1}{c}{0.783} & \multicolumn{1}{c}{0.82} & \multicolumn{1}{c}{1} \\
Levothyroxine, Tirosint & \multicolumn{1}{c}{0.8} & \multicolumn{1}{c}{0.57} & \multicolumn{1}{c}{0.81} \\
\bottomrule
\end{tabular}
\caption{\hspace{0.3em}Experiment on following trustworthy users.}
\label{tab:useFollower}
%\vspace*{-1.}
\end{table}

\subsection{Following Trustworthy Users}
\label{subsec:usecaseTrust}

In the second use-case experiment, we evaluate how well our model can identify trustworthy users in a community. We find the top-ranked users in the community given by their trustworthiness scores ($t_k$), for each of the drugs Alprazolam and Levothyroxine. As a baseline model, we consider the top-thanked contributors in the community. The moderators and facilitators of the community, listed by both models as top users, are removed from the ranked lists, in order to focus on the interesting, not obvious cases.
Two judges are asked to annotate the top-ranked users listed by each model
 as trustworthy or not, based on the users' postings on the target drug. The judges are asked to mark a user trustworthy if they would consider following the user in the community. Although this exercise may seem highly subjective, the Cohen's Kappa scores
show high inter-annotator agreement. The strength of agreement is considered to be
``very good'' for the user postings on Levothyroxine, and ``good'' for the Alprazolam users.

The baseline model performs poorly for Levothyroxine.
%, as the annotators disagreed on the top-most users of the list. In case of disagreement, a user was
%not considered trustworthy.
The CRF model outperforms the baseline in both cases.

%% file: main/chapter-credibility-analysis/experiments-news.tex
In this section, we present the first full-fledged analysis of credibility, trust, and expertise in
news communities; with data from {\tt \href{http://www.newstrust.net}{newstrust.net}}, one of
the most sophisticated news communities with a focus on quality
journalism. 

\subsection{Data}
\label{sec:newstrust-data}

We performed experiments with data from a typical news community: {\tt \href{http://www.newstrust.net}{newstrust.net}}\footnote{Code and data available at {\href{http://www.mpi-inf.mpg.de/impact/credibilityanalysis/}{http://www.mpi-inf.mpg.de/impact/credibilityanalysis/}}}.
This community is similar to {\tt \href{http://www.digg.com}{digg.com}} and {\tt \href{http://www.reddit.com}{reddit.com}}, but has more refined ratings and interactions. We chose NewsTrust 
because of the availability of {\em ground-truth} ratings for credibility analysis of news articles (i.e. postings);
such ground-truth is not available for the other communities.

We collected {\em stories} from NewsTrust from May, 2006 to May, 2014 on diverse topics ranging from sports, politics, environment to current affairs. Each such story features a {\em news article} (i.e. posting) from a source (E.g. BBC, CNN, Wall Street Journal) that is posted by a member, and reviewed by other members in the community, many of whom are {\em professional journalists} and {\em content experts}\footnote{\url{http://www.newstrust.net/help\#about\_newstrust}}. We crawled all the stories with their explicit topic tags and other associated meta-data.
We crawled all the {\em news articles} from their original sources that were featured in any NewsTrust story. The earliest story dates back to May 1, 1939 and the latest one is in May 9, 2014.

We collected all {\em member profiles} containing information about the demographics, occupation and expertise of the members along with their activity in the community in terms of the postings, reviews and ratings; as well as \textit{interaction} with other members. The members in the community can also rate each others' ratings. The earliest story rating by a member dates back to May, 2006 and the most recent one is in Feb, 2014.
In addition, we collected information on member evaluation of news sources, and other information (e.g., type of media, scope, viewpoint, topic specific expertise) about source from its {\em meta data}.

\begin{table}[htbp]
\small
\centering
\begin{tabular}{lr}
\toprule
\textbf{Factors} & \textbf{Count}\\\midrule
Unique news articles reviewed in NewsTrust & 62,064\\
NewsTrust stories on news articles & 84,704\\
NewsTrust stories with $\geq 1$ reviews & 43,107\\
NewsTrust stories with $\geq 3$ reviews & 18,521\\
%28,211, 18,521\\
NewsTrust member reviews of news articles & 134,407\\
\midrule
News articles extracted from original sources & 47,565\\
NewsTrust stories on extracted news articles & 52,579\\
\midrule
News sources & 5,658\\
Journalists who wrote news articles & 19,236\\
Timestamps (month and year) of posted news articles & 3,122\\
\midrule
NewsTrust members who reviewed news articles & 7,114\\
NewsTrust members who posted news articles & 1,580\\
\midrule
News sources reviewed by NewsTrust members & 668\\
\midrule
Explicit topic tags & 456\\
Latent topics extracted & 300\\
\bottomrule
\end{tabular}
\caption{\hspace{0.3em}Dataset statistics.}
\label{tab:data}
\end{table}

\noindent{\bf Crawled dataset:} Table~\ref{tab:data} shows the dataset statistics. In total $~62$K unique news articles were reviewed in NewsTrust in the given period, out of which we were able to extract $~47$K full articles from the original sources like New York Times, TruthDig, ScientificAmerican etc --- a total of $~5.6$K distinct sources. The remaining articles were not available for crawling. There are $~84.7$K stories featured in NewsTrust for all the above articles,
% (including a few re-submissions of news articles which are re-reviewed by different sections of the 
%community), 
out of which $~52.5K$ stories refer to the news articles we managed to extract from their original sources. The average number of reviews per story is $1.59$. For general analysis we use the entire dataset. 

For experimental evaluation of the CCRF and hypotheses testing, we use only those stories ($~18.5$K) with a {\em minimum of $3$ reviews} that refer to the news articles we were able to extract from original sources.

\noindent{\bf Generated graph:} Table~\ref{tab:graph} shows the statistics of the graph
%\footnote{Analysis done using {http://gephi.github.io/}} 
constructed by the method of Section~\ref{sec:newscom}. 
%The graph has high connectivity and overlapping modular network structure.

%{\tt GW: fix/trim table: keep only essential numbers, check weakly CC's (appears twice), fix decimal commas  (5,21,630  should probably be 521,630 - or is there a digit missing?}\\
\begin{table}[t]
\small
\centering
\begin{tabular}{lrlr}
\toprule
\textbf{Factors} & \textbf{Count} & \textbf{Factors} & \textbf{Count}\\\midrule
Nodes & 181,364 & No. of weakly connected components & 12\\
\myindent Sources & 1,704 & Diameter & 8\\
\myindent Members & 6,906 &  Average path length & 47 \\
\myindent News articles & 42,204 & Average degree & 6.641\\
\myindent Reviews & 130,550 & Average clustering coefficient & 0.884\\
Edges & 602,239 & Modularity & 0.516\\
Total triangles & 521,630 & &\\
\bottomrule
\end{tabular}
\caption{\hspace{0.3em}Graph statistics.}
\label{tab:graph}
\end{table}

\noindent{\bf Ground-Truth for evaluation}: The members in the community can rate the credibility of a news article on a scale from $1$ to $5$ regarding $15$ qualitative aspects like facts, fairness, writing style and insight, and popularity aspects like recommendation, credibility and views.
Members give an overall \emph{recommendation} for the article explained to them as:
\textit{``... Is this quality journalism? Would you recommend this story to a friend or colleague? ... This question is similar to the up and down arrows of popular social news sites like Digg and Reddit, but with a focus on quality journalism."}
Each article's aspect ratings by different members are weighted (and aggregated) by NewsTrust based on findings of \cite{lampe2007}, and the member expertise and member level (described below).
%Ratings weigh more if the reviewer is expert on a topic and receives positive ratings from other users.
This overall article rating is taken as the ground-truth for the article \emph{credibility} rating in our work.
A user's {\em member level} is calculated by NewsTrust
based on her community engagement, experience, other users' feedback on her ratings, profile transparency and validation by NewsTrust staff. This member level is taken as the proxy for user \emph{expertise} in our work.
Members rate news sources while reviewing an article. These ratings are aggregated for each source, and taken as a proxy for the source \emph{trustworthiness} in our work.

\noindent {\bf Training data:}
%{\tt GW: we need to explicitly say which parts of that data are used for training and which for testing. I suppose it's 90percent of everything for training, 10percent for testing, and then we do 10-fold cross-validation - correct? We can put this here, at the end of the Use Case section, or at the beginning of the experimental section, before 5.1 starts!!!!!}\\
We perform $10$-fold cross-validation {on the news articles}. During training on any $9$-folds of the data, the algorithm learns the user, source, language and topic models from user-assigned ratings to articles and sources present in the train split. We combine sources with less than $5$ articles and users with less than $5$ reviews into background models for sources and users, respectively.
This is to avoid modeling from sparse observations, and to reduce dimensionality of the feature space. However, while testing on the remaining {\em blind} $1$-fold we use {\em only the ids} of sources and users reviewing the article; we do not use any user-assigned ratings of sources or articles. For a new user and a new source, we draw parameters from the user or source background model. The results are averaged by $10$-fold cross-validation, and presented in the next section.

\noindent {\bf Experimental settings:} In the first two experiments we want to
find the power of the CCRF in predicting user rating behavior, and credibility
rating of articles. Therefore, the evaluation measure is taken as the
\textit{Mean Squared Error} (MSE) between the prediction and the actual
ground-rating in the community. For the latter experiments in finding expert
users (and trustworthy sources) there is no absolute measure for predicting
user (and, source) quality; it only makes sense to find the relative ranking of
users (and sources) in terms of their expertise (and, trustworthiness).
Therefore, the evaluation measure is taken as the {\em Normalized Discounted
Cumulative Gain} (NDCG) \cite{Jarvelin:TOIS2002} between the ranked list of users
(and sources) obtained from CCRF and their actual ranking in the
community.

\begin{table}
\small
\centering
\begin{tabular}{lll}
\toprule
\textbf{Model} & \textbf{MSE} &\\\midrule
\textbf{Latent Factor Models (LFM)} & &\\
\myindent Simple LFM~\cite{korenKDD2008} & 0.95 &\\
\myindent Experience-based LFM~\cite{mcauleyWWW2013} & 0.85 &\\
\myindent Text-based LFM~\cite{mcauleyrecsys2013} & 0.78 &\\
\textbf{Our Model: User SVR} & 0.60 &\\
\bottomrule
\end{tabular}
\caption{\hspace{0.3em}MSE comparison of models for predicting users' credibility rating
behavior with $10$-fold cross-validation. Improvements are statistically
significant with {\em P-value} < $0.0001$.}
\label{tab:MSE1}
\end{table}

\comment{
//all feautures- SVR - 0.84
//all feautures SVR + SVR - 0.29
explains why this is happening?
}

%%%%%%%%%%%%%%%%%%%%%%%%%%%%%%%%%%

\subsection{Predicting User Credibility Ratings of News Articles}

First we evaluate how good our model can predict the credibility ratings that users
assign to news articles using the \textit{Mean Squared Error} (MSE) between our prediction
and the actual user-assigned rating. 
%Table~\ref{tab:evaldata} shows the evaluation data statistics.

\noindent{\bf Baselines}: We consider the following baselines for comparison:

{\bf 1.} {\em Latent Factor Recommendation Model} (LFM)~\cite{korenKDD2008}: LFM considers the tuple $\langle userId,$ $itemId, rating \rangle$, and models each user and item as a vector of latent factors which are learned by minimizing the MSE between the rating and the product of the user-item latent factors. In our setting, each news article is considered an item, and rating refers to the credibility rating assigned by a user to an article.\\
{\bf 2.} {\em Experience-based LFM}~\cite{mcauleyWWW2013}: This model incorporates {\em experience} of a user in rating an item in the LFM. The model builds on the hypothesis that users at similar levels of experience have similar rating behaviors which evolve with {\em time}. The model has an extra dimension: the {\em time} of rating an item which is not used in our SVR model. Note the analogy between the {\em experience} of a user in this model, and the notion of user {\em expertise} in the SVR model. However, these models ignore the text of the reviews. \\
{\bf 3.} {\em Text-based LFM}~\cite{mcauleyrecsys2013}: This model incorporates text in the LFM by combining the latent factors associated to items in LFM with latent topics in text from topic models like LDA.\\\\ %Unlike our model, which learns the latent topics separately, the baseline model learns the ratings and latent topics jointly.\\
{\bf 4.} {\em Support Vector Regression} (SVR)~\cite{drucker97}: We train an SVR model $\text{SVR}_{u_k}$ for {\em each} user $u_k$ (refer to Section~\ref{subsec:SVR}) based on her reviews $\langle r_{j,k} \rangle$ with language and topic features $\langle F^L(r_{j,k}) \cup F^T(r_{j,k}) \rangle$, with the user's article ratings $\langle y_{j,k} \rangle$ as the response variable. We also incorporate the article language features and the topic features, as well as source-specific features to train the user model for this task. The other models ignore the stylistic features, and other fine-grained {\em user-item} interactions in the community.

Table~\ref{tab:MSE1} shows the MSE comparison between the different methods.
Our model (User SVR) achieved the lowest MSE and thus performed best.

%%%%%%%%%%%%%%%%%%%%%%%%%%%%%%%%%%

\begin{table}[t]
\small
\centering
\begin{tabular}{lll}
\toprule
\textbf{Model} & \textbf{Only Title} & \textbf{Title \& Text}\\
%& \textbf{Title} & \textbf{\& Text}\\
& \textbf{MSE} & \textbf{MSE}\\\midrule
\textbf{Language Model: SVR} & &\\
\myindent Language (Bias and Subjectivity) & 3.89 & 0.72\\
\myindent Explicit Topics & 1.74 & 1.74\\
\myindent Explicit + Latent Topics & 1.68 & 1.01\\
\myindent All Topics (Explicit + Latent) + Language & 1.57 & 0.61 \\\midrule

\textbf{News Source Features and Language Model: SVR} & & \\
\myindent News Source & 1.69 & 1.69\\
\myindent News Source + All Topics + Language & 0.91 & 0.46\\\midrule

\textbf{Aggregated Model: SVR} & & \\
\myindent Users + All Topics + Language + News Source & 0.43 & 0.41\\\midrule

\textbf{Our Model: CCRF+SVR} & & \\
\myindent User + All Topics + Language + News Source & 0.36 & 0.33\\
%User (Topic Preference) + All Topics + Language + News Source + User Features & 0.63\\
\bottomrule
\end{tabular}
\caption{\hspace{0.3em}MSE comparison of models for predicting aggregated article credibility rating with $10$-fold cross-validation. Improvements are statistically significant with {\em P-value} < $0.0001$.}
\label{tab:MSE}
%\end{minipage}
\end{table}

\subsection{Finding Credible News Articles}

As a second part of the evaluation, we investigate the predictive power of different models in order to find credible news articles based on the
{\em aggregated ratings from all users}. The above LFM models, unaware of the {\em user cliques}, cannot be used directly for this task, as each news article has multiple reviews from different users which need to be aggregated. 
%to get an overall credibility rating for the article.
%to predict the overall credibility.
We find the \textit{Mean Squared Error} (MSE) between the estimated overall article rating, and the ground-truth article rating. We consider stories with {\em at least 3 ratings} about a news article.
% we managed to extract from their original web-sources. 
We compare the CCRF against the following baselines:\\
{\bf 1.} {\em Support Vector Regression} (SVR)~\cite{drucker97}: We consider an SVR model with features on language (bag-of-all-words, subjectivity, bias etc.), topics (explicit tags as well as latent dimensions), and news-source-specific features. The language model uses all the lexicons and linguistic features discussed in Chapter~\ref{sec:language}.
%The topic model is similar to the one used in~\cite{mukherjee2013WWW}. 
The source model also includes topic features in terms of the top topics covered by the source, and its topic-specific expertise for a subset of the topics.\\
{\bf 2.} {\em Aggregated Model} (SVR)~\cite{drucker97}: As explained earlier, the user features cannot be directly used in the baseline model, which is agnostic of the user {\em cliques}. Therefore, we adopt a simple aggregation approach by taking the {\em average} rating of all the user ratings $\frac{\text{SVR}_{u_k}(d_j)}{|u_k|}$ for an article $d_j$ as a feature. 
Note that, in contrast to this simple average used here, our CCRF model learns the weights $\langle \alpha_u \rangle$ {\em per-user} to
combine their overall ratings for an article.

Table~\ref{tab:MSE} shows the MSE comparison of the different models.

\noindent{\bf MSE Comparison}: The first two models in Table~\ref{tab:MSE1} ignore the textual content of news articles, and reviews, and perform worse than the ones 
that incorporate full text. The text-based LFM considers title and text, and performs better than its predecessors. However, the User SVR model considers richer features and interactions, and attains $23\%$ MSE reduction over the best performing
LFM baselines.
%in the first task.

The baselines in Table~\ref{tab:MSE} show the model performance after incorporating different features in two different settings:
1) with news article {\em titles} only as text, and 
2) with titles and the {\em first few paragraphs} of an article.
The language model, especially the bias and subjectivity features, is less effective using only the article titles due to sparseness. 
On the other hand, using the entire article text may lead to very noisy features. 
So including the first few paragraphs of an article is the ``sweet spot''.
For this, we made an ad-hoc decision and included the first $1000$ characters
of each article.
With this setting, the language features made a substantial contribution to
reducing the MSE.
%{\tt GW: we may want to reconsider the wording here -- it should not sound like substantial tuning in order to get language features to work!!!!!!}\\

%Using language and topic features along with the {\em news source} specific features %attain a significant performance improvement over the news web-source features in isolation.
%%%GW: drop this, focus on primary message here

The aggregated SVR model further brings in the {\em user} features, and achieves
the lowest MSE among the baselines. 
This shows that a user-aware credibility model performs better than
user-independent ones.
Our CCRF model combines all features in a more sophisticated manner,
which results in $19.5\%$ MSE reduction over the most competitive baseline (aggregated SVR).
This is empirical evidence that the {\em joint} interactions between the different factors in a news community are indeed important to consider for identifying highly credible articles.
%\vspace{-0.5em}

%{\tt GW: shouldn't we have also NDCG scores for the top-k most credible articles? This would make this study perfectly analogous to the ones on trust and expertise!!!!!!}\\

\begin{table}[t]
\parbox{0.47\linewidth}{
\centering
\small
\begin{tabular}{lc}
\toprule
\textbf{Model} & \textbf{NDCG}\\
\midrule
Experience LFM~\cite{mcauleyWWW2013} & 0.80\\
PageRank & 0.83\\
CCRF & 0.86\\
\bottomrule
\end{tabular}
\caption{\hspace{0.3em}NDCG scores for ranking trustworthy sources.}
\label{tab:NDCGweb-sources}
}
\hfill
\parbox{0.47\linewidth}{
\centering
\small
\begin{tabular}{lc}
\toprule
\textbf{Model}  & \textbf{NDCG}\\
\midrule
Experience LFM~\cite{mcauleyWWW2013} & 0.81\\
Member Ratings & 0.85\\
CCRF & 0.91\\
\bottomrule
\end{tabular}
\caption{\hspace{0.3em}NDCG scores for ranking expert users.}
\label{tab:NDCGmembers}
}
\end{table}

%%%%%%%%%%%%%%%%%%%%%%%%%%%%%%%
\subsection{Finding Trustworthy Sources}
\label{subsec:expert-sources}

We shift the focus to two  use cases: 1) identifying the most trustworthy sources,
 and 2) identifying expert users in the community who can play the role of ``citizen journalists''.\\

Using the model of Section~\ref{sec:regression}, we rank all news sources in the community according to the learned $\langle \beta_{s_i} \rangle$ in Equation~\ref{eq.7}. The baseline is taken as the \emph{PageRank} scores of news sources in the Web graph. In the
experience-based LFM we can consider the sources to be users, and articles generated by them to be items. This allows us to obtain a ranking of the sources based on their 
%expertise
overall authority.
%{\tt GW: I would reserve "expertise" for individual users. Here it seems that the user expertise levels are aggregated by source. This leads to a notion similar to HITS-like authority IMO. May still have to discuss this!!!!!}
This is the second baseline against which we compare the CCRF.

We measure the quality of the ranked lists in terms of {\em NDCG} using the actual ranking of the news sources in the community as ground-truth.
%Kalervo Järvelin, Jaana Kekäläinen:
%Cumulated gain-based evaluation of IR techniques. ACM Trans. Inf. Syst. 20(4): 422-446 (2002)
NDCG gives geometrically decreasing weights to predictions at the various positions of the ranked list:

{
$NDCG_p = \frac{DCG_p}{IDCG_p}$ where
$DCG_p = rel_1 + \sum_{i=2}^p \frac{rel_i}{\log_2 i}$
}

Table \ref{tab:NDCGweb-sources} shows the NDCG scores for the different methods.
%{\tt GW: to be continued !!!!!}\\

\subsection{Finding Expert Users}
\label{subsec:expert-users}

Similar to news sources, we rank users according to the learned $\langle \alpha_{u_k} \rangle$ in Equation~\ref{eq.7}. The baseline is the average rating received by a user from other members in the community. We compute the NDCG score for the ranked lists of users
by our method.  We also compare against the ranked list of users from the experience-aware LFM~\cite{mcauleyWWW2013}. 
%another
%GW: ground-truth is not a baseline method, so LFM is the only baseline here
%baseline.
Table \ref{tab:NDCGmembers} shows the NDCG scores for different methods.

\begin{table}
\centering
\small
 \begin{tabular}{ll}
 \toprule
 \textbf{Factors} & \textbf{Corr.}\\\midrule
 \textbf{a) }{Stylistic Indicators} Vs. Article Credibility Rating & \\
 \myindent Insightful (Is it well reasoned? thoughtful?) & 0.77\\
 \myindent Fairness (Is it impartial? or biased?) & 0.75\\
 \myindent Style (Is this story clear? concise? well-written?) & 0.65\\
 \myindent Responsibility (Are claims valid, ethical, unbiased?) & 0.72\\
 \myindent Balance (Does this story represent diverse viewpoints?) & 0.49\\\midrule
 \textbf{b) }Influence of Politics Vs. Disagreement & 0.11\\
 \textbf{c) }Expertise (Moderate, High) Vs. Disagreement & -0.10, -0.31\\
 {Interactions} & \\
 \myindent \textbf{d) } User Expertise Vs. User-User Rating & 0.40\\
 \myindent \textbf{e) } Source Trustworthiness Vs. Article Credibility Rating & 0.47\\
 \myindent \textbf{f) } User Expertise Vs. MSE in Article Rating Prediction & -0.29\\
 \bottomrule
 \end{tabular}
 \caption{\hspace{0.3em}Pearson's product-moment correlation between various factors (with {\em P-value} $< 0.0001$ for each test).}
 \label{tab:corAspects}
\end{table}

%\subsection{Summarizing Findings}
%%%GW: "summarizing" diminishes the value of this part
\vspace{-.5em}
\subsection{Discussion}
\label{subsec:discussions}
\input{main/chapter-credibility-analysis/discussions-news}

%% file: main/chapter-credibility-analysis/discussions-news.tex
\vspace{-0.5em}
%{\tt GW: wirte this in a sequence of ca. 3 paragraphs, not in an itemized list - itemization emphasizes summary, whereas paragraphs suggested true discussion with added-value insight !!!!!}

\noindent{\bf Hypothesis Testing}: We test various hypotheses under the influence of the feature groups using explicit labels, and ratings available in the NewsTrust community. A summary of the tests is presented in Table~\ref{tab:corAspects} showing a {\em moderate} correlation between various factors which are put together in the CCRF to have a {\em strong} indicator for information credibility.
\vspace{-.5em}
%We present our experimental findings from Tables~\ref{tab:MSE1}-~\ref{tab:corAspects}.

\noindent{\bf Language}: The stylistic features (factor (a) in Table~\ref{tab:corAspects}) like \textit{assertives, hedges, implicatives, factives, discourse} and \textit{affective} play a significant role in credibility analysis, in conjunction with other language features like {\em topics}.
%%%GW: the table is very hard to interpret, as there is no detail given
%%%also: this seems to refer to explicit labels from NewsTrust, not to our model features ???
%%%I'm afraid the typical reader=reviewer will not get much out of this table
%Table~\ref{tab:corAspects} shows the correlation of various explicit stylistic indicators %rated by users with the overall posting credibility rating.

\noindent{\bf Topics}: Topics are an important indicator for credibility. We measured the influence of the {\em Politics} tag on other topics by their co-occurrence
frequency in the explicit tag sets over all the postings. 
%This allows us to obtain the overall influence of politics on any posting by averaging the %political association of each of its explicit topic tags. 
We found significant influence of Politics on all topics, with an average measure of association of $54\%$ to any topic, and $62\%$ for the overall posting.
%%%GW: the numbers are not really informative, as there is no comparison to other forums/media
The community gets polarized due to different perspectives on topical aspects of news. 
A moderate correlation (factor (b) in Table~\ref{tab:corAspects}) indicates a weak trend of disagreement, measured by the standard deviation in credibility rating (of postings) by users, increasing with its political content. In general, we find that community disagreement for different viewpoints are as follows: Right ($0.80$) > Left($0.78$) > Center($0.65$) > Neutral ($0.63$).

%{\tt GW: better trim this par -- it sounds a lot like very specific observations about NewsTrust!!!!!}\\

\noindent{\bf Users}: User engagement features are strong indicators of expertise.
%Users in the community have their own perspectives and expertise on various topics. Modeling user-specific topic perspectives and expertise helps to capture their user judgments on posting credibility.
Although credibility is ultimately subjective, experts show moderate agreement (factor (c) in Table~\ref{tab:corAspects}) on highly credible
postings. There is a moderate correlation (factor (d) in Table~\ref{tab:corAspects}) between feedback received by a user on his
ratings from community, and his expertise.

\noindent{\bf Sources}: Various traits of a source like viewpoint, format and topic expertise are strong indicators of trustworthiness.
In general,
science and technology websites (e.g., discovermagazine.com, nature.com, scientificamerican.com), investigative reporting and non-partisan sources (e.g., truthout.org, truthdig.com, cfr.org), book sites (e.g., nybooks.com, editorandpublisher.com), encyclopedia (e.g., Wikipedia)
and fact checking sites (e.g., factcheck.\\org) rank among the top trusted sources.
Table~\ref{tab:sampletopics} shows the most and least trusted sources on four sample topics.
Overall, sources are considered trustworthy with an average rating of $3.46$ and variance of $0.15$.
Tables \ref{tab:sampleview} and \ref{tab:samplemedia} show the most and least trusted sources on different viewpoints and media types respectively.
Contents from \emph{blogs} %(e.g., juancole.com, dailykos.com)
are most likely to be posted followed by newspaper, magazine %(e.g., rollingstone.com, nybooks.com)
and other online sources. Contents from \emph{wire service, TV} and \emph{radio} are deemed the most trustworthy, although they have the least subscription, followed by \emph{magazines}. %are also very much trustworthy and they constitute a good proportion of the posted postings.

\begin{table}
\center
\small
\begin{tabular}{p{2.5cm}p{2.5cm}p{2.5cm}p{2.5cm}}
\toprule
\multicolumn{1}{c}{\textbf{Money - Politics}} & \multicolumn{1}{c}{\textbf{War in Iraq}} & \multicolumn{1}{c}{\textbf{Media - Politics}} & \multicolumn{1}{c}{\textbf{Green Technology}} \\\midrule
\multicolumn{1}{c}{\textbf{Most Trusted}} &  &  &  \\\midrule
rollingstone.com & nybooks.com & consortiumnews & discovermagazine.com \\
truthdig.com & consortiumnews & thenation.com & nature.com \\
democracynow.org & truthout.org & thedailyshow.com & scientificamerican.com \\
%thenation.com & juancole.com & youtube.com & guardian.co.uk \\
%youtube.com & mcclatchydc & newyorker.com & bbc.co.uk \\
%alternet.org & commondreams & theatlantic.com & gigaom.com \\
%gregpalast.com & prospect.org & salon.com & popsci.com \\
%motherjones.com & npr.org & gregpalast.com & technologyreview.com \\
%dailykos.com & harpers.org & factcheck.org & newscientist.com \\
\toprule
\multicolumn{1}{c}{\textbf{Least Trusted}} &  &  &  \\\midrule
firedoglake.com & crooksandliars & rushlimbaugh.com &  \\
suntimes.com & timesonline & rightwingnews.com &  \\
trueslant.com & suntimes.com & foxnews.com &  \\
%thehill.com & iht.com & firedoglake.com &  \\
%money.cnn.com & trueslant.com & msnbc.msn.com &  \\
%startribune.com & abcnews.go & crooksandliars.com &  \\
\bottomrule
\end{tabular}
\caption{\hspace{0.3em}Most and least trusted sources on sample topics.}
\label{tab:sampletopics}
\end{table}

\begin{table}[!h]
\center
	\small
	\begin{tabular}{p{2.5cm}p{2.5cm}p{2.5cm}p{2.5cm}}
		\toprule
		\textbf{Left}&\textbf{Right}&\textbf{Center}&\textbf{Neutral}\\\midrule
		\textbf{Most} & \textbf{Trusted}\\\midrule
		democracynow, truthdig.com, rollingstone.com & courant.com, opinionjournal.com, townhall.com & armedforces- journal.com, bostonreview.net & spiegel.de,cfr.org, editorandpublisher.com
		\\\midrule
		\textbf{Least} & \textbf{Trusted}\\\midrule
		crooksandliars, suntimes.com, washingtonmonthly.com & rightwingnews, foxnews.com, weeklystandard.com & sltrib.com, examiner.com, spectator.org & msnbc.msn.com, online.wsj.com, techcrunch.com
		\\\bottomrule
	\end{tabular}
	\caption{\hspace{0.3em}Most and least trusted sources with different viewpoints.}
	\label{tab:sampleview}
\end{table}

\begin{table}
\center
\small
\begin{tabular}{p{2.5cm}p{2.5cm}p{2.5cm}p{2.5cm}}
\toprule
\multicolumn{1}{c}{\textbf{Magazine}} & \multicolumn{1}{c}{\textbf{Online}} & \multicolumn{1}{c}{\textbf{Newspaper}} & \multicolumn{1}{c}{\textbf{Blog}} \\
\toprule
\multicolumn{1}{c}{\textbf{Most Trusted Sources}} &  &  &  \\\midrule
rollingstone.com & truthdig.com & nytimes.com & juancole.com \\
nybooks.com & cfr.org & nola.com & dailykos.com \\
thenation.com & consortiumnews & seattletimes & huffingtonpost \\
%nature.com & truthout.org & guardian.co.uk & nytimes.com \\
%newyorker.com & youtube.com & csmonitor.com & scholarsandrog. \\
%theatlantic.com & propublica.org & baltimoresun.com & thinkprogress \\
%motherjones.com & theatlantic.com & ft.com & wired.com \\
%economist.com & alternet.org & citypaper.com & globalvoices. \\
%tnr.com & salon.com & miamiherald.com & glenngreenwald \\
\midrule
\multicolumn{1}{c}{\textbf{Least Trusted Sources}} &  &  &  \\
\midrule
weeklystandard.com & investigativevoice & suntimes.com & rightwingnews\\
commentarymagazine & northbaltimore & nydailynews.com & firedoglake.com \\
nationalreview.com & hosted.ap.org & dailymail.co.uk & crooksandliars \\
%money.cnn.com & online.wsj.com & online.wsj.com & techcrunch.com \\
%forbes.com & wbal.com & telegraph.co.uk & buzzmachine \\
%baltimoremagazine.net & nationalreview & iht.com &  \\
%city-journal.org & baltimore & thehill.com &  \\
% & trueslant.com & chicagotribune.com &  \\
% & thehill.com & startribune.com &  \\
 \bottomrule
\end{tabular}
\caption{\hspace{0.3em}Most and least trusted sources on different types of media.}
\label{tab:samplemedia}
\end{table}

\noindent{\bf Interactions}: In principle, there is a moderate correlation between \emph{trustworthy} sources generating \emph{credible} postings (factor (e) in Table~\ref{tab:corAspects}) identified by \emph{expert} users (factor (f) in Table~\ref{tab:corAspects}). A negative sign of correlation indicates decrease in disagreement or MSE with increase in expertise.
In a community, we can observe {\em moderate} signals of interaction between various factors that characterize users, postings, and sources. Our CCRF model brings all these features together to build a {\em strong} signal for credibility analysis.

%% file: main/chapter-credibility-analysis/conclusions.tex
In this chapter, we proposed a framework for credibility analysis of postings generated by users and sources in online communities (e.g., health and news). We analyzed the effect of different factors like \textit{writing style, topics}, and \textit{perspectives} of users and sources on ascertaining the credibility of postings. These factors and their mutual interactions are the features of two probabilistic graphical models --- specifically, i) a semi-supervised Conditional Random Field for credibility {\em classification}, and ii) a continuous Conditional Random Field for credibility {\em regression} --- for jointly capturing \textit{credibility} of postings, \textit{trustworthiness} of sources, and \textit{expertise} of users.
%, who can perform the role of citizen journalists in a news community.
%Topics, users and sources form a major component of this work, where we %exploit the preference, expertise and viewpoint of the entities to establish %credibility of the content. 

From an application perspective, we demonstrated that our method can reliably identify credible postings, trustworthy sources and expert users in online communities.  
In a novel use-case study in the healthforums, we show that our approach is effective in reliably extracting side-effects of drugs, and filtering out false information prevalent in the healthforums. We designed a user study to identify rare side-effects of drugs --- a scenario where large-scale non-expert data has the potential to complement expert knowledge,
and to identify trustworthy users in the community one would want to follow for certain topics. In the healthforum setting, we believe that our model can be a strong asset for possible in-depth analysis, like determining the specific conditions (age, gender, social group, life style, other medication, etc.) under which side-effects are observed. 

In another use-case study, we presented the first full-fledged analysis of credibility, trust, and expertise
in news communities, where our model identified expert users who can perform the role of {\em citizen journalists}. The proposed model can also be used for tasks like crowdsourcing aggregation, ensemble learning, and learning to rank --- where, we need to aggregate information from multiple sources (e.g., several weak learners, annotators) taking into account their mutual interactions, and weighing each source by its reliablility for the given task.

\comment{
Although our model achieves high accuracy in most of the test cases, 
it suffers from the usage of a simple Information Extraction (IE) machinery to identify possible side-effects in statements. The tool misses out on certain kinds of paraphrases (e.g. ``nightmares'' and ``unusual dream'' for Xanax) resulting in a drop in recall. We believe a more powerful IE approach can further boost the quality of our approach.
}

%% file: main/chapter-temporal-evolution/main.tex
\chapter{Temporal Evolution of Online Communities}
\label{chap:temporal}

\section{Introduction}
\input{main/chapter-temporal-evolution/introduction}

\section{Discrete Experience Evolution}
\label{sec:discrete}

\input{main/chapter-temporal-evolution/discrete-model/compmodel}
\input{main/chapter-temporal-evolution/discrete-model/features}
\input{main/chapter-temporal-evolution/discrete-model/relatedmodels}
\input{main/chapter-temporal-evolution/discrete-model/probinference}
\input{main/chapter-temporal-evolution/discrete-model/experiments}
%\input{main/chapter-temporal-evolution/discrete-model/usecases}

%\input{relatedwork}
%\input{conclusion}
%\pagebreak

\newpage
\section{Continuous Experience Evolution}
\label{sec:continuous}

\input{main/chapter-temporal-evolution/cont-model/compmodel}
\input{main/chapter-temporal-evolution/cont-model/inference}
\input{main/chapter-temporal-evolution/cont-model/experiments}

\input{main/chapter-temporal-evolution/usecases}

\input{main/chapter-temporal-evolution/conclusion}
%\input{relatedwork}
%\input{conclusion}

%% file: main/chapter-temporal-evolution/introduction.tex
Chapter~\ref{chap:framework} demonstrated the importance of modeling trustworthiness and expertise of users and sources for credibility analysis in online communities. Intuitively, postings from users and sources who are experts (or experienced) on a given topic are more reliable than those from amateur users. For instance, {\tt The Wall Street Journal} and {\tt National Geographic} are authoritative sources for postings related to economic policies and environmental matters, respectively. Similarly, experienced members in health and news communities can act as a proxy for medical experts and citizen journalists in the respective communities contributing credible information.  

{\em However, experience is not a static concept; instead it evolves over time. A user (or source) who was not an expert (or experienced) a few years back could have gained maturity over time}. 

In this chapter, we study the {\em temporal evolution of users' experience} in a collaborative filtering framework~\cite{korenKDD2008} in review communities (like, movies, beer, and electronics) --- where we recommend items to users based on their level of {\em maturity or experience} to consume them. Later (refer to Chapter~\ref{chap:helpful}) we propose an approach to exploit this notion of evolving user experience to extract {credible}, and {helpful} postings from online review communities.  

A simplistic way of mapping the task of item recommendation to our previous discussions on credibility analysis in Chapter~\ref{chap:framework} is the following. We can consider the side-effects ($Y$) of drugs ($X$) (SPO triples like $X$\_Causes\_$Y$) in health communities, and postings (i.e. articles) from sources in news communities to be {\em items} in a collaborative filtering framework~\cite{korenKDD2008}, on which {\em users} write {\em reviews} or assign {\em ratings} to {\em items} at different {\em timepoints}~\cite{KorenKDD2010}. Given such a setting, the objective can be to retrieve top-ranked items based on their credibility scores, top-ranked credible postings on any item, and top-ranked users based on their experience etc.
In the next section, we give further motivation for the temporal evolution of users' experience in online communities for recommendation tasks.

\section {Motivation and Approach}

\noindent{\bf State-of-the-Art and Its Limitations:} Collaborative filtering algorithms are at the heart of recommender systems
for items like movies, cameras, restaurants and beer.
Most of these methods exploit user-user and item-item similarities
in addition to the history of user-item ratings --- similarities being
based on latent factor models over user and item features~\cite{koren2011advances}, and more recently
%Yehuda Koren, Robert M. Bell: Advances in Collaborative Filtering. Recommender Systems Handbook 2011: 145-186
%\cite{xiongSDM2010, XiangKDD2010, KorenKDD2010, Gunnemann2014}.
%In addition, recent work has also considered 
on explicit links and interactions
among users~\cite{GuhaWWW2004,West-etal:2014}.
%GW: add a few citations here

%factor time
All these data evolve over {\em time}
leading to bursts in item popularity and other phenomena like anomalies\cite{Gunnemann2014}.
State-of-the-art recommender systems capture these temporal aspects
by introducing global bias components that reflect the evolution of
the user and community as a whole\cite{KorenKDD2010}.
A few models also consider changes in the social neighborhood
of users\cite{MaWSDM2011}.
What is missing in all these approaches, though, is the awareness of
how {\em experience} and {\em maturity} levels evolve in {\em individual users}.

%role of user experience
Individual experience is crucial in how users appreciate items, and thus
react to recommendations. For example, a mature cinematographer
would appreciate tips on art movies much more than
recommendations for new blockbusters. Also, the facets of an item that a user focuses on change with experience. For example, 
a mature user pays more attention to narrative, light effects, and style
rather than to actors or special effects.
Similar observations hold for ratings of wine, beer, food, etc.

%\cite{mcauleyWWW2013} considered only the {\em rating behavior} of users in harnessing their experience.
Our approach
advances state-of-the-art by tapping review texts,
modeling their properties as latent factors, and 
using them to
explain and predict item ratings as a function of a user's experience evolving over time.
Prior works considering review texts (e.g., \cite{mcauleyrecsys2013, wang2011, mukherjee2014JAST, lakkarajuSDM2011, wang2011})
did this only to learn topic similarities in a
static, snapshot-oriented manner, without considering time at all. The only prior work~\cite{mcauleyWWW2013}, considering time, ignores the text of user-contributed reviews in harnessing their experience. However, 
user experience and their interest in specific item facets
at different timepoints can often be observed only {\em indirectly}
through their ratings, and more {\em vividly} through her vocabulary and writing style in reviews.
%Julian J. McAuley, Jure Leskovec: Hidden factors and hidden topics: understanding rating dimensions with review text. RecSys 2013: 165-172

%%%%%%%%%%%%%%%%%%%%%%%%%%%%%%%%%

Consider the reviews and ratings by a user on a {\tt Canon DSLR} camera about the facet {\em lens} at two different timepoints in his lifecycle in the electronics review community.

\example{
[Posted on: August, 1997]: My first DSLR. Excellent camera, takes great pictures in HD, without a doubt it brings honor to its name. [Rating: 5]

[Posted on: October, 2012]: The EF 75-300 mm lens is only good to be used outside. The 2.2X HD lens can only be used for specific items; filters are useless if ISO, AP, ... The short 18-55mm lens is cheap and should have a hood to keep light off lens. [Rating: 3]
}

\noindent The user was clearly an amateur at the time of posting the first review; whereas, he is clearly more experienced a decade later while writing the second review, and more reserved about the lens quality of that camera model.

Future recommendations for this user should take into consideration her evolved maturity at the current timepoint.

As another example, consider the following reviews of Christopher Nolan movies where the facet of interest is the non-linear {\em narrative style}.
\example{
User 1 on Memento (2001): ``Backwards told is thriller noir-art empty ultimately but compelling and intriguing this.''
 
User 2 on The Dark Knight (2008): ``Memento was very complicated. The Dark Knight was flawless. Heath Ledger rocks!''

User 3 on Inception (2010): ``Inception is a triumph of style over substance. It is complex only in a structural way, not in terms of plot. It doesn't unravel in the way Memento does.''
}

\noindent The first user does not appreciate complex narratives, making
fun of it by writing her review backwards. The second user prefers simpler
blockbusters. The third user seems to appreciate the complex narration style
of Inception and, more of, Memento. We would consider this maturity level of the more experienced User 3 to generate future recommendations to her.
%We would consider User 3 to be more experienced
%and use this assessment when generating future recommendations to her.

%%%%%%%%%%%%%%%%%%%%%%%%%%%%%%

We model the joint evolution of {\em user experience}, interests in specific {\em item facets}, {\em rating behavior} and {\em writing style} (captured by her language model) in a community. As only item ratings and review texts are directly observed, we capture a user's experience and interests by a latent model learned from her reviews, and vocabulary. % and stylistic features.
All this is conditioned on {\em time}, considering the {\em maturing rate} of
a user. Intuitively, a user gains experience not only by writing many reviews, but she also needs to continuously improve the quality of her reviews. This varies for different users, as some enter the community being experienced.
%So our model is per timepoint and predicts future behavior.
This allows us to generate individual recommendations that
take into account the user's maturity level and interest in specific facets
of items, at different timepoints.

{\em We propose two approaches to model this evolving user experience, and her writing style: the first approach considers a user's experience to progress in a {\bf discrete} manner (refer to Section~\ref{sec:overview-discrete} for overview); whereas, the next approach (refer to Section~\ref{sec:overview-continuous} for overview) addresses several drawbacks of this discrete evolution, and proposes a natural and {\bf continuous} mode of temporal evolution of a user's experience, and her language model.}

\subsection{Discete Experience Evolution}
\label{sec:overview-discrete}

%more technical: generative model ...
\noindent{\bf Approach:} In the first approach, we assume that the user experience level is categorical with discrete levels (e.g., $[1,2,3,\cdots, E]$), and that users progress from each level to the next in a discrete manner. The experience level of each user is considered to be a {\em latent} variable that evolves over time conditioned on the user's progression in the community.

We develop a generative HMM-LDA model for a user's
evolution, where the Hidden Markov Model (HMM) traces her 
latent experience progressing over time, and the Latent Dirichlet Allocation (LDA) model captures her interests in specific item facets as a function of her (again, latent) experience level.
The only explicit input to our model is the ratings and review texts
upto a certain timepoint; everything else -- especially the user's
experience level -- is a latent variable.
The output is the predicted ratings for the user's reviews following
the given timepoint.
In addition, we can derive interpretations of a user's experience
and interests by salient words in the distributional vectors for
latent dimensions. 
Although it is unsurprising to see users writing sophisticated words with more experience, we observe something more interesting. For instance in  specialized communities like {\tt \href{www.beeradvocate.com}{beeradvocate.com}} and {\tt \href{www.ratebeer.com}{ratebeer.com}}, experienced users write more descriptive and {\em fruity} words to depict the beer taste (cf. Table~\ref{tab:beerTopics}).
Table~\ref{tab:facetWords} shows a snapshot of the words used by users at different experience levels to depict the facets {\em beer taste}, {\em movie plot}, and {\em bad journalism}, respectively.

%experimental results
\comment{
We apply our model to $12.7$ million ratings from $0.9$ million users on $0.5$ million items in five different communities on movies, food, beer, and news media, achieving an improvement of $5\%$ to $35\%$ for the mean squared
error for rating predictions over several competitive baselines.
%the baseline work \cite{mcauleyWWW2013}, and substantial improvement over other competitive baselines including content-based methods. 
We also show that users at the same (latent) experience level 
do indeed exhibit similar vocabulary, and facet interests.
%Finally, a use-case study with human judges assessing the experience levels of
%observed users, demonstrates that our model captures the
%maturity of users fairly well.
Finally, a use-case study in a news community to identify experienced {\em citizen journalists} demonstrates that our model captures user maturity fairly well.
}

\noindent{\bf Contributions:} This discrete-experience evolution model is discussed in-depth in Section~\ref{sec:discrete} that introduces the following novel contributions: 
\begin{itemize}
\item[a)] The first model (Section~\ref{sec:overview}) to consider the progression of user experience as expressed through the text of item reviews, thereby elegantly combining text and time.
\item[b)] An approach (Section~\ref{subsec:build},~\ref{sec:inference1}),to capture the natural {\em smooth} temporal progression in user experience factoring in the {\em maturing rate} of the user, as expressed through her writing. 
\item[c)] Offers interpretability by learning the vocabulary usage of users at different levels of experience.
\item[d)] A large-scale experimental study (Section~\ref{sec:experiments1}) in {\em five} real world datasets from different communities like movies, beer, and food.
\end{itemize}

%{\tt GW: if we want to place tables with salient words in the intro, here would be the right place !!!!!}
\begin{table}
\small
\centering
\begin{tabular}{p{2cm}p{2.5cm}p{2.5cm}p{3cm}}
\toprule
\bf{Experience} & \bf{Beer} &\bf{Movies} &\bf{News}\\
\midrule
Level 1 & bad, shit & stupid, bizarre & bad, stupid\\
Level 2 & sweet, bitter & storyline, epic & biased, unfair\\
Level 3 & caramel finish, coffee roasted & realism, visceral, nostalgic & opinionated, fallacy, rhetoric\\
\bottomrule
\end{tabular}
\caption{Vocabulary at different experience levels.}
\label{tab:facetWords}
\end{table}

\subsection{Continuous Experience Evolution}
\label{sec:overview-continuous}

{\bf Limitations of Discrete Evolution Models:} Section~\ref{sec:overview-discrete} gives the motivation for the evolution of user experience and how it affects ratings.
%has first been studied in~\cite{mcauleyWWW2013,Subho:ICDM2015}. 
However, the proposed approach and its precursor~\cite{mcauleyWWW2013} make the simplifying assumption that user experience is
{\em categorical} with discrete levels (e.g. $[1, 2, 3, \ldots, E]$), and that
users progress from one level to the next in a discrete manner. % resulting in a rigid model. 
As an artifact of this assumption, the experience level of a user changes abruptly by one transition. 
%%%GW: losing experience seems totally unintuitive and pointless, and our model doesn't capture it either , does it?
%Second, the models make the strict assumption that a user either stays at the current %(discrete) level of experience  or moves up to the next level -- losing experience, i.e. a %decrease in value, is not captured resulting in a very rigid model. 
%Third, 
Also, an undesirable consequence of the discrete model is that
all users at the same level of experience are treated similarly, although their maturity could still be far apart (if
we had a continuous scale of measuring experience). Therefore, the assumption of {\em exchangeability} of reviews --- for the latent factor model in the discrete approach --- for users at the same level of experience may not hold as the language model changes.

The prior work~\cite{mcauleyWWW2013} assumes user {\em activity} (e.g., number of reviews) to play a major role in experience evolution, which biases the model towards highly active users
(as opposed to an experienced person who posts only once in a while).
%In our work, instead, we model the progression of user experience through their language model evolution which implicitly rewards long-term contributors through vocabulary acquisition. 
In contrast, the discrete version of our own approach (refer to Section~\ref{sec:overview-discrete}) captures {\em interpretable} evidence for a user's experience level using her vocabulary, cast into a 
language model with latent facets.
However, this approach also exhibits the drawbacks of discrete levels of
experience, as discussed above. 

Therefore, we propose a {\em continuous} version of experience evolution that overcomes these limitations by modeling
the evolution of user experience, and the corresponding 
language model, as a {\em continuous-time} stochastic process. We model time {\em explicitly} in this work, in contrast to the prior works. 

{\noindent \bf Approach:} This is the first work to develop a continuous-time model of user experience 
and language evolution.
Unlike prior work, we do not rely on explicit features like ratings or number of reviews.  
Instead, 
%show that the number of posts written by a user does not have a strong influence on %experience progression, 
%unless they are written over a long period of time. 
we capture a user's experience by a latent language model learned from
the user-specific vocabulary in her review texts.
%more technical: generative model ...
We present a generative model where the user's experience and language model evolve according to a Geometric Brownian Motion (GBM) and Brownian Motion process, respectively. 
Analysis of the GBM trajectory of users offer interesting insights;
for instance, users who reach a high level of experience progress faster 
than those who do not, and also exhibit a comparatively higher variance.
Also, the number of reviews written by a user does not have a strong influence, 
unless they are written over a long period of time.

The facets in our model (e.g., narrative style, actor performance, etc. for  movies) are generated using Latent Dirichlet Allocation. 
%We show that our model is the continuous analog 
%of the discrete model in our prior work~\cite{XXX}. %\todo{is this still correct?}
%%%GW: I would not discuss the relation to our ICDM '15 work so explicitly
User experience and item facets are latent variables,
whereas the observables are
 {\em words}  at explicit {\em timepoints} 
in user reviews. 
%\todo{ time is made explicit in our work.!!}

%%%%%
%%%GW: I think this following paragraph is dispensable - it distracts a bit from the main flow of arguments in the intro
\comment{
As the language model is experience-aware and time is made explicit, 
%\todo{might sound too strong? }, 
we can trace how different words and community norms evolve
with user experience. 
For instance, ``aroma'' of a beer was the dominant norm used by experienced users in the early $2000$, and was later replaced by ``smell''. 
%, although both of their usage, in general, increased over time.
}
%%%%%

The parameter estimation and inference for our model are challenging since 
we combine discrete multinomial distributions (generating words per review) with a continuous Brownian Motion process for the language models' evolution, and a continuous Geometric Brownian Motion (GBM) process for the user experience. 
%%%GW: the earlier text only talked about inference - but isn't parameter estimation, to construct the model, equally challenging ???
%%%or do you mean by "inference" both model parameter estimation and the statistical inference when applied to new data? perhaps, with an unsupervised model, there is no difference

{\bf Contributions:}
To solve this technical challenge, we present an inference method consisting of three steps: a) estimation of user experience from a user-specific GBM using the 
Metropolis Hastings algorithm, 
b) estimation of the language model evolution by Kalman Filter, and c) estimation of latent facets using Gibbs sampling. Our experiments, with
real-life data from five different communities on movies, food, beer and
news media, show that the three components {\em coherently} work together and
yield a better fit of the data (in terms of log-likelihood) than the previously
best models with discrete experience levels.
%%%GW: instead of data likelihood, isn't it more usual to use perplexity for this purpose?
%
%experimental results
%We apply our model to $12.7$ million ratings from $0.9$ million users on $0.5$ million %items in five different communities on movies, food, beer, and news media. We show %that the generalized continuous version obtains a better fit over the data 
%in terms of log-likelihood. 
We also achieve an improvement of ca. $11\%$ to $36\%$ for the mean squared
error for predicting user-specific ratings of items compared to the baseline of~\cite{mcauleyWWW2013}, and the discrete version of the model (refer to Section~\ref{sec:overview-discrete} for overview).
%%%GW: what about our ICDM'15 paper being used as a baseline?
%Finally, a use-case study with human judges assessing the experience levels of
%observed users, demonstrates that our model captures the
%maturity of users fairly well.
%Finally, we present a use-case study with a news-media community, where experience-aware models can be used to identify experienced {\em citizen journalists} --- where our method performs well in capturing user maturity.

%\todo{should we add something like:
This continuous-experience evolution model is discussed in-depth in Section~\ref{sec:continuous} that introduces the following novel contributions:
%In summary, Section~\ref{sec:continuous} introduces the following novel contributions:
\begin{itemize}
%\squishlist
	\item [a)] Model: We devise a probabilistic model (Section~\ref{sec:cont-model}) for tracing {\em continuous} evolution of  {\em user experience}, combined with a language model for facets that explicitly captures smooth evolution over time.
	\item [b] Algorithm: We introduce an effective learning algorithm (Section~\ref{sec:inference}), that infers each users' experience progression, time-sensitive language models, 
and latent facets of each word. 
%%%GW: models, not model? one per time, constructed on demand? facets? a word can belong to multiple, right?
	\item [c)] Experiments: We perform extensive experiments (Section~\ref{sec:experiments}) with five real-word
 datasets, together comprising of $12.7$ million ratings from $0.9$ million users on $0.5$ million items, and demonstrate substantial improvements of our method over state-of-the-art baselines.
\end{itemize}
%}

As an interesting use-case application of our experience-evolution model, we perform an experimental study (Section~\ref{sec:usecases}) in a news community to identify {\em experienced} members who can play the role of {\em citizen journalists} in the community. This study is similar to Section~\ref{subsec:expert-users} for credibility analysis --- with the additional incorporation of temporal evolution.

%%%%%%%%%%%%%%%%%%%%%%%%%%%%%%%%%%%%
%GW: rest is leftover from prior paper ?

%% file: main/chapter-temporal-evolution/discrete-model/compmodel.tex
%\subsection{Overview}

\subsection{Model Dimensions}
\label{sec:overview}

Our approach is based on the intuition that there is a strong coupling between the \emph{facet preferences} of a user, her \emph{experience}, \emph{writing style} in reviews, and \emph{rating behavior}.
All of these factors jointly evolve with \emph{time} for a given user.
%For example, the preference of a user may change to the ``sci-fi'' genre with time, from the ``drama'' genre. Even within the same genre, the user may start appreciating intellectual movies like ``Gattaca'' and ``Inception'' more with time.

We model the user experience progression through discrete stages, so a state-transition model is natural. Once this decision is made, a Markovian model is the simplest, and thus natural choice. This is because the experience level of a user at the current instant $t$ depends on her experience level at the previous instant $t$-$1$. As experience levels are latent (not directly observable), a Hidden Markov Model is appropriate. Experience progression of a user depends on the following factors:\\\\

%In order to capture the \emph{experience progression} of a user with time, we consider a \emph{Hidden Markov Model} (HMM) where the experience level of the user is a \emph{latent} variable. The transition to an experience level at time $t$ depends on the experience level of the user at time $t$-$1$ and the following factors:
%\vspace{2em}

\begin{itemize}
\item \emph{Maturing rate} of the user which is modeled by her \emph{activity} in the community. The more engaged a user is in the community, 
the higher are the chances that she gains experience and advances in writing sophisticated reviews,
and develops taste to appreciate specific facets.
\item \emph{Facet preferences} of the user in terms of focusing on particular
facets of an item (e.g., narrative structure rather than special effects).
With increasing maturity, the taste for particular facets becomes more refined.
\item \emph{Writing style} of the user,  as expressed by the language model at her current level of experience. More sophisticated vocabulary and writing style indicates higher probability of progressing to a more mature level.
\item \emph{Time difference} between writing successive reviews. 
It is unlikely for the user's experience level to change from that of her last review in a short time span (within a few hours or days).
\item \emph{Experience level difference}: Since it is unlikely for a user to directly progress to say level $3$  from level $1$ without passing through level $2$, the model at each instant decides whether the user should stay at current level $l$, or progress to $l$+$1$.
%
%GW: I would drop the rest, as it is vague and not an explicit part of the our model anyway
\end{itemize}

In order to learn the \emph{facet preferences} and \emph{language model} of a user at different levels of experience, we use \emph{Latent Dirichlet Allocation} (LDA). In this work, we assume each review to
refer to exactly one item. Therefore, the facet distribution of items is 
expressed in the facet distribution of the review documents.

We make the following assumptions for the generative process of writing a review by a user at time $t$ at  experience level $e_t$:

\begin{itemize}
\item A user has a distribution over \emph{facets},
where the facet preferences of the user depend on her experience level $e_t$.
\item A facet has a distribution over \emph{words} where the words used to describe a facet depend on the user's vocabulary at experience level $e_t$.
%(cf.  Table~\ref{tab:beerTopics}). 
%Table~\ref{tab:beerTopics} in Section~\ref{sec:experiments} shows the words 
%chosen by users at different levels of experience 
%to describe the facet ``taste'' in a beer community. 
Table \ref{tab:amazonTopics} shows salient words for two facets of
Amazon movie reviews at different levels of user experience, automatically extracted by our latent model.
The facets are latent, but we can interpret them as {\em plot/script} and
{\em narrative style}, respectively.
\end{itemize}
%{\tt GW: todo: shrink Table , leave out some of the words !!!!! also need to 
%decide if table is better placed or in experiments section !!!!!}

\begin{table}
\small
\centering
\begin{tabular}{p{12cm}}
\toprule
\textbf{Level 1:} stupid people supposed wouldnt pass bizarre totally cant\\
\textbf{Level 2:}storyline acting time problems evil great times didnt money ended simply falls pretty\\
\textbf{Level 3:} movie plot good young epic rock tale believable acting\\
\textbf{Level 4:} script direction years amount fast primary attractive sense talent multiple demonstrates establish\\
\textbf{Level 5:} realism moments filmmaker visual perfect memorable recommended genius finish details defined talented visceral nostalgia\\
\midrule
\textbf{Level 1:} film will happy people back supposed good wouldnt cant\\
\textbf{Level 2:} storyline believable acting time stay laugh entire start funny\\
\textbf{Level 3 \& 4:} narrative cinema resemblance masterpiece crude undeniable admirable renowned seventies unpleasant myth nostalgic\\
\textbf{Level 5:} incisive delirious personages erudite affective dramatis nucleus cinematographic transcendence unerring peerless fevered\\
\bottomrule
\end{tabular}
\caption{Salient words for two facets at
five experience levels in movie reviews.}
\label{tab:amazonTopics}
\end{table}

\noindent As a sanity check for our assumption of the coupling between user {\em experience}, {\em rating behavior}, {\em language} and {\em facet preferences}, we perform experimental studies reported next.

%GW: moved this to the begin of the section on the joint model

%% file: main/chapter-temporal-evolution/discrete-model/features.tex
%\section{Experimental Study}
\label{sec:study}

\newpage

\subsection{Hypotheses and Initial Studies}

%{\tt GW: crop the first figure  so that "Heatmap - KL divergence" is removed - no need for this, the text says it all !!!!}

\noindent{\bf Hypothesis 1: Writing Style Depends on Experience Level.}

We expect users at different  experience levels to have divergent Language Models (LM's) --- with experienced users having a more sophisticated writing style and vocabulary than amateurs.
To test this hypothesis, we performed initial studies over two popular communities\footnote{Data available at \href{http://snap.stanford.edu/data/}{http://snap.stanford.edu/data/}}:
1) BeerAdvocate ({\tt \href{http://www.beeradvocate.com}{beeradvocate.com}}) with $1.5$ million reviews from $33,000$ users
and 2) Amazon movie reviews ({\tt \href{http://www.amazon.com}{amazon.com}}) with $8$ million reviews from $760,000$ users.
Both of these span a period of about 10 years.

In BeerAdvocate, a user gets \emph{points} on the basis of  likes received for her reviews, ratings from other 
users, number of posts written, diversity and number of beers rated, time in the community, etc. 
We use this points measure as a proxy for the user's \emph{experience}. 
In Amazon, reviews get \emph{helpfulness} votes from other users.  For {each user},
we aggregate these votes over all her reviews and take this as a proxy for her experience.

\begin{figure}[h]
        \centering
        \begin{subfigure}[b]{0.5\textwidth}
                \includegraphics[width=\textwidth]{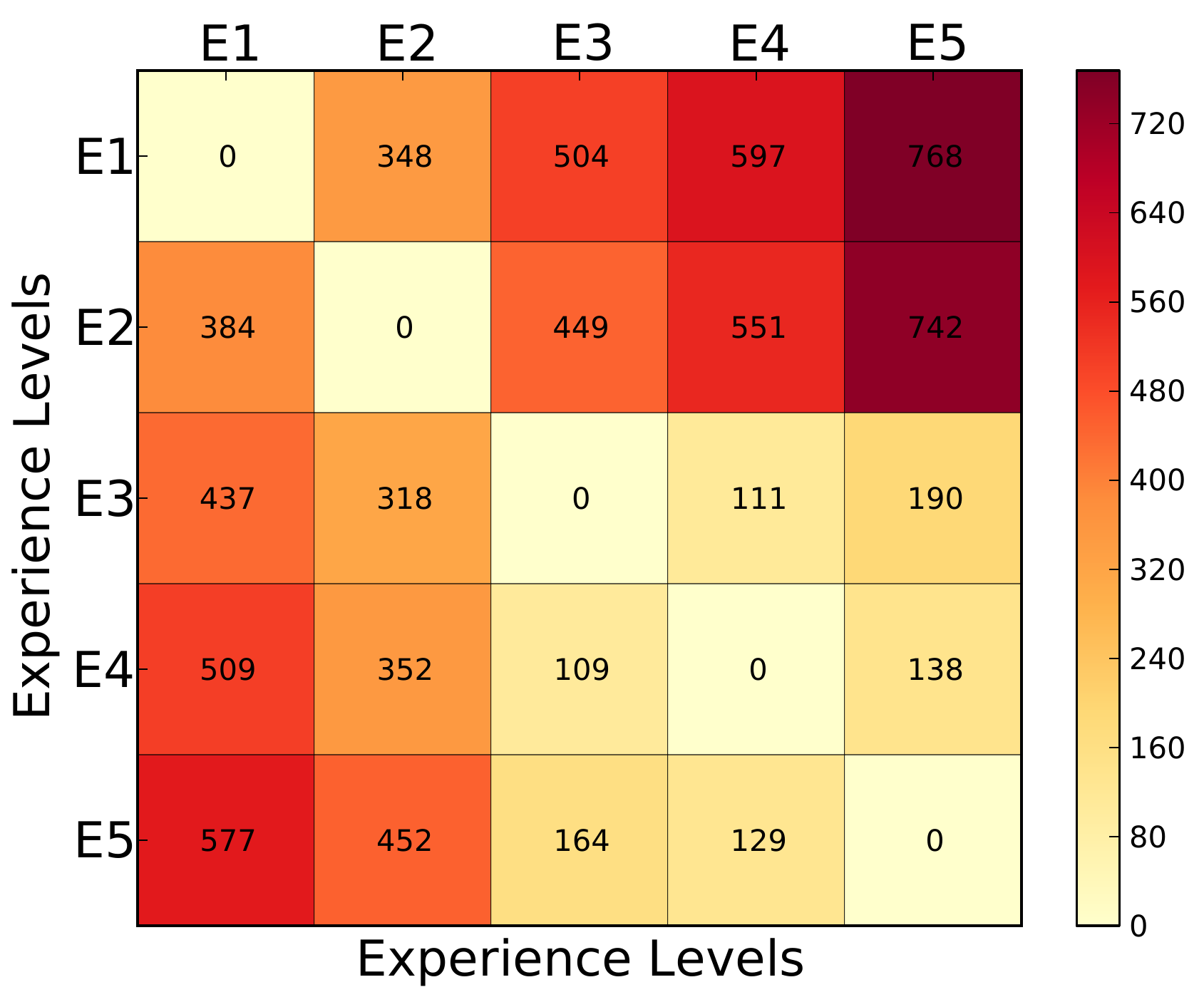}
 \caption{Divergence of language model\\ as a function of experience.}
 \label{fig:expertiseLangDivergence}
        \end{subfigure}%
        ~\hfill
        \begin{subfigure}[b]{0.5\textwidth}
                \includegraphics[width=\textwidth]{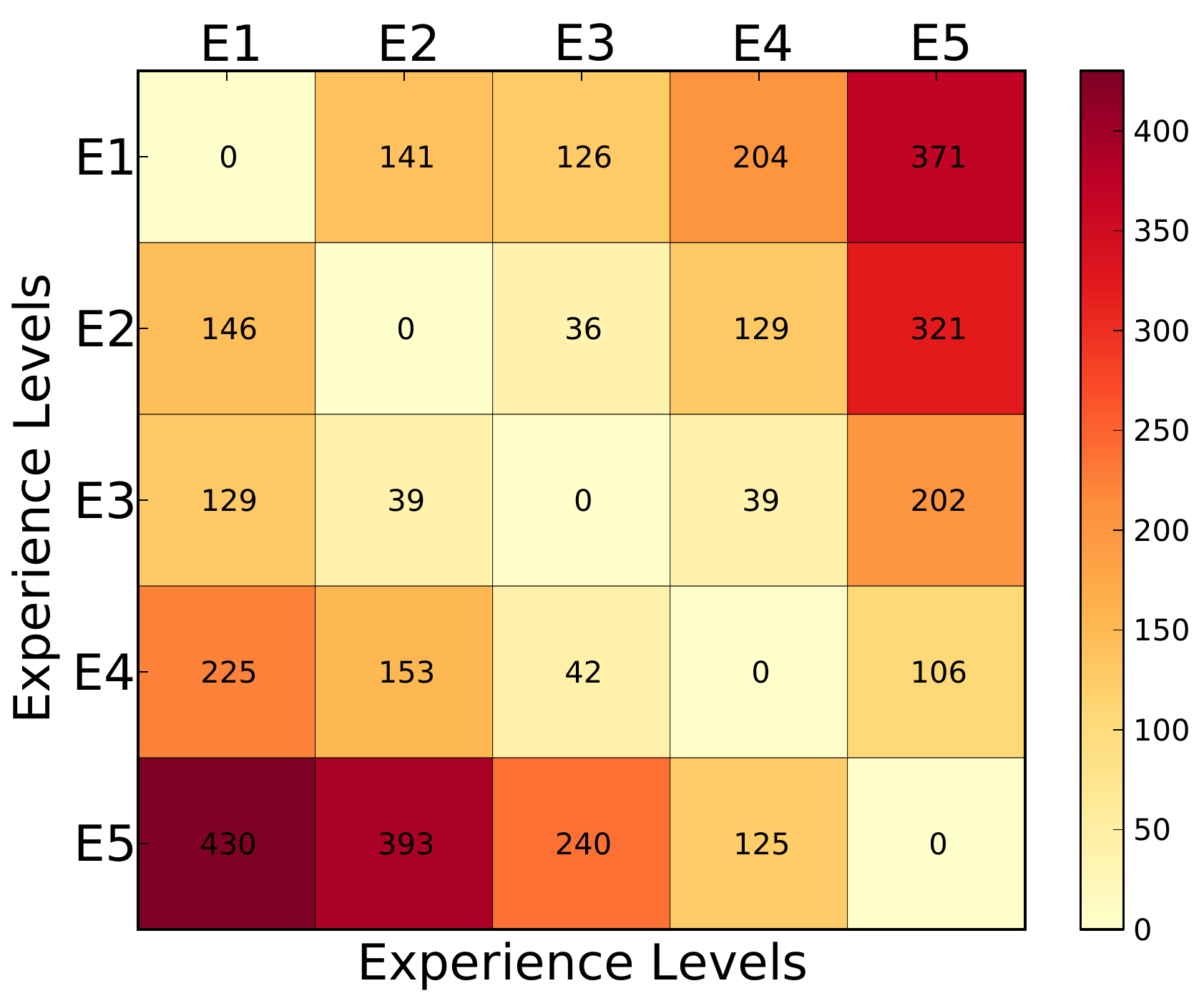}
 \caption{Divergence of facet preference\\ as a function of experience.}
 \label{fig:expertiseFacetDivergence}
        \end{subfigure}
         \caption{$KL$ Divergence as a function of experience.}
         \label{fig:expertiseDivergence}
\end{figure}

We partition the users into $5$ bins, based on the points / helpfulness votes received, each representing one of the experience levels. For each bin, we aggregate the review texts of all users in that bin and construct a unigram language model. 
The heatmap of Figure  \ref{fig:expertiseLangDivergence} shows the \emph{Kullback-Leibler} (KL) divergence between the LM's of different experience levels, for the BeerAdvocate case. 
The Amazon reviews lead to a very similar heatmap, which is omitted here.
The main observation is that the KL divergence is higher --- the larger the difference is between the experience
levels of two users. This confirms our hypothesis about the coupling of experience and user language.
%
%GW: the point is made, no need for detailed discussion here

\newpage
\noindent{\bf Hypothesis 2: Facet Preferences Depend on Experience Level.}

The second hypothesis underlying our work is that users at similar levels of experience have similar facet preferences.
%, and rating behavior.
In contrast to the LM's where words are {\em observed}, facets are {\em latent} so that validating or
falsifying the second hypothesis
is not straightforward.  We performed a three-step study:

\begin{itemize}
\item We use Latent Dirichlet Allocation (LDA)~\cite{Blei2003LDA} to compute
a latent facet distribution $\langle f_k \rangle$ of \emph{each review}. 
%As mentioned before, we assume each document to be associated to only one item. 
%Therefore the latent facet distribution of the document refers to that of the item associated to the document.
\item We run Support Vector Regression (SVR)~\cite{drucker97} for \emph{each user}. 
The user's item rating in a review is the response variable, with the facet proportions in the review given 
by LDA as features.
 The regression weight $w^{u_e}_k$ is then interpreted as the preference of user $u_e$ for facet $f_k$. 
%$e$ denotes the experience level of the user obtained on binning him on the basis of points / helpfulness votes.
\item Finally, we aggregate these facet preferences for each experience level $e$ to get the corresponding facet preference distribution given by $\ \ <\frac{\sum_{u_e} exp(w^{u_e}_k)}{\#u_e}>$.
\end{itemize}

Figure~\ref{fig:expertiseFacetDivergence} shows the $KL$ divergence between the facet preferences of users at different experience levels in BeerAdvocate. 
We see that the divergence clearly increases with the difference in user experience levels;
this confirms the hypothesis.
The heatmap for Amazon is similar and omitted. 

{\em Note} that Figure~\ref{fig:expertiseDivergence} shows how a {\em change} in the experience level can be detected. This is not meant to predict the experience level, which is done by the model in Section~\ref{sec:inference1}.% and performance evaluated by experiments in Section~\ref{sec:experiments}.
%The analysis of the heatmap is similar to that of the LM's. The entries near the diagonal are low indicating similar %facet preferences of users for close experience levels.

%Although, the trend is obvious from the figure the increase in divergence is not as smooth as that of the LM's. This is also related to the complexity of the task. In this case there are two sets of latent variables corresponding to \emph{experience} and \emph{topics}; whereas in case of LM's only topics are latent as \emph{words} are observed. This also affirms our notion of grounding the work based on \emph{language}, which is aided by other modules. Therefore the resulting distribution has the smoothness governed by that of the language model.

%% file: main/chapter-temporal-evolution/discrete-model/relatedmodels.tex
%\section{Related Work}
%%%GW: we need a real related work section at the end of the paper - perhaps brief, but explicit 

%\section{Basic Models}
\subsection{Building Blocks of our Model}
\label{subsec:build}

%{\tt GW: or should we call these "Building Blocks of our Model" ?????}

Our model, presented in the next section, builds on and compares itself
against various baseline models as follows.

\subsubsection{Latent Factor Recommendation}

According to the standard latent factor model (LFM)~\cite{korenKDD2008}, the rating assigned by a user $u$ to an item $i$ is given by:
\begin{equation}
\label{eq.1}
 rec(u, i) = \beta_g + \beta_u + \beta_i + \langle \alpha_u, \phi_i \rangle
\end{equation}

%{\tt GW: does <...,...> denote a scalar product ????? we need to be explicit here !!!!}
where $\langle .,. \rangle$ denotes a scalar product. $\beta_g$ is the average rating of all items by all users. $\beta_u$ is the offset of the average rating given by user $u$  from the global rating.
Likewise $\beta_i$ is the rating bias for item $i$. 
$\alpha_u$ and $\phi_i$ are the latent factors associated with user $u$ and item $i$, respectively. These latent factors are learned using gradient descent by minimizing the mean squared error ($MSE$) between observed ratings $r(u,i)$ and predicted ratings $rec(u,i)$:
$MSE = \frac{1}{|U|} \sum_{u,i \in U} (r(u,i) - rec(u, i))^2$

\subsubsection{Experience-based Latent Factor Recommendation}

The most relevant baseline for our work is the ``user at learned rate'' model of ~\cite{mcauleyWWW2013},
which exploits that users at the same experience level have similar rating behavior even if their ratings are temporarily far apart. 
Experience of each user $u$ for item $i$ is modeled as a latent variable $e_{u,i} \in \{1...E\}$.
Different recommenders are learned for different experience levels. Therefore Equation~\ref{eq.1} is parameterized as:
\begin{equation}
\label{eq.3}
 rec_{e_{u,i}}(u, i) = \beta_g(e_{u,i}) + \beta_u(e_{u,i}) + \beta_i(e_{u,i}) + \langle \alpha_u(e_{u,i}), \phi_i(e_{u,i}) \rangle
\end{equation}

The parameters are learned using Limited Memory BFGS
%, a quasi-Newton method for non-linear optimization of problems with many variables, 
with the additional constraint that experience levels should be non-decreasing over the reviews written by a user over time. 

However, this is significantly different from our approach. All of these models work on the basis of only user {\em rating behavior}, and ignore the review texts completely. Additionally, the {\em smoothness} in the evolution of parameters between experience levels is enforced via $L_2$ regularization, and does not model the {\em natural} user maturing rate (via HMM) as in our model. Also note that in the above parametrization, an experience level is estimated for each user-item pair. However, it is rare that a user reviews the same item multiple times. In our approach, we instead trace the evolution of users, and not user-item pairs. 

%Most importantly, these models do not use the text of reviews.

\subsubsection{User-Facet Model}
\vspace{-0.5em}

In order to find the facets of interest to a user, \cite{rosenzviUAI2004} extends Latent Dirichlet Allocation (LDA) to include authorship information. 
Each document $d$ is considered to have a distribution over authors. 
We consider the special case where each document has exactly one author
$u$ associated with a Multinomial distribution $\theta_u$ over facets $Z$ with a symmetric Dirichlet prior $\alpha$. The facets have a Multinomial distribution $\phi_z$ over words $W$ drawn from a vocabulary $V$ with a symmetric Dirichlet prior $\beta$. The generative process for a user writing a review is given by Algorithm~\ref{algo.1}.
%{\tt GW: do we really need - and want to keep - the algorithm here ?????}
Exact inference is not possible due to the intractable coupling between $\Theta$ and $\Phi$. Two ways for approximate inference are 
MCMC techniques like Collapsed Gibbs Sampling and Variational Inference.
The latter is typically much more complex and computationally expensive.
In our work, we thus use sampling.

%\comment{
\begin{algorithm}
\label{algo.1}
%\small
\SetAlgoLined
\DontPrintSemicolon
\For {each user $u = 1, ... U$} {choose $\theta_u \sim  Dirichlet(\alpha)$} \;
\For {each topic $z = 1, ... K$} {choose $\phi_z \sim Dirichlet(\beta)$} \;
\For {each review $d = 1, ... D$} {
  Given the user $u_d$ \;
  \For {each word $i = 1, ... N_d$} {
    Conditioned on $u_d$ choose a topic $z_{d_i} \sim Multinomial(\theta_{u_d})$ \;
    Conditioned on $z_{d_i}$ choose a word $w_{d_i} \sim Multinomial(\phi_{z_{d_i}})$ \;
  }
}
\caption{Generative Process for User-Facet Model}
\label{algo.1}
\end{algorithm}
%}

\vspace{-1em}
\subsubsection{Supervised User-Facet Model}
\vspace{-0.5em}

The generative process described above is unsupervised and does not take the ratings in reviews into account. Supervision is difficult to build into MCMC sampling where ratings are continuous values, as in communities like {\tt \href{http://www.newstrust.net}{newstrust.net}}.
%and cannot be modeled by a Multinomial distribution. 
For discrete ratings, a review-specific Multinomial rating distribution $\pi_{d,r}$ can be learned as in~\cite{linCIKM2009, ramageKDD2011}. 
Discretizing the continuous ratings into buckets bypasses the problem to some extent, but results in loss of information. Other approaches~\cite{lakkarajuSDM2011, mcauleyrecsys2013, mukherjee2014JAST} overcome this problem by learning the feature weights separately from the user-facet model.

%GW: I dismissed variational inference above, so we can drop this

A supervised version of the topic model using variational inference is proposed in~\cite{bleiNIPS2007}. It tackles the problem of coupling by removing some of the interactions altogether that makes the problem intractable; and learns a set of variational parameters that minimizes the $KL$ divergence between the approximate distribution and the true joint distribution. However, the flexibility comes at the cost of increasingly complex inference process.

\begin{figure}
\centering
 \includegraphics[scale=0.5]{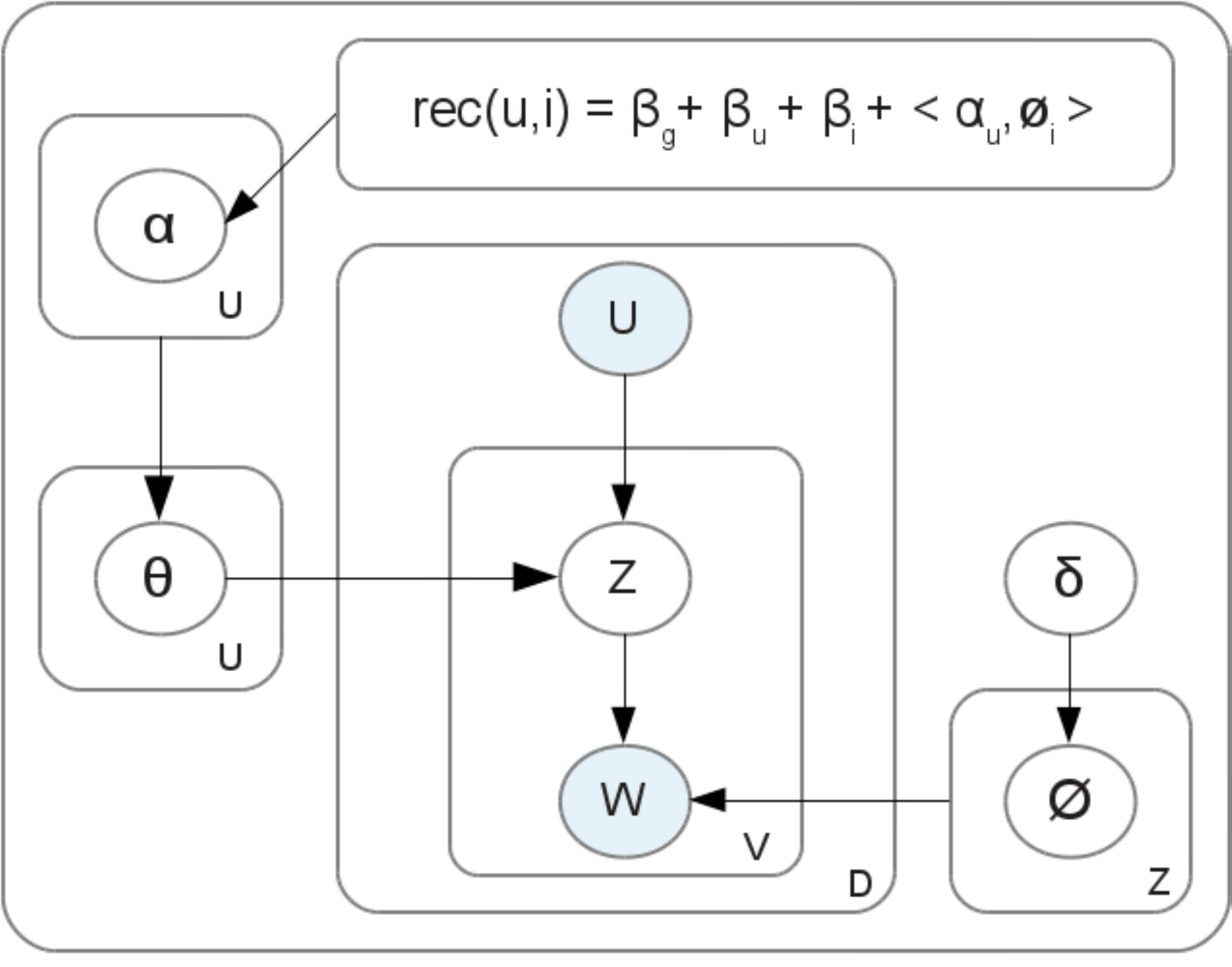}
  \caption{Supervised model for user facets and ratings.}
  \label{fig:1}
\end{figure}
 
An elegant approach using Multinomial-Dirichlet Regression is proposed in~\cite{mimnoUAI2008} to incorporate arbitrary types of observed continuous or categorical features. Each facet $z$ is associated with a vector $\lambda_z$ whose dimension equals the number of features. Assuming $x_d$ is the feature vector for document $d$, the Dirichlet hyper-parameter $\alpha$ for the document-facet Multinomial distribution $\Theta$ is parametrized as $\alpha_{d,z}=exp(x_d^T\lambda_z)$. The model is trained using stochastic \emph{EM} which alternates between 1) sampling facet assignments from the posterior distribution conditioned on words and features, and 2) optimizing $\lambda$ given the facet assignments using L-BFGS.
Our approach, explained in the next section, follows a similar approach to couple 
the User-Facet Model and the Latent-Factor Recommendation Model (depicted in Figure~\ref{fig:1}).
%depicted in Figure~\ref{fig:1}.
%GW: no point in making this forward reference here

%{\tt GW: we sometimes talk about user-facet models (formerly author-topic 
%models) and sometimes about document-facet models - these are the same, but
%a reader may get confused !!!???}

%% file: main/chapter-temporal-evolution/discrete-model/probinference.tex
\subsection{Joint Model: User Experience, Facet Preference, Writing Style}
\label{sec:inference1}

%\subsection{Approach}

\begin{figure}[t]
\centering
 \includegraphics[scale=0.5]{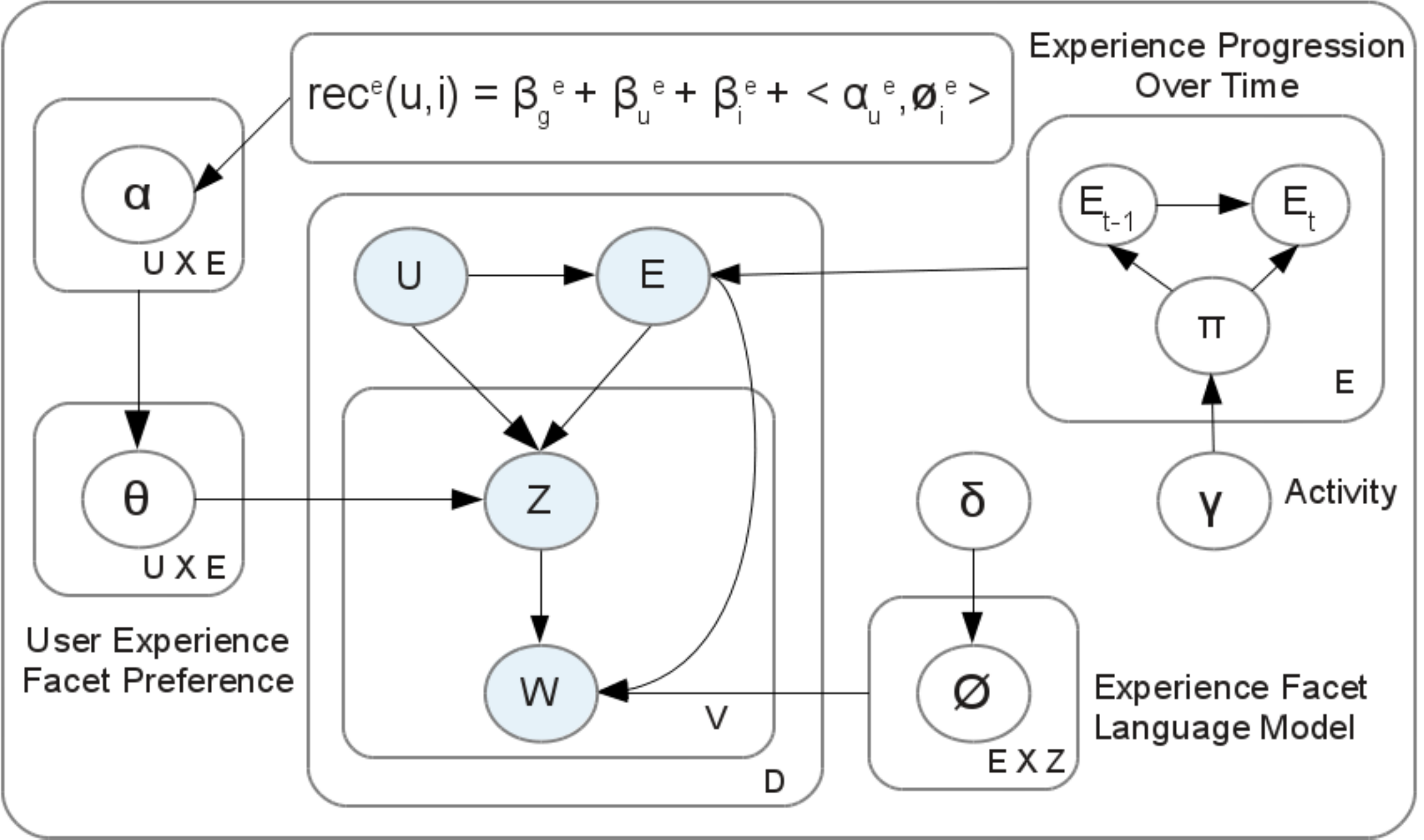}
  \caption{Supervised model for user experience, facets, and ratings.}
  \label{fig:2}
 \end{figure}

We start with a \emph{User-Facet Model (UFM)} (aka. Author-Topic Model~\cite{rosenzviUAI2004}) based on \emph{Latent Dirichlet Allocation} (LDA), where {users} have a distribution over {facets} and facets have a distribution over {words}. This is to determine the facets of interest to a user. These facet preferences can be interpreted as latent item factors in the traditional \emph{Latent-Factor Recommendation Model} (LFM)~\cite{korenKDD2008}. However, the LFM is supervised as opposed to the UFM. 
It is not obvious how to incorporate supervision into the UFM to predict ratings. 
The user-provided ratings of items can take continuous values (in some review communities), 
so we cannot incorporate them into a UFM with a Multinomial distribution of ratings.
%This problem has been tackeld using Variational Inference in case of 
%Supervised Topic Models~\cite{bleiNIPS2007}, 
%where the flexibility comes at the cost of increasingly intractable inference. 
%In our setting, 
We propose an \emph{Expectation-Maximization (EM)} approach to incorporate supervision, where the
latent facets are estimated in an \emph{E-Step} using \emph{Gibbs Sampling}, and \emph{Support Vector Regression} (SVR)~\cite{drucker97} is used in the \emph{M-Step} to learn the feature weights and predict ratings. Subsequently, we incorporate a layer for \emph{experience} in the UFM-LFM model, where the experience levels are drawn from a \emph{Hidden Markov Model} (HMM) in the \emph{E-Step}. The experience level transitions depend on the evolution of the user's \emph{maturing rate}, \emph{facet preferences}, and \emph{writing style} over \emph{time}. The entire process is a supervised generative process of generating a review based on the experience level of a user hinged on our HMM-LDA model.

%%%%%%%%%%%%%%%%%%%%%%%%%%

\subsubsection{Generative Process for a Review}
%by an Experienced User}

Consider a corpus with a set $D$ of review documents denoted by $\{d_1 \dots d_D\}$. For \emph{each user}, all her documents are ordered by timestamps $t$ when she wrote them, such that $t_{d_i}<t_{d_j}$ for $i<j$. Each document $d$ has a sequence of $N_d$ words denoted by $d=\{w_1 \dots .w_{N_d}\}$. 
Each word is drawn from a vocabulary $V$ having unique words indexed by $\{1 \dots V\}$. 
Consider a set of $U$ users involved in writing the documents in the corpus, 
where $u_d$ is the author of document $d$. 
Consider an ordered set of experience levels  $\{e_1,e_2,...e_E\}$ 
%to the document collection depicting the experience level at which the document was written,
where each  $e_i$ is from a set $E$,
and a set of facets $\{z_1,z_2,...z_Z\}$ where each $z_i$ is from a set $Z$ of possible facets. 
%GW: for experience levels the order matters, for facets it's a set, not a sequence - IMO
Each document $d$ is associated with a rating $r$ and an item $i$.

At the time $t_d$ of writing the review $d$, the user $u_d$ has experience level $e_{t_d} \in E$. We assume that her experience level transitions follow a distribution $\Pi$ with a Markovian assumption %$\pi_{e_{t_{d-1}}}$
and certain constraints.
This means the experience level of $u_d$ at time $t_d$ depends on her 
experience level when writing the previous document at time $t_{d-1}$. 

$\pi_{e_i}(e_j)$ denotes the probability of progressing to experience level $e_j$ from experience level $e_i$, with the constraint $e_j \in \{e_i, e_i+1\}$. This means at each instant the user can either stay at her current experience level, or move to the next one.

The experience-level transition probabilities
depend on the \emph{rating behavior}, \emph{facet preferences}, and \emph{writing style} of the user. 
%GW: this paragraph repeats some of the points made in the first paragraph of the section - before starting Subsection 4.1
The progression also takes into account the 1) \emph{maturing rate} of  $u_d$ modeled by the intensity of her activity in the community,
and 2) the \emph{time interval} between writing consecutive reviews.
We incorporate these aspects in 
%an asymmetric Dirichlet 
a prior
for the user's transition rates,
$\gamma^{u_d}$, defined as:
%GW: we need more explanation on how the prior captures these constraints !!!!!
\[
 \gamma^{u_d} = \frac{D_{u_d}}{D_{u_d} + D_{avg}} + {\lambda (t_d - t_{d-1})}
\]

$D_{u_d}$ and $D_{avg}$ denote the number of reviews written by ${u_d}$ and the average number of reviews per user in the community, respectively.
Therefore the first term
%in the above equation
%GW: "in the above equation" is correct, is it?
models the user activity with respect to the community average. The second term reflects the time interval between successive reviews.
The user experience is unlikely to change from the level when writing the previous review just a few hours or days ago.
$\lambda$ controls the effect of this time difference, and is set to a very small value. Note that if the user writes very infrequently, the second term may go up. But the first term which plays the \emph{dominating} role in this prior will be very small with respect to the community average in an active community, bringing down the influence of the entire prior.
Note that the constructed HMM encapsulates all the factors for experience progression outlined in Section~\ref{sec:overview}.

At experience level $e_{t_d}$, user $u_d$ has a Multinomial facet-preference distribution $\theta_{u_d, e_{t_d}}$.
From this distribution she draws a facet of interest $z_{d_i}$ for the $i^{th}$ word in her document.
For example, a user at a high level of experience may choose to write on the beer ``hoppiness'' or ``story perplexity'' in a movie.
The word that she writes depends on the facet chosen and the language model for 
her current experience level. Thus, she draws a word from the multinomial distribution $\phi_{e_{t_d}, z_{d_i}}$ with a symmetric Dirichlet prior $\delta$. 
For example, if the facet chosen is beer {\em taste} or movie {\em plot}, an experienced user may choose to use the words ``coffee roasted vanilla'' and ``visceral'', whereas an inexperienced user may use ``bitter'' and ``emotional'' respectively.

Algorithm~\ref{algo.3} describes this generative process for the review; Figure~\ref{fig:2}
depicts it visually in plate notation for graphical models.
We use \emph{MCMC} sampling for inference on this model.
%{\tt GW: not clear if this comprehensible - we may a bit more reader guidance here !!!!!}

\begin{algorithm}[h]
\label{algo.3}
\SetAlgoLined
\DontPrintSemicolon
\For {each facet $z = 1, ... Z$ and experience level $e = 1, ... E$} {choose $\phi_{e, z} \sim Dirichlet(\beta)$} \;
%\For {each experience level $e = 1, ... E$} {choose $\pi_{e} \sim Dirichlet(\gamma)$} \;
\For {each review $d = 1, ... D$} {
  Given user $u_d$ and timestamp $t_d$\;
  /*Current experience level depends on previous level*/ \;
  1. Conditioned on $u_d$ and previous experience $e_{t_{d-1}}$, 
  %(assuming documents arrive sorted in time)
  choose $e_{t_d} \sim \pi_{e_{t_{d-1}}}$ \;
  /*User's facet preferences at current experience level
are influenced by supervision via $\alpha$ $~$ -- $~$ scaled by
hyper-parameter $\rho$ controlling influence of supervision*/\;
  2. Conditioned on supervised facet preference $\alpha_{u_d, e_{t_d}}$ 
of $u_d$ at experience level $e_{t_d}$ scaled by $\rho$, choose $\theta_{u_d, e_{t_d}} \sim  Dirichlet(\rho \times \alpha_{u_d, e_{t_d}})$ \;
  \For {each word $i = 1, ... N_d$} {
    /*Facet is drawn from user's experience-based facet interests*/ \;
    3. Conditioned on $u_d$ and $e_{t_d}$ choose a facet $z_{d_i} \sim Multinomial(\theta_{u_d, e_{t_d}})$ \;
    /*Word is drawn from chosen facet and user's 
vocabulary at her current experience level*/ \;
    4. Conditioned on $z_{d_i}$ and $e_{t_d}$ choose a word $w_{d_i} \sim Multinomial(\phi_{e_{t_d}, z_{d_i}})$ \;
  }
  /*Rating computed via Support Vector Regression with \\
chosen facet proportions as input features to learn $\alpha$*/ \;
  5. Choose $r_d \sim F(\langle {\alpha_{u_d, e_{t_d}}}, \phi_{e_{t_d}, z_d} \rangle)$ \;
}
\caption{Supervised Generative Model for a User's Experience, Facets, and Ratings}
\label{algo.3}
\end{algorithm}
\vspace{-2em}

%%%%%%%%%%%%%%%%%%%%%%%%%%

\subsubsection{Supervision for Rating Prediction}

The latent item factors $\phi_i$ in Equation~\ref{eq.3} correspond to the latent facets $Z$ in Algorithm~\ref{algo.3}.
Assume that we have some estimation of the latent facet distribution $\phi_{e,z}$ of each document after one iteration of MCMC sampling, where $e$ denotes the experience level at which a document is written, and let $z$ denote a latent facet of the document. 
We also have an estimation of the preference of a user $u$ for facet $z$ 
at experience level $e$ given by $\theta_{u,e}(z)$.

For each user $u$, we compute a supervised regression function $F_u$
for the user's numeric ratings 
with the -- currently estimated -- experience-based facet distribution $\phi_{e,z}$ of her reviews as input features
and the ratings as output.

The learned feature weights $\langle\alpha_{u,e}(z)\rangle$ indicate the user's preference for facet $z$ at experience level $e$. These feature weights are used to modify $\theta_{u,e}$ to
 attribute more mass to the facet for which $u$ has a higher preference at level $e$. This is reflected in the next sampling iteration, when we draw a facet $z$ from the user's facet preference distribution $\theta_{u,e}$ smoothed by $\alpha_{u,e}$, and then draw a word from $\phi_{e,z}$. This
sampling process is repeated until convergence.

In any latent facet model, it is difficult to set the hyper-parameters.
%or select the optimal number of facets. 
Therefore, 
%researchers mostly work with
most prior work assume
symmetric Dirichlet priors with heuristically chosen concentration parameters. 
%Recent works~\cite{mccallumNIPS2009, mimnoUAI2008} show that optimizing 
%Dirichlet hyper-parameters for LDA results in improved performance by 
%decreasing the model's sensitivity to the number of latent dimensions, 
%resulting in lower model complexity and computational cost than 
%non-parametric models.
%\cite{mccallumNIPS2009} further shows that an asymmetric Dirichlet prior for the latent facet distribution has advantages over a symmetric prior, whereas an asymmetric prior for the facet-word distribution does not not provide any benefit.
%
Our approach is to {\em learn} the concentration parameter $\alpha$ of a {\em general} (i.e., asymmetric) Dirichlet prior for Multinomial distribution $\Theta$ -- where we optimize these hyper-parameters to learn user ratings for documents at a given experience level.

%{\tt GW: this whole paragraph -- and the previous one -- is awfully hard to 
%digest. Think about
%giving more guidance to non-expert readers here !!!!!}

%%%%%%%%%%%%%%%%%%%%%%%%%%%%%%%

\vspace{-1em}
\subsubsection{Inference}

We describe the inference algorithm to estimate the distributions $\Theta$, $\Phi$ and $\Pi$
from observed data.
For {each user}, we compute the conditional distribution over the set of hidden variables $E$ and $Z$ for all the words $W$ in a review. 
The exact computation of this distribution is intractable. 
We use {\em Collapsed Gibbs Sampling} \cite{Griffiths02gibbssampling} to estimate the conditional distribution for each hidden variable, which is computed over the current assignment for all other hidden variables, and integrating out other parameters of the model.

Let $U, E, Z$  and $W$ be the set of all users, experience levels, facets and words in the corpus. 
In the following, $i$ indexes a document and $j$ indexes a word in it.

The joint probability distribution is given by:
%\begin{equation}
% \begin{aligned}
\begin{multline}
%\centering
P(U, E, Z, W, \theta,\phi ,\pi;\alpha ,\delta ,\gamma) =
\prod_{u = 1}^U \prod_{e = 1}^E \prod_{i = 1}^{D_u} \prod_{z = 1}^Z \prod_{j = 1}^{{N_{d_u}}}\\
\{\underbrace{P(\pi_e;\gamma^u ) \times P(e_i|\pi_e)}_\text{experience transition distribution} \times \underbrace{P(\theta _{u,e};\alpha_{u,e}) \times P(z_{i,j}|\theta_{u,e_i})}_\text{user experience facet distribution}
\times \underbrace{P(\phi_{e,z};\delta) \times P(w_{i,j}|\phi_{e_i,z_{i,j}})}_\text{experience facet language distribution}\}
%\end{aligned}
\end{multline}
%\end{equation}

Let $n(u, e, d, z, v)$ denote the count of the word $w$
%indexed by the $v^{th}$ word in the vocabulary, 
occurring in document $d$ written by user $u$ at experience level $e$ belonging to facet $z$. In the following equation,
 $(.)$ at any position in a distribution indicates summation of the above counts  
for the respective argument.

Exploiting conjugacy of the Multinomial and Dirichlet distributions, 
we can integrate out $\Phi$ from the above distribution to obtain the posterior distribution
$P(Z|U,E; \alpha )$ of the latent variable $Z$ given by:

{
\[\prod_{u=1}^U \prod_{e=1}^E \frac{\Gamma(\sum_{z} \alpha_{u,e,z})\prod_{z} \Gamma(n(u, e, ., z, .)+ \alpha_{u,e,z})}{\prod_{z}{\Gamma(\alpha_{u,e,z})\Gamma(\sum_{z} n(u, e, ., z, .) + \sum_z \alpha_{u,e,z})}}\]\text{\normalsize \hspace{2.5em} where $\Gamma$ denotes the Gamma function.}
}

Similarly, by integrating out $\Theta$, $P(W|E,Z; \delta )$ is given by
{
\[\prod_{e=1}^E \prod_{z=1}^Z \frac{\Gamma(\sum_{v} \delta_v)\prod_{v} \Gamma(n(., e, ., z, v)+ \delta_v)}{\prod_{v}{\Gamma(\delta_v)\Gamma(\sum_{v} n(., e, ., z, v) + \sum_v \delta_v)}}\]
}

Let $m_{e_i}^{e_{i - 1}}$
denote the number of transitions from experience level $e_{i-1}$ to $e_i$ over {\em all} users in the community, with the constraint $e_i \in \{e_{i-1}, e_{i-1}+1\}$. Note that we allow self-transitions for staying at the same experience level. The counts capture the relative difficulty in progressing between different experience levels. For example, it may be easier to progress to level $2$ from level $1$ than to level $4$ from level $3$.

%experience level transitions from the
%${(i - 1)^{th}}$ state to the $i^{th}$ state over all users and documents. 
%

%
The state transition probability depending on the previous state, factoring in the user-specific activity rate, is given by:

$\hspace{3em}P(e_i|e_{i - 1}, u, e_{-i}) = \frac{m_{e_i}^{e_{i - 1}} + I(e_{i - 1} = e_i) + \gamma^u}{m_{.}^{e_{i-1}} + I(e_{i - 1} = e_i) + E \gamma^u}$

where $I(.)$ is an indicator function taking the value $1$ when the argument is true, and $0$ otherwise. The subscript $- i$ denotes the value of a variable excluding the data at the $i^{th}$ position. All the {\em counts} of transitions exclude transitions to and from $e_i$, when sampling a value for the current experience level $e_i$ during Gibbs sampling.
The conditional distribution for the experience level transition is given by:
\begin{equation}
\label{eq.6}
 P(E|U,Z,W) \propto P (E|U) \times P(Z|E, U) \times P(W|Z, E)
\end{equation}

Here the first factor models the rate of experience progression factoring in user activity; the second and third factor models the facet-preferences of user, and language model at a specific level of experience respectively. All three factors combined decide whether the user should stay at the current level of experience, or has matured enough to progress to next level.

In Gibbs sampling, the conditional distribution for each hidden variable is computed based on the current assignment of other hidden variables. The values for the latent variables are sampled repeatedly from this conditional distribution until convergence. In our problem setting we have two sets of latent variables corresponding to $E$ and $Z$ respectively.

We perform Collapsed Gibbs Sampling \cite{Griffiths02gibbssampling} in which we first sample a value for the experience level $e_i$ of the user for the current document $i$, keeping all facet assignments $Z$ fixed. In order to do this, we consider two experience levels $e_{i-1}$ and ${e_{i-1}+1}$. For each of these levels, we go through the current document and all the token positions to compute Equation~\ref{eq.6} --- and choose the level having the highest conditional probability.
Thereafter, we sample a new facet for each word $w_{i,j}$
of the document, keeping the currently sampled experience level of the user for the document fixed.

The conditional distributions for Gibbs sampling for the joint update of the latent variables $E$ and $Z$ are given by:
%\begin{equation}
%\hspace{-1em}
%\boxed{
\begin{multline}
\label{eq.4.0}
\textbf{E-Step 1: } P(e_i = e |e_{i-1}, u_i = u, \{z_{i,j}=z_j \}, \{w_{i,j}=w_j\}, e_{-i})  \propto \\
 P(e_i|u, e_{i-1}, e_{-i}) \times \prod_j P(z_j|e_i, u, e_{-i}) \times P(w_j|z_j, e_i, e_{-i}) \propto \\
\frac{m_{e_i}^{e_{i - 1}} + I(e_{i - 1} = e_i) + \gamma ^u}{m_{.}^{e_{i-1}} + I(e_{i - 1} = e_i) + E\gamma^u} \times
 \prod_j \frac{n(u, e, ., z_j, .) + \alpha_{u,e,z_j}}{\sum_{z_j} n(u, e, ., z_j, .) + \sum_{z_j} \alpha_{u,e,z_j}} \times \frac{n(., e, ., z_j, w_j) + \delta}{\sum_{w_j} n(., e, ., z_j, w_j) + V\delta}\\\\
\noindent \textbf{E-Step 2:\quad\quad} P(z_j = z|u_d = u, e_d=e, w_j =w, z_{-j}) \propto \\
\frac{n(u, e, ., z, .) + \alpha_{u,e,z}}{\sum_{z} n(u, e, ., z, .) + \sum_{z} \alpha_{u,e,z}} \times \frac{n(., e, ., z, w) + \delta}{\sum_{w} n(., e, ., z, w) + V\delta}\\
\end{multline}
%}
\\\\

The proportion of the $z^{th}$ facet in document $d$ with words $\{w_j\}$ written at experience level $e$ is given by:
\[
\centering
\phi_{e,z}(d) = \frac{\sum_{j=1}^{N_d} \phi_{e,z}(w_j)}{N_d}
\]

For each user $u$, we learn a regression model $F_u$ using these facet proportions in each document as features, along with the user and item biases (refer to Equation~\ref{eq.3}), 
with the user's item rating $r_d$ as the response variable. 
Besides the facet distribution of each document, the biases $<\beta_g(e), \beta_u(e), \beta_i(e)>$ also depend on the experience level $e$.

%The learned feature weights $\alpha_{u,e}$ for the facets are used to provide more mass to the facet for which the user has a higher preference in the next sampling iteration, when we use them as Dirichlet prior to $\theta_{u,e}$.

We formulate the function $F_u$ as Support Vector Regression~\cite{drucker97}, which forms the $M$-$Step$ in our problem:
%The primal form of the problem is given by:\\
\begin{equation*}
\begin{aligned}
\textbf{M-Step:    } \min_{\alpha_{u,e}} \frac{1}{2}  {\alpha_{u,e}} & ^T{\alpha_{u,e}} + C \times
 \sum_{d=1}^{D_u} (max(0, |r_d - {\alpha_{u,e}}^T<\beta_g(e), & \beta_u(e), \beta_i(e), \phi_{e,z}(d)>| - \epsilon))^2
 \end{aligned}
  \label{eq.5}
\end{equation*}

The total number of parameters learned is $[E \times Z + E \times 3] \times U$.
%We can define $\alpha_{u,e} = [1\ 1\ 1\ \alpha_{u,e}]$ to incorporate the different biases in the feature vector.
Our solution may generate a mix of positive and negative real numbered weights. In order to ensure that 
the concentration parameters of the Dirichlet distribution are positive reals, we take $exp(\alpha_{u,e})$. The learned $\alpha$'s are typically very small, whereas the value of $n(u, e, ., z, .)$ in Equation~\ref{eq.4.0} is very large. Therefore we scale the $\alpha$'s by a hyper-parameter $\rho$ to control the influence of supervision.
$\rho$ is tuned
using a validation set by varying it from $\{10^0, 10^1... 10^5\}$. In the \emph{E-Step} of the next iteration, we choose $\theta_{u,e} \sim Dirichlet(\rho \times \alpha_{u,e})$. We use the LibLinear\footnote{\href{http://www.csie.ntu.edu.tw/~cjlin/liblinear}{http://www.csie.ntu.edu.tw/~cjlin/liblinear}} package for Support Vector Regression.

%% file: main/chapter-temporal-evolution/discrete-model/experiments.tex
\subsection{Experiments}
\label{sec:experiments1}

\subsubsection{Setup: Data and Baselines}

\noindent {\bf Data:} We perform experiments with data from five communities in different domains:\\
BeerAdvocate ({\tt \href{http://www.beeradvocate.com}{beeradvocate.com}}) and RateBeer ({\tt \href{http://www.ratebeer.com}{ratebeer.com}}) for beer reviews, Amazon ({\tt \href{http://www.amazon.com}{amazon.com}}) for movie reviews, Yelp ({\tt \href{http://www.yelp.com}{yelp.com}}) for food and restaurant reviews, and\\ NewsTrust ({\tt \href{http://www.newstrust.net}{newstrust.net}}) for reviews of news media. Table~\ref{tab:statistics1} gives the dataset statistics\footnote{\href{http://snap.stanford.edu/data/}{http://snap.stanford.edu/data/}, \href{http://www.yelp.com/dataset\_challenge/}{http://www.yelp.com/dataset\_challenge/}}. 
We have a total of $12.7$ million reviews from $0.9$ million users from all of the five communities combined. The first four communities are used for product reviews, from where we extract the following quintuple for our model $<userId, itemId, timestamp, rating, review>$. \\
NewsTrust is a special community, which we discuss in Section~\ref{sec:usecases}.

For all models, we used the three most recent reviews of each user as withheld test data. All experience-based models consider the \emph{last} experience level reached by each user, and corresponding learned parameters for rating prediction. In all the models, we group {\em light} users with less than $50$ reviews in {\em training} data into a background model, treated as a single user, to avoid modeling from sparse observations. We do not ignore any user. During the {\em test} phase for a light user, we take her parameters from the background model. We set $Z=20$ for BeerAdvocate, RateBeer and Yelp facets; and $Z=100$ for Amazon movies and NewsTrust which have much richer latent dimensions. For experience levels, we set $E=5$ for all. However, for NewsTrust and Yelp datasets our model categorizes users to belong to one of {\em three} experience levels.

\begin{table}[htbp]
\centering
\begin{tabular}{lrrr}
\toprule
\bf{Dataset} & \bf{\#Users} & \bf{\#Items} & \bf{\#Ratings}\\
\midrule
\bf{Beer (BeerAdvocate)} & 33,387 & 66,051 & 1,586,259\\
\bf{Beer (RateBeer)} & 40,213 & 110,419 & 2,924,127\\
\bf{Movies (Amazon)} & 759,899 & 267,320 & 7,911,684\\
\bf{Food (Yelp)} & 45,981 & 11,537 & 229,907\\
\bf{Media (NewsTrust)} & 6,180 & 62,108 & 134,407\\
\midrule
\bf{TOTAL} & 885,660 & 517,435 & 12,786,384\\
\bottomrule
\end{tabular}
\caption{Dataset statistics.}
\label{tab:statistics1}
\end{table}

\noindent {\bf Baselines:} We consider the following baselines for our work, and use the available code\footnote{\url{http://cseweb.ucsd.edu/~jmcauley/code/}} for experimentation.
%(refer to Figure~\ref{fig:baselines}).\\
\begin{itemize}
\item[a)]\emph{LFM}: A standard latent factor recommendation model~\cite{korenKDD2008}.
\item[b)]\emph{Community at uniform rate}: Users and products in a community evolve using a single ``global clock''~\cite{KorenKDD2010}\cite{xiongSDM2010}\cite{XiangKDD2010}, where the different stages of the community evolution appear at uniform time intervals. So the community prefers different products at different times.
\item[c)]\emph{Community at learned rate}: This extends (b) by learning the rate at which the community evolves with time, eliminating the uniform rate assumption.
\item[d)]\emph{User at uniform rate}: This extends (b) to consider individual users, by modeling the different stages of a user's progression based on preferences and experience levels evolving over time. 
The model assumes a uniform rate for experience progression. 
\item[e)]\emph{User at learned rate}: This extends (d) by allowing each user to evolve on a ``personal clock'', so that the time to reach certain experience levels depends on the user\\ \cite{mcauleyWWW2013}.
\end{itemize}
\pagebreak
f) \emph{Our model with past experience level}: In order to determine how well our model captures {\em evolution of user experience over time},
%or in other words, if the \emph{last} experience level attained 
%by the user is a good predictor of the user's current preferences, 
we consider another baseline
%Instead of considering the \emph{last} experience level of each user, for 
%predicting his ratings for the \emph{most} recent reviews --- we consider a 
where we
\emph{randomly sample} the experience level reached by users at some timepoint {\em previously} in their lifecycle, who may have evolved thereafter. 
We learn our model parameters from the data up to this time, and again predict the user's most recent three item ratings. Note that this baseline considers textual content of user contributed reviews, unlike other baselines that ignore them. Therefore it is better than vanilla content-based methods, with the notion of past evolution, and is the strongest baseline for our model.

%Note that this experience level is not truly random in the sense it has 
%already been reached by the user at some point in his lifecycle, after which %he may have progressed to a new level. 
%Table~\ref{tab:userExperienceMSE} shows the MSE results for this setup.

\subsubsection{Quantitative Comparison}

\begin{table}[t]
\centering
\begin{tabular}{p{5cm}p{1.4cm}p{1.2cm}p{1.2cm}p{1.2cm}p{1.2cm}}
\toprule
\bf{Models} & \bf{Beer} & \bf{Rate} & \bf{News} & \bf{Amazon} & \bf{Yelp}\\
 & \bf{Advocate} & \bf{Beer} & \bf{Trust} & &\\
\midrule
Our model & 0.363 & 0.309 & 0.373 & 1.174 & 1.469\\
(most recent experience level) & & & & &\\
{\bf f)} Our model & 0.375 & 0.362 & 0.470 & 1.200 & 1.642\\
(past experience level) & & & & &\\
{\bf e)} User at learned rate & 0.379 & 0.336 & 0.575 & 1.293 & 1.732\\
{\bf c)} Community at learned rate & 0.383 & 0.334 & 0.656 & 1.203 & 1.534\\
{\bf b)} Community at uniform rate & 0.391 & 0.347 & 0.767 & 1.203 & 1.526\\
{\bf d)} User at uniform rate & 0.394 & 0.349 & 0.744 & 1.206 & 1.613\\
{\bf a)} Latent factor model & 0.409 & 0.377 & 0.847 & 1.248 & 1.560\\
\bottomrule
\end{tabular}
\caption{MSE comparison of our model versus baselines.}
\label{fig:MSE1}
\end{table}

\begin{figure}
\centering
 \includegraphics[scale=0.8]{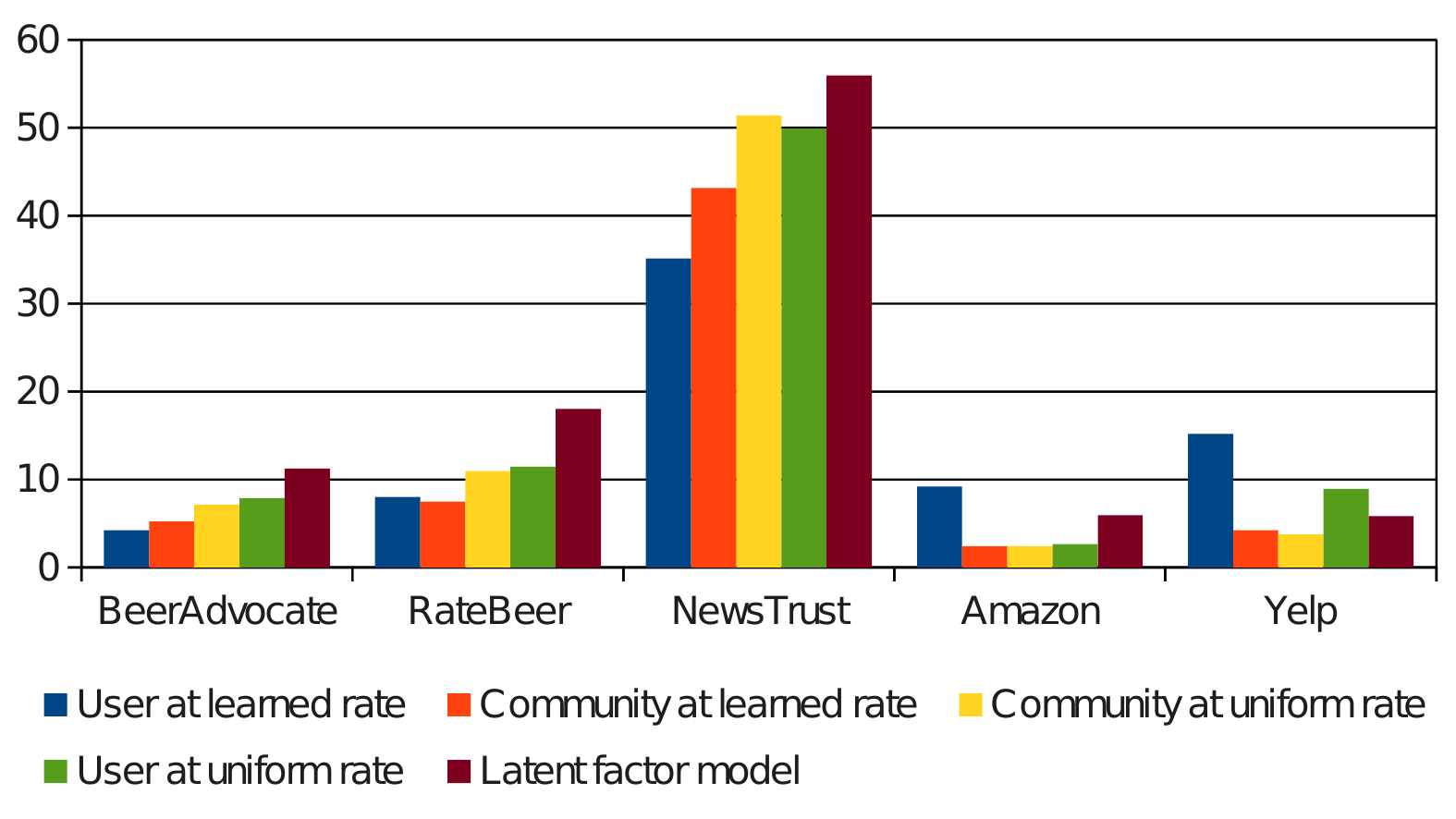}
 \caption{MSE improvement (\%) of our model over baselines.}
  \label{fig:improvement}
  \vspace{-1em}
\end{figure}

\noindent {\bf Discussions:} Table~\ref{fig:MSE1} compares the \emph{mean squared error (MSE)} for rating predictions, generated by our model versus the six baselines. Our model consistently outperforms all baselines, reducing the MSE by ca. $5$ to $35\%$. Improvements of our model over baselines are statistically significant at p-value $<0.0001$.

Our performance improvement is most prominent for the NewsTrust community, which exhibits strong language features, and topic polarities in reviews. The lowest improvement (over the best performing baseline in any dataset) is achieved for Amazon movie reviews. A possible reason is that the community is very diverse with a very wide range of movies and that review texts heavily mix statements about movie plots with the actual review aspects like praising or criticizing certain facets of a movie. The situation is similar for the food and restaurants case. Nevertheless, our model always wins over the best baseline from {\em other} works, which is typically the ``user at learned rate" model.

\noindent {\bf Evolution effects:} We observe in Table~\ref{fig:MSE1} that our model's predictions degrade when applied to the users' {\em past} experience level, compared to their {\em most recent} level. This signals that the model captures user evolution past the previous timepoint. Therefore the last (i.e., most recent) experience level attained by a user is most informative for generating new recommendations.
%Additionally, we also perform an experiment where we switch off all time-related components in our method's software. The results are slighlty worse than that of the past experience level in Table~\ref{fig:MSE}. 

%This is contrary to the finding in~\cite{mcauleyWWW2013} in which a random experience level in the user evolution model \emph{often} performed better than that of the most recent one.

%%%%%%%%%%%%%%%%%%%%%%%%%%%%%%%%%%%%%%%%%%%%%%%%%%%%
\begin{table}[t]
\centering
\small
\begin{tabular}{p{10.5cm}}
\toprule
\textbf{Experience Level 1:} drank, bad, maybe, terrible, dull, shit
\\\midrule
\textbf{Experience Level 2:} bottle, sweet, nice hops, bitter, strong light, head, smooth, good, brew, better, good
\\\midrule
\textbf{Expertise Level 3:} sweet alcohol, palate down, thin glass, malts, poured thick, pleasant hint, bitterness, copper hard
%expected palate, bubbles pale quickly, short snifter, attractive set leaving robust impression, deep feel satisfying, high quality hop, cloudy fruit mouthfeel, initial tongue, fizzy work, mouthfeel, burst orangey, brewers pint
\\\midrule
\textbf{Experience Level 4:} smells sweet, thin bitter, fresh hint, honey end, sticky yellow, slight bit good, faint bitter beer, red brown, good malty, deep smooth bubbly, damn weak
%green murky, fizzy syrupy, spicey, tasty american average lager, strange watery, weird mouthfeel, cheap grain, foam stuff, hop, kinda english, macro carmel, cask gold, definately head, aroma follows
\\\midrule
\textbf{Experience Level 5:} golden head lacing, floral dark fruits, citrus sweet, light spice, hops, caramel finish, acquired taste, hazy body, lacing chocolate, coffee roasted vanilla, creamy bitterness, copper malts, spicy honey
%slight faint bitter coffee, chocolate alcohol, fruity bubbly, amber pine, dried cinnamon, toffee, tart reddish cherries, molasses, translucent malt, candi ruby lemon, tropical acidity, smoked fruitiness
\\
\bottomrule
\end{tabular}
\caption{Experience-based facet words for the {\em illustrative} beer facet {\em taste}.}
\label{tab:beerTopics}
\vspace{-1em}
\end{table}

\subsubsection{Qualitative Analysis}

\noindent{\bf Salient words for facets and experience levels}: We point out typical word clusters, with {\em illustrative} labels, to show the variation of language for users of different experience levels and different facets. Tables \ref{tab:amazonTopics} and \ref{tab:beerTopics} show salient words to describe the beer facet {\em taste} and movie facets {\em plot} and {\em narrative style}, respectively -- at different experience levels. Note that the facets being latent, their labels are merely our interpretation. Other similar examples can be found in Tables~\ref{tab:facetWords} and~\ref{tab:newstrustTopics}.

BeerAdvocate and RateBeer are very focused communities; so it is easier for our model to characterize the user experience evolution by vocabulary and writing style in user reviews. We observe in Table~\ref{tab:beerTopics} that users write more descriptive and {\em fruity} words to depict the beer taste as they become more experienced.

For movies, the wording in reviews is much more diverse and harder to track.
Especially for blockbuster movies, which tend to dominate this data, the reviews mix all kinds of aspects.
A better approach here could be to focus on specific kinds of movies (e.g., by genre or production studios) that may better distinguish experienced users from amateurs or novices in terms of their refined taste and writing style.

%placed in Section 2

\noindent{\bf MSE for different experience levels}: We observe a weak trend that the MSE decreases with increasing experience level. Users at the highest level of experience almost always exhibit the lowest MSE, and, therefore, more preditable in their behavior. So we tend to better predict the rating behavior for the most mature users than for the remaining user population. This in turn enables generating better recommendations for the ``connoisseurs" in the community.\\

\begin{figure}[t]
\centering
 \includegraphics[scale=0.6]{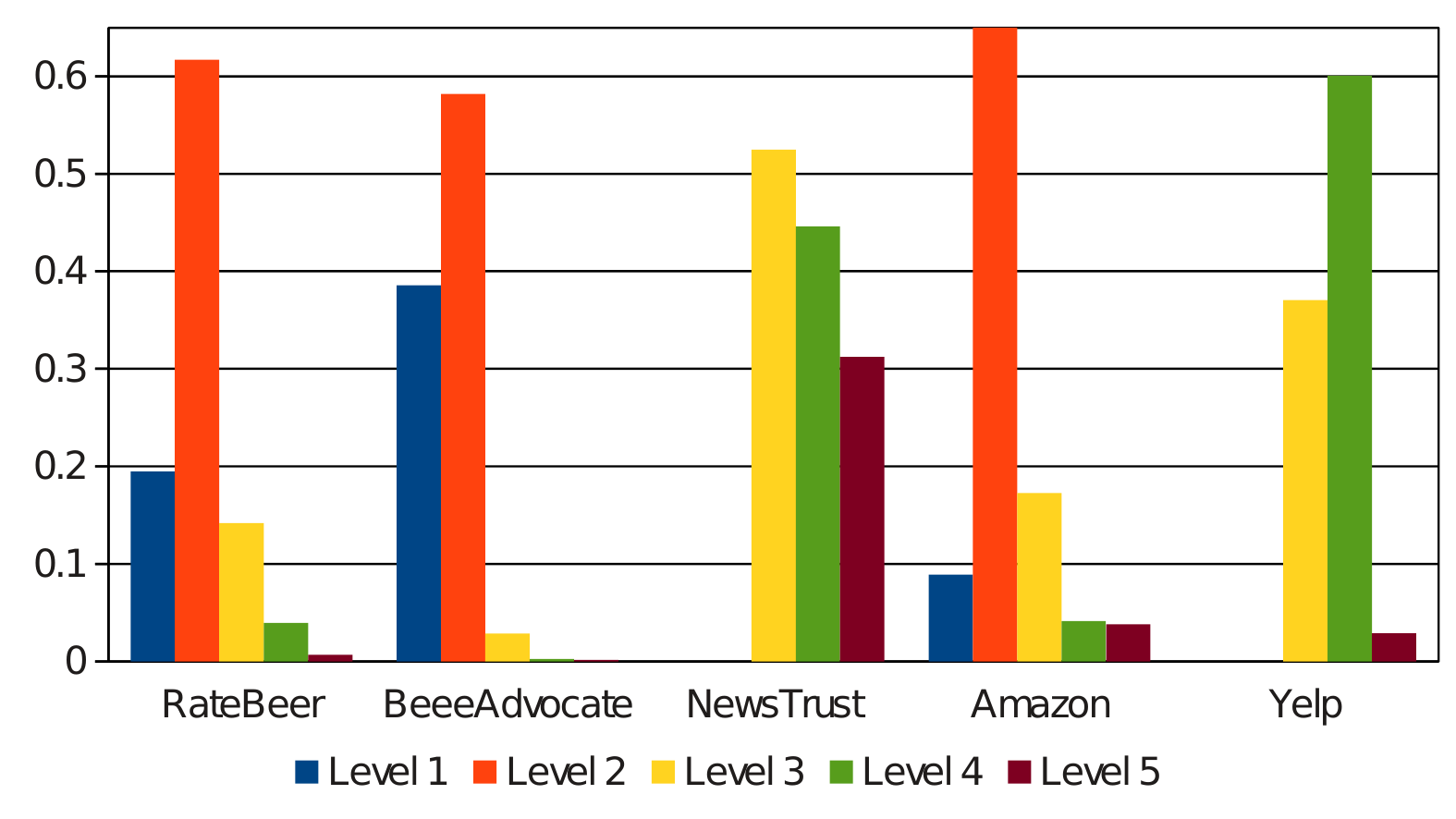}
 \caption{Proportion of reviews at each experience level of users.}
  \label{fig:progression}
\end{figure}

\noindent{\bf Experience progression}: Figure~\ref{fig:progression} shows the proportion of reviews written by community members at different experience levels right before advancing to the next level. Here we plot users with a minimum of $50$ reviews, so they are
certainly not ``amateurs".
A large part of the community progresses from level $1$ to level $2$. However, from here only few users move to higher levels, leading to a skewed distribution. We observe that the majority of the population stays at level $2$.

%GW: how does this phenomenon fit with the statement that we considered only somewhat advanced users -- with at least 50 reviews ?????

%This is in contrast to the finding in~\cite{mcauleyWWW2013} that most users progress through all levels and spend their longest time in their final experience level.

%subho: somehow in their model output more than 90% of the users come out to be experts. they do not state this in the paper,but their model output show this.
%GW: any explanation ??????
%or should we better omit this comment about contradicting the McAuley work ?????

\begin{table}[!h]
\centering
 \begin{tabular}{lccccc}
 \toprule
\bf{Datasets} &	\bf{e=1} & \bf{e=2} & \bf{e=3} & \bf{e=4} &\bf{e=5}\\
\midrule
BeerAdvocate & 0.05 & 0.59 & 0.19 & 0.10 & 0.07\\
RateBeer & 0.03 & 0.42 & 0.35 & 0.18 & 0.02\\
NewsTrust & - & - & 0.15 & 0.60 & 0.25\\
Amazon & - & 0.72 & 0.13 & 0.10 & 0.05\\
Yelp & - & - & 0.30 & 0.68 & 0.02\\
\bottomrule
 \end{tabular}
 \caption{Distribution of users at different experience levels.}
 \label{tab:userExperienceDistr}
\end{table}

\pagebreak

\noindent{\bf User experience distribution}:
Table~\ref{tab:userExperienceDistr} shows the number of users per experience level in each domain, for users with $>50$ reviews. The distribution also follows our intuition of a highly skewed distribution.
Note that almost all users with $<50$ reviews belong to levels 1 or 2.

\begin{figure*}
    \centering
    \begin{subfigure}[b]{\textwidth}
        \centering
        \includegraphics[width=0.96\columnwidth]{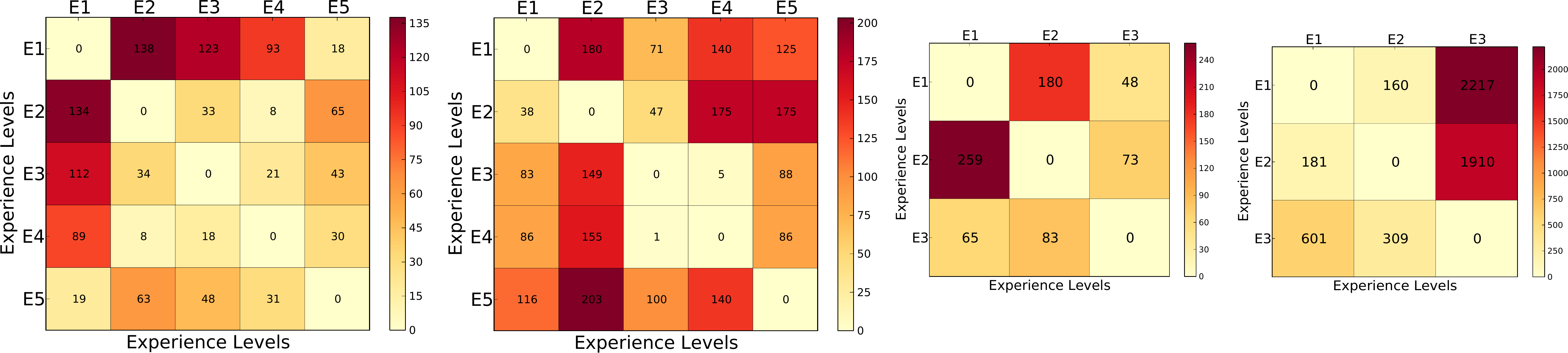}
        \caption{User at learned rate~\cite{mcauleyWWW2013}: Facet preference divergence with experience.}
        \label{fig:baselineFacet}
    \end{subfigure}
    %\vfill
    \begin{subfigure}[b]{\textwidth}
        \centering
        \includegraphics[width=0.96\columnwidth]{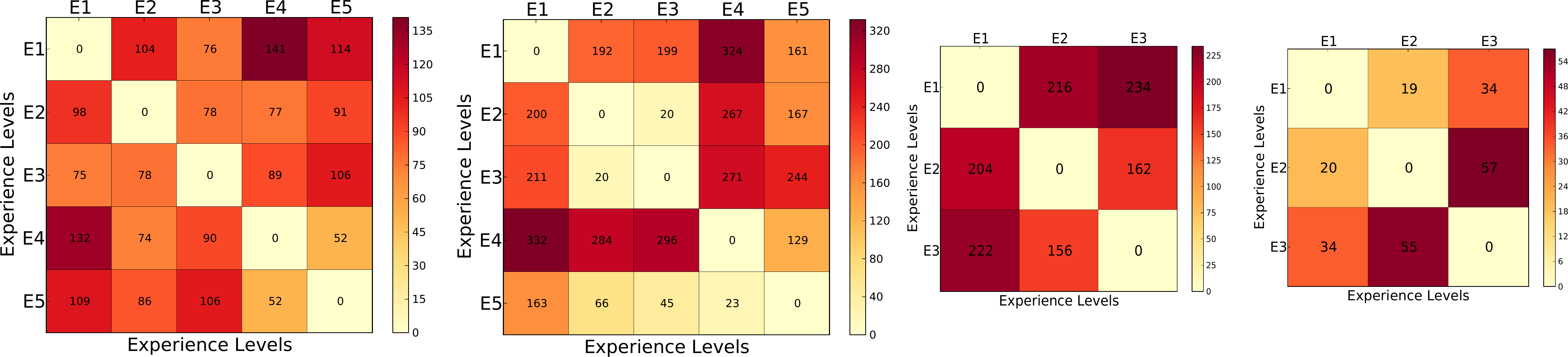}
        \caption{Our model: Facet preference divergence with experience.}
        \label{fig:modelFacet}
    \end{subfigure}
    %\vfill
    \begin{subfigure}[b]{\textwidth}
        \centering
        \includegraphics[width=0.96\columnwidth]{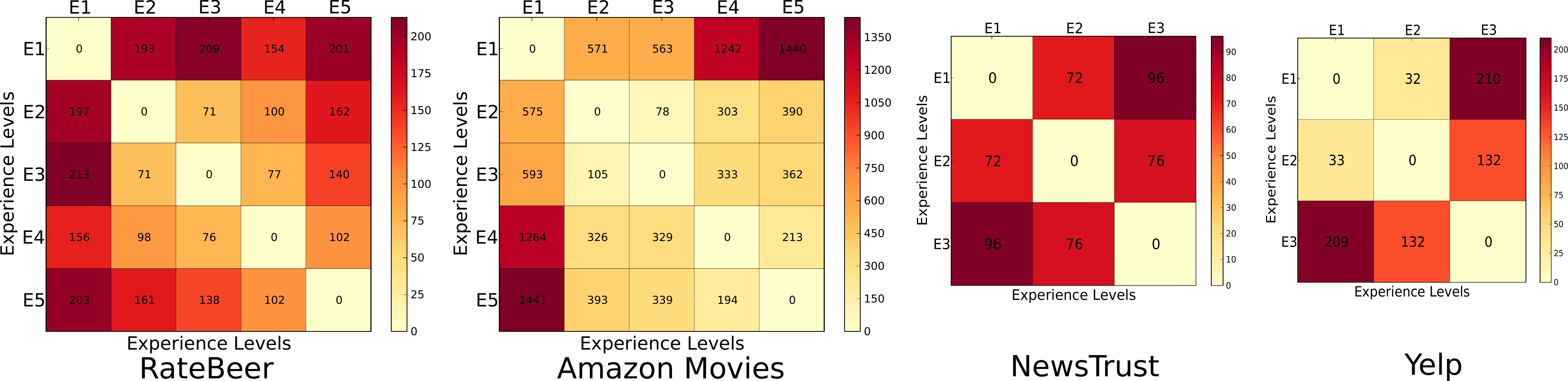}
        \caption{Our model: Language model divergence with experience.}
        \label{fig:modelLang}
    \end{subfigure}
    \caption{Facet preference and language model $KL$ divergence with experience.}
    \label{fig:heatmaps}
\end{figure*}

\noindent{\bf Language model and facet preference divergence}: Figure~\ref{fig:modelFacet} and~\ref{fig:modelLang} show the $KL$ divergence for facet-preference and language models of users at different experience levels, as computed by our model.
%All the heatmaps follow a similar distribution as we observed in our 
%experimental studies (refer to Section~\ref{sec:study}). 
The facet-preference divergence increases with the gap between experience levels, but not as {\em smooth} and prominent as for the language models. 
On one hand, this is due to the complexity of {\em latent} facets vs. {\em explicit} words.
On the other hand, this also affirms our notion of grounding the model on \emph{language}. 
%The resulting distribution has the smoothness governed by that of the language model.

\noindent{\bf Baseline model divergence}: Figure~\ref{fig:baselineFacet} shows the facet-preference divergence of users at different experience levels computed by the baseline model ``user at learned rate''~\cite{mcauleyWWW2013}.
The contrast between the heatmaps of our model and the baseline is revealing. The increase in divergence with increasing gap between experience levels is very {\em rough} in the baseline model, although the trend is obvious. 

%% file: main/chapter-temporal-evolution/cont-model/compmodel.tex
In the previous section, we presented an approach to model the experience evolution of users in online communities. However, the proposed model has several assumptions, and resulting drawbacks. In the following, we propose a {\em generalized} model that captures the evolution of user experience as is commonly observed in the Nature.

\subsection{Model Components}
\label{sec:cont-model}
%In the following, we will introduce the major components of our probabilistic model. We start with a discussion about the importance of time.

\subsubsection{Importance of Time}\label{sec:time}
Previous approaches [Section~\ref{sec:discrete}]~\cite{mcauleyWWW2013} on experience evolution  
model time only {\em implicitly} by assuming the (discrete) latent experience to progress from one review to the next. 
%In this work, in contrast, there are two components that are directly affected by time: 1) the experience evolution, and 2) the language model evolution.
%
%In these works the (discrete) experience is simply considered as an additional dimension in the language model. While this approach might be suitable %for modeling experience as a {\em discrete} random variable, making it {\em continuous} results in an infinite dimensional language model. While in %principle approximate inference is still possible, such a solution would neither be computationally efficient nor practically useful since the vocabulary %does not evolve at a super-fine granularity, as experience might do.
%
%Therefore, instead of considering experience as a dimension of the language model, it is more suitable to directly use {\em time}. As the reviews' %timestamps are observed, time is considered as an explicit dimension in the language model; since our purpose is to obtain an experience-aware %language model, this process is conditioned on the experience progression of the users. By using time as an explicit dimension, we are able to trace the %evolution of vocabulary and trends {\em jointly} on the temporal and experience dimension.
%
%Therefore we allow the language model to evolve temporally with the {\em variance} conditioned on the experience progression of a user. The idea
In contrast, we now model time {\em explicitly}, and allow experience to {\em continuously} evolve over time --- so that we are able to trace the joint evolution of experience, and vocabulary. This is challenging as the discrete Multinomial distribution based language model (to generate words) needs to be combined with a continuous stochastic process for experience evolution.

We use {two} levels of temporal granularity. Since experience is naturally {continuous}, it is beneficial to model its evolution at a very fine resolution (say, minutes or hours). 
On the other hand, the language model has a much coarser granularity (say, days, weeks or months). 
We show in Section~\ref{sec:inference} how to smoothly merge the two granularities using continuous-time models. 
Our model for language evolution is motivated by the seminal work of Wang and Blei et al.~\cite{BleiCTM},
%on Continuous Time Dynamic Topic Models, 
with major differences and extensions. 
%While the variance of the language model in their work grows with the {\em time} difference at which topics occur --- in our case it evolves with the %{\em experience} value difference and time. That is, it captures the change in the underlying users' experience evolution. In principle, our language %model can be made as fine as possible, although the resolution has to be pre-defined.
%
In the following subsections, we formally introduce the two components affected by time: the experience evolution and the language model evolution.

%We can set the granularity of time in the language model to days, month or years. We still need to map the experience value which evolves at a fine granularity (say, seconds) to the language model evolving at a course granularity.

% \subsection{Continuous State Experience Evolution}
\subsubsection{Continuous Experience Evolution}

\begin{figure*}[htbp]
%\centering
\begin{subfigure}{.5\textwidth}
  \centering
  \includegraphics[width=\linewidth]{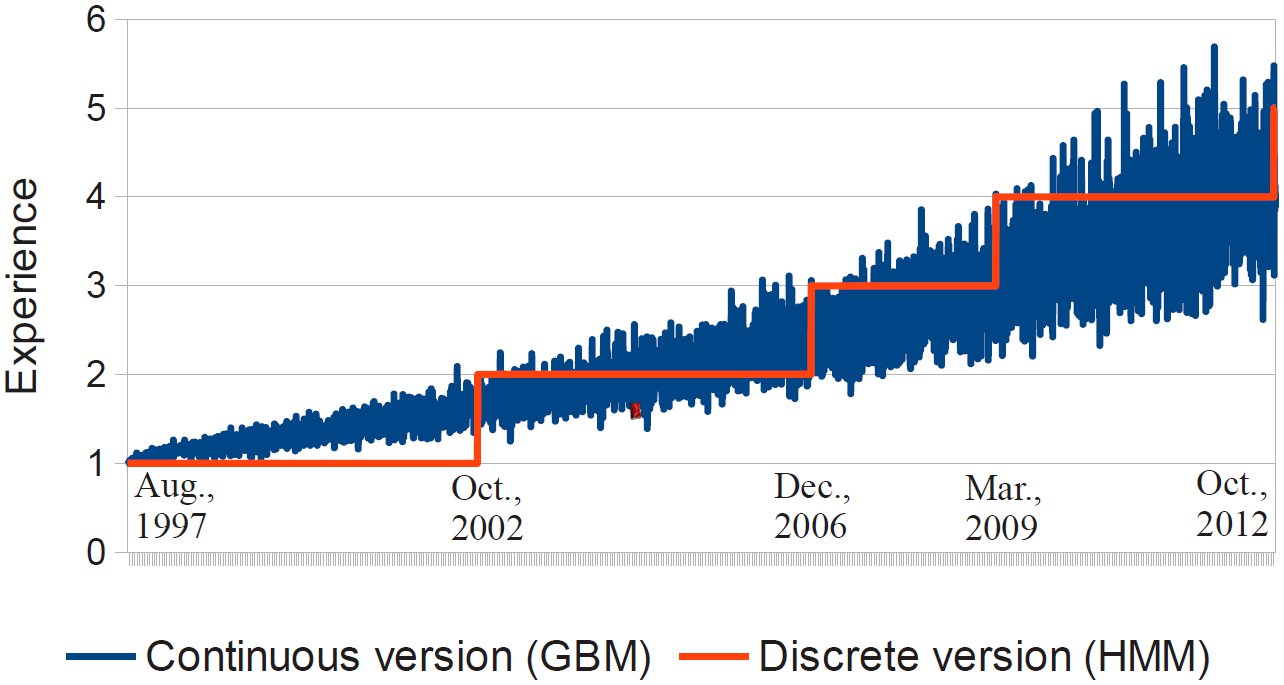}
  \caption{Evolution of an experienced user.}
  \label{fig:exp1}
\end{subfigure}%
\begin{subfigure}{.5\textwidth}
  \centering
  \includegraphics[width=1.05\linewidth]{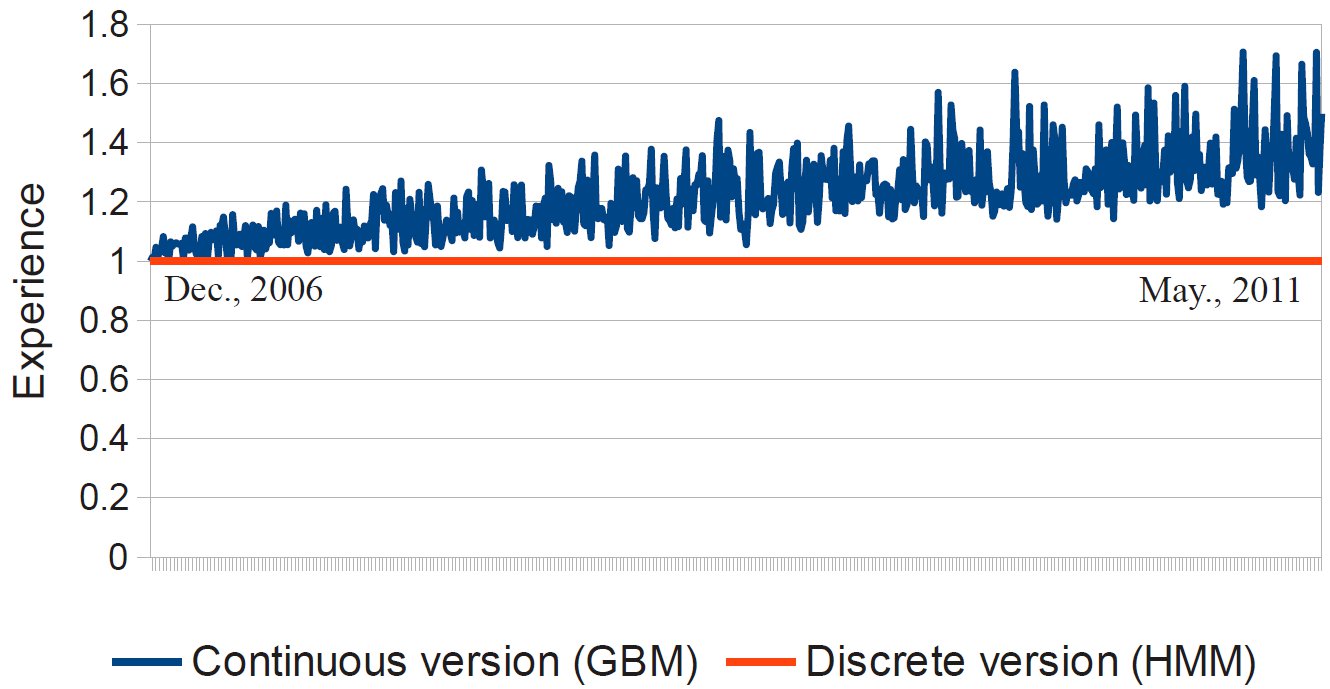}
  \caption{Evolution of an amateur user.}
  \label{fig:exp2}
\end{subfigure}
\caption{Discrete state and continuous state experience evolution of some typical users from the BeerAdvocate community.}
\label{fig:exp}
\end{figure*}

Prior approaches [Section~\ref{sec:discrete}]~\cite{mcauleyWWW2013} model experience as a discrete random variable . At each timepoint, a user is allowed to stay at level $l$, or move to level $l+1$. As a result the transition is abrupt when the user switches levels. Also, the model does not distinguish between users at the same level of experience, (or even for the same user at beginning or end of a level) even though their experience can be quite far apart (if measured in a continuous scale). For instance, in Figure~\ref{fig:exp2} the language model uses the same set of parameters as long as the user stays at level $1$, although the language model changes. 

In order to address these issues, our goal is to develop a continuous experience evolution model with the following requirements:

\newpage

\begin{itemize}
 \item The experience value is always positive.
 \item Markovian assumption for the continuous-time process: The experience value at any time $t$ depends only on the value at the {\em most recent 
%%%GW: added "observed" to make it meaningful; otherwise "most recent" is not well-defined for continuous time
observed 
time prior to $t$}.
 \item Drift: It has an overall {\em trend} to increase over time.
 \item Volatility:  The evolution may not be smooth with occasional volatility. For instance, an experienced user may write a series of expert reviews, followed by a sloppy one.
 %i.e. we allow experience value to go down resulting in a flexible model, unlike the previous ones.
\end{itemize}

To capture all of these aspects, we model each user's experience as a {\em Geometric Brownian Motion} (GBM) process (also known as Exponential Brownian Motion). 

GBM is a natural continuous state alternative to the discrete-state space based Hidden Markov Model (HMM) used in our previous approach (refer to Section~\ref{sec:discrete}). Figure~\ref{fig:exp} shows a real-world example of the evolution of an experienced and amateur user in the {\tt \href{http://www.beeradvocate.com}{BeerAdvocate}} community, as traced by our proposed model --- along with that of its discrete counterpart from our previous approach. The GBM is a stochastic process used to model population growth, financial processes like stock price behavior (e.g., Black-Scholes model) with random noise. It is a continuous time stochastic process, where the logarithm of the random variable (say, $X_t$) follows Brownian Motion with a {\em volatility} and {\em drift}.
Formally, a stochastic process $X_t$, with an arbitrary initial value $X_0$, for $t \in [0,\infty)$ is said to follow Geometric Brownian Motion, if it satisfies the following Stochastic Differential Equation (SDE)~\cite{gbm}:
\begin{equation}
\label{eq:sde}
 dX_t = \mu X_t dt + \sigma X_t dW_t
\end{equation}
where, $W_t$ is a Wiener process (Standard Brownian Motion); $\mu \in \mathbb{R}$ and $\sigma \in (0, \infty)$ are constants called the {\em percentage trend} and {\em percentage volatility} respectively. The former captures deterministic trends, whereas the latter captures unpredictable events occurring during the motion.

\pagebreak

In a Brownian Motion trajectory, $\mu X_tdt$ and $\sigma X_t dW_t$ capture the ``trend'' and ``volatility'', as is required for experience evolution. However, in real life communities  each user might show a different experience evolution; therefore our model considers a {\em multivariate} version of this GBM -- we model one trajectory {\em per-user}. Correspondingly, during the inference process we learn $\mu_u$ and $\sigma_u$ for each user $u$.

\emph{Properties:} A straightforward application of {\em It\^{o}'s} formula yields the following analytic solution to the above SDE (Equation~\ref{eq:sde}):
\begin{equation}
\begin{aligned}
 X_t &= X_0 \ exp\big((\mu - \frac{\sigma^2}{2})t + \sigma W_t\big)\\
 \end{aligned}
\end{equation}

Since $log(X_t)$ follows a Normal distribution, $X_t$ is Log-Normally distributed with mean $\big( log(X_0) + (\mu - \frac{\sigma^2}{2})t \big)$ and variance $\sigma \sqrt{t}$. The probability density function $f_t(x)$, for $x \in (0, \infty)$, is given by:
\begin{equation}
\label{eq:ln}
 f_t(x) = \frac{1}{\sqrt{2 \pi t}\sigma x} exp \bigg(-\frac{\big(log(x) - log\ (x_0) - (\mu - \frac{\sigma^2}{2})t\big)^2}{2\sigma^2t}\bigg)
\end{equation}

It is easy to show that GBM has the Markov property. Consider $U_t = (\mu - \frac{\sigma^2}{2})t + \sigma W_t$.
\begin{equation}
\begin{aligned}
 X_{t+h} &= X_0 exp(U_{t+h})\\
&= X_0 exp (U_t + U_{t+h} - U_t)\\
&= X_0 exp (U_t)exp(U_{t+h} - U_t)\\
&= X_t exp (U_{t+h} - U_t)
\end{aligned}
\end{equation}

Therefore, future states depend only on the future increment of the Brownian Motion, which satisfies our requirement for experience evolution. Also, for $X_0 > 0$, the GBM process is always positive. {\em Note} that the start time of the GBM of {\em each} user is relative to her first review in the community.

\subsubsection{Experience-aware Language 
%Model 
Evolution}

Once the experience values for each user are generated from a Log-Normal distribution (more precisely: the experience of the user at the times when she wrote each review), we develop the language model whose parameters evolve according to the Markov property for experience evolution.

As users get more experienced, they use more sophisticated words to express a concept. For instance, 
experienced cineastes refer to a movie's ``protagonist'' whereas amateur movie lovers talk about the ``hero''.
Similarly, in a Beer review community (e.g., BeerAdvocate, RateBeer) experts use more {\em fruity} words to describe a beer like ``caramel finish, coffee roasted vanilla'', and ``citrus hops''. Facet preferences of users also evolve with experience. For example, users at a high level of experience prefer ``hoppiest'' beers which are considered too ``bitter'' by amateurs~\cite{mcauleyWWW2013}. 
Encoding explicit time in our model allows us to trace the evolution of vocabulary and trends {\em jointly} on the temporal and experience dimension.

{\noindent \bf Latent Dirichlet Allocation (LDA):} In the traditional LDA process \cite{Blei2003LDA}, a document is assumed to have a distribution over $Z$ facets (a.k.a. topics) $\beta_{1:Z}$, and each of the facets has a distribution over words from a fixed vocabulary collection. The per-facet word (a.k.a topic-word) distribution $\beta_{z}$ is drawn from a Dirichlet distribution, and words $w$ are generated from a Multinomial($\beta_z$).

The process assumes that documents are drawn {\em exchangeably} from the same set of facets. However, this process neither takes experience nor the evolution of the facets over {\em time} into account. 

{\noindent \bf Discrete Experience-aware LDA:} Our previous approach (refer to Section~\ref{sec:discrete}) incorporates a layer for \emph{experience} in the above process. The user experience is manifested in the set of facets that the user chooses to write on, and the vocabulary and writing style used in the reviews. The experience levels were drawn from a \emph{Hidden Markov Model} (HMM). %, where the experience transition depended on the evolution of the user's \emph{maturing rate}, \emph{facet preferences}, and \emph{writing style}. 
The reviews were assumed to be exchangeable for a user at the same level of experience -- an assumption which generally may not hold; since the language model of a user at the same discrete experience level may be different at different points in time (refer to Figure~\ref{fig:exp2}) (if we had a continuous scale for measuring experience). The process considers {\em time} only {\em implicitly} via the transition of the latent variable for experience.

{\noindent \bf Continuous Time LDA:} The seminal work of \cite{BleiDTM,BleiCTM} capture evolving content, for instance, in scholarly journals and news articles where the themes evolve over time, by considering time {\em explicitly} in the generative LDA process. Our language model evolution is motivated by their Continuous Time Dynamic Topic Model~\cite{BleiDTM}, with the major difference that the facets, in our case, evolve over both {\em time} and {\em experience}. 

{\noindent \bf Continuous Experience-aware LDA (this work):} Since the assumption of exchangeability of documents at the same level of experience of a user may not hold, we want the language model to explicitly evolve over experience and time.
To incorporate the effect of changing experience levels, our goal is to condition the parameter evolution of $\beta$ on the experience progression.
%
%In the previous work, we generate the experience level $e$ of a user from a discrete state-space HMM. Thereafter we generate the facet $k$, and a word $w$ conditioned on $e$ from the tensor $\beta_{e,z}$ that models per-experience, per-facet word distribution. But now $e$ being continuous, such a tensor cannot be constructed directly. Alternatively,
%
%
%However, experience being continuous is difficult to encode explicitly as a dimension in the language model as in our earlier work. Therefore we allow the language model to evolve {\em temporally} (with observed timestamps) such that its parameters evolve according to the Markov property for experience evolution.
%

In more detail, for the language model evolution, we desire the following properties: 
\begin{itemize}
 \item It should {\em smoothly} evolve over time preserving the Markov property of experience evolution.
 \item Its variance should {\em linearly increase} with the  {\em experience change} between successive timepoints. This entails that if the experience of  a user does not change between successive timepoints, the language model remains almost the same. 
\end{itemize}

%In the traditional Latent Dirichlet Allocation process, the facet-word (a.k.a. topic-word) distribution $\beta_{z}$ is drawn from a Dirichlet distribution, and words are generated from a Multinomial ($\beta_z$). 

To incorporate the temporal aspects of data, in our model, we use multiple distributions $\beta_{t,z}$ for each time $t$ and facet $z$. Furthermore, to capture the smooth temporal evolution of the facet language model, we need to chain the different distributions to sequentially evolve over time $t$: the distribution $\beta_{t,z}$ should affect the distribution $\beta_{t+1,z}$.

Since the traditional parametrization of a Multinomial distribution via its mean parameters is not amenable to sequential modeling, and inconvenient to work with in gradient based optimization -- since any gradient step requires the projection to the feasible set, the simplex --- we follow a similar approach as \cite{BleiCTM}: instead of operating on the mean parameters, we consider the natural parameters of the Multinomial. The natural parameters are unconstrained and, thus, enable an easier sequential modeling.

From now on, we denote with $\beta_{t,z}$ the natural parameters of the Multinomial at time $t$ for facet $z$. For {\it identifiability} one of the parameters $\beta_{t,z,w}$ needs to be fixed at zero. By applying the following mapping we can obtain back the mean parameters that are located on the simplex:
\begin{equation}
\label{eq:pi}
\pi(\beta_{t,z,w}) = \frac{exp(\beta_{t,z,w})}{1 + \sum_{w=1}^{V-1} exp(\beta_{t,z,w})}
\end{equation}

Using the natural parameters, we can now define the facet-model evolution:
The underlying idea is that strong changes in the users' experience can lead to strong changes in the language model, while low changes should lead to only few changes. To capture this effect, let $l_{t,w}$ denote the average experience of a word $w$ at time $t$ (e.g. the value of $l_{t,w}$ is high if many experienced users have used the word). That is, $l_{t,w}$ is given by the average experience of all the reviews $D_t$ containing the word $w$ at time $t$.
 
 \begin{equation}
 \label{eq:word-exp}
 l_{t,w} =  \frac{\sum_{d\in D_t: w \in d} e_d}{|D_t|}
 \end{equation}
 where, $e_d$ is the experience value of review $d$ (i.e. the experience of user $u_d$ at the time of writing the review).
 
 %\todo{what if a word accours mutiple times? should we count it then multiple times?}

The language model evolution is then modeled as:
 \begin{equation}
 \label{eq:beta}
 \beta_{t,z,w} \sim Normal (\beta_{t-1,z,w}, \sigma \cdot |l_{t,w} - l_{t-1,w}|) 
 \end{equation}
Here, we simply follow the idea of a standard dynamic system with Gaussian noise, where the mean is the value at the previous timepoint, and the variance increases linearly with increasing change in the experience. Thereby, the desired properties of the language model evolution are ensured.

%In the next section, we explain how to combine all of these components to 
%obtain the full generative process, and how to perform approximate inference.

%However, the language model does not evolve at the fine granularity at which experience evolves (where, timestamps are measured in seconds). Therefore, we fix a separate granularity for the language model evolution (say, days, months, or years), which, in principle, can be made as fine as possible. 

%% file: main/chapter-temporal-evolution/cont-model/inference.tex
\subsection{Joint Model for Experience-Language Evolution}
\label{sec:inference}

\subsubsection{Generative Process}
Consider a corpus $D=\{d_1,\ldots,d_D\}$ of review documents written by a set of users $U$ at timestamps $T$. For each review $d\in D$, we denote $u_d$ as its user, $t'_d$ as the fine-grained timestamp of the review (e.g. minutes or seconds; used for experience evolution) and with $t_d$ the timestamp of coarser granularity (e.g. yearly or monthly; used for language model evolution). 
The reviews are assumed to be ordered by timestamps, i.e. $t{'}_{d_i}<t{'}_{d_j}$ for $i<j$. We denote with $D_t=\{d\in D \mid t_d=t\}$ all reviews written at timepoint $t$.
 Each review $d \in D$ consists of a sequence of $N_d$ words denoted by $d=\{w_1,\ldots ,w_{N_d}\}$, where each word is drawn from a vocabulary $V$ having unique words indexed by $\{1 \dots V\}$. The number of facets corresponds to $Z$.
 
 Let $e_d \in (0, \infty)$ denote the experience value of review $d$. Since each review $d$ is associated with a unique timestamp $t'_d$ and unique user $u_d$, the experience value of a review refers to the experience of the user at the time of writing it.
 %
  %Consider a set of facets $\{z_1,z_2,...z_K\}$ where each $z_i$ is from a set $K$ of possible facets. 
%
%Each review $d$ is associated with a rating $r$ and an item $i$.
%
In our model, each user $u$ follows her own Geometric Brownian Motion trajectory -- starting time of which is relative to the first review of the user in the community --  parametrized by the mean $\mu_u$, variance $\sigma_u$, and her {\em starting experience} value $s_{0,u}$. As shown in Equation~\ref{eq:ln}, the analytical form of a GBM translates to a Log-Normal distribution with the given mean and variance. We use this user-dependent distribution to generate an experience value $e_d$ for the review $d$ written by her at timestamp $t'_d$. 

Following standard LDA, the facet proportion $\theta_d$ of the review is drawn from a Dirichlet distribution with concentration parameter $\alpha$, and the facet $z_{d,w}$ of each word $w$ in $d$ is drawn from a Multinomial($\theta_d$).
%
%Each word $w$ that is going to be used at time $t$ is also associated with an experience value $l_{t,w}$ --- which depends on the experience $e_{D_t}$ of the reviews $D_t$ written at time $t$ (for instance, many sophisticated words are expected in the review of experienced users). We simply draw $l_{t,w} \sim \text{Normal} (e_{D_t}, \sigma_1)$.

Having generated the experience values, we can now generate the language model and individual words in the review.
Here, the language model $\beta_{t,z,w}$ uses the state-transition Equation~\ref{eq:beta}, and the actual word $w$ is based on its facet $z_{d,w}$ and timepoint $t_d$ according to a Multinomial($\pi(\beta_{t_d,z_{d,w}})$), where the transformation $\pi$ is given by Equation~\ref{eq:pi}.

Note that technically, the distribution $\beta_t$ and word $w$ have to be generated simultaneously: for $\beta_t$ we require the terms $l_{t,w}$, which depend on the experience and the words. Thus, we have a joint distribution $P(\beta_t,w|\ldots)$. Since, however, words are \emph{observed} during inference, this dependence is not crucial, i.e.\ $l_{t,w}$ can be computed once the experience values are known using Equation~\ref{eq:word-exp}. 

\begin{figure}
 \centering
 \includegraphics[scale=0.5]{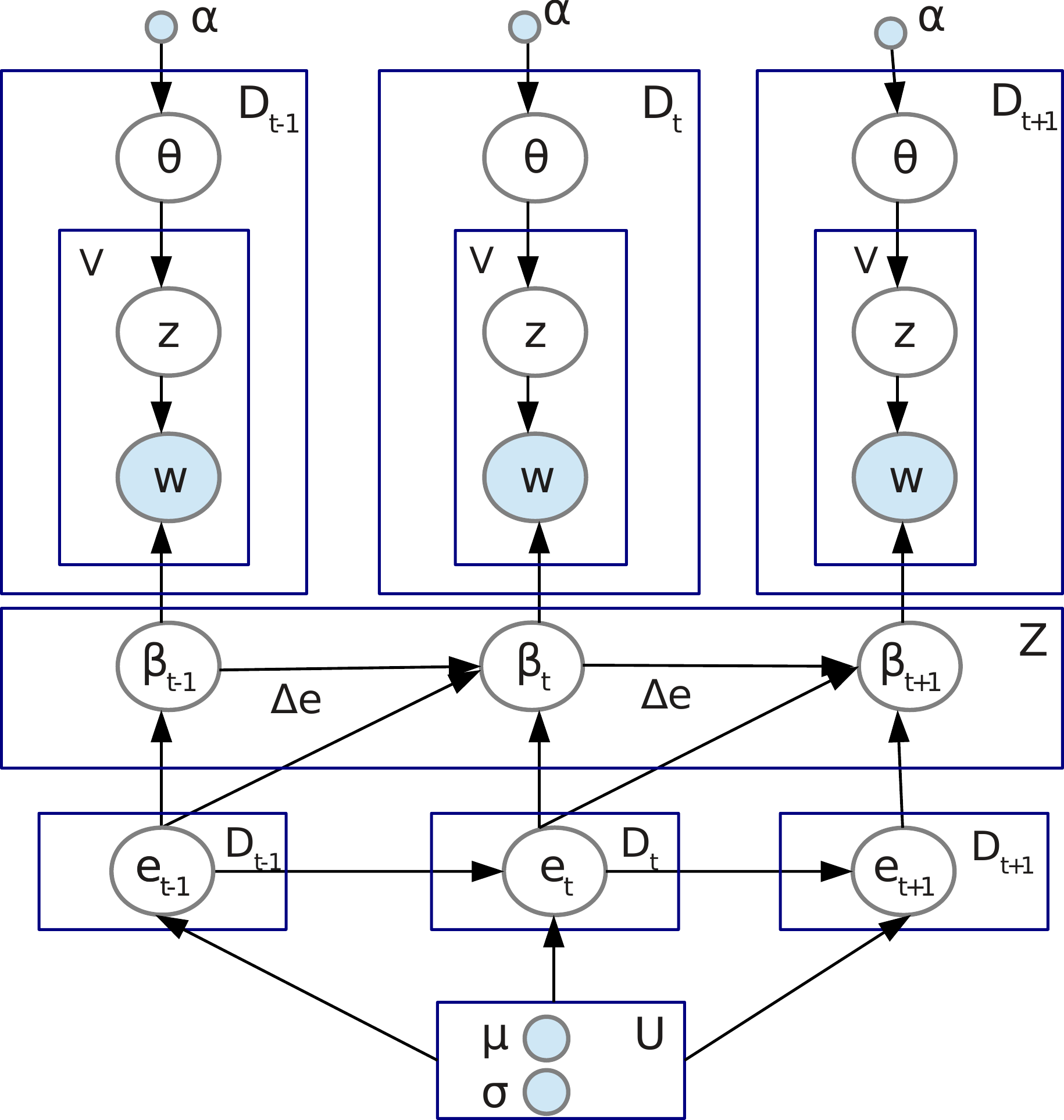}
  \caption{Continuous experience-aware language model. Words (shaded in blue), and timestamps (not shown for brevity) are observed.}
  %\todo{remove the l; keep directed edges! e --> b, b-->b, b--> w}} 
  %\todo{needs to be fixed! beta are outside. and parameters in boxed were wrong. should we include the brownian motion parameters? we should also make the alpha as dots since they are hyperparameter. does the z influence the b!?!? same question  as before!}}
  \label{fig:2}
 \end{figure}

We use this observation to simplify the notations and illustrations of Algorithm~\ref{algo:1}, which outlines the generative process, and Figure~\ref{fig:2}, which depicts it visually in plate notation for graphical models.

% \todo{should we somewhere add the formal definition for $l_tw$ since i have removed it from the inference -- either here or probably even better to the previous chapter where we explain the l anyways.}
%\todo{actually i think this is not correct; since we still dont know which words are effected. probably we shouldn't do it per word then.... hmmm.. it might becomes an undirected graphical model...}
%

 \begin{algorithm}
\SetAlgoLined
\DontPrintSemicolon
{
1. {Set granularity $t$ for language model evolution (e.g., years, months, days)}\;
2. {Set granularity for experience evolution, timestamp $t'$ (e.g., minutes, seconds)}\;

\For {each coarse timepoint t} {
\For {each review $d\in D_t$} {
	// retrieve user $u=u_d$ and fine-grained timepoint $t'=t'_d$\;
3. Draw $e_d \sim \text{Log-Normal}((\mu_u-\frac{\sigma_u^2}{2})t{'} +log(s_{0,u}), \sigma_u \sqrt{t{'}})$\;
4. Draw $\theta_d \sim \text{Dirichlet}(\alpha)$\;
\For {each word $w$ in $d$} {
5. Draw $z_{d,w} \sim \text{Multinomial}(\theta_d$)\;
}
}
%6. Draw $l_{t,w} \sim \text{Normal}(e_{D_t}, \sigma_1)$\;
6. Draw $\beta_{t,z,w} \sim \text{Normal}(\beta_{t-1,z,w}, \sigma \cdot |l_{t,w} - l_{t-1,w}|)$\;
\For {each review $d\in D_t$} {
	\For {each word $w$ in $d$} {
7. Draw $w \sim \text{Multinomial}(\pi(\beta_{t_d,z_{d,w}}))$
}
}
}
	}

%\For {each user $u$} {
%	\For {each of her (sorted) review $d$ at timestamp $t^{'} \in t$} {
%		3. Draw $e_d \sim \text{Log-Normal}((\mu_u-\frac{\sigma_u^2}{2})t^{'} +log(s_{0,u}), \sigma_u \sqrt{t^{'}})$\;
%		4. Draw $\theta_d \sim \text{Dirichlet}(\alpha)$\;
%		\For {each word $w$ in $d$} {
%			5. Draw $l_t \sim \text{Normal}(e_d, \sigma_1 \cdot I)$\;
%			6. Draw $z \sim \text{Multinomial}(\theta_d$)\;
%			7. Draw $\beta_{z,t} \sim \text{Normal}(\beta_{z,t-1}, \sigma \cdot |l_t - l_{t-1}|)$\;
%			8. Draw $w \sim \text{Multinomial}(\beta_{z,t})$
%		}
%	}
%}

\caption{Generative model for continuous experience-aware language model.}
\label{algo:1}
\end{algorithm}

\subsubsection{Inference}

Let $E, L, Z, T$  and $W$ be the set of experience values of all reviews, experience values of words, facets, timestamps and words in the corpus, respectively. 
In the following, $d$ denotes a review and $j$ indexes a word in it. %Each review $i$, is associated with a unique user $u$, and timestamp $t{'}$. Each word is associated with a set of timestamps $t$ (e.g., $t = Year(t^{'})$). 
$\theta$ denotes the per-review facet distribution, and $\beta$ the language model respectively. 

The joint probability distribution is given by:

{
\begin{multline}
 P(E,L, Z,W, \theta, \beta | U, T; \alpha, \langle \mu \rangle, \langle \sigma \rangle) \propto \\
 \prod_{t \in T} \prod_{d\in D_t}  P(e_d; s_{0,u_d}, \mu_{u_d}, \sigma_{u_d})
 \boldsymbol{\cdot} \bigg( P(\theta_d ; \alpha) \cdot \prod_{j=1}^{N_d} P(z_{d,j} | \theta_d)   \cdot P(w_{d,j} | \pi(\beta_{z_{d,j},t})) \bigg) \\
 \boldsymbol{\cdot}  \bigg( \prod_{z \in Z} \prod_{w \in W} P (l_{t,w}; e_d)\cdot P(\beta_{t,z,w}; \beta_{t-1,z,w}, \sigma \cdot |l_{t,w} - l_{t-1,w}|) \bigg)\\
\end{multline}
}

The exact computation of the above distribution is intractable, and we have to resort to approximate inference. 

Exploiting conjugacy of the Multinomial and Dirichlet distributions, we can integrate out $\theta$ from the above distribution. Assuming $\theta$ has been integrated out, we can decompose the joint distribution as:

{
\begin{equation}
  P(Z, \beta, E, L | W, T) \propto P(Z, \beta | W, T) \cdot P( E | Z, \beta, W, T) \cdot P(L | E, W, T)
\end{equation}
}

%\todo{I am a bit confused here.  Could you please verify this again.}
The above decomposition makes certain conditional independence assumptions in line with our generative process.

\noindent{\bf Estimating Facets $Z$: }
We use Collapsed Gibbs Sampling~\cite{Griffiths02gibbssampling}, as in standard LDA,
to estimate the conditional distribution for each of the latent facets $z_{d,j}$,
which is computed over the current assignment for all other hidden
variables, after integrating out $\theta$. Let $n(d, z)$ denote the count of the topic $z$ appearing in review $d$.
In the following equation,
 $n(d,.)$  indicates the summation of the above counts over all possible $z\in Z$.  
%for the respective review. 
The subscript $-j$ denotes the value
of a variable excluding the data at the $j^{th}$ position.

The posterior distribution
$P(Z| \beta, W, T; \alpha )$ of the latent variable $Z$ is given by:
\begin{equation}
\centering
\begin{aligned}
 & P(z_{d,j} = k | z_{d,-j}, \beta, w_{d,j}, t, d; \alpha) \\
 & \propto \frac{n(d, k) + \alpha}{n(d, .) + Z \cdot \alpha} \boldsymbol{\cdot} P(w_n = w_{d,j}| \beta, t, z_n =k, z_{-n}, w_{-n}) & \\
 & = \frac{n(d, k) + \alpha}{n(d, .) + Z \cdot \alpha} \boldsymbol{\cdot} \pi(\beta_{t,k,w_n})\\
\end{aligned} 
\label{eq:gibbs}
%\todo{is it $\alpha_k$?}
\end{equation}
\hspace{2.5em} where, the transformation $\pi$ is given by Equation~\ref{eq:pi}.

\noindent{\bf Estimating Language Model $\beta$: }
In contrast to $\theta$, the variable $\beta$ cannot be integrated out by the same process, as Normal and Multinomial distributions are not conjugate. Therefore, we refer to  another approximation technique to estimate $\beta$.

In this work, we use {\em Kalman Filter}~\cite{kalman} to model the sequential language model evolution. It is widely used to model linear dynamic systems from a series of observed measurements over time, containing statistical noise, that produces robust estimates of unknown variables over a single measurement. It is a continuous analog to the Hidden Markov Model (HMM), where the state space of the latent variables is continuous (as opposed to the discrete state-space HMM); and the observed and latent variables evolve with Gaussian noise. 

We want to estimate the following state-space transition model:
\begin{equation}
\begin{aligned}
\beta_{t,z,w} | \beta_{t-1,z, w} & \sim N (\beta_{t-1,z, w}, \sigma \cdot |l_{t,w} - l_{t-1, w}|)\\
w_{d,j} | \beta_{t,z, w} & \sim Mult(\pi(\beta_{t,z,w})) \text {\hspace{1em}where, } z=z_{d,j}, t=t_d.\\
\end{aligned} 
\end{equation}

However, unlike standard Kalman Filter, we do not have any {\em observed} measurement of the variables --- due to the presence of {\em latent} facets $Z$. Therefore, we resort to {\em inferred} measurement from the Gibbs sampling process.  

Let $n(t, z, w)$ denote the number of times a given word $w$ is assigned to a facet $z$ at time $t$ in the corpus. Therefore,
\begin{equation}
 \beta^{inf}_{t,z,w} = \pi^{-1} \bigg( \frac{n(t, z, w) + \gamma}{n(t, z, .) + V \cdot \gamma} \bigg)
\end{equation}
where, we use the inverse transformation of $\pi$ given by Equation~\ref{eq:pi}, and 
%\todo{probably not beta; since we dont do the transformation here. just call it with some other greek letter; or we do the inverse pi transformation here. this would be the alternative}
$\gamma$ is used for smoothing.

\noindent{\bf Update Equations for Kalman Filter: }
%\todo{shouldn't it be $x_{t,z,w}$ too? otherwise it is confusing}
Let $p_t$ and $g_t$ denote the {\em prediction error}, and {\em Kalman Gain} at time $t$ respectively. The variance of the process noise and measurement is given by the difference of the experience value of the word observed at two successive timepoints. Following standard Kalman Filter calculations~\cite{kalman}, predict equations are given by:
\begin{equation}
\label{eq:kalman-predict}
\begin{aligned}
\widehat{\beta}_{t,z,w} & \sim N(\beta_{{t-1},z,w}, \sigma \cdot |l_{t,w} - l_{t-1, w}|)\\
\widehat{p}_t & = p_{t-1} + \sigma \cdot |l_{t-1,w} - l_{t-2, w}|
\end{aligned} %\todo{is it $\sim$ or $=$?}
\end{equation}
and the update becomes:
\begin{equation}
\label{eq:kalman-update}
\begin{aligned}
g_t & = \frac{\widehat{p}_t}{ \widehat{p}_t + \sigma \cdot |l_{t,w} - l_{t-1, w}|}\\
\beta_{t,z,w} & = \widehat{\beta}_{t,z,w} + g_t \cdot (\beta^{inf}_{t,z,w} - \widehat{\beta}_{t,z,w}) \\
p_t & = (1 - g_t) \cdot \widehat{p}_t
\end{aligned} 
\end{equation}
Thus, the new value for $\beta_{t,z,w}$ is given by Eq. \ref{eq:kalman-update}.

If the experience does not change much between two successive timepoints, i.e. the variance is close to zero, the Kalman Filter just emits the counts as estimated by Gibbs sampling (assuming, $P_0 = 1$). This is then similar to the Dynamic Topic Model~\cite{BleiDTM}. Intuitively, the Kalman Filter is smoothing the estimate of Gibbs sampling taking the experience evolution into account.

\noindent{\bf Estimating Experience $E$: }
The experience value of a review depends on the user and the language model $\beta$. Although we have the state-transition model of $\beta$, the previous process of estimation using Kalman Filter cannot be applied in this case, as there is no observed or inferred value of $E$. Therefore, we resort to
Metropolis Hastings sampling. 
%It is a Markov Chain Monte Carlo (MCMC) method for obtaining sequences of random variables from a distribution for which direct sampling is difficult.
Instead of sampling the $E$'s from the complex true distribution, we use a proposal distribution for sampling the random variables --- followed by an acceptance or rejection of the newly sampled value.
% , whose density is {\em proportional} to the target distribution so that exact computation is not required.
 That is, at each iteration, the algorithm samples a value of a random variable --- where the current estimate depends only on the previous estimate, thereby, forming a Markov chain. %The decision to accept the candidate depends on how close the proposal density is to the target one.

 Assume all reviews $\{\cdots d_{i-1}, d_i, d_{i+1} \cdots\}$ from all users are sorted according to their timestamps.
 As discussed in Section~\ref{sec:time}, for computational feasibility,
 %tractability 
 we use a coarse granularity for the language model $\beta$. For the inference of $E$, however, we need to operate at the fine temporal resolution of the reviews' timestamps (say, in minutes or seconds).
Note that the process defined in Eq.~\eqref{eq:beta} represents the aggregated language model over multiple fine-grained timestamps. Accordingly, its corresponding fine-grained counterpart is $\beta_{t'_{d_i},z,w} \sim Normal (\beta_{t'_{d_{i-1}},z,w}, \sigma \cdot |e_{d_i} - e_{d_{i-1}}|) $ --- now operating on $t'$ and the review's individual experience values.
Since the language model is given (i.e. previously estimated) during the inference of $E$ , we can now easily refer to this fine-grained definition for the Metropolis Hastings sampling.

 %Note that during inference we do not materialize the values $l_{d,w}$, instead we directly 
 
% Note that when sampling an new value for $e_d$, we directly re-estimate the values of $l_{t,w}$
% its $l_{d,j}$ via its maximum likelihood estimate.

%At a fine temporal resolution, the variance $\Delta l$ of the language model at successive timepoints is proportional to the experience value difference $\Delta e$  of the reviews at those timepoints.

As the proposal distribution for the experience of review $d_i$ at time $t'_{d_i}$ , we select the corresponding user's GBM ($u=u_d$) and sample a new experience value $\widehat{e_{d_i}}$ for the review:
$$\widehat{e}_{d_i} \sim \text{Log-Normal}((\mu_u-\frac{\sigma_u^2}{2})t'_{d_i} +log(s_{0,u}), \sigma_u \sqrt{t'_{d_i}})$$

%The true experience value $e_t$ of $d_t$ depends on its user $u$, and the set of words and facets used in the review, i.e. the language model.
 
\noindent The language model $\beta_{t'_{d_i}}$ at time $t'_{d_i}$ depends on the language model $\beta_{t'_{d_{i-1}}}$ at time $t'_{d_{i-1}}$, and experience value difference $|e_{d_i} - e_{d_{i-1}}|$ between the two timepoints. Therefore, a change in the experience value at any timepoint affects the language model at the \emph{current} and {next} timepoint, i.e. $\beta_{t^{'}_{d_{i+1}}}$ is affected by $\beta_{t^{'}_{d_{i}}}$, too.

 %Each review $d_t$, associated to a user $u$, consists of a sequence of words $\{w\}$ with the corresponding ``best'' sequence of facets $\{z\}$ estimated during Gibbs sampling. 
Thus, the acceptance ratio of the Metropolis Hastings sampling becomes:
\begin{multline}
\label{eq:prop}
Q = \prod_{w,z} \bigg[\frac{N(\beta_{t'_b,z,w}; \beta_{t'_a, z, w}, \sigma \cdot | \widehat{e_b} - e_{a}|)}{N(\beta_{t'_b,z,w}; \beta_{t'_a, z, w}, \sigma \cdot | e_b - e_{a}|)}
\boldsymbol{\cdot} \frac{N(\beta_{t'_c,z,w}; \beta_{t'_b, z, w}, \sigma \cdot | e_{c} - \widehat{e_b}|)}{N(\beta_{t'_c,z,w}; \beta_{t'_b, z, w}, \sigma \cdot | e_{c} - e_b|)}\bigg]
\end{multline}
where $a=d_{i-1}$, $b=d_{i}$ and $c=d_{i+1}$.
The numerator accounts for the modified distributions affected by the updated experience value, and the denominator discounts the old ones. Note that since the GBM has been used as the proposal distribution, its factor cancels out in the term $Q$.

Overall, the Metropolis Hastings algorithm iterates over the following steps:
{\setlength{\leftmargini}{5mm}
\begin{enumerate}
 \item Randomly pick a review $d$ at time $t'=t'_d$ by user $u=u_d$ with experience $e_d$
 \item Sample $\widehat{e_d} \sim \text{Log-Normal}\bigg((\mu_u-\frac{\sigma_u^2}{2})t' +log(s_{0,u}), \sigma_u \sqrt{t'}\bigg)$
 \item Accept $\widehat{e_d}$ as the new experience with probability $P$\,$=$\,$min(1, Q)$
\end{enumerate}}

%Once the experience value of all reviews have been estimated, it is straightforward to estimate the experience $l_{t,w}$ of a word $w$ at time $t$ via its maximum likelhood estimate: Since $l_{t,w}$ is drawn from a Normal distribution with mean $e_{D_t}$, the estimate is given by the average experience of all the reviews $\langle d_t \rangle$ containing the word $w$ at time $t$.
 
 %\begin{equation}
 %\label{eq:word-exp}
 %l_{t,w} =  \frac{\sum_{d\in D_t: w \in d} e_d}{|D_t|}
 %\end{equation}\todo{what if a word accours mutiple times? should we count it then multiple times?}

\noindent{\bf Estimating Parameters for the Geometric Brownian Motion: }
For each user $u$, the mean $\mu_u$ and variance $\sigma_u$ of her GBM trajectory are estimated from the sample mean and variance. 

Consider the set of all reviews $\langle d_t \rangle$ written by $u$, and $\langle e_t \rangle$ be the corresponding experience values of the reviews. 

Let $\widehat{m}_u=\frac{\sum_{d_t} log(e_t)}{|d_t|}$, and $\widehat{s}_u^2 = \frac{\sum_{d_t} (log(e_t)-\widehat{m}_u)^2}{|d_t-1|}$. 

Furthermore, let $\Delta$ be the average length of the time intervals for the reviews of user $u$. 
%\todo{what does this mean? average length? for certain reviews? indexed, i.e. $\Delta_t$?}

Now, $log(e_t) \sim N \big((\mu_u-\frac{\sigma_u^2}{2})\Delta +log(s_{0,u}), \sigma_u \sqrt{\Delta}\big)$.

From the above equations we can obtain the following estimates using Maximum Likelihood Estimation (MLE):
%\todo{do we have a reference for this?}:
\begin{equation}
 \begin{aligned}
 \widehat{\sigma}_u &= \frac{\widehat{s}_u}{\sqrt{\Delta}}\\ 
 \widehat{\mu}_u &= \frac{\widehat{m}_u - log(s_{0,u})}{\Delta} + \frac{{\widehat{\sigma}_u}^2}{2}\\
  &= \frac{\widehat{m}_u - log(s_{0,u})}{\Delta} + \frac{{\widehat{s}_u}^2}{2\Delta}
 \end{aligned}
\end{equation}

\pagebreak

\noindent{\bf Overall Processing Scheme: } Exploiting the results from the above discussions, the overall inference is an iterative process consisting of the following steps:
\begin{enumerate}\setlength{\itemsep}{0pt}
 \item Estimate facets $Z$ using Equation~\ref{eq:gibbs}.
 \item Estimate $\beta$ using Equations~\ref{eq:kalman-predict} and~\ref{eq:kalman-update}.
 \item Sort all reviews by timestamps, and estimate $E$ using Equation~\ref{eq:prop} and the Metropolis Hastings algorithm, for a random subset of the reviews.
 \item Once the experience values of all reviews have been determined, estimate $L$ using Equation~\ref{eq:word-exp}.
\end{enumerate}

%% file: main/chapter-temporal-evolution/cont-model/experiments.tex
\subsection{Experiments}
\label{sec:experiments}

We perform experiments with data from five communities in different domains:
\begin{itemize}
 \item BeerAdvocate ({\tt \href{http://www.beeradvocate.com}{beeradvocate.com}}) and RateBeer ({\tt \href{http://www.ratebeer.com}{ratebeer.com}}) for beer reviews
 \item Amazon ({\tt \href{http://www.amazon.com}{amazon.com}}) for movie reviews
 \item Yelp ({\tt \href{http://www.yelp.com}{yelp.com}}) for food and restaurant reviews
 \item NewsTrust ({\tt \href{http://www.newstrust.net}{newstrust.net}}) for reviews of news media
\end{itemize}

Table~\ref{tab:statistics} gives the dataset statistics\footnote{\url{http://snap.stanford.edu/data/},\url{http://www.yelp.com/dataset\_challenge/}, \url{http://resources.mpi-inf.mpg.de/impact/credibilityanalysis/data.tar.gz}}. 
We have a total of $12.7$ million reviews from $0.9$ million users over $16$ years from all of the five communities combined. The first four communities are used for product reviews, from where we extract the following quintuple for our model $<userId, itemId, timestamp, rating, review>$. 
NewsTrust is a special community, which we discuss in Section~\ref{sec:usecases}.

\begin{table}
\centering
 \setlength{\tabcolsep}{1.4mm}
\begin{tabular}{lrrrc}
\toprule
\bf{Dataset} & \bf{\#Users} & \bf{\#Items} & \bf{\#Ratings} & \bf{\#Years}\\
% & & & & \bf{(Years)}\\
\midrule
\bf{Beer (BeerAdvocate)} & 33,387 & 66,051 & 1,586,259 & 16\\
\bf{Beer (RateBeer)} & 40,213 & 110,419 & 2,924,127 & 13\\
\bf{Movies (Amazon)} & 759,899 & 267,320 & 7,911,684 & 16\\
\bf{Food (Yelp)} & 45,981 & 11,537 & 229,907 & 11\\
\bf{Media (NewsTrust)} & 6,180 & 62,108 & 89,167 & 9\\
\midrule
\bf{TOTAL} & 885,660 & 517,435 & 12,741,144 & -\\
\bottomrule
\end{tabular}
\caption{Dataset statistics.}
\label{tab:statistics}
\end{table}

\subsubsection{Data Likelihood, Smoothness and Convergence}

Inference of our model is quite involved with different Markov Chain Monte Carlo methods. It is imperative to show that the resultant model is not only stable, but also improves the log-likelihood of the data. Although there are several measures to evaluate the quality of facet models, we report the following from~\cite{wallach}:\\
$LL = \sum_d \sum_{j=1}^{N_d} log\ P(w_{d,j} | \beta; \alpha)$. A higher likelihood indicates a better model.

\begin{figure*}
\centering
 \includegraphics[width=\linewidth, height=3.5cm]{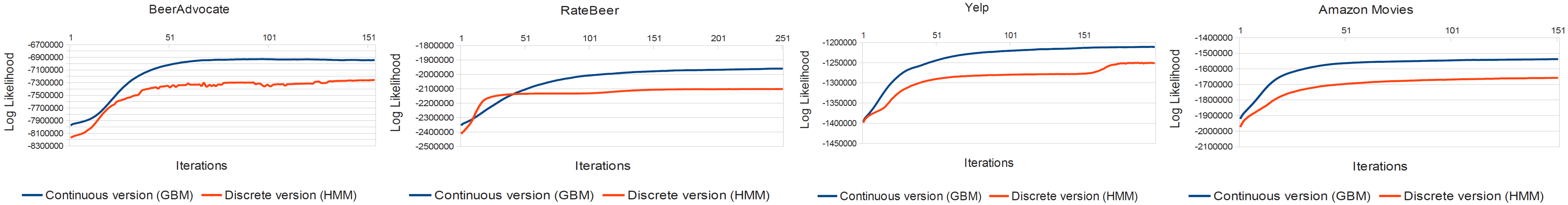}
 \caption{Log-likelihood per iteration of discrete model (refer to Section~\ref{sec:discrete}) vs. continuous experience model (this work).}
  \label{fig:log-likelihood}
\end{figure*}

Figure~\ref{fig:log-likelihood} contrasts the log-likelihood of the data from the continuous experience model and its discrete counterpart (refer to Section~\ref{sec:discrete}). We find that the continuous model is stable and has a {\em smooth} increase in the data log-likelihood {\em per iteration}. This can be attributed to how smoothly the language model evolves over time, preserving the Markov property of experience evolution. Empirically our model also shows a fast convergence, as indicated by the number of iterations. 

On the other hand, the discrete model not only has a worse fit, but is also less smooth. It exhibits abrupt state transitions in the Hidden Markov Model, when the experience level changes (refer to Figure~\ref{fig:exp}). This leads to abrupt changes in the language model, as it is coupled to experience evolution. %Even for the same user, at the same level of experience, the language model may change 

%to perform much better than  in improving the log likelihood of data to obtain a better fit. Empirically, it also shows a faster and smoother convergence.

\subsubsection{Experience-aware Item Rating Prediction}

In the first task, we show the effectiveness of our model for item rating prediction. Given a user $u$, an item $i$, time $t$, and review $d$ with words $\langle w \rangle$ --- the objective is to predict the rating the user would assign to the item based on her {\em experience}. 

For prediction, we use the following features: The experience value $e$ of the user is taken as the last experience attained by the user during training. %The language model $\beta$ is also learned during inference. 
Based on the learned language model $\beta$, we construct the language feature vector $\langle F_w = log(max_z (\beta_{t,z,w})) \rangle$ of dimension $V$ (size of the vocabulary). That is, for each word $w$ in the review, we consider the value of $\beta$ corresponding to the best facet $z$ that can be assigned to the word at the time $t$. We take the log-transformation of $\beta$ which empirically gives better results.

Furthermore, as also done in the baseline works~\cite{mcauleyWWW2013} and the discrete version of our model (refer to Section~\ref{sec:discrete}), we consider: $\gamma_g$, the average rating in the community;  $\gamma_u$, the offset of the average rating given by user $u$ from the global average; and $\gamma_i$, the rating bias for item $i$. 

Thus, combining all of the above, we construct the feature vector
\noindent $\langle \langle F_w \rangle, e, \gamma_g, \gamma_u, \gamma_i \rangle$ for each review with the user-assigned ground rating for training. We use Support Vector Regression~\cite{drucker97}, with the same set of default parameters as used in our discrete model (refer to Section~\ref{sec:discrete}), for rating prediction. %Table~\ref{fig:MSE} shows the mean-squared error comparison of our model with all the baselines for this prediction task.

\begin{table*}[ht]
\centering
\begin{tabular}{lccccc}
\toprule
\bf{Models} & \bf{BeerAdvocate} & \bf{RateBeer} & \bf{NewsTrust} & \bf{Amazon} & \bf{Yelp}\\
\midrule
{\bf Continuous experience model} & {\bf 0.247} & {\bf 0.266} & {0.494} & {\bf 1.042} & {\bf 0.940}\\
(this work) & & & & &\\
Discrete experience model & 0.363 & 0.309 & 0.{\bf 464} & 1.174 & 1.469\\
(Section~\ref{sec:discrete}) & & & & &\\
User at learned rate & 0.379 & 0.336 & 0.575 & 1.293 & 1.732\\
\cite{mcauleyWWW2013} & & & & &\\
Community at learned rate  & 0.383 & 0.334 & 0.656 & 1.203 & 1.534\\
\cite{mcauleyWWW2013} & & & & &\\
Community at uniform rate  & 0.391 & 0.347 & 0.767 & 1.203 & 1.526\\
\cite{mcauleyWWW2013} & & & & &\\
User at uniform rate & 0.394 & 0.349 & 0.744 & 1.206 & 1.613\\
\cite{mcauleyWWW2013} & & & & &\\
Latent factor model & 0.409 & 0.377 & 0.847 & 1.248 & 1.560\\
\cite{koren2011advances} & & & & &\\
\bottomrule
\end{tabular}
\caption{Mean squared error (MSE) for rating prediction. Our model performs better than competing methods.}
\label{fig:MSE}
\end{table*}

\pagebreak

\noindent{\bf Baselines: }
We consider the following baselines [b -- e] from~\cite{mcauleyWWW2013}, and use their code\footnote{Code available from \url{http://cseweb.ucsd.edu/~jmcauley/code/}} for experiments. Baseline (f) is our prior discrete experience model (refer to Section~\ref{sec:discrete}).
%(refer to Figure~\ref{fig:baselines}).\\

\begin{itemize}
\item[a)]\emph{LFM}: A standard latent factor recommendation model~\cite{korenKDD2008}.
\item[b)]\emph{Community at uniform rate}: Users and products in a community evolve using a single ``global clock''~\cite{KorenKDD2010, xiongSDM2010, XiangKDD2010}, where the different stages of the community evolution appear at uniform time intervals. %So the community prefers different products at different times.
\item[c)]\emph{Community at learned rate}: This extends b) by learning the rate at which the community evolves with time, eliminating the uniform rate assumption.\\
\item[d)]\emph{User at uniform rate}: This extends b) to consider individual users, by modeling the different stages of a user's progression based on preferences and experience levels evolving over time. 
The model assumes a uniform rate for experience progression. 
\item[e)]\emph{User at learned rate}: This extends d) by allowing the {\em experience} of each user to evolve on a ``personal clock'', where the time to reach certain ({\em discrete}) experience levels depends on the user \cite{mcauleyWWW2013}. This is reportedly the best version of their experience evolution models.
\item[f)]\emph{Discrete experience model}: This is our prior approach (refer to Section~\ref{sec:discrete}) for the discrete version of the experience-aware language model, where the experience of a user depends on the evolution of the user's maturing rate, facet preferences, and writing
style.
\end{itemize}

%Only the last baseline (i.e. discrete model) uses textual features for user evolution. \todo{questions: by the text feature is used for all models? the sentence is a bit unclear}
%\vspace{-1em}

\subsubsection{Quantitative Results}

Table~\ref{fig:MSE} compares the \emph{mean squared error (MSE)} for rating predictions in this task, generated by our model versus the six baselines. Our model outperforms all baselines --- except in the NewsTrust community, performing slightly worse than our prior discrete model (discussed in Section~\ref{sec:usecases}) --- reducing the MSE by ca. $11\%$ to $36\%$. Our improvements over the baselines are statistically significant at $99\%$ level of confidence determined by {\em paired sample t-test}.

For all models, we used the three most recent reviews of each user as withheld test data. All experience-based models consider the \emph{last} experience value reached by each user during training, and the corresponding learned parameters for rating prediction. Similar to the setting in~\cite{mcauleyWWW2013}, we consider users with a minimum of $50$ reviews. Users with less than $50$ reviews are grouped into a background model, and treated as a single user. We set $Z=5$ for BeerAdvocate, RateBeer and Yelp facets; and $Z=20$ for Amazon movies and $Z=100$ for NewsTrust which have richer latent dimensions. All {\em discrete} experience models consider $E=5$ experience levels. In the continuous model, the experience value $e \in (0, \infty)$. We initialize the parameters for our joint model as: $s_{0,u} = 1, \alpha=50/Z, \gamma=0.01$. Our performance improvement is strong for the {\em BeerAdvocate} community due to large number of reviews per-user for a long period of time, and low for NewsTrust for the converse.

\begin{figure*}
\centering
 \includegraphics[width=\linewidth, height=3.8cm]{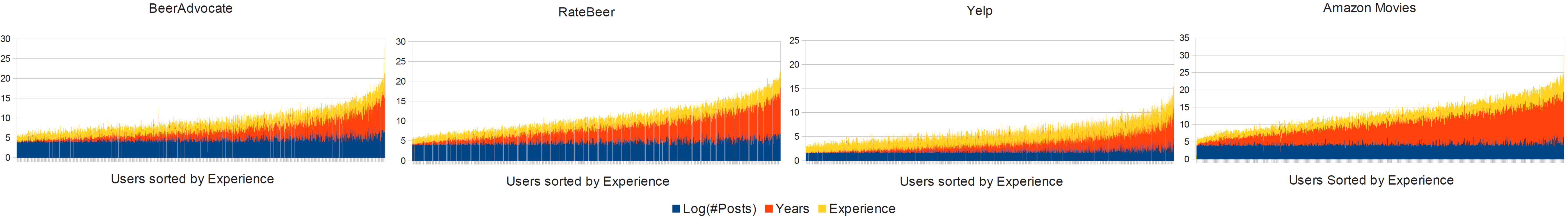}
 \caption{\hspace{0.2em}Variation of {\em experience} ($e$) with {\em years} and {\em reviews} of each user. Each bar in the above stacked chart corresponds to a user with her most recent experience, number of years spent, and number of reviews posted in the community.}
  \label{fig:user-activity}
\end{figure*}

\begin{figure*}
\centering
 \includegraphics[width=\linewidth, height=3.8cm]{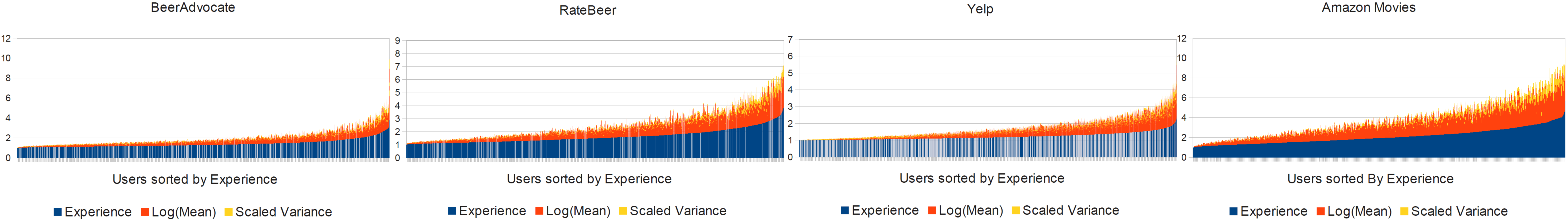}
 \caption{\hspace{0.2em}Variation of {\em experience} ($e$) with {\em mean} ($\mu_u$) and {\em variance} ($\sigma_u$) of the GBM trajectory of each user ($u$). Each bar in the above stacked chart corresponds to a user with her most recent experience, mean and variance of her experience evolution.}
  \label{fig:user-mean}
\end{figure*}

\begin{figure*}
\centering
 \includegraphics[width=\linewidth, height=3.8cm]{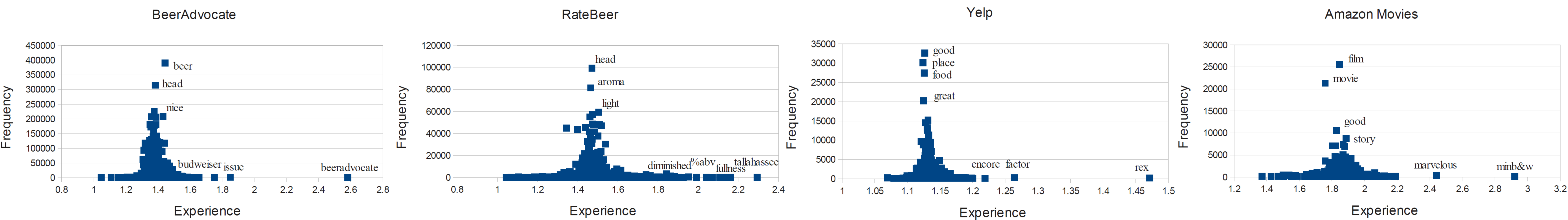}
 \caption{\hspace{0.2em}Variation of {\em word frequency} with {\em word experience}. Each point in the above scatter plot corresponds to a word ($w$) in ``2011'' with corresponding frequency and experience value ($l_{t\text{=2011},w}$).}
  \label{fig:word-exp-freq}
\end{figure*}

\begin{figure*}
\centering
 \includegraphics[width=\linewidth, height=3.8cm]{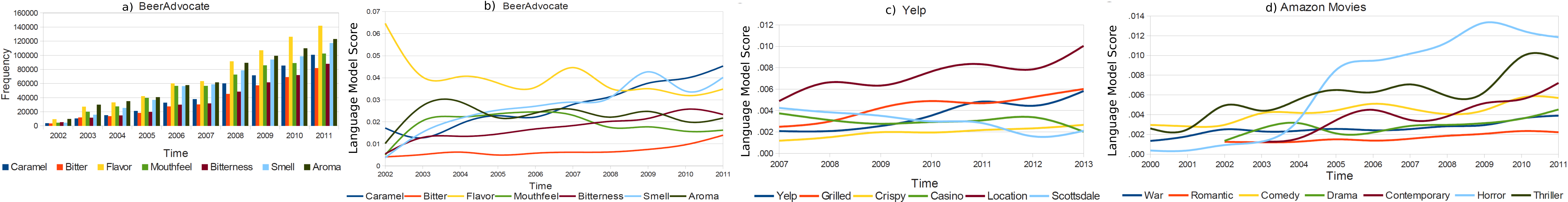}
 \caption{\hspace{0.2em}{\em Language model} score ($\beta_{t,z,w} \boldsymbol{\cdot} l_{t,w}$) variation for sample words with {\em time}. Figure a) shows the count of some sample words over time in BeerAdvocate community, whose evolution is traced in Figure b). Figures c) and d) show the evolution in Yelp and Amazon Movies.}
  \label{fig:word-evol}
\end{figure*}

%\paragraph{Qualitative Results}
\subsubsection{Qualitative Results}

{\noindent \bf User experience progression:} Figure~\ref{fig:user-activity} shows the variation of the users' {\em most recent} experience (as learned by our model), along with the number of reviews posted, and the number of years spent in the community. As we would expect, a user's experience increases with the amount of {\em time} spent in the community. On the contrary, number of reviews posted does not have a strong influence on experience progression. Thus, if a user writes a large number of reviews in a short span of time, her experience does not increase much; in contrast to if the reviews are written over a long period of time. 

Figure~\ref{fig:user-mean} shows the variation of the users' {\em most recent} experience, along with the mean $\mu_u$ and variance $\sigma_u$ of her Geometric Brownian Motion (GBM) trajectory --- all learned during inference. We observe that users who reach a high level of experience progress faster (i.e. a higher value of $\mu_u$) than those who do not. Experienced users also exhibit comparatively higher variance than amateur ones. This result also follows from using the GBM process, where the mean and variance tend to increase with time.

{\noindent \bf Language model evolution:} Figure~\ref{fig:word-exp-freq} shows the variation of the frequency of a word --- used in the community in ``2011'' --- with the {\em learned} experience value $l_{t,w}$ associated to each word. The plots depict a bell curve. Intuitively, the experience value of a word does not increase with general usage; but increases if it has been used by experienced users. Highlighted words in the plot give some interesting insights. For instance, the words ``beer, head, place, food, movie, story'' etc. are used with high frequency in the beer, food or movie community, but have an average experience value. On the other hand specialized words like ``beeradvocate, budweiser, \%abv, fullness, encore, minb\&w'' etc. have high experience value. 

Table~\ref{tab:sample-words} shows some top words used by {\em experienced} users and amateur ones in different communities, as learned by our model. Note that this is a ranked list of words with numeric values (not shown in the table). %\todo{tab 3 has alread been mentioned; merge it?}
We see that experienced users are more interested about fine-grained facets like the mouthfeel, ``fruity'' flavors, and texture of food and drinks; narrative style of movies, as opposed to popular entertainment themes; discussing government policies and regulations in news reviews etc.

The word ``rex'' in Figure~\ref{fig:word-exp-freq} in Yelp, appearing with low frequency and high experience, corresponds to a user ``Rex M.'' with ``Elite'' status who writes humorous reviews with {\em self} reference.

Figure~\ref{fig:word-evol} shows the evolution of some sample words over {\em time} and experience (as given by our model) in different communities. The score in the {\em y-axis} combines the language model probability $\beta_{t,z,w}$ with experience value $l_{t,w}$ associated to each word $w$ at time $t$. 

Figure~\ref{fig:word-evol} a) illustrates the frequency of the words in BeerAdvocate, while their evolution is traced in Figure~\ref{fig:word-evol} b). It can be seen that the overall usage of each word increases over time; but the evolution path is different for each word. For instance, the ``smell'' convention started when ``aroma'' was dominant; but the latter was less used by {\em experienced} users over time, and slowly replaced by (increasing use of) ``smell''. This was also reported in~\cite{DanescuWWW2013} in a different context. Similarly ``caramel'' is likely to be used more by {\em experienced} users, than ``flavor''. Also, contrast the evolution of ``bitterness'', which is used more by 
experienced users, compared to ``bitter''. %\todo{where exactly do we see this?}

In Yelp, we see certain food trends like ``grilled'' and ``crispy'' increasing over time; in contrast to a decreasing feature like ``casino'' for restaurants. For Amazon movies, we find certain genres like ``horror, thriller'' and ``contemporary'' completely dominating other genres in recent times.

\begin{table}[t]
%\centering
\small
 \setlength{\tabcolsep}{2mm}
\begin{tabular}{p{7cm}p{7cm}}
\toprule
{\bf Most Experience} & {\bf Least Experience}\\\midrule
{\noindent \bf BeerAdvocate} & \\
chestnut\_hued near\_viscous rampant\_perhaps faux\_foreign cherry\_wood sweet\_burning bright\_crystal faint\_vanilla boned\_dryness woody\_herbal citrus\_hops mouthfeel& originally flavor color didnt favorite dominated cheers tasted review doesnt drank version poured pleasant bad bitter sweet\\ \midrule
{\noindent \bf Amazon} & \\
aficionados minimalist underwritten theatrically unbridled seamless retrospect overdramatic diabolical recreated notwithstanding oblivious featurettes precocious & viewer entertainment battle actress tells emotional supporting evil nice strong sex style fine hero romantic direction superb living story\\\midrule
{\bf Yelp} &\\
rex foie smoked marinated savory signature contemporary selections bacchanal delicate grits gourmet texture exotic balsamic & mexican chicken salad love better eat atmosphere sandwich local dont spot day friendly order sit \\\midrule
{\bf NewsTrust} &\\
health actions cuts medicare oil climate major jobs house vote congressional spending unemployment citizens events & bad god religion iraq responsibility questions clear jon led meaningful lives california powerful\\
\bottomrule
\end{tabular}
\caption{\hspace{0.2em}Top words used by experienced and amateur users.}
\label{tab:sample-words}
\end{table}

%% file: main/chapter-temporal-evolution/usecases.tex
\section{Use-Case Study}
\label{sec:usecases}

Sections~\ref{sec:discrete} and~\ref{sec:continuous} discuss the evolution of user experience in online communities --- with applications focused on recommending items (like beers or movies) to users based on their maturity. As another application use-case, we switch to a different kind of items -- newspapers and news articles -- tapping into the NewsTrust online community ({\tt \href{www.newstust.net}{newstrust.net}}). NewsTrust features news stories posted and reviewed by members, 
many of whom are professional journalists and content experts. Stories are reviewed based on their objectivity, rationality, and general quality of language to present an unbiased and balanced narrative of an event. The focus is on \emph{quality journalism}. Unlike the other datasets, NewsTrust contains expertise of members that can be used as ground-truth for evaluating our model-generated {\em experience} values of users. Previously in Section~\ref{sec:newstrust-data}, we had discussed several characteristics of this community that were employed for credibility analysis therein.

In our framework of item recommendation, each story is an item, which is rated and reviewed by a user. The facets are the underlying topic distribution of reviews, with (latent) topics being {\em Healthcare, Obama Administration, NSA}, etc. The facet preferences can be mapped to the (political) polarity of users in the news community. 

\subsection{Recommending News Articles}

Our first objective is to recommend news to readers catering to their facet preferences, viewpoints, and experience. We apply our joint model to this task, and compare the predicted ratings with the ones observed for withheld reviews in the NewsTrust community. 

The mean squared error (MSE) results for this task were reported in Table~\ref{fig:MSE}. Our continuous model clearly outperforms most of the baselines; it performs only slightly worse regarding our prior discrete model (discussed in Section~\ref{sec:discrete}) in this task --- possibly due to high rating / data sparsity in face of a large number of model parameters and less number of reviews per-user. 

%For the {\em discrete} version of our model (refer to Section~\ref{sec:discrete}), the mean squared error (MSE) results were reported in Table~\ref{fig:MSE1} in Section~\ref{sec:experiments1}. 
Table~\ref{tab:newstrustTopics} shows salient examples of the vocabulary by users at different (discrete) experience levels on the topic {\em US Election} as generated by the {\em discrete} version of our model (refer to Section~\ref{sec:discrete}).

\begin{table}
\small
\centering
\begin{tabular}{p{12cm}}
\toprule
\textbf{Level 1:} bad god religion iraq responsibility\\
\textbf{Level 2:} national reform live krugman questions clear jon led meaningful lives california powerful safety impacts\\
\textbf{Level 3:} health actions cuts medicare nov news points oil climate major jobs house high vote congressional spending unemployment strong taxes citizens events failure\\
\bottomrule
\end{tabular}
\caption{\hspace{0.2em}Salient words for the {\em illustrative} NewsTrust topic {\em US Election} used by users at different levels of experience.}
\label{tab:newstrustTopics}
%\vspace{-1em}
\end{table}

%For the {\em continuous} version of our model (refer to Section~\ref{sec:continuous}), 

%\paragraph{Identifying Experienced Users}

\subsection{Identifying Experienced Users}

Our second task is to find experienced members of this community, who have the potential of being \emph{citizen journalists}. In order to evaluate the quality of the ranked list of experienced users generated by our model, we consider the following proxy measure for user experience. In NewsTrust, users have {\em Member Levels} determined by the NewsTrust staff based on  community engagement, time in the community, other users' feedback on reviews, profile transparency, and manual validation.

\pagebreak

We use these member levels to categorize users as {\em experienced} or {\em inexperienced}. This is treated as the ground truth for assessing the ranking quality of our model against the baseline models~\cite{mcauleyWWW2013}, and the discrete version of our prior work (discussed in Section~\ref{sec:discrete}) --- considering top $100$ users from each model ranked by experience. Here we consider the top-performing baseline models from the previous task.

%\todo{this is a bit unclear; we rank our users according to experience. and then? we use top k? how many?}

%Table~\ref{tab:ndcgUsers} shows the $F_1$ scores of these two competitors.

We report the \emph{Normalized Discounted Cumulative Gain (NDCG)} and the {\em Normalized Kendall Tau Distance} for the ranked lists of users generated by all the models. NDCG gives geometrically decreasing weights to predictions at the various positions of the ranked list:

$NDCG_p = \frac{DCG_p}{IDCG_p}$, \hspace{1em} where
$DCG_p = rel_1 + \sum_{i=2}^p \frac{rel_i}{\log_2 i}$

Here, $rel_i$ is the relevance ($0$ or $1$) of a result at position $i$.

The better model should exhibit higher {\em NDCG}, and lower {\em Kendall Tau Distance}. 

As Table~\ref{tab:ndcgUsers} shows, the {\em continuous} version of our model performs better than its discrete counterpart, which, in turn, outperforms~\cite{mcauleyWWW2013} in capturing user maturity.

%We also report the {\em Normalized Kendall Tau Distance} of the ranked lists given by the models, against the ground truth rank list.

\begin{table}

\begin{center}

%\centering

 \setlength{\tabcolsep}{2mm}

 \begin{tabular}{p{6cm}p{2cm}p{4cm}}

 \toprule

\bf{Models} &\bf{NDCG} & \bf{Kendall Tau}\\

\bf{} & \bf{} & \bf{Normalized Distance}\\

\midrule 

{\bf Continuous experience model}\newline (this work) & {\bf 0.917} & {\bf 0.113} \\

Discrete experience model\newline (refer to Section~\ref{sec:discrete}) & 0.898 & 0.134 \\

User at learned rate~\cite{mcauleyWWW2013} & 0.872 & 0.180 \\

\bottomrule

 \end{tabular}

\caption{\hspace{0.2em}Performance on identifying experienced users.}

\label{tab:ndcgUsers}

\end{center}

\end{table}

%% file: main/chapter-temporal-evolution/conclusion.tex
\section{Conclusion}

In this chapter, we propose models to capture the temporal evolution of users in online communities. These can be used to identify users who were not experienced when they joined the community, but could have evolved into a matured user now. Current recommender systems do not consider the temporal dynamics of user experience when generating recommendations. We propose experience-aware recommendation models --- that can adapt to the changing preferences and maturity of users in a community --- to recommend items that she will appreciate at her current maturity level. We exploit the coupling between the \emph{facet preferences} of a user, her \emph{experience}, \emph{writing style} in reviews, and \emph{rating behavior} to capture the user's temporal evolution. Our model is the first work that considers the progression of users' experience as expressed in the text of item reviews.

%We propose a generative HMM-LDA model for tracing a user's evolution, where %the Hidden Markov Model (HMM) traces her latent experience progressing over %time, and the Latent Dirichlet Allocation (LDA) model captures her interest %in specific item facets as a function of her (again latent) experience level %and writing style. The model considers user assigned ratings to items and 

%review text till a given timepoint as observables, and outputs the rating 

%the user would assign to an item at a later timepoint based on her evolved 

%experience.

\pagebreak

Furthermore, we develop an experience-aware language model that can trace the {\em continuous} evolution of a user's experience and her language explicitly over {\em time}. 
We combine principles of Geometric Brownian Motion, Brownian Motion, and Latent Dirichlet Allocation to model a smooth temporal progression of user experience, and language model over time.
This is also the first work to develop a continuous and generalized version of user experience evolution. 

We derive interesting insights from the evolution trajectory of users, and their vocabulary usage with change in experience. For instance, experienced users progress faster than amateurs, with the progression depending more on their time spent in the community than on activity. Experienced users also show a more predictable behavior, and have a distinctive writing style and facet preferences --- for example, experienced users in the Beer community use more ``fruity'' words to depict the smell and taste of a beer; and users in the News community are more interested about policies and regulations than amateurs who are more interested in polarizing topics.

%We propose a generative HMM-LDA model for tracing a user's evolution, where %the Hidden Markov Model (HMM) traces her latent experience progressing over %time, and the Latent Dirichlet Allocation (LDA) model captures her interest %in specific item facets as a function of her (again latent) experience level %and writing style. The model considers user assigned ratings to items and 
%review text till a given timepoint as observables, and outputs the rating 
%the user would assign to an item at a later timepoint based on her evolved 
%experience.

Our experiments -- with data from domains like beer, movies, food, and news -- demonstrate that our model effectively exploits user experience for item recommendation that
substantially reduces the mean squared error for predicted ratings, compared to the state-of-the-art baselines. %~\cite{mcauleyWWW2013,Subho:ICDM2015}.  
This shows our method can generate better recommendations than those models.

We further demonstrate the utility of our model in a use-case study on identifying experienced members in the NewsTrust community, where these users would be top candidates for being citizen journalists. Another similar use-case for our model can be to detect experienced medical professionals in the health community who can contribute valuable medical knowledge.

%Overall in this work we propose an experience-aware recommendation model 
%that can adapt to the changing preferences and maturity of users in a community.

%% file: main/chapter-credible-applications/main.tex
\chapter{Credibility Analysis of Product Reviews}
\label{chap:applications}

\section{Introduction}
\input{main/chapter-credible-applications/introduction}

\section{Exploring Latent Semantic Factors to Find Useful Product Reviews}
\label{chap:helpful}

%\footnote{This is a joint work with Sourav Dutta from Max Planck Institute for Informatics, Saarbruecken, Germany.}

%\input{main/chapter-credible-applications/helpful-reviews/introduction}

\input{main/chapter-credible-applications/helpful-reviews/features}

\input{main/chapter-credible-applications/helpful-reviews/joint-model}

\input{main/chapter-credible-applications/helpful-reviews/experiments}

\section{Finding Credible Reviews with Limited Information using Consistency Features}

\input{main/chapter-credible-applications/consistency-analysis/credibilityanalysis}

\input{main/chapter-credible-applications/consistency-analysis/experiments}

\input{main/chapter-credible-applications/consistency-analysis/discussions}
%\input{main/chapter-credible-applications/consistency-analysis/conclusions}

\pagebreak
\section{Conclusions}
\input{main/chapter-credible-applications/conclusions}

%% file: main/chapter-credible-applications/introduction.tex
Chapters~\ref{chap:framework} and~\ref{chap:temporal} develop probabilistic graphical models for credibility analysis in online communities and their temporal evolution, respectively. In the current chapter, we use the principles and models developed therein for some related tasks that have been of serious concern for product review communities in recent times.

With the rapid growth in e-Commerce, product reviews have become a crucial component for the business nowadays. As consumers cannot test the functionality of a product prior to purchase, these reviews help them make an informed decision to buy the product or not. As per the survey conducted by Nielsen Corporations, $40\%$ of online consumers have indicated that they would not buy electronics without consulting online reviews first~\cite{nielsen}. Due to the increasing dependency on user-generated reviews, it is crucial to understand their quality --- that can widely vary from being an excellent-detailed opinion to superficial criticizing or praising, to spams in the worst case. %Without any indication of the review quality, it is overwhelming for consumers to browse through a multitude of reviews. 
Unfortunately, review forums such as TripAdvisor, Yelp, Amazon, and others are being increasingly game to manipulative and deceptive reviews: fake (to promote or demote some item), 
incompetent (rating an item based on irrelevant aspects), or biased (giving a distorted and inconsistent view of the item). For example, recent studies depict that $20\%$ of Yelp reviews might be fake and Yelp internally rejects $16\%$ of user submissions~\cite{Luca} as ``not-recommended''. 

Recent research has proposed approaches to identify helpful reviews and spams automatically, but they suffer from major drawbacks: most of these approaches are geared towards active users and items in the community with a lot of reviews and activity information, and, therefore, not suitable for ``long-tail'' users and items with limited data. Most importantly, these works --- based on crude user behavioral, and shallow textual features --- do not provide any interpretable explanation as to why a review should be deemed helpful, or non-credible.

\pagebreak

In order to address the above issues, we propose probabilistic approaches based on analyzing reviews on several aspects like consistency, (latent) semantics, and temporal dynamics for {\em two} tasks in online review communities: (i) finding useful product reviews that are {\em helpful} to the end consumers, and (ii) finding credible reviews with {\em limited} information about users and items, specifically, for the ``long-tail'' ones using {\em consistency} features. We provide user-interpretable explanations for our verdict for both the tasks.

%In order to address these credibility issues in product review communities, we focus on the following applications in this section:

\section{Motivation and Approach}

\subsection{Finding Useful Product Reviews}

\noindent{\bf Motivation:} Online reviews provided by consumers are a valuable asset for e-Commerce platforms, influencing potential consumers in making purchasing decisions. However, without any indication of the review quality, it is overwhelming for consumers to browse through a multitude of reviews. In order to help consumers in finding useful reviews, most of the e-Commerce platforms nowadays allow users to vote whether a product review is helpful or not. For instance, any {\em Amazon} product review is accompanied with information like $x$ out of $y$ users found the review helpful. This {\em helpfulness score} ($x/y$) can be considered as a proxy for the review quality and its usefulness to the end consumers. In this task, we aim to automatically find the helpfulness score of a review based on certain consistency, and semantic aspects of the review like: whether the review is written by an expert, what are the important facets of the product outlined in his review, what do other 
experts have to say about the given product, timeliness of the review etc. --- that are automatically mined as latent factors from review texts.

%These external votes about the review helpfulness assist new consumers in focusing on the helpful reviews. 
%In this work, we aim to determine the helpfulness +of a review based on various semantic aspects.\\

\noindent {\bf State-of-the-Art and its Limitations:} Prior works on predicting review helpfulness mostly operate on shallow syntactic textual features like bag-of-words, part-of-speech tags, and tf-idf (term, and inverse document frequency) statistics~\cite{Kim:2006:AAR:1610075.1610135,Lu:2010:ESC:1772690.1772761}. These works, and other related works on finding review spams~\cite{Liu2008, Liu2013} classify extremely opinionated reviews as not helpful. Similarly, other works exploiting rating \& activity features like frequency of user posts, average ratings of users and items~\cite{O'Mahony:2009:LRH:1639714.1639774,Lu:2010:ESC:1772690.1772761, liu-EtAl:2007:EMNLP-CoNLL2007} consider extreme ratings and deviations as indicative of unhelpful reviews. Some recent works incorporate additional information like community-specific characteristics (who-voted-whom) with explicit user network~\cite{Tang:2013:CRH:2507157.2507183,Lu:2010:ESC:1772690.1772761}, and item-specific meta-data like {\em explicit} item facets 
and product brands~\cite{icdm2008,Kim:2006:AAR:1610075.1610135}. Apart from the requirement of a large number of meta-features that restrict the generalizability of many of these models to any arbitrary domain, these shallow features do not 
analyze what the review is {\em about}, and, therefore, cannot {\em explain} why it should be helpful for a given product. Some of these works~\cite{O'Mahony:2009:LRH:1639714.1639774,icdm2008} identify {\em expertise} of a review's author as an important feature. However, in absence of suitable modeling techniques, they consider prior reputation features like user activity, and low rating deviation as a proxy for user expertise.\\

The work closest to our approach is~\cite{icdm2008} --- where the authors identify syntactic features, user expertise, and timeliness of a review as important indicators of its quality. However, even in this case, the authors use part-of-speech tags as syntactic features, and user preferences for {\em explicit} item facets (pre-defined genres of IMDB movies in their work) as proxy for user expertise. In contrast, we explicitly model user expertise as a function of their writing style, rating style, and preferences for (latent) item facets --- all of which are jointly learned from user-contributed reviews --- going beyond the usage of shallow syntactic features, and the requirement for additional item meta-data.

\noindent {\bf Problem Statement:} Our work aims to overcome the limitations of prior works by exploring the {\em semantics} and {\em consistency} of a review to predict its {\em helpfulness score} for a given item. Unlike prior works, all of these features can be harnessed from only the information of a {\em user reviewing an item at an explicit timepoint}, making our approach fairly general for all communities and domains. We also provide {\em interpretable} explanation in terms of latent word clusters that gives interesting insights as to what makes the review helpful.

%\noindent {\bf Example:}\\
\noindent {\bf Approach:} The first step towards understanding the {\em semantics} of a review is to uncover the facet descriptions of the target item outlined in the review. We treat these facets as {\em latent} and use Latent Dirichlet Allocation (LDA) to discover them as topic clusters. The second step is to find the {\em expertise} of the users who wrote the review, and their description of the different (latent) facets of the item. 
Our approach in 
modeling user expertise is similar to that outlined in Chapter~\ref{chap:temporal}. However, there are   
%similar to~\cite{mukherjee2015jertm}, 
significant differences and modifications (discussed in Section~\ref{subsec:diff}) in modeling the joint interactions between several factors, where our proposed model has a better coupling between the factors, all of which are learned directly from the review helpfulness.

We make use of {\em distributional hypotheses} (outlined in Section~\ref{subsec:distrhyp}) like: expert users agree on what are the important facets of an item, and their description (or, writing style) of those facets 
influences the helpfulness of a review. We also derive several {\em consistency} 
features --- all from the given quintuple $\langle$\textit{userId, itemId, rating, reviewText, timepoint}$\rangle$ --- like prior user reputation, item prominence, and timeliness of a review, that are used in conjunction with the semantic features. Finally, we leverage the interplay between all of the above factors in a {\em joint} setting to predict the review helpfulness.

For interpretable explanation, we derive interesting insights from the latent word clusters used by experts --- for instance, reviews describing the underlying ``theme and storytelling'' of {\em movies} and {\em books}, the ``style'' of {\em music}, and ``hygiene'' of {\em food} are considered most helpful for the respective domains.

\noindent{\bf Contributions:}  The salient contributions of this work can be summarized as:
\vspace{-2em}
\begin{itemize}
\item[a)] {\bf Model:} We propose an approach to leverage the {\em semantics} and {\em consistency} of reviews to predict their helpfulness. We propose a Hidden Markov Model -- Latent Dirichlet Allocation (HMM-LDA) based model that jointly learns the (latent) item facets, (latent) user expertise, and his writing style from {\em observed} words in reviews at explicit timepoints.
\item[b)] {\bf Algorithm:} We introduce an effective learning algorithm based on an iterative stochastic optimization process that reduces the mean squared error of the predicted helpfulness scores with the ground scores, as well as maximizes the log-likelihood of the data.
\item[c)] {\bf Experiments:} We perform large-scale experiments with real-world datasets from {\em five} different domains in {\em Amazon}, together comprising of $29$ million reviews from $5.7$ million users on $1.9$ million items, and demonstrate substantial improvement over state-of-the art baselines for {\em prediction} and {\em ranking} tasks.
\end{itemize}

%\noindent The rest of the paper is organized as follows: Section~\ref{sec:features} introduces the different semantic and consistency factors to model review helpfulness, along with a preliminary study. Section~\ref{sec:model} presents the HMM-LDA based model to {\em jointly} infer the item facets, user expertise, and review helpfulness, along with an inference method based on Gibbs sampling and regression. Section~\ref{sec:experiments} presents the experimental evaluation of our approach, comparing it against several baselines, followed by related works and conclusion in Section~\ref{sec:relWork} and Section~\ref{sec:conc}, respectively.

\subsection{Finding Credible Reviews with Limited Information}

{\noindent \bf Motivation:} Starting with the work of \cite{Liu2008}, research efforts have been undertaken to automatically detect non-credible reviews. In parallel, industry (e.g., stakeholders such as Yelp) has 
developed its own standards\footnote{\url{officialblog.yelp.com/2009/10/why-yelp-has-a-review-filter.html}} to filter out ``illegitimate'' reviews. Although details are not disclosed, studies suggest that these filters tend to be fairly 
crude~\cite{Liu2013a}; for instance, exploiting user activity like the number of reviews posted, and treating users whose ratings show high deviation from the mean/majority ratings as suspicious. Such a policy seems to 
over-emphasize trusted long-term contributors and suppress outlier opinions off the mainstream. Moreover, these filters also employ several aggregated metadata, and are thus hardly viable for 
new items that initially have very few reviews --- often by not so active users or newcomers in the community.

%state-of-the-art research methods and their limitations
{\noindent \bf State-of-the-Art and Its Limitations:}
Research on this topic has cast the problem of review credibility into a binary classification task: a review is either credible or deceptive.
To this end, supervised and semi-supervised methods have been developed that largely rely on features about users and their activities as well as statistics about item ratings. 
Most techniques also consider spatio-temporal patterns of user activities like 
IP addresses or user locations (e.g., \cite{Liu2014,Li2015}), 
burstiness of posts on an item or an item group (e.g., \cite{Fei2013}), 
and further 
correlation measures across users and items as discussed in Chapter~\ref{chap:framework}. 
However, the classifiers built this way are mostly geared for popular items, and the meta-information about user histories and activity correlations are not always available. For example, 
someone interested in opinions on a new art film or a ``long-tail'' bed-and-breakfast in a rarely visited town, is not helped at all by the above methods. 
%for review credibility classification.
%
%now one short par on text-based approaches
Several existing works~\cite{Mihalcea2009,Ott2011,Ott2013} consider the textual content of user reviews for tackling opinion spam by using word-level unigrams or bigrams as features, 
along with specific lexicons (e.g., LIWC~\cite{liwc} psycholinguistic lexicon, WordNet Affect~\cite{WNAffect}), to learn latent topic models and classifiers (e.g., \cite{Ott2013a}).
Although these methods achieve high classification accuracy for various gold-standard datasets, they do not provide any interpretable evidence as to why a certain review is classified as 
non-credible.

\pagebreak

{\noindent \bf Problem Statement:}
This task focuses on detecting credible reviews {\em with limited information}, namely, in the absence of rich data about user histories, community-wide correlations, and for ``long-tail'' items. 
In the extreme case, we are provided with only the review texts and ratings for an item. Our goal is then to analyze various inconsistencies that may exist within the reviews --- using which we can compute a {\em credibility score} and provide  
{\em interpretable evidence} for explaining why certain reviews have been categorized as non-credible.

{\noindent \bf Approach:}
Our proposed method to this end is to learn a model based on {\em latent topic models} and combining them with limited metadata to provide a novel notion of {\em consistency features} 
characterizing each review. We use the LDA-based Joint Sentiment Topic model (JST)~\cite{linCIKM2009} to cast the user review texts into a number of informative facets. We do this per-item, 
aggregating the text among all reviews for the same item, and also per-review. This allows us to identify, score, and highlight inconsistencies that may appear between a review and 
the community's overall characterization of an item. We perform this for the item as a whole, and also for each of the latent facets separately.
Additionally, we learn inconsistencies such as discrepancy between the contents of a review and its rating, and temporal ``bursts'' --- where a number of reviews are written in a short span 
of time targeting an item. We propose five kinds of inconsistencies that form the key assets of our credibility scoring model, fed into a Support Vector Machine for classification, or for ordinal ranking.

{\noindent \bf Contributions:} In summary, our contributions are summarized as:

\begin{itemize}
 \item {\em Model:} We develop a novel {\em consistency model} for credibility analysis of reviews that works with limited information, with particular attention to ``long-tail'' items, and offers interpretable evidence for reviews classified as non-credible.
 \item {\em Tasks:} We investigate how credibility scores affect the overall ranking of items. To address the scarcity of labeled training data, we transfer the learned model from Yelp to Amazon to rank top-selling items based on (classified) {\em credible} user reviews. In the presence of proxy labels for item ``goodness'' (e.g., item sales rank), we develop a better ranking model for domain adaptation.% in the community.
 \item {\em Experiments:} We perform extensive experiments in TripAdvisor, Yelp, and Amazon to demonstrate the viability of our method and its advantages over state-of-the-art baselines in dealing with ``long-tail'' items and providing interpretable evidence.
\end{itemize}

\vspace{10em}

%% file: main/chapter-credible-applications/helpful-reviews/features.tex
%\section{Experimental Study}

\subsection{Review Helpfulness Factors}\label{sec:features3}
In this section, we outline the components of our model that analyze the {\em semantics} and {\em consistency} features of reviews, and show how these can help in predicting the review helpfulness. 

\subsubsection{Item Facets}

Given a review on an item, it is essential to understand the different facets of the item described in the review. For instance, a camera review can focus on different facets like ``resolution", ``zoom", ``price", ``size'', or a movie review can focus on ``narration", ``cinematography", ``acting", ``direction'' etc. However, not all facets are equally important for an item. For example, a review downrating a camera for ``late delivery'' by the seller is not as helpful to the general consumer as opposed to downrating it due to ``grainy resolution'' or ``shaky zoom''. Therefore, a helpful review should focus on the {\em important facets} of an item. Another important aspect of a detailed review is to consider a {\em wide range of facets} of an item, rather than harping on a specific facet~\cite{Mudambi:2010:MHO:2017447.2017457, Kim:2006:AAR:1610075.1610135,liu-EtAl:2007:EMNLP-CoNLL2007}. 

Prior works~\cite{liu-EtAl:2007:EMNLP-CoNLL2007,Lu:2010:ESC:1772690.1772761,Kim:2006:AAR:1610075.1610135} consider the length distribution (like the number of words, sentences, or paragraphs in the review), and the overlap of {\em explicit} facets from the product description (including brand names, categories, specifications etc.) in the review as a proxy of how detailed it is.

In contrast, we model facets as {\em latent} variables, similar to that of a topic model~\cite{Blei2003LDA}. The latent facet distribution of an item in the review text is indicative of how {\em detailed} and {\em diverse} the review is. 

\subsubsection{Review Writing Style}

Similar to the importance of the facets outlined in a review, the {\em words} used to describe the facets play a crucial role in making the review readable, and useful to the consumers. Due to diverse background of the reviewers with different language skills, the writing style too varies widely. An important aspect of an expert writing style is to use {\em precise, domain-specific} vocabulary to describe a facet in {\em details}, rather than using generic words. For instance, contrast this {\em expert} camera review: 

\example{60D focus screen is `grainy'. It is the `precision matte' surface that helps to increase contrast and minimize depth of field for manual focusing. The Ef-s screen is even more so for use with fast primes. The T1i focus screen is smoother and brighter to compensate for the dimmer pentamirror design and typical economy f/3.5-5.6 zooms, but gives less precise manual focus.} 

with this {\em amateur} one: 

\example{This camera is pure garbage. This is the worst 
camera I have ever owned. I bought it last xmas on a deal and I have thrown it away and replaced it with a decent camera.}

%{\small \tt ``Very disappointed! Takes for ever to focus on stills, the moment is over by the 
%time it right. Won't focus good on the video portion either. Thinking of sending it back. I am not usually a complainer, but this is awful, don't buy it.''}

Another important factor to observe here is the {\em balance} in the reviewer's opinion on an item. An expert review depicts a detailed judgment about the item, rather than just criticizing or praising it. Therefore, it is essential to distinguish the writing style of an experienced user from an amateur one.

Prior works~\cite{Liu2008,Kim:2006:AAR:1610075.1610135,Lu:2010:ESC:1772690.1772761,liu-EtAl:2007:EMNLP-CoNLL2007} capture the writing style from syntactic features like bag of words, part-of-speech tags, and use sentiment lexicons to find the distribution of positive and negative sentiment words in the review. In contrast, we learn a language model from the latent facets and user expertise that uncovers the hidden semantics in a review. %\todo{It would be nice to explain this one line further}

%\todo{Insert a table of excellent, good, and bad reviews: maybe in experiments}
\vspace{-1em}

\subsubsection{Reviewer Expertise}\label{subsec:revExp}

Previous works~\cite{Liu2008,O'Mahony:2009:LRH:1639714.1639774} in this domain attempted to harness a user's expertise in writing a review under the hypothesis that expert reviews are positively correlated to review helpfulness. However, none of them explicitly modeled the users' expertise. Instead, they considered the following proxy features for user reputation, namely:\\
{\noindent \bf Activity:} Number of posts written by the user in the community.\\
{\noindent \bf Rating deviation:} Deviation of the user rating from the community rating on an item.\\
{\noindent \bf Prior user reputation:} Average number of helpfulness votes received by the user from her previous reviews.

In this work, we explicitly model user expertise adopting a similar approach as outlined in Section~\ref{sec:discrete}. However, we make substantial modifications (outlined in Section~\ref{subsec:diff}) in modeling and learning the joint distributions conditioned on expertise --- where all the distributions are explicitly learned from the review helpfulness scores as observables. 

Unlike other factors in the model, expertise is not static, but evolves over time. A user who was not an expert at the time of entering the community, may have become an expert now contributing helpful reviews.

We model {\em expertise} as a {\em latent} variable that evolves over time, exploiting the hypothesis that users at similar levels of expertise have similar rating behavior, facet preferences, and writing style. The facets discovered in the previous step, and the writing style would therefore help us in finding a reviewer's expertise. Once we figure out the reviewer's expertise, we can find out the important facets of the item that he is concerned about, as well as the domain-specific vocabulary for describing the facets --- thereby forming an effective feedback loop between {\em facets, writing style,} and {\em expertise}.

\vspace{-1em}

\subsubsection{Distributional Hypotheses} 
\label{subsec:distrhyp}
Once we identify the (latent) item facets, (latent) expertise and preferences of different users, we can make use of the following hypotheses to capture the helpfulness of reviews:
\squishlist
\item[i)] If the past reviews of a given user on an item (with a certain facet distribution) have been deemed helpful, then an incoming review by the given user on a similar kind of item (i.e. similar facet distribution) is also likely to be helpful. 
For instance, in the movie domain, if a user's past reviews in the ``drama'' genre have been found to be helpful, and the movie under preview is also from the same genre, then its review is likely to be helpful.
\item[ii)] If the past reviews of users with certain characteristics (like, specific facet preferences and expertise) have been deemed helpful, and the given user has tastes and expertise similar to those users, then her current review is also likely to be helpful. For instance, assume we have learned how ``expert'' reviews in the ``drama'' genre looks like, and the current review text indicates the user to be an expert in the ``drama'' genre, then her review is likely to be helpful.
\squishend

\noindent Note that in traditional collaborative filtering approaches for recommender systems, (i) and (ii) are similar to item-item and user-user similarities, respectively.

\subsubsection{Consistency}

Users and items do not gain reputation overnight. %\todo{This is not completely true as we see some of the cases in movie reviews} 
Therefore prior reputation of users and items are good indicators of the associated reviews' helpfulness. In this work, we use the following consistency features that are used to guide our model to learn the {\em latent} distributions conditioned on the reviews' helpfulness and ratings.

{\noindent \bf Prior user reputation:} Average helpfulness votes received by the user's past reviews from other users.\\ %\todo{This is same as Prior Helpfulness in last Section-C}\\
{\noindent \bf Prior item prominence:} Average helpfulness votes received by the item's past reviews from other users, which is also indicative of the {\em prominence} of the item. \\
{\noindent \bf User rating deviation:} Absolute deviation between the user's rating on an item, and the average rating assigned by the user over all other items. This captures the mean user rating behavior, and, therefore, scenarios where the user is too dis-satisfied (or, otherwise) with an item.\\
{\noindent \bf Item rating deviation:} Absolute deviation between a user's rating on the item, and average rating received by the item from all other users. This captures the scenario where a user unnecessarily criticizes or praises the item, that the community does not agree with.\\
{\noindent \bf Global rating deviation:} Absolute deviation between the user's rating on an item, and the average rating of all items by all users in the community. This captures the scenario where the user rating deviates from the general rating behavior of the community.

\subsubsection{Timeliness or ``Early-bird'' bias}
%{\noindent \bf Timeliness or ``Early-bird'' bias:} \todo{This could be a different subsection} 
Prior work~\cite{icdm2008} has shown a positive influence of a review's publication date on the number of helpfulness votes received by it. The reason being that early and ``timely'' reviews are more useful to the consumers when the item is launched, so that they can make an informed decision about the item. Also, early reviews are exposed to consumers for a longer period of time which allows them to garner more votes over time, compared to recent reviews. The timestamp of the {\em first} review on a given item $i$ is considered to be the reference timepoint (say, $t_{i,0}$). Therefore, the timeliness of any other review on the item at time $t_i$ is computed as: $exp^{-(t_i - t_{i,0})}$.

\subsubsection{Preliminary Study of Feature Significance}
In order to understand the significance of different consistency features in predicting review helpfulness, we find correlation between various features described in the previous section and helpfulness scores of reviews. We consider reviews from 
{\em five} real-world datasets from Amazon in the domains namely, {\em food, movies, music} and {\em electronics}. We selected reviews that received a minimum of $five$ votes to maintain the robustness of the task.

\begin{table}
\centering
\small
\begin{tabular}{p{4cm}p{1cm}p{1cm}p{1cm}p{1cm}p{1cm}}
\toprule
{\bf Factors} & {\bf Elect.} & {\bf Foods} & {\bf Music} & {\bf Movies} & {\bf Books}\\
\midrule
Item rating deviation & -0.364 & -0.539 & -0.596 & -0.519 & -0.516\\
Global rating deviation & -0.295 & -0.507 & -0.526 & -0.439 & -0.443\\
User rating deviation & -0.292 & -0.429 & -0.477 & -0.267 & -0.327\\
%Negative sentiment & -0.069 & -0.113 & -0.200 & -0.143 & -0.136\\
%Objective sentiment & 0.026 & -0.023 & 0.069 & 0.012 & -0.030\\
User activity & -0.056 & 0.002$^{*}$ & -0.074 & 0.032 & -0.033\\
Timeliness & 0.036 & 0.083 & 0.102 & 0.114 & 0.137\\
%Positive sentiment & 0.021 & 0.149 & 0.064 & 0.074 & 0.076\\
Prior user reputation & 0.062 & 0.191 & 0.353 & 0.525 & 0.386\\
%Review length & 0.286 & 0.193 & 0.282 & 0.260 & 0.255\\
Prior item prominence & 0.221 & 0.251 & 0.343 & 0.303 & 0.343\\
\bottomrule
\end{tabular}
\caption{Pearson correlation between different features and helpfulness scores of reviews in the domains {\em electronics, foods, music, movies,} and {\em books}. All factors (except the one marked with $*$) are statistically significant with {\em p-value} $< 2e-16$.}
\label{tab:correlation}
\end{table}

%From Table~\ref{tab:correlation}, we observe that negative (and, in some cases, neutral) sentiment, and rating deviations --- where the user diverges with the community rating on items, and her prior rating history --- negatively impact helpfulness; whereas the prior reputation of users \& items, timeliness, review details, and positive sentiment have a positive impact. 

From Table~\ref{tab:correlation}, we observe that rating deviations --- where the user diverges with the community rating on items, and her prior rating history --- negatively impact helpfulness; whereas the prior reputation of users \& items, and timeliness
%, and review details 
have a positive impact.

We also find that user activity alone does not have a significant impact on review helpfulness. In some cases (e.g., food domain) it is non-signifcant, or even has a negative impact on review helpfulness (e.g., electronics, music, and books domains). In order to find out if this feature fires in unison with other features, we use {\em linear regression} to predict the review helpfulness considering all of these features together. From the corresponding {\em f-statistic} we find user activity to be statistically significant with {\em p-value} $<2e-16$ in all cases (including the food domain) with a moderate {\em positive} weight. This feature, therefore, is used later in our expertise evolution model as a hyper-parameter that controls the rate of user progression. 

Similarly, all of the other factors are reinforced in a {\em joint} setting, even though the correlations are quite low (for many of the features) in this study. 

In the following section, we propose an approach to model all of these factors {\em jointly} to 
predict review helpfulness.

%% file: main/chapter-credible-applications/helpful-reviews/joint-model.tex
\subsection{Joint Model for Review Helpfulness}\label{sec:model3}

\subsubsection{Incorporating Consistency Factors} \label{subsec:consist}
Let $u \in U$ be a user writing a review at time $t \in T$ on an item $i \in I$. Let $d=\{w_1, w_2,... w_{|N_d|}\}$ be the corresponding review text with a sequence of words $\langle w \rangle$, and rating $r \in R$. Each such review is associated with a helpfulness score $h \in [0-1]$. Let $b_t$ be the corresponding timeliness of the review computed as  $exp^{-(t - t_{i,0})}$, where $t_{i,0}$ is the {\em first} review on the item $i$.

Let $\beta_u$ be the average helpfulness score of user $u$ over all the reviews written by her (capturing user reputation), and $\beta_i$ be the average helpfulness score of all reviews for item $i$ (capturing item prominence). Let $\overline{r}_u$ be the average rating assigned by the user over all items, $\overline{r}_i$ be the average rating assigned to the item by all users, and $\overline{r}_g$ be the average global rating over all items and users. {\em Consistency} features include prior item and user reputation, deviation features, and burst. 

Let $\xi$ be a tensor of dimension $E \times Z$, where $E$ is the number of expertise levels of the users, and $Z$ is the number of latent facets of the items. $\xi_{e,z}$ depicts the opinion of users at (latent) expertise level $e \in E$ about the (latent) facet $z \in Z$. Therefore, the distributional hypotheses (outlined in the previous section) are intrinsically integrated in $\xi$ that is estimated from the reviews' text, conditioned on the helpfulness score of the reviews. 

%We want to {\em jointly minimize the mean squared error, and maximize the likelihood of the data (in terms of the latent factors $\xi$)}. Therefore, 
The estimated helpfulness score $\widehat{h}(u,i)$ of a review by user $u$ on item $i$ is a function $f$ of the following {\em consistency} and {\em latent} factors, parametrized by $\Psi$:

\begin{equation}
\label{eq:features}
 \widehat{h}(u,i) = f(\beta_u, \beta_i, |r-\overline{r_u}|, |r-\overline{r_i}|, |r-\overline{r_g}|, b_t, \xi; \Psi)
\end{equation}

Here, $f$ can be a polynomial, radial basis, or a simple linear function for combining the features. 
The objective is to estimate the parameters $\Psi$ (of dimension: $6 + E \times Z$) that reduces the {\em mean squared error} of the predicted helpfulness scores with the ground scores:

\begin{equation}
\label{eq:mse}
 \Psi^{*} = argmin_\Psi \frac{1}{|U|} \sum_{u,i \in U,I} (h(u,i) - \widehat{h}(u, i))^2 + \mu ||\Psi||^2_2
\end{equation}
\noindent where, we use $L_2$ regularization for the parameters to penalize complex models.

There are several ways to estimate the parameters like alternate least squares, gradient-descent, and Newton based approaches.

\subsubsection{Incorporating Latent Facets}
We use principles of {\em Latent Dirichlet Allocation} (LDA)~\cite{Blei2003LDA} to learn the latent facets associated to an item. Each review $d$ on an item is assumed to have a Multinomial distribution $\theta$ over facets $Z$ with a symmetric Dirichlet prior $\alpha$. Each facet $z$ has a Multinomial distribution $\phi_z$ over words drawn from a vocabulary $W$ with a symmetric Dirichlet prior $\delta$. 
Exact inference is not possible due to the intractable coupling between $\Theta$ and $\Phi$. 

Two popular ways for approximate inference are 
MCMC techniques like Collapsed Gibbs Sampling and Variational Inference.
%The latter is typically much more complex and computationally expensive. Therefore, we use sampling in this work.

\subsubsection{Incorporating Latent Expertise}

Expertise influences both the facet distribution $\Theta$, as users at different levels of expertise have different facet preferences, and the language model $\Phi$ as the writing style is also different for users at different levels of expertise. Therefore, we parametrize both of these distributions with user expertise similar to our approach in Chapter~\ref{sec:discrete}, with some major modifications (discussed in the next section).

Consider $\Theta$ to be a tensor of dimension $E \times Z$, and $\Phi$ to be a tensor of dimension $E \times Z \times W$, where $\theta_{e,z}$ denotes the preference for facet $z \in Z$ for users at expertise level $e \in E$, and $\phi_{e,z,w}$ denotes the probability of the word $w \in W$ being used to describe the facet $z$ by users at expertise level $e$. 

Now, expertise changes as users evolve over time. However, the transition should be {\em smooth}. Users cannot abruptly jump from expertise level $1$ to $4$ without passing through expertise levels $2$ and $3$. Therefore, at each timepoint $t+1$ (of posting a review), we assume a user at expertise level $e_t \in E$ to stay at $e_t$, or move to $e_t + 1$ (i.e.
expertise level is monotonically non-decreasing). This progression depends on how the writing style (captured by $\Phi$), and facet preferences (captured by $\Theta$) of the user is evolving {\em with respect to} other expert users in the community, as well as the rate of {\em activity} of the user. User activity is used as a proxy for expertise in many of the prior works~\cite{O'Mahony:2009:LRH:1639714.1639774,Lu:2010:ESC:1772690.1772761, liu-EtAl:2007:EMNLP-CoNLL2007}. However, we find it to play a weak role during our preliminary study. Therefore, we use it only as a hyper-parameter for controlling the rate of progression. Let $\gamma_u$, the activity rate of user $u$ be defined as: $\gamma_u = \frac{D_
u}{D_u + D_{avg}}$,
where $D_u$ and $D_{avg}$ denote the number of posts written by $u$, and the average number of posts written by any user in the community, respectively.

Let $\Pi$ be a tensor of dimension $E \times E$ with hyper-parameters $\langle \gamma_u \rangle$ of dimension $U$, where $\pi_{e_i,e_j}$ denotes the probability of moving to expertise level $e_j$ from $e_i$ with the constraint $e_j \in \{e_i, e_i + 1\}$. However, not all users start at the same level of expertise, when they enter the community; some may enter already being an expert. The algorithm figures this out during the inference process. We assume all users to 
%be {\em weak learners}\todo{verify this term} 
start at expertise level $1$ during parameter initialization.

During inference, we want to learn the parameters $\Psi, \xi, \Theta, \Phi, \Pi$ jointly for predicting review helpfulness. 

\vspace{-1em}
\subsubsection{Difference with Prior Works for Modeling Expertise} \label{subsec:diff}

The generative process of user expertise has the following differences with our previous approach in Chapter~\ref{sec:discrete}:

\begin{itemize}
 \item [i)] Previously we had learned {\em user-specific} preferences for personalized recommendation. However, we assume users at the same level of expertise to have similar facet preferences. Therefore, the facet distribution $\Theta$ is conditioned {\em only} on the user {\em expertise}, and not the user explicitly, unlike the prior works. This helps us to reduce the dimensionality of $\Theta$, and exploit the correspondence between $\Theta$ and $\xi$ to {\em tie} the parameters of the consistency and latent factor models together for joint inference.
 \item [ii)] Our previous approach incorporates supervision, for predicting ratings, {\em only indirectly} via optimizing the Dirichlet hyper-parameters $\alpha$ of the Multinomial facet distribution $\Theta$ --- and cannot guarantee an increase in the data log-likelihood over iterations. In contrast, we exploit (i) to learn the expertise-facet distribution $\Theta$ {\em directly} from the review helpfulness scores by minimizing the {\em mean squared error}  during inference. This is also tricky as the parameters of the distribution $\Theta$, for an unconstrained optimization, are not guaranteed to lie on the simplex --- for which we do certain transformations, discussed during inference. Therefore, the parameters are {\em strongly} coupled in our model, not only reducing the mean squared error, but also leading to a near smooth increase in the data log-likelihood over iterations (refer to Figure~\ref{fig:logL}).
\end{itemize}

\vspace{-1em}
\subsubsection{Generative Process}

Consider a corpus $D=\{d_1,\ldots,d_D\}$ of reviews written by a set of users $U$ at timestamps $T$. For each review $d\in D$, we denote $u_d$ as its user, $t_d$ as the timestamp of the review.
The reviews are assumed to be ordered by timestamps, i.e., $t_{d_i}<t_{d_j}$ for $i<j$. 
 Each review $d \in D$ consists of a sequence of $N_d$ words denoted by $d=\{w_1,\ldots ,w_{N_d}\}$, where each word is drawn from a vocabulary $W$ having unique words indexed by $\{1 \dots W\}$. Number of facets correspond to $Z$.
 
 Let $e_d \in \{1, 2, ..., E\}$ denote the expertise value of review $d$. Since each review $d$ is associated with a unique timestamp $t_d$ and unique user $u_d$, the expertise value of a review refers to the expertise of the user at the time of writing it. Following Markovian assumption, the user's expertise level transitions follow a distribution $\Pi$ with the Markovian assumption $e_{u_d} \sim \pi_{e_{u_{d-1}}}$ i.e. the expertise level of $u_d$ at time $t_d$ depends on her 
expertise level when writing the previous review at time $t_{d-1}$. 

\begin{figure}
	%\vspace{-1.5em}
	\centering
	\includegraphics[scale=0.5]{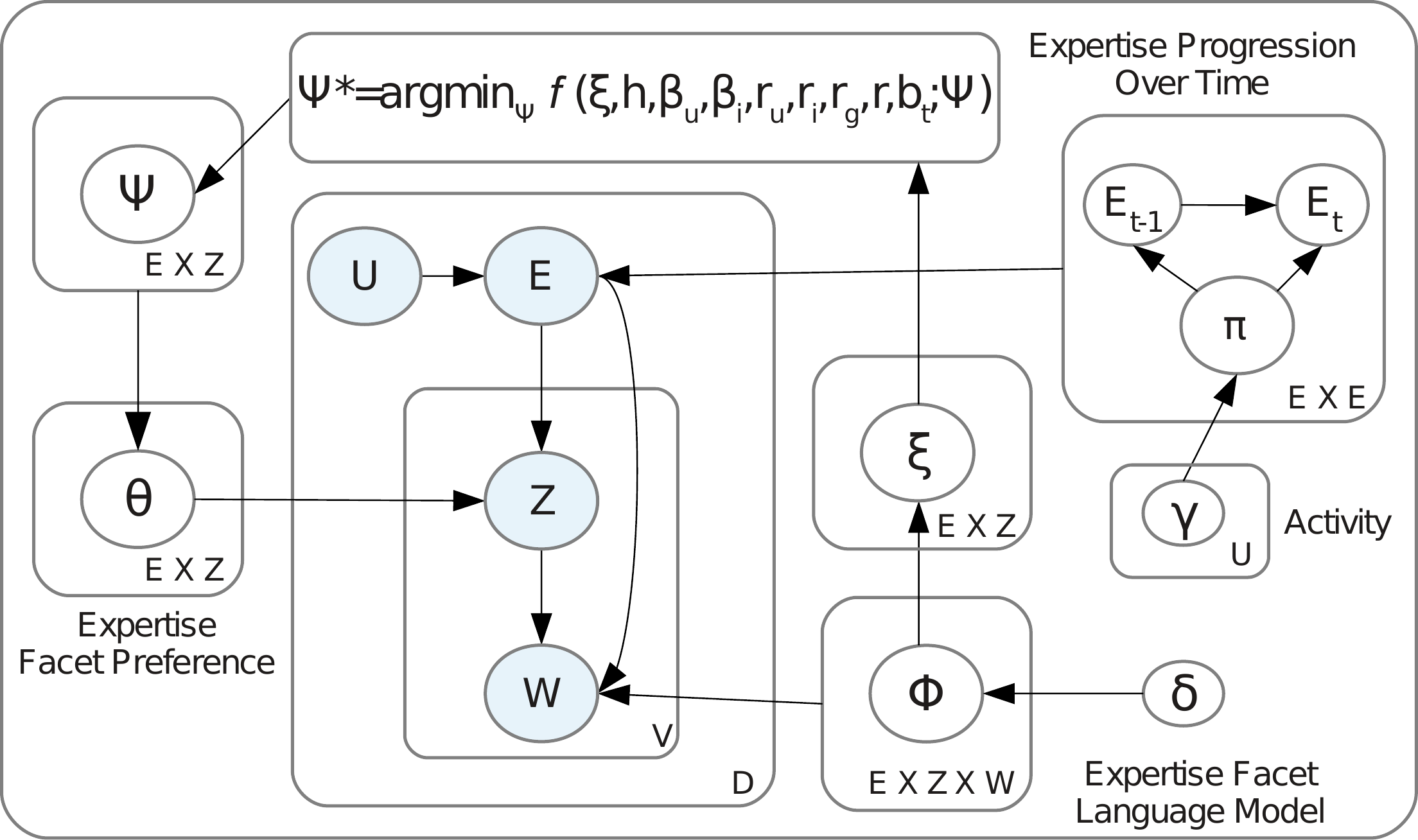}
	\caption{Generative process for helpful product reviews.}
	\label{fig:plate}
\end{figure}

Once the expertise level $e_{d}$ of the user $u_d$ for review $d$ is known, her facet preferences are given by $\theta_{e_d}$. Thereafter, the facet $z_{d,w}$ of each word $w$ in $d$ is drawn from a Multinomial ($\theta_{e_{d}}$). 
Now that the expertise level of the user, and her facets of interest are known, we can generate the language model $\Phi$ and individual words in the review ---  where the user draws a word from the Multinomial distribution $\phi_{e_{d}, z_{d,w}}$ with a symmetric Dirichlet prior $\delta$. Refer Figure~\ref{fig:plate} for the generative process.

%
%Each word $w$ that is going to be used at time $t$ is also associated with an expertise value $l_{t,w}$ --- which depends on the expertise $e_{D_t}$ of the reviews $D_t$ written at time $t$ (for instance, many sophisticated words are expected in the review of expert users). We simply draw $l_{t,w} \sim \text{Normal} (e_{D_t}, \sigma_1)$.

%\pagebreak

The joint probability distribution is given by:

{
\begin{multline}
\label{eq:joint}
 P(E,Z,W, \Theta, \Phi | U; \langle \gamma_u \rangle, \delta) \propto
 \prod_{u \in U} \prod_{d\in D_u}  P(\pi_{e_d}; \gamma_u) \boldsymbol{\cdot} P(e_d | \pi_{e_d}) \\ 
 \boldsymbol{\cdot} \bigg(\prod_{j=1}^{N_d} P(z_{d,j} | \theta_{e_d}) \cdot P(\phi_{e_d, z_{d,j}}; \delta) \cdot P(w_{d,j} | \phi_{e_d,z_{d,j}}) \bigg) 
\end{multline}
}

\subsubsection{Inference}

Given a corpus of reviews indexed by $\langle${\em userId, itemId, rating, reviewText, timepoint}$\rangle$, with corresponding helpfulness scores, our objective is to learn the parameters $\Psi$ that minimizes the mean squared error given by Equation~\ref{eq:mse}. 

In case $\xi$ was known, we could have directly plugged in its values (other features being {\em observed}) in Equation~\ref{eq:features} to learn a model (e.g., using regression) with parameters $\Psi$. However, the dimensions of $\xi$, corresponding to both facets and user expertise, are {\em latent} that need to be inferred from text. Now, the parameter weight $\psi_{e,z}$ corresponding to $\xi_{e,z}$ from Equation~\ref{eq:mse} depicts the importance of the facet $z$ for users at expertise level $e$ for predicting review helpfulness. We want to exploit this observation to infer the latent dimensions from text.

During the generative process of a review document, for a user at expertise level $e$, we want to draw her facet of interest $z$ with probability $\theta_{e,z} \propto \psi_{e,z}$. However, we cannot directly replace $\Theta$ with $\Psi$ due to the following reason. The traditional parametrization of a Multinomial distribution ($\Theta$ in this case) is via its mean parameters. Any unconstrained optimization will take the parameters out of the feasible set, i.e. they may not lie on the simplex. Hence, it is easier to work with the natural parameters instead. If we consider the unconstrained parameters $\langle \psi_{e,z} \rangle$ (learned from Equation~\ref{eq:mse}) to be the natural parameters of the Multinomial distribution $\Theta$, we need to transform the natural parameters to the mean parameters that lie on the simplex (i.e. $\sum_z \theta_{e,z} = 1$). In this work, we follow the same principle as in Equation~\ref{eq:pi} in Chapter~\ref{sec:continuous} to do this transformation:

\begin{equation}
\label{eq:transform}
 \theta_{e,z} = \frac{exp(\psi_{e,z})}{\sum_z exp(\psi_{e,z})}
\end{equation}
\noindent 
where, $\psi_{e,z}$ corresponds to the learned parameter for $\xi_{e,z}$.

%Let $n(u, e, d, z, v)$ denote the count of the word $w$
%indexed by the $v^{th}$ word in the vocabulary, 
%occurring in document $d$ written by user $u$ at expertise level $e$ belonging to facet $z$. In the following equation,
% $(.)$ at any position in a distribution indicates summation of the above counts  
%for the respective argument.

Exploiting conjugacy of the Multinomial and Dirichlet distributions, 
we can integrate out $\Phi$ from the joint distribution in Equation~\ref{eq:joint} to obtain the posterior distribution
$P(W|Z,E; \delta )$ given by:

{
\[\prod_{e=1}^E \prod_{z=1}^Z \frac{\Gamma(\sum_{w} \delta)\prod_{w} \Gamma(n(e, z, w)+ \delta)}{\prod_{w}{\Gamma(\delta)\Gamma(\sum_{w} n(e, z, w) + \sum_w \delta)}}\]
} 

\noindent where, $\Gamma$ denotes the Gamma function, and $n(e,z,w)$ is the number of times the word $w$ is used for facet $z$ by users at expertise level $e$.

We use Collapsed Gibbs Sampling~\cite{Griffiths02gibbssampling}, as in standard LDA,
to estimate the conditional distribution for each of the latent facets $z_{d,j}$,
which is computed over the current assignment for all other hidden
variables, after integrating out $\Phi$. In the following equation,
 $n(e,z,.)$  indicates the summation of the counts over all possible $w\in W$.  
%for the respective review. 
The subscript $-j$ denotes the value
of a variable excluding the data at the $j^{th}$ position.

The posterior distribution
$P(Z| \Phi, W, E)$ of the latent variable $Z$ is given by:

\begin{equation}
\begin{aligned}
\label{eq:facets}
 & P(z_{d,j} = k | z_{d,-j}, \Phi, w_{d,j}=w, e_d=e, d) \\
 & \propto \theta_{e,k} \boldsymbol{\cdot} \frac{n(e, k, w) + \delta}{n(e,k, .) + W \cdot \delta} \\
 & = \frac{exp(\psi_{e,k})}{\sum_z exp(\psi_{e,z})} \boldsymbol{\cdot} \frac{n(e, k, w) + \delta}{n(e,k, .) + W \cdot \delta} \\
\end{aligned} 
%\todo{is it $\alpha_k$?}
\end{equation}

Similar to the above process, we use Collapsed Gibbs Sampling \cite{Griffiths02gibbssampling} also to sample the expertise levels, keeping all facet assignments $Z$ fixed. 

Let $n(e_{i-1}, e_i)$ denote the number of transitions from expertise level $e_{i-1}$ to $e_i$ over {all} users in the community, with the Markovian constraint $e_i \in \{e_{i-1}, e_{i-1}+1\}$. 
%The state transition probability depending on the previous state, factoring in the user-specific activity rate, is given by:

\begin{equation}
\label{eq:transition}
P(e_i|e_{i - 1}, e_{-i}, u; \gamma_u) = \frac{n(e_{i - 1}, e_i) + I(e_{i - 1} = e_i) + \gamma_u}{n(e_{i-1},.) + I(e_{i - 1} = e_i) + E \boldsymbol{\cdot}\gamma_u} 
\end{equation}
\noindent where $I(.)$ is an indicator function taking the value $1$ when the argument is true (a self-transition, in this case, where the user has the same expertise level over subsequent reviews), and $0$ otherwise. The subscript $- i$ denotes the value of a variable excluding the data at the $i^{th}$ position. Note that the transition function is similar to prior works in Hidden Markov Model -- Latent Dirichlet Allocation (HMM-LDA) based models~\cite{rosenzviUAI2004},~\cite{mukherjee2014JAST}.
%All the {\em counts} of transitions exclude transitions to and from $e_i$, when sampling a value for the current expertise level $e_i$ during Gibbs sampling.

The conditional distribution for the expertise level transition is given by:

\begin{equation}
\label{eq:expertise}
 P(E|U,Z,W; \langle \gamma_u \rangle) \propto P (E|U; \langle \gamma_u \rangle) \boldsymbol{\cdot} P(Z|E) \boldsymbol{\cdot} P(W|Z, E)\\
\end{equation}
Using Equations~\ref{eq:facets},~\ref{eq:transition},~\ref{eq:expertise}, we obtain the conditional distribution for updating latent variables $E$ as:
\begin{equation}
\begin{aligned}
\label{eq:exp-gibbs}
& P(e_{u_d} = e_i |e_{u_{d-1}}=e_{i-1}, u_d = u, \{z_{i,j}=z_j \}, \{w_{i,j}=w_j\}, e_{-i}) \\
& \propto \frac{n(e_{i - 1}, e_i) + I(e_{i - 1} = e_i) + \gamma_u}{n(e_{i-1}, .) + I(e_{i - 1} = e_i) + E \boldsymbol{\cdot}\gamma_u} \\
& \boldsymbol{\cdot} \bigg( \prod_j  \frac{exp(\psi_{e_i,z_j})}{\sum_z exp(\psi_{e_i,z})} \boldsymbol{\cdot} \frac{n(e_i, z_j, w_j) + \delta}{n(e_i,z_j, .) + W \cdot \delta} \bigg) \\
\end{aligned}
\end{equation}

Consider a document $d$ containing a sequence of words $\{w_j\}$ with corresponding facets $\{z_j\}$. The first factor models the probability of the user $u_d$ reaching expertise level $e_{u_d}$ for document $d$; whereas the second and third factor models the probability of the facets $\{z_j\}$ being chosen at the expertise level $e_{u_d}$, and the probability of observing the words $\{w_j\}$ with the facets $\{z_j\}$ and expertise level $e_{u_d}$, respectively. Following the Markovian assumption, we only consider the expertise levels $e_{u_d}$ and $e_{u_d} + 1$ for sampling, and select the one with the highest conditional probability.

Samples obtained from Gibbs sampling are
used to approximate the expertise-facet-word distribution
$\Phi$:

\begin{equation}
\label{eq:phi}
 \phi_{e,k,w} = \frac{n(e, k, w) + \delta}{n(e,k, .) + W \cdot \delta}
\end{equation}

Once the generative process for a review $d$ with words $\{w_j\}$ is over, we can estimate $\xi$ from $\Phi$ as the proportion of the $z^{th}$ facet in the document written at expertise level $e$ as:

\begin{equation}
\label{eq:xi}
\xi_{e,z} \propto \sum_{j=1}^{N_d} \phi_{e, z, w_j}
\end{equation}

In summary, $\xi$, $\Phi$, and $\Theta$ are linked via $\Psi$:

\begin{itemize}
\item[i)] $\Psi$ generates $\Theta$ via Equation~\ref{eq:transform}.
\item[ii)] $\Theta$ and $\Phi$ are coupled in Equations~\ref{eq:joint},~\ref{eq:facets}.
\item[iii)] $\Phi$ generates $\xi$ using Equation~\ref{eq:xi}.
\item[iv)] $\Psi$ is learned via regression (with $\xi$ as latent features) using Equations~\ref{eq:features},~\ref{eq:mse}, so as to minimize the mean squared error for predicting review helpfulness. 
\end{itemize}

\noindent {\bf Overall Processing Scheme:} Exploiting results from the above discussions, the overall inference is an iterative stochastic optimization process consisting of the following steps:
\vspace{-1em}
\begin{itemize}
\item[i)] Sort all reviews by timestamps, and estimate $E$ using Equation~\ref{eq:exp-gibbs}, by Gibbs sampling. During this process, consider all facet assignments $Z$ and $\Psi$, from the earlier iteration fixed.
\item[ii)] Estimate facets $Z$ using Equation~\ref{eq:facets}, by Gibbs sampling, keeping the expertise levels $E$ and $\Psi$, from the earlier iteration fixed.
\item[iii)] Estimate $\xi$ using Equations~\ref{eq:phi} and~\ref{eq:xi}.
\item[iv)] Learn $\Psi$ from $\xi$ and other consistency factors using Equations~\ref{eq:features},~\ref{eq:mse}, by regression.
\item[v)] Estimate $\Theta$ from $\Psi$ using Equation~\ref{eq:transform}.
\end{itemize}

\noindent{\bf Regression:} For regression, we use the fast and scalable Support Vector Regression implementation from LibLinear\footnote{\url {www.csie.ntu.edu.tw/~cjlin/liblinear}} that uses trust region Newton method for learning the parameters $\Psi$.

\noindent{\bf Test:} Given a test review with $\langle${\em user=$u$, item=$i$, words=$\{w_j \}$, rating=$r$, timestamp=$t$}$\rangle$, we find its helpfulness score by plugging in the consistency features, and latent factors in Equation~\ref{eq:features} with the parameters $\langle \Psi, \beta_u, \beta_i, \overline{r_u}, \overline{r_i}, \overline{r_g} \rangle$ having been learned from the training data. $\xi$ is computed over the words $\{w_j\}$ using Equation~\ref{eq:xi}, where the counts are estimated over all the documents and words in the training dataset.

%\todo{Table of Notations}
%\todo{histograms for helpfulness}
%\todo{train mse graphs}

%% file: main/chapter-credible-applications/helpful-reviews/experiments.tex
\subsection{Experiments}
\label{sec:experiments3}

\subsubsection{Setup: Data} 

We perform experiments with data from {\em Amazon} in {\em five} different domains:
(i) movies, (ii) music, (iii) food, (iv) books, and (v) electronics. The statistics of the dataset\footnote{Data available from \url {snap.stanford.edu/data/}} is given in Table~\ref{tab:statistics3}. 
In total, we have $29$ million reviews from $5.6$ million users on $1.8$ million items from all of the five domains combined. We extract the following quintuple for our model $\langle${\em userId, itemId, timestamp, rating, review, helpfulnesVotes}$\rangle$ from each domain. For the average number of votes per review in Table~\ref{tab:statistics3}, we consider those reviews that received non-zero number of votes.

During {\em training}, for movies, books, music, and electronics, we consider only those reviews for which at least $y \geq 20$ users have voted about their helpfulness (including for, and against) to have a robust dataset (similar to the setting in~\cite{icdm2008, O'Mahony:2009:LRH:1639714.1639774}) for learning. Since the food dataset has less number of reviews, we lowered this threshold to {\em five}. 

For {\em test}, we used the $3$ most recent reviews of each user as withheld test data (similar to our setting in Chapter~\ref{chap:temporal}), that received atleast {\em five} votes (including for, and against). The same data is used for all the models for comparison.

We group {\em long-tail} users with less than $10$ reviews in {\em training} data into a background model, treated as a single user, to avoid modeling from sparse observations. We do not ignore any user. During the {\em test} phase for a ``long-tail'' user, we take her parameters from the background model. We set the number of facets as $Z=50$, and number expertise levels as $E=5$, for all the datasets. 

\begin{table}
	\centering
	 \small
	\begin{tabular}{lrrrrr}
		\toprule
		\bf{Factors} & \bf{Books} & \bf{Music} & \bf{Movie} & \bf{Electronics} & \bf{Food}\\
		\midrule
		\#Users & 2,588,991 & 1,134,684 & 889,176 & 811,034 & 256,059\\
		\#Items & 929,264 & 556,814 & 253,059 & 82,067 & 74,258\\
		\#Reviews & 12,886,488 & 6,396,350 & 7,911,684 & 1,241,778 & 568,454\\
		\midrule
		$\frac{\textbf{\#Reviews}}{\textbf{\#Users}}$ & 4.98 & 5.64 & 8.89 & 1.53 & 2.22 \\\midrule
		$\frac{\textbf{\#Reviews}}{\textbf{\#Items}}$ & 13.86 & 11.48 &	31.26 &	15.13 &	7.65 \\\midrule
		$\frac{\textbf{\#Votes}}{\textbf{\#Reviews}}$ & 9.71 & 5.95 & 7.90 & 8.91 & 4.24\\
		\bottomrule
	\end{tabular}
	\caption{Dataset statistics. Votes indicate the total number of helpfulness votes (both, for and against) cast for a review. Total number of users $=5,679,944$, items $=1,895,462$, and reviews $=29,004,754$.}
	\label{tab:statistics3}
\end{table}

\pagebreak

\subsubsection{Tasks and Evaluation Measures}

We use all the models for the following tasks:
\vspace{-1em}
\begin{itemize}
\item[1)] {\bf Prediction:} Here the objective is to predict the helpfulness score of a review as $x/y$, where $x$ is the number of users who voted the review as helpful out of $y$ number of users. We use the following evaluation measures:
\begin{itemize}
 \item[i)] {\em Mean squared error:}  The mean squared error of the predicted scores with the ground helpfulness scores is obtained using Equation~\ref{eq:mse}.
 \item[ii)] {\em Squared correlation coefficient ($R^2$):} The $R^2$ statistic gives an indication of the goodness of fit of a model, i.e., how well the regression function approximates the real data points, with $R^2 =1$ indicating a perfect fit. In linear least squares regression, $R^2$ is given by the square of the Pearson correlation between the observed and predicted values.
\end{itemize}
\item[2)] {\bf Ranking:} A more suitable way of evaluation is to compare the ranking of the reviews from different models based on their helpfulness scores --- where the reviews at the top of the rank list should be more helpful than the ones below them. We use the predicted helpfulness scores from each model to rank the reviews, and compute {\em rank correlation} with the gold rank list --- obtained by ranking all the reviews by their ground-truth helpfulness scores ($x/y$) --- using the following measures:
\begin{itemize}
 \item[i)] {\em Spearman correlation ($\rho$):} This assesses how well the relationship between two variables can be described using a {\em monotonic} function, unlike Pearson correlation that only indicates a {\em linear} relationship between the variables. $\rho$ can be computed by the Pearson correlation between the {\em rank} values of the variables in the rank list.
 \item[ii)] {\em Kendall-Tau correlation ($\tau$):} This measures the number of concordant and discordant pairs, to find whether the ranks of two elements agree or not based on their scores, out of the total number of combinations possible. Unlike Spearman correlation, Kendall-Tau is not affected by the distance between the ranks, but only depends on whether they agree or not.
\end{itemize}
\end{itemize}

\subsubsection{Baselines} 

We consider the following baselines to compare our work:

\squishlist
%(refer to Figure~\ref{fig:baselines}).\\
\item [a)] \cite{O'Mahony:2009:LRH:1639714.1639774} use several rating based features as proxy for reviewer reputation and sentiment; review length and letter cases for content; and review count statistics for social features to classify if the review is helpful or not %{\em Note} that our consistency features (except timeliness) are a subset of their rating features. 
\item [b)] \cite{Lu:2010:ESC:1772690.1772761} use syntactic features (part-of-speech tags of words), sentiment (using a lexicon to find word polarities), review length and reviewer rating statistics to predict the quality of a review. We ignore the social network related features in their work, in absence of user-user links in our dataset. Similar kinds of syntactic and semantic features are also used in the next baseline.
\item [c)] \cite{Kim:2006:AAR:1610075.1610135} use structural (review length statistics), lexical (tf-idf), syntactic (part-of-speech tags), semantic (explicit product features, and sentiment of words), and meta-data related features to rank the reviews based on their helpfulness. We ignore the explicit product-specific (meta-data) features that are absent in our dataset.
\item [d)] \cite{icdm2008} predict the helpfulness of reviews on IMDB based on three factors: \textit{reviewer expertise}, \textit{syntactic features}, and \textit{timeliness} of a review. The authors use reviewer preferences for explicit facets (pre-defined genres of movies in IMDB) as proxy for their expertise, part-of-speech tags of words for the syntactic features, and review publication dates to compute timeliness of reviews. This baseline is the closest to our work as we attempt to model similar factors. However, we model reviewer expertise {\em explicitly}, and the facets as {\em latent} --- therefore not relying on any additional item meta-data (like, genres).
%Since there are no explicit facets for items in our dataset, the reviewer expertise (as modeled in this baseline) could not be implemented.
\squishend

For all of the above baselines, we use all the features from their works that are supported by our dataset for a fair comparison.
		\vspace{-1em}

\subsubsection{Quantitative Comparison}

\begin{table*}
	\centering
	\small
		\begin{tabular}{p{3.2cm}p{0.8cm}p{0.6cm}p{0.6cm}p{0.6cm}p{0.8cm}|p{0.8cm}p{0.6cm}p{0.6cm}p{0.6cm}p{0.5cm}}
			\toprule
			\multirow{2}{*}{\bf{Models}} & \multicolumn{5}{c|}{\textbf{Mean Squared Error (MSE)}} &  \multicolumn{5}{c}{\textbf{Squared Correlation Coefficient ($R^2$)}} \\ \cmidrule{2-11}
			& \bf{Movies} & \bf{Music} & \bf{Books} & \bf{Food}  & \bf{Elect.} & \bf{Movies} & \bf{Music} & \bf{Books} & \bf{Food}  & \bf{Elect.}\\
			\midrule
			Our model & {\bf 0.058}  & {\bf 0.059} & {\bf 0.055} & {\bf 0.053}  & {\bf 0.050} & {\bf 0.438} & {\bf 0.405} & {\bf 0.397} & {\bf 0.345}  & {\bf 0.197}\\
			{\bf a)} \cite{O'Mahony:2009:LRH:1639714.1639774} & 0.067 & 0.069 &  0.069 & 0.060 & 0.064 & 0.325 & 0.295 & 0.249 & 0.312 & 0.134\\
			{\bf b)} \cite{Lu:2010:ESC:1772690.1772761} & 0.093 & 0.087 & 0.077 & 0.072 & 0.071 & 0.111 & 0.128 & 0.139 & 0.134 & 0.056 \\
			{\bf c)} \cite{Kim:2006:AAR:1610075.1610135} & 0.107 & 0.125 & 0.094 & 0.073 & 0.161 & 0.211 & 0.025 & 0.211 & 0.309 & 0.065 \\
			{\bf d)} \cite{icdm2008} & 0.091 & 0.091 & 0.082 & 0.075 & 0.063 & 0.076 & 0.053 & 0.076 & 0.039 & 0.043 \\
			\bottomrule
		\end{tabular}
		\vspace{-0.5em}
	\caption{{\em Prediction Task:} Performance comparison of our model versus baselines. Our {improvements} over the baselines are statistically significant at {\em p-value} $<2.2e-16$ using {\em paired sample t-test}.}
			\vspace{-0.5em}
	\label{fig:MSE3}
\end{table*}

\begin{table*}
	\centering
	\small
	        \begin{tabular}{p{3.2cm}p{0.8cm}p{0.6cm}p{0.6cm}p{0.6cm}p{0.8cm}|p{0.8cm}p{0.6cm}p{0.6cm}p{0.6cm}p{0.6cm}}
			\toprule
			\multirow{2}{*}{\bf{Models}} & \multicolumn{5}{c|}{\textbf{Spearman ($\rho$)}} &  \multicolumn{5}{c}{\textbf{Kendall-Tau ($\tau$)}} \\ \cmidrule{2-11}
			& \bf{Movies} & \bf{Music} & \bf{Books} & \bf{Food}  & \bf{Elect.} & \bf{Movies} & \bf{Music} & \bf{Books} & \bf{Food} & \bf{Elect.}\\
			\midrule
			Our model & {\bf 0.657} & {\bf 0.610} &  {\bf 0.603} & 0.533  & {\bf 0.394} & {\bf 0.475} & {\bf 0.440} & {\bf 0.435} & 0.387  & {\bf 0.280}\\
			{\bf a)} \cite{O'Mahony:2009:LRH:1639714.1639774} & 0.591 & 0.554  & 0.496 & 0.541 & 0.340 & 0.414 & 0.390 & 0.347 & 0.398 & 0.237\\
			{\bf b)} \cite{Lu:2010:ESC:1772690.1772761} & 0.330  & 0.349 & 0.334 & 0.367 & 0.205 & 0.224 & 0.242 & 0.230 & 0.259 & 0.144 \\
			{\bf c)} \cite{Kim:2006:AAR:1610075.1610135} & 0.489 & 0.166 & 0.474 & \textbf{0.551} & 0.261 & 0.342 & 0.114 & 0.334 & \textbf{0.414}  & 0.184 \\
			{\bf d)} \cite{icdm2008} & 0.268 & 0.232 & 0.258 & 0.199 & 0.159 & 0.183 & 0.161 & 0.178 & 0.141 & 0.112 \\
			\bottomrule
		\end{tabular}
	\caption{{\em Ranking Task:} Correlation comparison between the ranking of reviews and gold rank list --- our model versus baselines. Our {\em improvements} over the baselines are statistically significant at {\em p-value} $<2.2e-16$ using {\em paired sample t-test}.}
	\label{fig:rank}
\end{table*}

%\begin{figure}
%	\vspace{-1.5em}
%	\centering
%	\includegraphics[scale=0.45]{blank}
%	\caption{MSE improvement (\%) of our model over baselines.}
%	\label{fig:improvement}
%	\vspace{-1.5em}
%\end{figure}

Table~\ref{fig:MSE3} shows the comparison of the \emph{Mean Squared Error (MSE)} and \textit{Squared Correlation Coefficient ($R^2$)} for review helpfulness predictions, as  generated by our model with the four baselines. Our model consistently outperforms all baselines in reducing the MSE.
%by ca. $??\%$. 
Table~\ref{fig:rank} shows the comparison of the \textit{Spearman ($\rho$)} and \textit{Kendall-Tau ($\tau$)} correlation between the rank list of helpful user reviews, as generated by all the models, and the gold rank list.

The most competitive baseline for our model is~\cite{icdm2008}. Since there is a high overlap in the consistency features of our model with this baseline, the performance improvement of our model can be attributed to the incorporation of the {\em latent} factors in our model. We perform {\em paired sample t-tests}, and find that our performance improvement over all the baselines is statistically significant at {\em p-value} $<2e-16$.

We observe that our model's performance, for the ranking task, is better for the domains {\em movies, music,} and {\em books} with average number of reviews  per-user $\geq 5$; and worse for {\em food}, and, especially, {\em electronics} with very few number of reviews per-user at $2.2$ and $1.5$ respectively --- although we still outperform the baseline models that perform worse. The poor performance of our model in the last two datasets can be attributed to data sparsity due to which user maturity could not be captured well.

\begin{figure*}[t]
	\centering
	\includegraphics[scale=0.75]{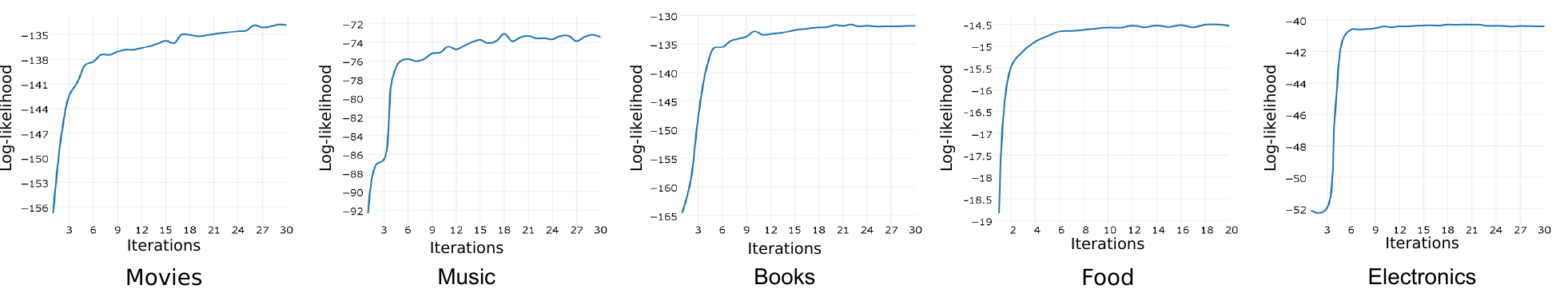}
	\caption{Increase in log-likelihood (scaled by $10e+07$) of the data {\em per-iteration} in the five domains.}
			\vspace{-1em}
	\label{fig:logL}
\end{figure*}
\begin{figure*}
	\centering
	\begin{subfigure}{\textwidth}
	\includegraphics[scale=0.75]{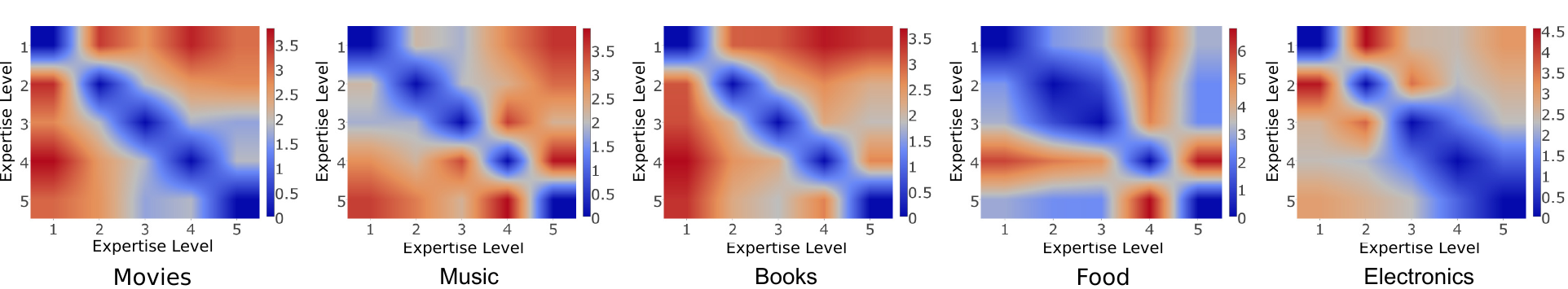}
	\caption{Our model: Facet preference divergence with expertise learned from review helpfulness.}
	\label{fig:heatMap_facet}
	\end{subfigure}
	\begin{subfigure}{\textwidth}
	\includegraphics[scale=0.75]{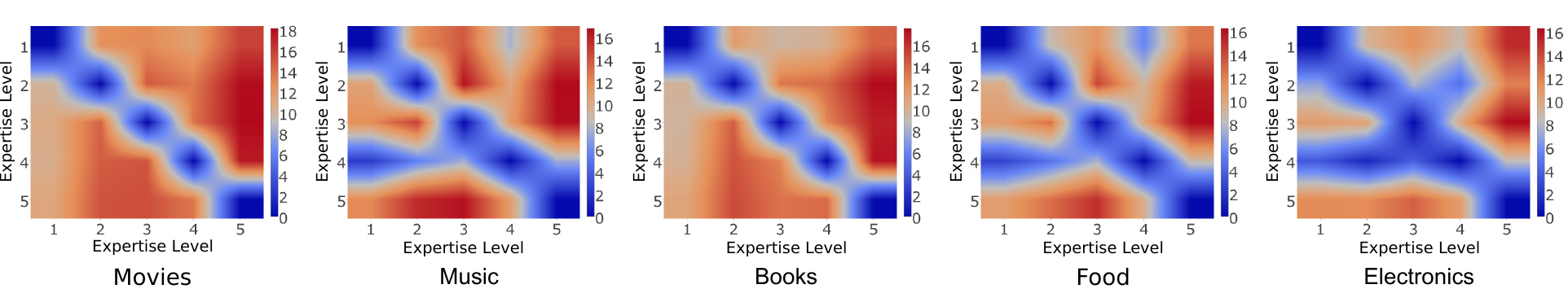}
	\caption{Our model: Language model divergence with expertise learned from review helpfulness.}
	\label{fig:heatMap_lang}
	\end{subfigure}
			\vspace{-0.5em}
	\caption{Facet preference and language model KL divergence with expertise.}
			\vspace{-1em}
\end{figure*}

\subsubsection{Qualitative Comparison}

{\noindent \bf Log-likelihood of data and convergence:} The inference of our model is quite involved with the coupling between several variables, and the alternate stochastic optimization process. Figure~\ref{fig:logL} shows the increase in the data log-likelihood of our model per-iteration for each of the five datasets. We observe that the model is stable, and achieves a near smooth increase in the data log-likelihood per-iteration. It also converges quite fast between $20-30$ iterations depending on the complexity of the dataset. For {\em electronics} the convergence is quite rapid as the data is quite sparse, and the model does not find sufficient evidence for categorizing users to different expertise levels; this behavior is reflected in all the experiments involving the {\em electronics} dataset.

{\noindent \bf Language model and facet preference divergence:} From our initial hypothesis of the joint interaction between review helpfulness, reviewer expertise, facet preferences, and writing style: we expect users at different expertise levels to have divergent facet preferences and Language Models (LM's) --- with expert users having a more sophisticated writing style and vocabulary than amateurs.

Figures~\ref{fig:heatMap_facet} and~\ref{fig:heatMap_lang} show the heatmaps of the Kullback-Leibler (KL) divergence for facet preferences and language models of users at different expertise levels, as computed by our model conditioned on review {\em helpfulness} --- given by $D_{KL}(\theta_{e_i}||\theta_{e_j})$ and $D_{KL}(\phi_{e_i}||\phi_{e_j})$ respectively, where $\Theta$ and $\Phi$ are given by Equations~\ref{eq:transform} and~\ref{eq:phi}, respectively. 

The main observation is that the {\em KL} divergence is higher --- the larger the difference is between the expertise levels of two users. This confirms our hypothesis. We also note that the increase in divergence with the increase in gap between expertise levels is not smooth for {\em food} and {\em electronics} --- due to the sparsity of {\em per-user} data.

{\noindent \bf Interpretable explanation by salient words used by experts for helpful reviews:} Table~\ref{tab:examples} shows a snapshot of the latent word clusters, as used by experts and amateurs, for helpful reviews and otherwise, as generated by our model. Once the model parameters are estimated, for each dataset, we consider the expertise-facet pairs $\{e^{+},z^{+}\}$ and $\{e^{-},z^{-}\}$ for which the learned feature weights $\langle \psi_{e,z} \rangle$ are maximum and minimum, respectively. Now, given the language model $\Phi$, we rank the {\em top} words from $\phi_{e^{+},z^{+}}$ and $\phi_{e^{-},z^{-}}$ as the words contributing most to helpful reviews, and least helpful reviews, respectively.

We observe that the most helpful reviews pertaining to {\em music} talk about the essence and style of music; for {\em books} they describe the theme and writing style; for {\em movies} they write about screenplay and storytelling; for {\em electronics} they discuss about specific product features --- note that earlier works~\cite{icdm2008,Kim:2006:AAR:1610075.1610135} used {\em explicit} product descriptions as features, that we were able to automatically discover as latent features from textual reviews; whereas for {\em food} reviews these are mostly concerned about hygiene and allergens. The least helpful reviews mostly describe some generic concepts in the domain, praise or criticize an item without going in depth about the facets, and are generally quite superficial in nature.

\begin{table*}[!h]
	\centering
	\small
	\begin{tabular}{p{\linewidth}}
	\toprule
	{\bf Top words used by experts in {\em most} helpful reviews.}\\
	\midrule
	  {\bf Music:} album, lyrics, recommend, soundtrack, touch, songwriting, features, rare, musical, ears, lyrical, enjoy, absolutely, musically, individual, bland, soothing, released, inspiration, share, mainstream, deeper, flawless, wonderfully, eclectic, heavily, critics, presence, popularity, brilliantly, inventive%, sensual, classical, moods, worn, baritone, groovy, contributions, conviction, undoubtedly, floyd, vinyl 
	  \\\midrule
	  {\bf Books:} serious, complex, claims, content, illustrations, picture, genre, beautifully, literary, witty, critics, complicated, argument, premise, scholarship, talented, divine, twists, exceptional, obsession, commentary, landscape, exposes, influenced, accomplished, oriented, exploration, styles, storytelling%, overwhelming, humour, implications, contradictions, sequence, doctrines, elaborate, complexities%, horrific, angst, vignettes 
	  \\\midrule
	  {\bf Movies:} scene, recommend, screenplay, business, depth, justice, humanity, packaging, perfection, flicks, sequels, propaganda, anamorphic, cliche\&acute, pretentious, goofy, ancient, marvelous, perspective, outrageous, intensity, mildly, immensely, bland, subplots, anticipation, rendered, atrocious%, admire, continually, macho, vague, futuristic, uncover, context, ensues, scripted, inept, fuzzy, revisit%, mediocrity
	  \\\midrule
	  {\bf Electronics:} adapter, wireless, computer, sounds, camera, range, drives, mounted, photos, shots, packaging, antenna, ease, careful, broken, cards, distortion, stick, media, application, worthless, clarity, technical, memory, steady, dock, items, cord, systems, amps, skin, watt, monitors, arms, pointed%, instructions, directional, updates, responsive, inkjet, warranty, handheld, refund, interfere, sharpness%, processor
	  \\\midrule
	  {\bf Food:} expensive, machine, months, clean, chips, texture, spicy, odor, inside, processed, robust, packs, weather, sticking, alot, press, poured, swallow, reasonably, portions, beware, fragrance, basket, volume, sweetness, terribly, caused, scratching, serves, sensation, sipping, smelled, italian, sensitive, suffered%, stayed, superb, crappy, chewy, dumplings, exquisite, venture, ranks, gulp, barbecue, touted
	  \\\toprule
	  	{\bf Top words used by amateurs in {\em least} helpful reviews.}\\
	\midrule
	  {\bf Music:} will, good, favorite, cool, great, genius, earlier, notes, attention, place, putting, superb, style, room, beauty, realize, brought, passionate, difference, god, fresh, save, musical, grooves, consists, tapes, depressing, interview, short, rock, appeared, learn, brothers, considering, pitched, badly, adding, kiss%, shape, main, rolling, shut, mark, violinist, suspect, murder, rousing, absolute, albeit, suites, gloomy%, looked, checked
	  \\\midrule
	  {\bf Books:} will, book, time, religious, liberal, material, interest, utterly, moves, movie, consistent, false, committed, question, turn, coverage, decade, novel, understood, worst, leader, history, kind, energy, fit, dropped, current, doubt, fan, books, building, travel, sudden, fails, wanted, ghost, presents, honestly%, mission, reaching, motivation, blacks, themes, marriages, chosen, prefer, games, blank, baseball%, attempt, evaluate
	  \\\midrule
	  {\bf Movies:} movie, hour, gay, dont, close, previous, features, type, months, meaning, wait, boring, absolutely, truth, generation, going, fighting, runs, fantastic, kids, quiet, kill, lost, angles, previews, crafted, teens, help, believes, brilliance, touches, sea, hardcore, continue, album, formula, listed, drink, text%, author, scripts, requires, animation, seek, flag, stronger, legal, reminds, light, exactly, till, seeking%, mythology
	  \\\midrule
	  {\bf Electronics:} order, attach, replaced, write, impressed, install, learn, tool, offered, details, turns, snap, price, digital, well, buds, fit, problems, photos, hear, shoot, surprisingly, continue, house, card, sports, writing, include, adequate, nice, programming, protected, mistake, response, situations, effects%, systems, ear, auto, directions, easily, album, existing, spending, considerably, antennas, steal, rate, block%, respond	  
	  \\\midrule
	  {\bf Food:} night, going, haven, sour, fat, avoid, sugar, coffee, store, bodied, graham, variety, salsa, reasons, favorite, delicate, purpose, brands, worst, litter, funny, partially, sesame, handle, excited, close, awful, happily, fully, fits, effects, virgin, salt, returned, powdery, meals, matcha, great, bites, table, pistachios%, liver
	  \\\bottomrule
	\end{tabular}
	\caption{Snapshot of latent word clusters as used by experts and amateurs for most and least helpful reviews in different domains.} \label{tab:examples}
\end{table*}

%% file: main/chapter-credible-applications/consistency-analysis/credibilityanalysis.tex
\subsection{Review Credibility Analysis}

Unlike prior works --- in opinion spam and fake review detection --- leveraging crude user behavioral and shallow textual features of reviews for credibility classification, we delve deep into the semantics of the reviews to under the inconsistencies that can be used to explain why the review is non-credible, or otherwise using (latent) facet models.

\subsubsection{Facet Model}

Given review snippets like ``{\tt the hotel offers free wi-fi}'', we now aim to find the different facets present in the reviews along with their corresponding sentiment polarities. Since the aim of this work is to present a model requiring limited prior information, we extract the {\em latent} facets from the review text, without the help of any explicit facet or seed words. The ideal machinery should map ``wi-fi'' to a latent facet cluster like ``network, Internet, computer, access, ...''. We also want to extract the sentiment expressed in the review about the facet. Interestingly, although ``free'' does not 
have a polarity of its own, in the above example ``free'' in conjunction with ``wi-fi'' expresses a positive sentiment of a service being offered without charge. The hope is that although ``free'' does not have an individual polarity, it appears in the neighborhood of words that have known polarities (from lexicons). This helps in the joint discovery of facets and sentiment labels, as ``free wi-fi'' and ``internet without extra charge'' should ideally map to the same facet cluster with similar polarities using their co-occurrence with similar words with positive polarities. In this work, we use the Joint Sentiment Topic Model approach (JST)~\cite{linCIKM2009} to jointly discover the latent facets along with their expressed polarities.

%Although JST is unsupervised, we modify the formulation to make it supervised using the user-assigned ratings to reviews. This leads to better sentiment prediction of reviews, for instance in the TripAdvisor dataset we obtain $8.4\%$ improvement in accuracy over the unsupervised version.

Consider a set of reviews $\langle D \rangle$ written by users $\langle U \rangle$ on a set of items $\langle I \rangle$, with $r_d \in R$ being the rating assigned to review $d \in D$. Each review document $d$ consists of a sequence of words $N_d$ denoted by $\{w_1, w_2, ... w_{N_d}\}$, and each word is drawn from a vocabulary $V$ indexed by $1,2,..V$. Consider a set of facet assignments $z=\{z_1, z_2, ... z_K\}$ and sentiment label assignments $l = \{l_1, l_2, ... l_L\}$ for $d$, where each $z_i$ can be from a set of $K$ possible facets, and each label $l_i$ is from a set of $L$ possible sentiment labels. 

JST adds a layer of sentiment in addition to the topics as in standard LDA~\cite{Blei2003LDA}. It assumes each document $d$ to be associated with a multinomial distribution $\theta_d$ over facets $z$ and sentiment labels $l$ with a symmetric Dirichlet prior $\alpha$. $\theta_d(z,l)$ denotes the probability of occurrence of facet $z$ with polarity $l$ in document $d$. Topics have a multinomial distribution $\phi_{z,l}$ over words drawn from a vocabulary $V$ with a symmetric Dirichlet prior $\beta$. $\phi_{z,l}(w)$ denotes the probability of the word $w$ belonging to the facet $z$ with polarity $l$. In the generative process, a sentiment label $l$ is first chosen from a document-specific rating distribution $\pi_d$ with a symmetric Dirichlet prior $\gamma$ . 
%In case of a supervised process this distribution is known {\em apriori}. 
Thereafter, one chooses a facet $z$ from $\theta_d$ conditioned on $l$, and subsequently a word $w$ from $\phi$ conditioned on $z$ and $l$. Algorithm~\ref{algo.3} outlines the generative process. Exact inference is not possible due to 
intractable coupling between $\Theta$ and $\Phi$, and thus we use Collapsed Gibbs Sampling for approximate inference.

Let $n(d, z, l, w)$ denote the count of the word $w$ occurring in document $d$ belonging to the facet $z$ with polarity $l$. The conditional distribution for the latent variable $z$ (with components $z_1$ to $z_K$) and $l$ (with components $l_1$ to $l_L$) is given by:

\begin{equation}
\label{eq.3.1}
\begin{aligned}
 P(z_i=k, l_i=j| &w_i=w, z_{-i}, l_{-i}, w_{-i}) \propto\\
  \frac{n(d, k, j, .) + \alpha}{\sum_{k}n(d, k, j, .) + K \alpha} &\times \frac{n(., k, j, w) + \beta}{\sum_{w}n(., k, j, w) + V \beta} \times \frac{n(d,.,j,.) + \gamma}{\sum_j n(d, ., j,.) + L \gamma}
 \end{aligned}
\end{equation}

In the above equation, the operator $(.)$ in the count indicates marginalization, i.e.,  summing up the counts over all values for the corresponding position in $n(d,z,l,w)$, and 
the subscript $-i$ denotes the value of a variable excluding the data at the $i^{th}$ position.

\begin{algorithm}[t]
\small
\label{algo.3}
\SetAlgoLined
\DontPrintSemicolon
\For {each document $d$} {choose a distribution $\pi_d \sim Dir(\gamma)$} \;
\For {each sentiment label $l$ under document $d$} {choose a distribution $\theta_{d,l} \sim Dir(\alpha)$} \;
\For {each word $w_i$ in document $d$} {
  Choose a sentiment label $l_i \sim \pi_d$ \;
  Choose a topic $z_i \sim \theta_{d,l_i}$ \;
  Choose a word $w_i$ from the distribution over words $\phi_{l_i, z_i}$ \;
}
\caption{Joint sentiment topic model~\cite{linCIKM2009}.}
\label{algo.3}
\end{algorithm}

\subsubsection{Consistency Features}
\label{subsec:cons}

We extract the following features from the latent facet model enabling us to detect {\em inconsistencies} in reviews and ratings of items for credibility analysis.
%1) {\em Users}, 2) {\em Items}, 3) {\em Facets}, 4) {\em Reviews}, 5) {\em Ratings}, and 6) {\em Timestamps} 

\noindent 1. {\bf User Review -- Facet Description:} The facet-label distribution of different items differ; for some items, certain facets (with their polarities) are more important than 
other dimensions. For instance, the ``battery life'' and ``ease of use'' for consumer electronics are more important than ``color''; for hotels, certain services are available for free (e.g., wi-fi) 
which may be charged elsewhere. Similarly, user reviews involving less relevant facets of the item under discussion, e.g., downrating hotels for ``not allowing pets'' should also be detected.

Given a review $d(i)$ on an item $i \in I$ with a sequence of words $\{w\}$ and previously learned $\Phi$, its facet label distribution $\Phi^{'}_d(i)$ with dimension $K \times L$ is given by:
\begin{equation}
 \phi^{'}_{k,l} = \sum_{w: l^{*} = argmax_{l}\ \phi_{k,l}(w)} \phi_{k, l^{*}}(w)
\end{equation}
For each word $w$ and each latent facet dimension $k$, we consider the sentiment label $l^{*}$ that maximizes the facet-label-word distribution $\phi_{k,l}(w)$, and aggregate this over all the words.
This facet-label distribution of the review $\Phi^{'}_d(i)$ of dimension $K \times L$ is used as a feature vector to a classifier to figure out the importance of the different latent dimensions that 
also captures {\em domain-specific} facet-label importance.

\noindent 2. {\bf User Review --- Rating:} The rating assigned by a user to an item should be consistent to her opinion expressed in the review about the item. For instance, it is unlikely that the user will assign an average or poor rating to an item when she has expressed positive opinion about all the important facets of the item in the review. 
The inferred rating distribution $\pi^{'}_d $ (with dimension $L$) of a review $d$ consisting of a sequence of words $\{w\}$ and learned $\Phi$ is computed as:
\begin{equation}
 \pi^{'}_l = \sum_{w,k:\{k^{*}, l^{*}\} = argmax_{k,l}\ \phi_{k,l}(w)} \phi_{k^{*}, l^{*}}(w)
\end{equation}
For each word, we consider the facet and label that jointly maximizes the facet-label-word distribution, and aggregate over all the words and facets. The absolute deviation (of dimension $L$) between the user-assigned 
rating $\pi_d$, and estimated rating $\pi^{'}_d$ from user text is taken as a component in the overall feature vector.

\noindent 3. {\bf User Rating:} Prior works~\cite{Ott2011,Sun2013,Hu2012} dealing with opinion spam and fake reviews found that these kinds of reviews tend to express overtly positive or overtly negative opinions. Therefore, we also 
use $\pi^{'}_d$ as a component of the overall feature vector to detect cues from such extreme ratings.%from text.

\noindent 4. {\bf Temporal Burst:} This is typically observed in {\em group spamming}, where a number of reviews are posted targeting an item in a short span of time. Consider a set of reviews 
$\{d_j\}$ at timepoints $\{t_j\}$ posted for a {\em specific} item. The temporal burstiness of review $d_i$ for the given item is given by $\big( \sum_{j, j\neq i} \frac{1}{1 + e^{t_i-t_j}} \big)$. 
Here, exponential decay is used to weigh the temporal proximity of reviews to capture the burst.

\noindent 5. {\bf User Review -- Item Description:} In general, the description of the facets outlined in a user review about an item should not differ  markedly from that of the majority. %Here, we exploit the fact that non-credible reviews form only a small fraction of all the reviews on an item. 
For instance, if the user review says ``internet is charged'', and majority says the ``hotel offers free wi-fi''  --- this presents a possible inconsistency. For the facet model this corresponds to word 
clusters having the same facet label but different sentiment labels. During experiments, however, we find this feature to play a weak role in the presence of other inconsistency features.\\

We aggregate the {\em per-review} facet distribution $\phi^{'}_{k,l}$ over all the reviews $d(i)$ on the item $i$ to obtain the facet-label distribution $\Phi^{''}(i)$ of the item. We use the Jensen-Shannon divergence, a symmetric and smoothed version of the Kullback-Leibler divergence as a feature. This depicts how much the facet-label distribution in the given review diverges from the general opinion of other people about the item.
\begin{equation}
 JSD(\Phi^{'}_d(i)\ ||\ \Phi^{''}(i)) = \frac{1}{2} (D(\Phi^{'}_d(i)\ ||\ M) + D(\Phi^{''}(i)\ ||\ M))
\end{equation}
where, $M = \frac{1}{2}(\Phi^{'}_d(i)+\Phi^{''}(i))$, and $D$ represents Kullback-Leibler divergence.

\noindent{\bf Feature vector construction}: For each review $d_j$, all the above {\em consistency features} are computed, and a facet feature vector $\langle F^T(d_j) \rangle$ of dimension 
$2 + K \times L + 2L$ is created for subsequent processing.

\subsubsection{Additional Language and Behavioral Features}
\label{subsec:lang}

In addition to the above consistency features, we also use limited language and user behavioral features. We later show during experiments that all these features, in conjunction, perform better than the individual feature classes.

In order to capture the distributional difference in the words of deceptive and authentic reviews, we consider {\em unigram and bigram} language features that have been shown to outperform other fine-grained linguistic features using psycholinguistic features (e.g., LIWC lexicon) and Part-of-Speech tags~\cite{Ott2011}. Chapter~\ref{sec:language} discusses in-depth the various linguistic features effective for distinguishing credible reviews, from non-credible ones.

\noindent {\bf Language feature vector construction}: Consider a vocabulary $V$ of unique unigrams and bigrams in the corpus (after removing stop words). For each token type $f_i \in V$ 
and each review $d_j$, we compute the presence/absence of words, $w_{ij}$, of type $f_i$ occurring in $d_j$, thus constructing a feature vector $F^L(d_j) = \langle w_{ij}=I({w_{ij} = f_i}) ~ / ~length(d_j) \rangle, \forall i$, with $I(.)$ denoting an indicator function (notations used are presented in Table~\ref{tab:notation}).

%\subsection{Behavioral Model}
%\label{subsec:user}

Earlier works~\cite{Liu2007,Liu2008,Liu2010} on review spam show that user-dependent models detecting user-preferences and biases perform well in credibility analysis. However, such information 
is not always available, especially for newcomers, and not so active users in the community.
%We used the Joint Author Sentiment Topic Model~\cite{mukherjee2014JAST} which puts a layer of {\em users} over JST to have a generative model of the facet-label distribution {\em per-user}, taking her preferences and writing style into account. The model performs well when there are multiple reviews per-user. However, in our setting we have very few reviews per-user. 
Besides, ~\cite{Liu2012,Liu2013} show that spammers tend to open multiple fake accounts to write reviews for malicious activities --- using each of those accounts sparsely to avoid detection. 
%Therefore the user model learned very sparse distributions, and did not perform well in the task of credible review classification.
Therefore, instead of relying on extensive user history, we use simple proxies for user activity that are easier to aggregate from the community:

\begin{itemize}
\item {\bf User Posts:} number of posts written by the user in the community.
\item {\bf Review Length:} length of the reviews --- longer reviews tend to frequently go off-topic with high emotional digression.
\item {\bf User Rating Behavior:} absolute deviation of the review rating from the mean and median rating of the user to other items, as well as the first three moments of the user rating distribution 
--- capturing the scenario where the user has a {\em typical rating behavior} across all items.
\item {\bf Item Rating Pattern:} absolute deviation of the item rating from the mean and median rating obtained from other users captures the extent to which the user disagrees with other users about the item quality; the first three moments of the item rating distribution captures the general item rating pattern.
\item {\bf User Friends:} number of friends of the user.
\item {\bf User Check-in:} if the user checked-in the hotel --- first hand experience of the user adds to the review credibility.
\item {\bf Elite:} elite status of the user in the community.
\item {\bf Review helpfulness:} number of helpfulness votes received by the user post --- captures the quality of user postings.
\end{itemize}

Note that user rating behavior and item rating pattern are also captured {\em implicitly} using the consistency features in the latent facet model. 

Also, note that some of these consistency features are also used in the earlier task on detecting helpful product reviews.

%We test various statistical measures for measuring deviation in ratings, where the Median Absolute Deviation performs the best. One possible reason is that it is robust, captures typical rating behavior, and is not affected by outliers (common in this task) like arithmetic mean.

Since our aim is to detect credible reviews in the case of limited information, we further split the above activity or behavioral features into two components: (a) $Activity^{-}$ using features $[1-4]$ 
that can be straightforward obtained from the tuple $\langle userId, itemId, review, rating \rangle$ and are easily available even for ``long-tail'' items and newcomers; and (b) $Activity^{+}$ using all 
the listed features. However the latter requires additional information (features $[5-8]$) that might not always be available, or takes long time to aggregate for new items/users.

\noindent{\bf Behavioral feature vector construction}: For each review $d_j$ by user $u_k$, we construct a behavioral feature vector $\langle F^B(d_j) \rangle$ using the above features.

%\subsection{Applicability Tasks}
\subsection{Tasks}

%\subsubsection{Credible Review Classification:}

\subsubsection{Credible Review Classification}
In the first task, we {\em classify} reviews as {\em credible} or not. For each review $d_j$ by user $u_k$, we construct the joint feature vector $F(d_j) = F^L(d_j) \cup F^T(d_j) \cup F^B(d_j)$, and 
use Support Vector Machines (SVM)~\cite{Cortes1995} for classification of the reviews. 
%SVM maps the examples (using Kernels) to a high dimensional space, and constructs a hyperplane to 
%separate the two categories of examples. Although there can be an infinite number of such hyperplanes possible, SVM constructs the one with the largest %functional margin given by the distance of the 
%nearest point to the hyperplane on each side of it. New points are mapped to the same space and classified to a category based on which side of the hyperplane it lies. We use a linear 
%kernel %(seen to perform better than other kernels like polynomial, radial basis, and sigmoid) 
%which has been shown to perform the best for text classification tasks. 

We use the $L_2$ regularized $L_2$ loss SVM with dual formulation from the LibLinear package\footnote{\url {csie.ntu.edu.tw/cjlin/liblinear}}~\cite{LibLinear} with other default parameters. We report classification accuracy 
with $10$-fold cross-validation on ground-truth from TripAdvisor and Yelp.

\subsubsection{Item Ranking and Evaluation Measures}
%\noindent {\bf Item Ranking:}
Due to the scarcity of ground-truth data pertaining to review credibility, a more suitable way to evaluate our model is to examine the {\em effect} of non-credible reviews on the relative 
{\em ranking} of items in the community. For instance, in case of popular items with large number of reviews, even if a fraction of it were non-credible, its effect would not be so severe 
as would be on ``long-tail'' items with fewer reviews. 

A simple way to find the ``goodness'' of an item is to aggregate ratings of all reviews -- using which we also obtain a ranking of items. We use our model to filter 
out non-credible reviews, aggregate ratings of credible reviews, and re-compute the item ranks.

\noindent {\em \bf Evaluation Measures} --
We use the {\em Kendall-Tau Rank Correlation Co-efficient} ($\tau$) to find effectiveness of the rankings, against a {\em reference ranking} --- for instance, the {\em sales rank} of items 
in Amazon.
$\tau$ measures the number of concordant and discordant pairs, to find whether the ranks of two elements agree or not based on their scores, out of the total number of combinations possible. 
Given a set of observations $\{x, y\}$, any pair of observations $(x_i, y_i)$ and $(x_j, y_j)$, where $i \not= j$, are said to be {\em concordant} if either $x_i > x_j$ and $y_i > y_j$, or 
$x_i < x_j$ and $y_i < y_j$, and {\em discordant} otherwise. If $x_i=x_j$ or $y_i=y_j$, the ranks are tied --- neither discordant, nor concordant.

We use the {\em Kendall-Tau-B} measure ($\tau_b$) which allows for rank adjustment. Consider $n_c$, $n_d$, $t_x$, and $t_y$ to be the number of concordant, discordant, tied pairs on $x$, and tied 
pairs on $y$ respectively, whereby Kendall-Tau-B is given by: $\frac{n_c-n_d}{\sqrt{(n_c + n_d + t_x)(n_c + n_d + t_y)}}$. 

However, this is a conservative estimate as multiple items --- typically the top-selling ones in Amazon --- have the same rating (say, $5$). Therefore, we use a second estimate 
(say, {\em Kendall-Tau-M} ($\tau_m$)) which considers non-zero tied ranks to be concordant. Note that, an item can have a zero-rank if all of its reviews are classified as non-credible. A high positive (or, negative) value of Kendall-Tau indicates the two series are positively (or, negatively) correlated; whereas a value close to zero indicates they are independent. 

\subsubsection{Domain Transfer from Yelp to Amazon}

A typical issue in credibility analysis task is the scarcity of labeled training data. In the first task, we use labels from the Yelp Spam Filter (considered to be the industry standard) to train our model. However, such ground-truth labels are not available in Amazon. Although, in principle, we can train a model $M_\text{Yelp}$ on Yelp, and use it to filter out non-credible reviews in Amazon.

Transferring the learned model from Yelp to Amazon (or other domains) entails using the learned weights of {\em features} in Yelp that are analogous to the ones in Amazon. However, this process 
encounters the following issues:

\begin{itemize}
\item Facet distribution of Yelp (food and restaurants) is different from that of Amazon (products such as software, and consumer electronics). Therefore, the facet-label distribution and the 
corresponding learned feature weights from Yelp cannot be directly used, as the latent dimensions are different. 
\item Additionally, specific metadata like check-in, user-friends, and elite-status are missing in Amazon.
\end{itemize}

However, the learned weights for the following features can still be directly used:
\begin{itemize}
\item Certain unigrams and bigrams, especially those depicting opinion, that occur in both domains.
\item Behavioral features like user and item rating patterns, review count and length, and usefulness votes.
\item Deviation features derived from {\em Amazon-specific} facet-label distribution that is obtained using the JST model on Amazon corpus:
\begin{itemize}
\item Deviation (with dimension $L$) of the user assigned rating from that inferred from review content.
\item Distribution (with dimension $L$) of positive and negative sentiment as expressed in the review.
\item Divergence, as a unary feature, of the facet-label distribution in the review from the aggregated distribution over other reviews on a given item.
\item Burstiness, as a unary feature, of the review.
\end{itemize}
\end{itemize}

Using the above components, that are common to both Yelp and Amazon, we {\em first} re-train the model $M_{\text{Yelp}}$ from Yelp to remove the non-contributing features for Amazon. 

Now, a direct transfer of the model weights from Yelp to Amazon assumes the distribution of credible to non-credible reviews, and corresponding feature importance, to be the same in both domains --- 
which is not necessarily true. In order to boost certain features to better identify non-credible reviews in Amazon, we tune the {\em soft margin parameter} $C$ in the SVM. $C^{+}$ and $C^{-}$ are regularization parameters for positive and negative class (credible and deceptive), respectively. We use 
{\em C-SVM}~\cite{chen04}, with slack variables, that optimizes:

\begin{align*}
min_{\vec{w}, b, \xi_i \ge 0} \frac{1}{2}\vec{w}^T\vec{w}+C^{+}\sum_{y_i=+1}\xi_i+C^{-}\sum_{y_i=-1}\xi_i \\
\text{subject\ to}\ \forall \{(\vec{x_i}, y_i)\}, y_i(\vec{w}^{T}\vec{x}_i + b) \ge 1- \xi_i
\end{align*}

The parameters $\{C\}$ provide a trade off as to how wide the margin can be made by moving around 
certain points which incurs a penalty of $\{C\xi_i\}$. A high value of $C^{-}$, for instance, places a large penalty for mis-classifying instances from the negative class, and therefore 
boosts certain features from that class. As the value of $C^{-}$ increases, the model starts classifying more reviews as non-credible. In the worse case, all the reviews of an item are classified 
as non-credible, leading to the aggregated item rating being zero.

\begin{figure}[t]
\centering
\includegraphics[width=0.7\linewidth, height=5cm]{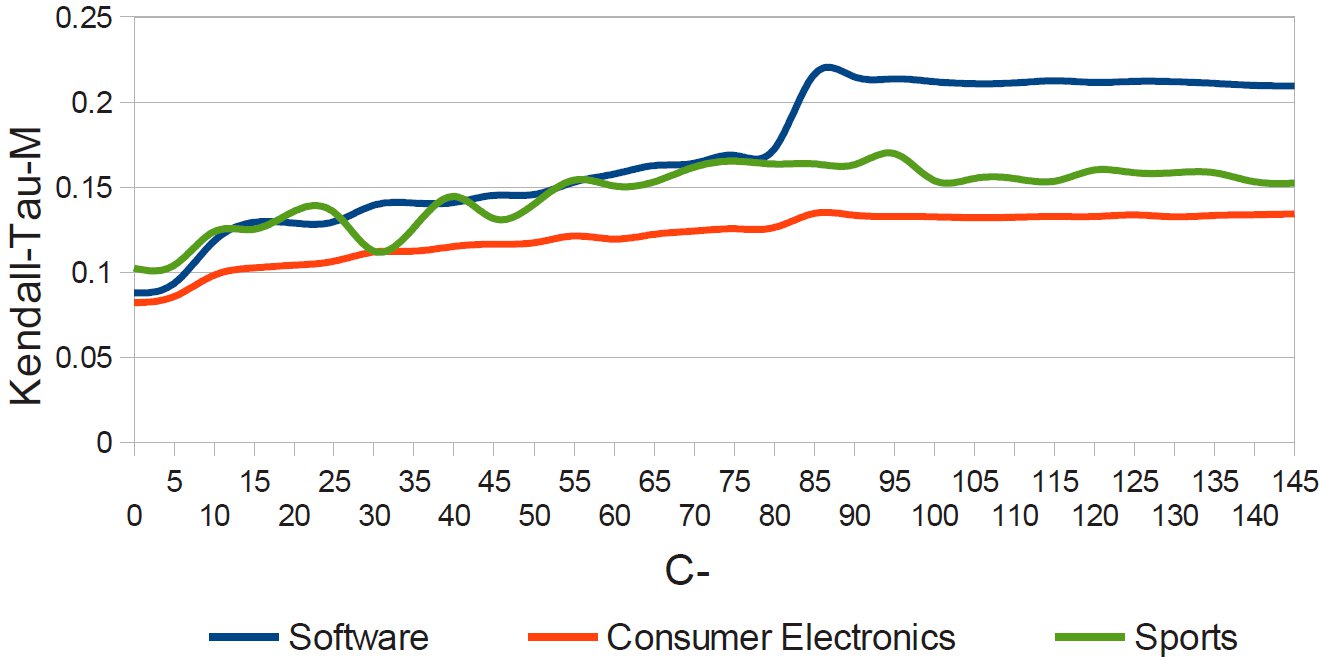}
\caption{Variation of Kendall-Tau-M ($\tau_m$) on different Amazon domains with parameter $C^{-}$ variation (using model $M_\text{Yelp}$ trained in Yelp and tested in Amazon).}
\vspace{-3em}
\label{fig:parameter}
\end{figure}

We use $\tau_m$ to find the optimal value of $C^{-}$ by varying it in the interval $C^{-} \in \{0, 5, 10, 15, ... 150\}$ using a {\em validation set} from Amazon as shown in 
Figure~\ref{fig:parameter}. We observe that as $C^{-}$ increases, $\tau_m$ also increases till a certain point as more and more non-credible reviews are filtered out, after which it stabilizes.

\begin{table}[t]
\centering
\begin{tabular}{c|l}
\toprule
{\bf Notation} & {\bf Description} \\ \midrule
%$\langle U \rangle, \langle D \rangle, \langle I \rangle$ & set of users, reviews, and items resp. \\
$U, D, I$ & set of users, reviews, and items resp. \\
$d, r_d$ & review text and associated rating \\
$V, f$ & unigrams and bigrams vocab. \& token types\\
$w_{ij}$ & word of token type $f_i$ in review $d_j$ \\
$I(\cdot)$ & indicator fn. for presence/absence of words \\
$z, l$ & set of facets and sentiment labels resp. \\
$K, L$ & cardinality of facets and sentiment labels \\
$\theta_d(z,l)$ & multinom. prob. distr. of facet $z$ \\
& $\quad$ with sentiment label $l$ in document $d$ \\
$\phi_{z,l}(w)$ & multinom. prob. distr. of word $w$ belonging \\
& $\quad$ to facet $z$ with sentiment label $l$ \\
$\Phi', \Phi''$ & facet-label distr. of review and item resp.\\
$\alpha, \beta, \gamma$ & Dirichlet priors \\
$\pi, \pi'$ & review rating distr. \& inferred rating distr. \\
$n(\cdot)$ & word count in reviews \\
$F^x(d_j)$ & feature vec. of review $d_j$ using lang. (x=L), \\
& $\quad$ consistency (x=T), and behavior (x=B) \\
$C^+, C^-$ & C-SVM regularization parameters \\
\bottomrule
\end{tabular}
\caption{List of variables and notations used with corresponding description.}
\label{tab:notation}
\end{table}

\subsubsection{Ranking SVM}

Our previous approach uses the model $M_\text{Yelp}$ trained on Yelp, with the reference ranking (i.e., sales ranking) in Amazon being used only for evaluating the item ranking using the Kendall-Tau measure. 
As the objective is to obtain a good item ranking based on credible reviews, we can have a model $M_\text{Amazon}$ that directly optimizes for Kendall-Tau using the reference ranking as training 
labels. This allows us to use the entire feature space available in Amazon, including the explicit facet-label distribution and the full vocabulary, which could not be used earlier. The feature 
space is constructed similarly to that of Yelp.

The goal of Ranking SVM~\cite{joachims02} is to learn a ranking function which is concordant with a given ordering of items. The objective is to learn $\vec{w}$ such that $\vec{w}\cdot \vec{x_i} > \vec{w} \cdot \vec{x_j}$ for most data pairs $\{(\vec{x_i}, \vec{x_j}): y_i > y_j \in R\}$. Although the problem is known to be NP-hard, it is approximated using SVM techniques with pairwise slack variables $\xi_{i,j}$. The optimization problem is equivalent to that of classifying SVM, but now operating on {\em pairwise difference vectors} $(\vec{x_i} - \vec{x_j})$ with corresponding labels $+1 /-1$ indicating which one should be ranked ahead. We use the implementation\footnote{\url{https://www.cs.cornell.edu/people/tj/svm\_light/svm\_rank.html}} of~\cite{joachims02} that maximizes the empirical Kendall-Tau by minimizing the number of discordant pairs.

Unlike the classification task, where labels are {\em per-review}, the ranking task requires labels {\em per-item}. Consider $\langle f_{i,j, k} \rangle$ to be the feature vector for the $j^{th}$ review of an item $i$, with $k$ indexing an element of the feature vector. We aggregate these feature vectors element-wise over all the reviews on item $i$ to obtain its feature vector 
$\langle \frac{\sum_j f_{i,j, k}}{\sum_j \mathbf{1}} \rangle$.

%% file: main/chapter-credible-applications/consistency-analysis/experiments.tex
\subsection{Experiments}

\subsubsection{Setup and Data} 

\noindent {\bf Parameter initialization:} The sentiment lexicon from~\cite{Hu2004} consisting of $2006$ positive and $4783$ negative polarity bearing words is used to initialize the 
review text based facet-label-word tensor $\Phi$ prior to inference. We consider the number of topics, $K=20$ for Yelp, and $K=50$ for Amazon with the review sentiment labels $L=\{+1, -1\}$ (corresponding to positive and negative rated reviews) initialized randomly. 

The symmetric Dirichlet priors are set to $\alpha = 50/K$, $\beta = 0.01$, and $\gamma = 0.1$. 
%We consider a supervised version of the Joint Sentiment Topic Model where the distribution $\pi_{d,l}$ denoting the probability of review $d$ having sentiment label $l$ is known apriori. For a review with positive sentiment, we set $\pi_{d,+1}=1-\gamma$ and $\pi_{d,-1}=\gamma$, where the smoothness hyper-parameter $\gamma$ is set to a small value ($0.10$). In case of negative sentiment, the corresponding distribution is $\{\pi_{d,-1}=1-\gamma,\ \pi_{d,+1}=\gamma\}$.

\noindent{\bf Datasets and Ground-Truth:} In this work, we consider the following datasets (refer to Table~\ref{tab:data1} and~\ref{tab:data2}) with available ground-truth information.

\begin{table}
\centering
\small
\begin{tabular}{p{2cm}p{4cm}p{3cm}p{1cm}p{1.5cm}}
\toprule
\bf{Dataset} & \bf{Non-Credible Reviews} & \bf{Credible Reviews} & \bf{Items} & \bf{Users}\\
\midrule
TripAdvisor & 800 & 800 & 20 & -\\
Yelp & 5169 & 37,500 & 273 & 24,769\\
Yelp$^{*}$ & 5169 & 5169 & 151 & 7898\\
\bottomrule
\end{tabular}
\caption{Dataset statistics for review classification (Yelp$^{*}$ denotes balanced dataset using random sampling).}
\label{tab:data1}
\end{table}

\begin{table}
\small
\begin{tabular}{lrr|rrrrrrr}
\toprule
\textbf{Domain} & \multicolumn{1}{l}{\textbf{\#Users}} & \multicolumn{1}{l|}{\textbf{\#Reviews}} & \multicolumn{ 7}{c}{\textbf{\#Items with reviews per-item}} \\\midrule 
 & \multicolumn{1}{l}{} & \multicolumn{1}{l|}{} & \multicolumn{1}{l}{$\leq$5} & \multicolumn{1}{l}{$\leq$10} & \multicolumn{1}{l}{$\leq$20} & \multicolumn{1}{l}{$\leq$30} & \multicolumn{1}{l}{$\leq$40} & \multicolumn{1}{l}{$\leq$50} & \multicolumn{1}{l}{Total} \\\midrule
%{\bf Baby Products} & 22,676 & 31,913 & 562 & 708 & 818 & 868 & 908 & 937 & 1,072 \\ 
{\bf Electronics} & 94,664 & 121,234 & 14,797 & 16,963 & 18,350 & 18,829 & 19,053 & 19,187 & 19,518 \\ 
%{\bf Musical Instruments} & 87 & 94 & 60 & 60 & 60 & 60 & 60 & 60 & 60 \\ 
{\bf Software} & 21,825 & 26,767 & 3,814 & 4,354 & 4,668 & 4,767 & 4,807 & 4,828 & 4,889 \\ 
{\bf Sports} & 656 & 695 & 202 & 226 & 233 & 235 & 235 & 235 & 235 \\ 
\bottomrule
\end{tabular}
\caption{Amazon dataset statistics for item ranking, with cumulative \#items and varying \#reviews.}
\label{tab:data2}
\end{table}

\noindent $\bullet$ The {\em TripAdvisor Dataset}~\cite{Ott2011,Ott2013} contains reviews on $20$ most popular Chicago hotels. The data consists of $1600$  reviews with positive ($5$ star) and negative ($1$ star) sentiment, with $20$ credible and $20$ non-credible reviews on each of the hotels. The authors crawled the {\em credible} reviews from online review portals like TripAdvisor; whereas the {\em non-credible} ones were generated by users in Amazon Mechanical Turk. The dataset has only the review text and sentiment label (positive/negative ratings) with corresponding hotel names, with no other information on users or items. 

\noindent $\bullet$ The {\em Yelp Dataset} contains reviews on $273$ restaurants in Chicago. The data consists of $37.5K$ recommended (i.e., {\em credible}) reviews, and $5K$ non-recommended (i.e., {\em non-credible}) reviews given by the Yelp filtering algorithm. The annotated labels (recommended, or not-recommended) for the reviews by the Yelp filter are considered as ground-truth in our work. \cite{Liu2013a} found that the Yelp spam filter primarily relies on linguistic, behavioral, and social networking features. Additionally, we extract the following information for each review: $\langle userId, itemId$ $, timestamp, rating, review, metadata \rangle$. The meta-data consists of some user activity information as outlined in Section~\ref{subsec:lang}. 

%The reviews marked as ``not recommended'' by the Yelp spam filter are considered to be the ground-truth for comparing the accuracy for credible review detection for our 
%proposed model. 
%\todo{talk about the Yelp ground truth about Yelp spam filter in short}

\noindent $\bullet$ The {\em Amazon Dataset} used in~\cite{Liu2008} consists of around $149K$ reviews from $117K$ users on $25K$ items from three domains, namely, Consumer Electronics, Software, and Sports. For each review, we gather the same information tuple as that from Yelp. However, the metadata in this dataset is not as rich as in Yelp, consisting only of helpfulness votes of the reviews. 

Further, there exists no explicit ground-truth characterizing the reviews as credible or deceptive in Amazon. To this end, we re-rank the items using our approaches, filtering 
out possible deceptive reviews (based on the feature vectors), and then compare the ranking to the {\em item sales rank} considered as the pseudo ground-truth.
%\todo{talk about the Amazon ground truth for item sales rank in short}

\subsubsection{Baselines} 

We use the following state-of-the-art baselines (given the full set of features that fit with their model) for comparison with our proposed model. 

\noindent{\em (1) Language Model Baselines:} We consider the unigram and bigram language model baselines from~\cite{Ott2011,Ott2013} that have been shown to outperform other baselines using psycholinguistic 
features, part-of-speech tags, information gain, etc. We take the best baseline from their work which is a combination of unigrams and bigrams. Our proposed model (N-gram+Facet) enriches it by using 
length normalization, presence or absence of features, latent facets, etc. 
The recently proposed {\em doc-to-vec} model based on Neural Networks, overcomes the weakness of bag-of-words models by taking the context of words into account, and learns a dense vector representation for each document~\cite{doc2vec}. We train the doc-to-vec model in our dataset as a baseline model.  
In addition, we also consider readability (ARI) and review sentiment scores~\cite{Hu2012} under the hypothesis that writing styles would be random because of diverse customer background. ARI measures the reader's ability to comprehend a text and is measured as a function of 
the total number of characters, words, and sentences present, while review sentiment tries to capture the fraction of occurrences of positive/negative sentiment words to the total 
number of such words used.
%Results in Table~\ref{tab:classification} exhibit that our proposed model performs better than the competing baselines on both the TripAdvisor data (containing only review text) and Yelp data.

\noindent{\em (2) Activity \& Rating Baselines:} We extract all activity, rating and behavioral features of users as proposed in~\cite{Liu2007,Liu2008,Liu2010,Liu2011,Liu2012,Liu2013,Liu2013a,Liu2014} from the tuple $\langle userId, itemId, rating, review,$ $metadata \rangle$ in the Yelp dataset. Specifically, we utilize the number of helpful feedbacks, review title 
length, review rating, use of brand names, percent of positive and negative sentiments, average rating, and rating deviation as features for classification. Further, based on the recent 
work of~\cite{Rahman2015}, we also use the user check-in and user elite status information as additional features for comparison.

%{\tt To Do ... give a short description on the different features based on ARI, Ratio of positive and negative words, review length etc. Split the baselines into the works by Bing Liu's group aggregating all their features in a single model, Rahman's work, and the work on ARI-ReviewLengthSentimentProportion as 3 or more different baselines.}

\subsubsection{Quantitative Analysis} 

Our experimental setup considers the following evaluations:

\noindent {\em (1) Credible review classification:} We study the performance of the various approaches in distinguishing a {\em credible} review from a {\em non-credible} one. Since this 
forms a binary classification task, we consider a balanced dataset containing equal proportion of data from each of the two classes. On the Yelp dataset, for each item we randomly sample an equal number 
of credible and non-credible reviews (to obtain Yelp$^*$); while the TripAdvisor dataset is already balanced. Table~\ref{tab:classification} shows the $10$-fold cross validation accuracy results for the different models on the two datasets. We observe that our proposed {\em consistency and behavioral features} exhibit around $15\%$ improvement in Yelp$^{*}$ for classification accuracy over the best performing baselines (refer to Table~\ref{tab:classification}). Since the TripAdvisor dataset has {\em only} review text, the user/activity models could {\em not} be used there. This experiment could not be performed on Amazon as well, as the ground-truth for credibility labels of reviews is absent.

\begin{table}
\centering
\small
\begin{tabular}{p{3cm}p{5cm}p{2cm}p{1cm}}
\toprule
\bf{Models} & \bf{Features} & \bf{TripAdvisor} & \bf{Yelp$^{*}$} \\
\midrule
\multirow{2}{*}{\bf Deep Learning} & Doc2Vec & 69.56 & 64.84 \\
& Doc2Vec + ARI + Sentiment & 76.62 & 65.01 \\
\multirow{2}{*}{\bf Activity \& Rating} & Activity+Rating & - & 74.68 \\
& Activity+Rating+Elite+Check-in & - & 79.43 \\
\midrule
\multirow{2}{*}{\bf Language} & Unigram + Bigram& 88.37 & 73.63\\
& Consistency & 80.12 & 76.5\\
\midrule
\multirow{2}{*}{\bf Behavioral} & Activity Model$^{-}$ & - & 80.24\\
& Activity Model$^{+}$ & - & 86.35\\
\midrule
\multirow{6}{*}{\bf Aggregated} & N-gram + Consistency & {\bf 89.25} & 79.72\\
& N-gram + Activity$^{-}$ & - & 82.84\\
& N-gram + Activity$^{+}$ & - & 88.44\\
& N-gram + Consistency + Activity$^{-}$ & - & 86.58\\
& N-gram + Consistency + Activity$^{+}$ & - & {\bf 91.09}\\
& $M_\text{Yelp}$  & - & 89.87\\
\bottomrule
\end{tabular}
\caption{Credible review classification accuracy with $10$-fold cross validation. TripAdvisor dataset contains only review texts and no user/activity information.}
\label{tab:classification}
\end{table}

\noindent{\em (2) Item Ranking:} In this task we examine the effect of non-credible reviews on the ranking of items in the community. This experiment is performed {\em only} on Amazon using 
the item {\em sales rank} as ground or reference ranking, as Yelp does not provide such item rankings. The sales rank provides an indication as to how well a product is selling on 
Amazon.com and highlights the item's rank in the corresponding category\footnote{\url{www.amazon.com/gp/help/customer/display.html?nodeId=525376}}.

The baseline for the item ranking is based on the aggregated rating of all reviews on an item. The first model $M_\text{Yelp}$ (C-SVM) trained on Yelp filters out the non-credible reviews, before 
aggregating review ratings on an item. The second model $M_\text{Amazon}$ (SVM-Rank) is trained on Amazon using SVM-Rank with the reference ranking as training labels. $10$-fold cross-validation results 
are reported on the two measures of Kendall-Tau ($\tau_b$ and $\tau_m$) in Table~\ref{tab:model-ranking} with respect to the reference ranking. $\tau_b$ and $\tau_m$ for SVM-Rank are the same since there are no ties. Our first model performs substantially better than the baseline, which, in turn, is outperformed by our second model.

In order to find the effectiveness of our approach in dealing with ``long-tail'' items, we perform an additional experiment with our best performing model i.e., $M_\text{Amazon}$ (SVM-Rank). 

We use the model to find Kendall-Tau-M ($\tau_m$) rank correlation (with the reference ranking) of items having less than (or equal to) $5, 10, 20, 30, 40,$ and $50$ reviews in different domains in Amazon (results reported in Table~\ref{tab:long-tail} with $10$-fold cross validation). We observe that our model performs substantially well even with items having as few as {\em five} reviews, with the performance progressively getting better with more reviews per-item.

\begin{table}[t]
\small
\centering
\begin{tabular}{l|p{1.5cm}c|p{1.5cm}p{2.2cm}|p{2.2cm}}
\toprule
\textbf{Domain} & \multicolumn{2}{c|}{\textbf{Kendall-Tau-B ($\tau_b$)}} & \multicolumn{2}{c|}{\textbf{Kendall-Tau-M ($\tau_m$)}} & \multicolumn{1}{c}{\textbf{Kendall-Tau ($\tau_b = \tau_m$)}}\\ 
\midrule
 & \multicolumn{1}{l}{\textbf{Baseline}} & \multicolumn{1}{l|}{\textbf{$M_\text{Yelp}$ (C-SVM)}} & \multicolumn{1}{l}{\textbf{Baseline}} & \multicolumn{1}{l|}{\textbf{$M_\text{Yelp}$ (C-SVM)}} & \multicolumn{1}{l}{\textbf{$M_\text{Amazon}$ (SVM-Rank)}} \\ 
 \midrule
%Baby & 0.026 & 0.179 & 0.093 & 0.204 & 0.382 \\
CE & 0.011 & 0.109 & 0.082 & 0.135 & 0.329 \\ 
Software & 0.007 & 0.184 & 0.088 & 0.216 & 0.426 \\ 
Sports & 0.021 & 0.155 & 0.102 & 0.170 & 0.325 \\ 
\bottomrule
\end{tabular}
\caption{\hspace{0.4em}Kendall-Tau correlation of different models across domains.}
\label{tab:model-ranking}
\end{table}

\begin{table}
\centering
\small
\begin{tabular}{p{1.5cm}p{1.2cm}p{1.2cm}p{1.2cm}p{1.2cm}p{1.2cm}p{1.2cm}p{1.2cm}}
\toprule
\textbf{Domain} & \multicolumn{7}{c}{\textbf{$\tau_m$ with \#reviews per-item}} \\ 
\midrule
 & $\leq$5 & $\leq$10 & $\leq$20 & \multicolumn{1}{l}{$\leq$30} & \multicolumn{1}{l}{$\leq$40} & \multicolumn{1}{l}{$\leq$50} & \multicolumn{1}{l}{Overall} \\ \midrule
%Baby & - & - & - & 0.017 & 0.047 & 0.073 & 0.204 \\ 
%\textbf{$M_\text{Yelp}$} & CE & - & \multicolumn{1}{l}{0.020} & \multicolumn{1}{l}{0.069} & 0.092 & 0.104 & 0.112 & 0.135 \\ 
%\textbf{(C-SVM)} & Software & \multicolumn{1}{l}{0.039} & \multicolumn{1}{l}{0.102} & \multicolumn{1}{l}{0.161} & 0.184 & 0.195 & 0.200 & 0.216 \\ 
%& Sports & \multicolumn{1}{l}{0.056} & \multicolumn{1}{l}{0.122} & \multicolumn{1}{l}{0.155} & 0.170 & 0.170 & 0.170 & 0.170 \\ \midrule
CE & 0.218 & \multicolumn{1}{l}{0.257} & \multicolumn{1}{l}{0.290} & 0.304 & 0.312 & 0.317 & 0.329 \\ 
Software & \multicolumn{1}{l}{0.353} & \multicolumn{1}{l}{0.375} & \multicolumn{1}{l}{0.401} & 0.411 & 0.417 & 0.419 & 0.426 \\ 
Sports & \multicolumn{1}{l}{0.273} & \multicolumn{1}{l}{0.324} & \multicolumn{1}{l}{0.310} & \multicolumn{1}{l}{0.325} & \multicolumn{1}{l}{0.325} & \multicolumn{1}{l}{0.325} & \multicolumn{1}{l}{0.325} \\
\bottomrule
\end{tabular}
\caption{\hspace{0.4em}Variation of Kendall-Tau-M ($\tau_m$) correlation with \#reviews with \textbf{$M_\text{Amazon}$ (SVM-Rank)}.}
\label{tab:long-tail}
\end{table}

\begin{table}[t]
\centering
\small
\begin{tabular}{p{7cm}p{6.8cm}}
\toprule
\textbf{Credible Reviews} & \textbf{Non-Credible Reviews}\\\midrule
not, also, really, just, like, get, perfect, little, good, one, space, pretty, can, everything, come\_back, still, us, right, definitely, enough, much, super, free, around, delicious, no, fresh, big, favorite, lot, selection, sure, friendly, way, dish, since, huge, etc, menu, large, easy, last, room, guests, find, location, time, probably, helpful, great, now, something, two, nice, small, better, sweet, though, loved, happy, love, anything, actually, home & dirty, mediocre, charged, customer\_service, signature\_lounge, view\_city, nice\_place, hotel\_staff, good\_service, never\_go, overpriced, several\_times, wait\_staff, signature\_room, outstanding, establishment, architecture\_foundation, will\_not, long, waste, food\_great, glamour\_closet, glamour, food\_service,  love\_place, terrible, great\_place, wonderful, atmosphere, bill, will\_never, good\_food, management, great\_food, money, worst, horrible, manager, service, rude \\
\bottomrule
\end{tabular}
\caption{\hspace{0.4em}Top n-grams (by feature weights) for credibility classification.}
\label{tab:ngrams}
\end{table}

%% file: main/chapter-credible-applications/consistency-analysis/discussions.tex
\subsubsection{Qualitative Analysis}

{\noindent \bf Language Model:} The bigram language model performs very well (refer to Table~\ref{tab:classification}) on the TripAdvisor dataset due to  its artificial creation. Workers in Amazon Mechanical Turk were asked to study all the hotel amenities in their websites, and then write fake reviews about them. As a result, the reviews closely follow the actual hotel descriptions, and, therefore it is quite difficult for the facet model to find contradictions or mismatch in facet descriptions. Consequently, the facet model gives marginal improvement when combined with the language model.

However, the bigram language model and doc-to-vec do not perform so well on the real-world, and naturally noisy Yelp dataset, as they do in the previous one. The facet model also does not perform well in isolation. However, all the components put together give significant performance improvement over the ones in isolation (around $8\%$).

Incorporating writing style using ARI and sentiment measures improves performance of doc-to-vec in the TripAdvisor dataset. However, the improvements are not significant in the real-world Yelp data.\\

We rank all the features in the {\em joint model} for credibility classification by their weights --- as given by the C-SVM --- and show a snapshot of the top unigrams and bigrams in Table~\ref{tab:ngrams}. We observe that credible reviews mostly contain a mix of function and content words, balanced opinions, and a lot of informative unigrams. Non-credible reviews, on the other hand, contain extreme opinions, less function words, and more of sophisticated content words, like, a lot of signature bigrams, to catch the readers' attention. 

{\noindent \bf Behavioral Model:} We find the activity based model to perform the best in isolation (refer to Table~\ref{tab:classification}). Combined with language and consistency features, the joint model exhibits around $5\%$ improvement in performance. Additional meta-data like the user elite and check-in status improves the performance of activity based baselines, which are not typically available for newcomers in the community. Our model using limited information ({\em N-gram+Consistency+Activity$^{-}$}) performs better than the activity baselines using fine-grained information about items (like brand description) and user history. Incorporating additional user features ({\em Activity$^{+}$}) further boosts its performance.

{\noindent \bf Consistency Features:} We perform ablation tests (refer to Table~\ref{tab:classification}) to find the effectiveness of the facet based consistency features. We remove the consistency model from the aggregated model, and see significant performance degradation of $3-4\%$ for the Yelp$^{*}$ dataset. In the TripAdvisor dataset the performance reduction is less compared to Yelp due to reasons outlined before.

%We also find the facet based model inefficient in isolation. It works well in conjunction with other features, and improves the overall model performance.

Table~\ref{tab:revcon} shows a snapshot of the non-credible reviews, with corresponding (in)consistency features in Yelp and Amazon. We observe inconsistencies like: ratings of deceptive reviews not corroborating with the textual description, irrelevant facets influencing the rating of the target item, contradictions between users, expressing extreme opinions without explanation, depicting temporal ``burst'' in ratings, etc. In principle, these features can also be used to detect other anomalous phenomena like group-spamming (one of the principal indicators of which is temporal burst), which is out of scope of this work.

%Table~\ref{tab:model-ranking} shows the Kendall-Tau correlation ($\tau_b$ and $\tau_m$) of different models for the ranking task in Amazon. 
{\noindent \bf Ranking Task:} For the ranking task in Amazon (refer to Table~\ref{tab:model-ranking}), the first model $M_\text{Yelp}$ --- trained on Yelp and tested on Amazon using C-SVM --- performs much better than the baseline exploiting various consistency features. The second model $M_\text{Amazon}$ --- trained on Amazon using SVM-Rank --- outperforms the former exploiting the power of the entire feature space and domain-specific proxy labels unavailable to the former.

{\noindent \bf ``Long-Tail'' Items:} Table~\ref{tab:long-tail} shows the gradual degradation in performance of the second model $M_\text{Amazon}$ (SVM-Rank) in dealing with items with lesser number of reviews. Nevertheless, we observe it to give a substantial Kendall-Tau correlation ($\tau_m$) with the reference ranking, with as few as {\em five} reviews per-item, demonstrating the effectiveness of our model in dealing with ``long-tail'' items.

%Even though only a small proportion of the reviews per-item are non-credible in Yelp, we show that an efficient filtering algorithm can still improve the overall item ranking (Table~\ref{tab:KendallTau}) in the community.

\begin{table}[t]
\centering
\small
\begin{tabular}{p{2.4cm}|p{4.5cm}|p{5.5cm}}
\toprule
{\bf Inconsistency Features} & {\bf Yelp} Review \& [Rating] & {\bf Amazon} Review \& [Rating] \\
\midrule
{\bf user review -- rating} {\em (promotion/demotion)}:  & 
\underline{never been inside James.} \underline{never checked in.} \underline{never visited bar.} yet, one of my favorite hotels in Chicago. James has dog friendly area. my dog loves it there. [5] & 
Excellant product-alarm zone, technical support is almost non-existent because of this \underline{i will look to another product.} \underline{this is unacceptible}. [4]\\
{\bf user review -- facet description} {\em (irrelevant)}: &
you will learn that they are actually \underline{EVANGELICAL CHRISTIANS} working to proselytize the coffee farmers they buy from. [2] &
DO NOT BUY THIS. I used turbo tax since 2003, it never let me down until now. I can't file because Turbo Tax doesn't have software updates from the IRS \underline{``because of Hurricane Katrina''}. [1] \\
{\bf user review -- item description} {\em (deviation from community)}: & 
{internet is charged} in a $300$ dollar hotel! [3] &
The book Amazon offers is a joke! All it provides is the forward which is not written by Kalanithi. I don't have any \underline{sample of} \underline{HIS writing} to know if it appeals. [1] \\
{\bf extreme user rating}: &
{GREAT!!!i give 5 stars!!!Keep it up.} [5] &
GREAT. This camera takes pictures. [1]\\
\multirow{4}{*}{\bf temporal bursts\footnotemark:} &
\multicolumn{2}{|l}{Dan's apartment was beautiful and a great downtown location... \underline{(3/14/2012)} [5]} \\
& \multicolumn{2}{|l}{I highly recommend working with Dan and NSRA... \underline{(3/14/2012)} [5]} \\
& \multicolumn{2}{|l}{Dan is super friendly, demonstrating that he was confident... \underline{(3/14/2012)} [5]} \\
& \multicolumn{2}{|l}{my condo listing with no activity, Dan really stepped in... \underline{(4/18/2012)} [5]} \\
\bottomrule
\end{tabular}
\caption{\hspace{0.4em}Snapshot of non-credible reviews (reproduced verbatim) with inconsistencies.}
\label{tab:revcon}
\end{table}

\footnotetext{These reviews have also been flagged by the Yelp Spam Filter as not-recommended (i.e., non-credible).}

%\todo{explain negative sign of correlation}

%% file: main/chapter-credible-applications/conclusions.tex
In this section, we apply the principles and methods developed earlier for credibility analysis for two tasks in product review communities.

For the first task, 
we propose an approach to predict helpful product reviews by exploiting the {\em joint interaction} between user expertise, writing style, timeliness, and review consistency using Hidden Markov Model -- Latent Dirichlet Allocation. Unlike prior works exploiting a variety of syntactic and domain-specific features, our model uses {\em only} the information of a user reviewing an item at an explicit timepoint to perform this task --- making our approach generalizable across all communities and domains. Additionally, we provide {\em interpretable explanation} as to why a review is helpful, in terms of salient words from latent word clusters --- that are used by experts to describe important facets of the item under consideration. 

Thereafter, for the second task, we harness various {\em (in)consistency features} from the latent facet models to analyze (in)consistencies between review description, facets, ratings, and timestamps to find credible product reviews with limited information. Additionally, these features help in providing interpretable explanations as to why a review has been deemed as non-credible. 

Our approach works well for ``long-tail'' items or newcomers in the community with limited prior information / history. We develop multiple models for domain transfer and adaptation, where our model performs very well in the ranking tasks involving ``long-tail'' items, with as few as {\em five} reviews per-item.

We perform extensive experiments on real-world reviews from different domains in Amazon (like books, movies, music, food, and electronics), Yelp and TripAdvisor that demonstrate the effectiveness of our approach over state-of-the-art baselines.

%% file: main/chapter-conclusions/main.tex
\chapter{Conclusions}
\label{chap:conclusions}

\section{Contributions}

The {\em first contribution} of this dissertation is to develop novel forms of probabilistic graphical models, namely, Conditional Random Fields (CRF), for credibility analysis in online communities. These models {\em jointly} leverage the context, structure, and interactions between sources, users, postings, and statements in online communities to ascertain the credibility of user-contributed information. They capture the complex interplay between several factors: the writing style (e.g., subjectivity and rationality, attitude and emotions), (latent) trustworthiness and (latent) expertise of users and sources, (latent) topics of postings, user-user and user-item interactions etc. 

We first develop a {\em semi-supervised CRF} model for {\em credibility classification} of postings and statements that is partially supervised by expert knowledge. We apply this framework to the {\em healthcare} domain to extract rare or unobserved side-effects of drugs from user-contributed postings in online healthforums. This is one of the problems where large-scale non-expert data 
has 
the potential to complement expert medical knowledge. Furthermore, we develop a {\em continuous CRF} model for fine-grained {\em credibility regression} in online communities to deal with user-assigned numeric ratings to items. We demonstrate its usefulness for {\em news communities} that are plagued with misinformation, bias, and polarization induced by the fairness and style of reporting, and political perspectives of media sources and users. We use the model to {\em jointly} identify objective news articles, trustworthy media sources, expert users and their credible postings.

The {\em second contribution} deals with the temporal evolution and dynamics of online communities where, users join and leave, adapt to evolving trends, and mature over time. We study this temporal evolution in a collaborative filtering framework to recommend items to users based on their experience or maturity to consume them. To this end, we develop two models for {\em experience evolution} of users in online communities. The first one models the users to evolve in a {\em discrete} manner employing Hidden Markov Model -- Latent Dirichlet Allocation that captures the change in writing style and vocabulary usage with change in users' (latent) experience level. The second one addresses several drawbacks of this discrete evolution with a natural and {\em continuous} evolution model of users' experience, and their corresponding language model employing Geometric Brownian Motion, Brownian Motion, and Latent Dirichlet Allocation. We, thereafter, develop efficient probabilistic inference techniques using 
Metropolis Hastings, Kalman Filter, and Gibbs Sampling that are empirically shown to smoothly and continuously increase data log-likelihood over time, as well as have a fast convergence. Experimentally, we show that such experience-aware user models can perform item recommendation better than other state-of-the-art algorithms in communities like beer, movies, food, and news. We also use this model to find {\em useful} product reviews that are helpful to the end-users in communities like Amazon.% --- by identifying experienced users, their preferences, and writing style.   

The {\em third contribution} is a method to perform credibility analysis with limited information, especially for ``long-tail'' items and users with limited history of activity information. We develop methods leveraging latent topic models that analyze inconsistencies between review texts, their ratings and facet descriptions, and temporal bursts to identify non-credible reviews. All these methods for product review communities operate only on the information of {\em a user reviewing an item at an explicit timepoint} --- making our approach generalizable across all communities and domains. We also propose approaches for domain transfer to deal with missing ground-truth information in one domain, by transferring learned models from other domains.

The {\em fourth contribution} deals with providing user-interpretable explanations from probabilistic graphical models that can be used to explain their verdict. To this end, we show (latent) distributional word clusters that demonstrate the usage of words by users with varying experience and trustworthiness, discourse and affective norms of credible vs. non-credible postings, evolution traces of how the users evolve over time and acquire community norms, etc.

\section{Outlook}

%This dissertation developed computational models to harness collective user intelligence, denoise and debias them, by jointly leveraging context, structure, and interactions in online communities. We used these models to perform a myriad of tasks ranging from credibility analysis (e.g., finding credible content, trustworthy sources and users, etc.), expert-finding, subjectivity analysis for sentiment analysis and opinion mining, collaborative filtering for item recommendation etc. 
Some future applications and extensions of our model to related tasks are the following.

The proposed models, especially, the ones for product review communities --- operating only on user-user, user-item, and item-item interactions --- are fairly generic in nature, and easily applicable to other communities and domains. For instance, these can be applied to Question-Answering forums (e.g., Quora) to find reliable and expert answers to queries, and experts one would want to follow for certain topics. These can also be used in other crowdsourcing applications to find reliability of user-contributed information. These models can also be used to analyze inconsistencies between credible and non-credible (i.e. abnormal) behavior to detect anomalies and frauds in networks and systems.

Our proposed continuous Conditional Random Field model --- for aggregating information from multiple users and sources (e.g., several weak learners or annotators) taking into account their expertise and interactions --- can be used for learning to rank and ensemble learning. For instance, these can be used in Amazon Mechanical Turk to assess annotator reliability, and gold answer for certain query types.

%Large-scale Knowledge Bases (KBs) like Yago~\cite{Yago} DBpedia~\cite{DbPedia}, and Freebase~\cite{Freebase} mostly harness information from {\em structured} data (e.g., Wikipedia infoboxes, categorization information, geo-coordinates, etc.). Additionally, they also require manual curation to maintain quality and consistency of the KB. Consequently, they have a high precision, but low coverage: whereby, they store information mostly about the prominent entities. On the contrary, crowd-sourced information, being noisy and unstructured, have a high coverage but low precision. 

%The proposed models can be used to 

Prior works on Knowledge Base (KB) construction (e.g., Yago \cite{Yago}, DBpedia \cite{DbPedia}, Freebase \cite{Freebase}) mostly leverage structured information like Wikipedia infoboxes, category information, etc. Additionally, they also require manual curation to maintain quality and consistency of the KB. Consequently, they have a high precision, but low coverage: whereby, they store information mostly about the prominent entities. On the contrary, crowd-sourced information, being noisy and unstructured, have a high coverage but low precision. To bring these together, our proposed models --- specifically, the semi-supervised Conditional Random Field model that learns from partial expert knowledge --- can be used to automatically construct KBs (and curate them) from large-scale, structured and unstructured Web content, and structured KBs. Recently, an approach for knowledge fusion using a similar approach has been proposed in \cite{DBLP:conf/kdd/0001GHHLMSSZ14}.

Many of the language features for capturing the subjectivity and rationality of information in user postings have been manually identified using bias and affective lexicons, discourse relations, etc. Due to the recent advances in representation learning and deep learning, and correspondence between graphical models and neural networks --- a natural extension of our work is to automatically learn these linguistic cues and patterns for credibility analysis from the joint embeddings of context and structure of communities using neural networks.% without requiring any hand-crafted features or lexicons.

Most of the prior works on truth-finding and data fusion operate over structured data. Although this dissertation relaxes many of these assumptions, it is mostly geared for online communities with user and item interactions. Therefore, future research should be to address the case of arbitrary
textual claims that are expressed freely in an open-domain
setting, without making any assumptions on the structure
of the claim, or characteristics of the community or website
where the claim is made.

%% file: tail/biblio.tex
\cleardoublepage
\phantomsection
\bibliographystyle{tail/mybibstyle}
\bibliography{tail/credibility-analysis}
\addcontentsline{toc}{chapter}{Bibliography}

%% file: credibility-analysis.bbl
\begin{thebibliography}{xxxx}

\bibitem[Adler~2007]{Adler2007}
B.~Thomas Adler and Luca de~Alfaro.
\newblock {\em A content-driven reputation system for the wikipedia}.
\newblock In Proceedings of the 16th International Conference on World Wide
  Web, {WWW} 2007, Banff, Alberta, Canada, May 8-12, 2007, pages 261--270,
  2007.

\bibitem[Agarwal~2009]{DBLP:reference/db/AgarwalL09}
Nitin Agarwal and Huan Liu.
\newblock {\em Trust in Blogosphere}.
\newblock In Encyclopedia of Database Systems, pages 3187--3191. 2009.

\bibitem[Auer~2007]{DbPedia}
S{\"{o}}ren Auer, Christian Bizer, Georgi Kobilarov, Jens Lehmann, Richard
  Cyganiak and Zachary~G. Ives.
\newblock {\em DBpedia: {A} Nucleus for a Web of Open Data}.
\newblock In The Semantic Web, 6th International Semantic Web Conference, 2nd
  Asian Semantic Web Conference, {ISWC} 2007 + {ASWC} 2007, Busan, Korea,
  November 11-15, 2007., pages 722--735, 2007.

\bibitem[Baltrusaitis~2014]{tadas14}
Tadas Baltrusaitis, Peter Robinson and Louis{-}Philippe Morency.
\newblock {\em Continuous Conditional Neural Fields for Structured Regression}.
\newblock In Computer Vision - {ECCV} 2014 - 13th European Conference, Zurich,
  Switzerland, September 6-12, 2014, Proceedings, Part {IV}, pages 593--608,
  2014.

\bibitem[Bj{\"{o}}rne~2010]{DBLP:journals/bioinformatics/BjorneGPTS10}
Jari Bj{\"{o}}rne, Filip Ginter, Sampo Pyysalo, Jun'ichi Tsujii and Tapio
  Salakoski.
\newblock {\em Complex event extraction at PubMed scale}.
\newblock Bioinformatics {[ISMB]}, vol.~26, no.~12, pages 382--390, 2010.

\bibitem[Blei~2001]{Blei2003LDA}
David~M. Blei, Andrew~Y. Ng and Michael~I. Jordan.
\newblock {\em Latent Dirichlet Allocation}.
\newblock In Advances in Neural Information Processing Systems 14 [Neural
  Information Processing Systems: Natural and Synthetic, {NIPS} 2001, December
  3-8, 2001, Vancouver, British Columbia, Canada], pages 601--608, 2001.

\bibitem[Blei~2006]{BleiDTM}
David~M. Blei and John~D. Lafferty.
\newblock {\em Dynamic topic models}.
\newblock In Machine Learning, Proceedings of the Twenty-Third International
  Conference {(ICML} 2006), Pittsburgh, Pennsylvania, USA, June 25-29, 2006,
  pages 113--120, 2006.

\bibitem[Blei~2007]{bleiNIPS2007}
David~M. Blei and Jon~D. McAuliffe.
\newblock {\em Supervised Topic Models}.
\newblock In Advances in Neural Information Processing Systems 20, Proceedings
  of the Twenty-First Annual Conference on Neural Information Processing
  Systems, Vancouver, British Columbia, Canada, December 3-6, 2007, pages
  121--128, 2007.

\bibitem[Blei~2012]{Blei:2012:PTM:2133806.2133826}
David~M. Blei.
\newblock {\em Probabilistic Topic Models}.
\newblock Communications of the ACM, vol.~55, no.~4, pages 77--84, April 2012.

\bibitem[Bohannon~2012]{Bohannon2012}
Philip Bohannon, Nilesh~N. Dalvi, Yuval Filmus, Nori Jacoby, Sathiya Keerthi
  and Alok Kirpal.
\newblock {\em Automatic web-scale information extraction}.
\newblock In Proceedings of the {ACM} {SIGMOD} International Conference on
  Management of Data, {SIGMOD} 2012, Scottsdale, AZ, USA, May 20-24, 2012,
  pages 609--612, 2012.

\bibitem[Bollacker~2008]{Freebase}
Kurt~D. Bollacker, Colin Evans, Praveen Paritosh, Tim Sturge and Jamie Taylor.
\newblock {\em Freebase: a collaboratively created graph database for
  structuring human knowledge}.
\newblock In Proceedings of the {ACM} {SIGMOD} International Conference on
  Management of Data, {SIGMOD} 2008, Vancouver, BC, Canada, June 10-12, 2008,
  pages 1247--1250, 2008.

\bibitem[Bundschus~2008]{DBLP:journals/bmcbi/BundschusDSTK08}
Markus Bundschus, Math{\"{a}}us Dejori, Martin Stetter, Volker Tresp and
  Hans{-}Peter Kriegel.
\newblock {\em Extraction of semantic biomedical relations from text using
  conditional random fields}.
\newblock {BMC} Bioinformatics, vol.~9, 2008.

\bibitem[Canini~2011]{canini2011}
Kevin~Robert Canini, Bongwon Suh and Peter Pirolli.
\newblock {\em Finding Credible Information Sources in Social Networks Based on
  Content and Social Structure}.
\newblock In PASSAT/SocialCom 2011, Privacy, Security, Risk and Trust (PASSAT),
  2011 {IEEE} Third International Conference on and 2011 {IEEE} Third
  International Conference on Social Computing (SocialCom), Boston, MA, USA,
  9-11 Oct., 2011, pages 1--8, 2011.

\bibitem[Castillo~2011a]{DBLP:conf/www/CastilloMP11}
Carlos Castillo, Marcelo Mendoza and Barbara Poblete.
\newblock {\em Information credibility on twitter}.
\newblock In Proceedings of the 20th International Conference on World Wide
  Web, {WWW} 2011, Hyderabad, India, March 28 - April 1, 2011, pages 675--684,
  2011.

\bibitem[Castillo~2011b]{castillo2011}
Carlos Castillo, Marcelo Mendoza and Barbara Poblete.
\newblock {\em Information credibility on twitter}.
\newblock In Proceedings of the 20th International Conference on World Wide
  Web, {WWW} 2011, Hyderabad, India, March 28 - April 1, 2011, pages 675--684,
  2011.

\bibitem[Chen~2004]{chen04}
Di-Rong Chen, Qiang Wu, Yiming Ying and Ding-Xuan Zhou.
\newblock {\em Support Vector Machine Soft Margin Classifiers: Error Analysis}.
\newblock Journal of Machine Learning Research, vol.~5, pages 1143--1175, 2004.

\bibitem[Cline~2001]{cline2001consumer}
Rebecca~JW Cline and Katie~M Haynes.
\newblock {\em Consumer health information seeking on the Internet: the state
  of the art}.
\newblock Health education research, vol.~16, no.~6, 2001.

\bibitem[Coates~1987]{Coates1987}
Jennifer Coates.
\newblock {\em Epistemic Modality and Spoken Discourse}.
\newblock Transactions of the Philological Society, vol.~85, pages 100--131,
  1987.

\bibitem[Cortes~1995]{Cortes1995}
Corinna Cortes and Vladimir Vapnik.
\newblock {\em Support-Vector Networks}.
\newblock Machine Learning, vol.~20, no.~3, pages 273--297, 1995.

\bibitem[Danescu{-}Niculescu{-}Mizil~2013]{DanescuWWW2013}
Cristian Danescu{-}Niculescu{-}Mizil, Robert West, Dan Jurafsky, Jure Leskovec
  and Christopher Potts.
\newblock {\em No country for old members: user lifecycle and linguistic change
  in online communities}.
\newblock In 22nd International World Wide Web Conference, {WWW} '13, Rio de
  Janeiro, Brazil, May 13-17, 2013, pages 307--318, 2013.

\bibitem[Dave~2003]{Dave2003}
Kushal Dave, Steve Lawrence and David~M. Pennock.
\newblock {\em Mining the Peanut Gallery: Opinion Extraction and Semantic
  Classification of Product Reviews}.
\newblock In Proceedings of the 12th International Conference on World Wide
  Web, WWW '03, pages 519--528, New York, NY, USA, 2003. ACM.

\bibitem[de Alfaro~2011]{DBLP:journals/cacm/AlfaroKPA11}
Luca de~Alfaro, Ashutosh Kulshreshtha, Ian Pye and B.~Thomas Adler.
\newblock {\em Reputation systems for open collaboration}.
\newblock Commun. {ACM}, vol.~54, no.~8, pages 81--87, 2011.

\bibitem[Despotovic~2009]{DBLP:reference/db/Despotovic09a}
Zoran Despotovic.
\newblock {\em Trust and Reputation in Peer-to-Peer Systems}.
\newblock In Encyclopedia of Database Systems, pages 3183--3187. 2009.

\bibitem[Dong~2009]{Dong:2009:ICD:1687627.1687690}
Xin~Luna Dong, Laure Berti-Equille and Divesh Srivastava.
\newblock {\em Integrating Conflicting Data: The Role of Source Dependence}.
\newblock Proceedings of VLDB Endowment, vol.~2, no.~1, pages 550--561, 2009.

\bibitem[Dong~2013]{Dong:2013:CED:2488388.2488422}
Xin~Luna Dong and Divesh Srivastava.
\newblock {\em Compact explanation of data fusion decisions}.
\newblock In 22nd International World Wide Web Conference, {WWW} '13, Rio de
  Janeiro, Brazil, May 13-17, 2013, pages 379--390, 2013.

\bibitem[Dong~2014]{DBLP:conf/kdd/0001GHHLMSSZ14}
Xin Dong, Evgeniy Gabrilovich, Geremy Heitz, Wilko Horn, Ni~Lao, Kevin Murphy,
  Thomas Strohmann, Shaohua Sun and Wei Zhang.
\newblock {\em Knowledge vault: a web-scale approach to probabilistic knowledge
  fusion}.
\newblock In The 20th {ACM} {SIGKDD} International Conference on Knowledge
  Discovery and Data Mining, {KDD} '14, New York, NY, {USA} - August 24 - 27,
  2014, pages 601--610, 2014.

\bibitem[Dong~2015]{Dong:2015:KTE:2777598.2777603}
Xin~Luna Dong, Evgeniy Gabrilovich{\em  et~al.}
\newblock {\em Knowledge-based Trust: Estimating the Trustworthiness of Web
  Sources}.
\newblock Proceedings of VLDB Endowment, vol.~8, no.~9, pages 938--949, May
  2015.

\bibitem[Drucker~1996]{drucker97}
Harris Drucker, Christopher J.~C. Burges, Linda Kaufman, Alexander~J. Smola and
  Vladimir Vapnik.
\newblock {\em Support Vector Regression Machines}.
\newblock In Advances in Neural Information Processing Systems 9, NIPS, Denver,
  CO, USA, December 2-5, 1996, pages 155--161, 1996.

\bibitem[Einhorn~1977]{einhorn1977}
Hillel~J. Einhorn, Robin~M. Hogarth and Eric Klempner.
\newblock {\em Quality of group judgment}.
\newblock Psychological Bulletin, 1977.

\bibitem[Ernst~2014]{Ernst2014}
Patrick Ernst, Cynthia Meng, Amy Siu and Gerhard Weikum.
\newblock {\em KnowLife: {A} knowledge graph for health and life sciences}.
\newblock In {IEEE} 30th International Conference on Data Engineering, Chicago,
  {ICDE} 2014, IL, USA, March 31 - April 4, 2014, pages 1254--1257, 2014.

\bibitem[Esuli~2006]{Sentiwordnet}
Andrea Esuli and Fabrizio Sebastiani.
\newblock {\em SENTIWORDNET: A Publicly Available Lexical Resource for Opinion
  Mining}.
\newblock In In Proceedings of the 5th Conference on Language Resources and
  Evaluation (LREC'06, pages 417--422, 2006.

\bibitem[Fan~2008]{LibLinear}
Rong{-}En Fan, Kai{-}Wei Chang, Cho{-}Jui Hsieh, Xiang{-}Rui Wang and
  Chih{-}Jen Lin.
\newblock {\em {LIBLINEAR:} {A} Library for Large Linear Classification}.
\newblock Journal of Machine Learning Research, vol.~9, pages 1871--1874, 2008.

\bibitem[Fang~2014]{fang2014}
Hui Fang, Jie Zhang and Nadia Magnenat{-}Thalmann.
\newblock {\em Subjectivity grouping: learning from users' rating behavior}.
\newblock In International conference on Autonomous Agents and Multi-Agent
  Systems, {AAMAS} '14, Paris, France, May 5-9, 2014, pages 1241--1248, 2014.

\bibitem[Fei~2013]{Fei2013}
Geli Fei, Arjun Mukherjee, Bing Liu, Meichun Hsu, Mal{\'{u}} Castellanos and
  Riddhiman Ghosh.
\newblock {\em Exploiting Burstiness in Reviews for Review Spammer Detection}.
\newblock In Proceedings of the Seventh International Conference on Weblogs and
  Social Media, {ICWSM} 2013, Cambridge, Massachusetts, USA, July 8-11, 2013.,
  2013.

\bibitem[Feng~2012]{Feng2012}
Song Feng, Ritwik Banerjee and Yejin Choi.
\newblock {\em Syntactic Stylometry for Deception Detection}.
\newblock In The 50th Annual Meeting of the Association for Computational
  Linguistics, Proceedings of the Conference, July 8-14, 2012, Jeju Island,
  Korea - Volume 2: Short Papers, pages 171--175, 2012.

\bibitem[Fogg~2003]{fogg2003}
B.~J. Fogg.
\newblock {\em Prominence-interpretation theory: explaining how people assess
  credibility online}.
\newblock In Extended abstracts of the 2003 Conference on Human Factors in
  Computing Systems, {CHI} 2003, Ft. Lauderdale, Florida, USA, April 5-10,
  2003, pages 722--723, 2003.

\bibitem[Fox~2013]{Fox:PewInternetAndAmericanLifeProject:2013}
Susannah Fox and Maeve Duggan.
\newblock {\em Health online 2013}.
\newblock Pew Internet and American Life Project, 2013.

\bibitem[Galland~2010]{Galland:2010:CID:1718487.1718504}
Alban Galland, Serge Abiteboul, Am{\'{e}}lie Marian and Pierre Senellart.
\newblock {\em Corroborating information from disagreeing views}.
\newblock In Proceedings of the Third International Conference on Web Search
  and Web Data Mining, {WSDM} 2010, New York, NY, USA, February 4-6, 2010,
  pages 131--140, 2010.

\bibitem[Gallup.com~]{gallop}
Gallup.com.
\newblock {\em Americans' Confidence in Newspapers Continues to Erode}.
\newblock {http://www.gallup.com/poll/163097/americans-confidence-
  newspapers-continues-erode.aspx}.
\newblock {Accessed: 2015-05-07}.

\bibitem[Greene~2009]{greene2009}
Stephan Greene and Philip Resnik.
\newblock {\em More than Words: Syntactic Packaging and Implicit Sentiment}.
\newblock In Human Language Technologies: Conference of the North American
  Chapter of the Association of Computational Linguistics, Proceedings, May 31
  - June 5, 2009, Boulder, Colorado, {USA}, pages 503--511, 2009.

\bibitem[Griffiths~2002]{Griffiths02gibbssampling}
Tom Griffiths.
\newblock {\em Gibbs sampling in the generative model of Latent Dirichlet
  Allocation}.
\newblock Technical report, 2002.

\bibitem[Guha~2004a]{DBLP:conf/www/GuhaKRT04}
Ramanathan~V. Guha, Ravi Kumar, Prabhakar Raghavan and Andrew Tomkins.
\newblock {\em Propagation of trust and distrust}.
\newblock In Proceedings of the 13th international conference on World Wide
  Web, {WWW} 2004, New York, NY, USA, May 17-20, 2004, pages 403--412, 2004.

\bibitem[Guha~2004b]{GuhaWWW2004}
Ramanathan~V. Guha, Ravi Kumar, Prabhakar Raghavan and Andrew Tomkins.
\newblock {\em Propagation of trust and distrust}.
\newblock In Proceedings of the 13th international conference on World Wide
  Web, {WWW} 2004, New York, NY, USA, May 17-20, 2004, pages 403--412, 2004.

\bibitem[G{\"{u}}nnemann~2014]{Gunnemann2014}
Stephan G{\"{u}}nnemann, Nikou G{\"{u}}nnemann and Christos Faloutsos.
\newblock {\em Detecting anomalies in dynamic rating data: a robust
  probabilistic model for rating evolution}.
\newblock In The 20th {ACM} {SIGKDD} International Conference on Knowledge
  Discovery and Data Mining, {KDD} '14, New York, NY, {USA} - August 24 - 27,
  2014, pages 841--850, 2014.

\bibitem[Gupta~2012]{agupta2012}
Aditi Gupta and Ponnurangam Kumaraguru.
\newblock {\em Credibility Ranking of Tweets During High Impact Events}.
\newblock In Proceedings of the 1st Workshop on Privacy and Security in Online
  Social Media, PSOSM '12, pages 2:2--2:8, New York, NY, USA, 2012. ACM.

\bibitem[Gupta~2013]{DBLP:conf/www/0003LKJ13}
Aditi Gupta, Hemank Lamba, Ponnurangam Kumaraguru and Anupam Joshi.
\newblock {\em Faking Sandy: characterizing and identifying fake images on
  Twitter during Hurricane Sandy}.
\newblock In 22nd International World Wide Web Conference, {WWW} '13, Rio de
  Janeiro, Brazil, May 13-17, 2013, Companion Volume, pages 729--736, 2013.

\bibitem[Hang~2013]{Hang2013}
Chung{-}Wei Hang, Zhe Zhang and Munindar~P. Singh.
\newblock {\em Shin: Generalized Trust Propagation with Limited Evidence}.
\newblock {IEEE} Computer, vol.~46, no.~3, pages 78--85, 2013.

\bibitem[Howard~2011]{howard2011}
Philip~N. Howard, Aiden Duffy, Deen Freelon, Muzammil Hussain, Will Mari and
  Marwa Mazaid.
\newblock {\em Opening Closed Regimes: What Was the Role of Social Media During
  the Arab Spring?}
\newblock 2011.

\bibitem[Hu~2004]{Hu2004}
Minqing Hu and Bing Liu.
\newblock {\em Mining and summarizing customer reviews}.
\newblock In Proceedings of the Tenth {ACM} {SIGKDD} International Conference
  on Knowledge Discovery and Data Mining, Seattle, Washington, USA, August
  22-25, 2004, pages 168--177, 2004.

\bibitem[Hu~2012]{Hu2012}
Nan Hu, Indranil Bose, Noi~Sian Koh and Ling Liu.
\newblock {\em Manipulation of online reviews: An analysis of ratings,
  readability, and sentiments}.
\newblock Decision Support Systems, vol.~52, no.~3, pages 674--684, 2012.

\bibitem[IMS~Institute~2014]{IMS2014}
Healthcare~Informatics IMS~Institute.
\newblock {\em Engaging Patients through Social Media}.
\newblock \url{http://www.theimsinstitute.org/}, 2014.

\bibitem[J{\"{a}}rvelin~2002]{Jarvelin:TOIS2002}
Kalervo J{\"{a}}rvelin and Jaana Kek{\"{a}}l{\"{a}}inen.
\newblock {\em Cumulated gain-based evaluation of {IR} techniques}.
\newblock {ACM} Trans. Inf. Syst., vol.~20, no.~4, pages 422--446, 2002.

\bibitem[Jindal~2007]{Liu2007}
Nitin Jindal and Bing Liu.
\newblock {\em Analyzing and Detecting Review Spam}.
\newblock In Proceedings of the 7th {IEEE} International Conference on Data
  Mining {(ICDM} 2007), October 28-31, 2007, Omaha, Nebraska, {USA}, pages
  547--552, 2007.

\bibitem[Jindal~2008]{Liu2008}
Nitin Jindal and Bing Liu.
\newblock {\em Opinion spam and analysis}.
\newblock In Proceedings of the International Conference on Web Search and Web
  Data Mining, {WSDM} 2008, Palo Alto, California, USA, February 11-12, 2008,
  pages 219--230, 2008.

\bibitem[Jindal~2013]{DBLP:conf/ijcai/JindalR13}
Prateek Jindal and Dan Roth.
\newblock {\em End-to-End Coreference Resolution for Clinical Narratives}.
\newblock In {IJCAI} 2013, Proceedings of the 23rd International Joint
  Conference on Artificial Intelligence, Beijing, China, August 3-9, 2013,
  pages 2106--2112, 2013.

\bibitem[Joachims~2002]{joachims02}
Thorsten Joachims.
\newblock {\em Optimizing search engines using clickthrough data}.
\newblock In Proceedings of the Eighth {ACM} {SIGKDD} International Conference
  on Knowledge Discovery and Data Mining, July 23-26, 2002, Edmonton, Alberta,
  Canada, pages 133--142, 2002.

\bibitem[Jordan~2002]{jordan2002probabilistic}
M.I. Jordan and Y.~Weiss.
\newblock {\em {Probabilistic inference in graphical models}}.
\newblock Handbook of neural networks and brain theory, 2002.

\bibitem[Kalman~1960]{kalman}
R.~E. Kalman.
\newblock {\em {A New Approach to Linear Filtering and Prediction Problems}}.
\newblock Transactions of the ASME --- Journal of Basic Engineering, no.~82
  (Series D), pages 35--45, 1960.

\bibitem[Kamvar~2003]{DBLP:conf/www/KamvarSG03}
Sepandar~D. Kamvar, Mario~T. Schlosser and Hector Garcia{-}Molina.
\newblock {\em The Eigentrust algorithm for reputation management in {P2P}
  networks}.
\newblock In Proceedings of the Twelfth International World Wide Web
  Conference, {WWW} 2003, Budapest, Hungary, May 20-24, 2003, pages 640--651,
  2003.

\bibitem[Kang~2012]{kang2012}
Byungkyu Kang, John O'Donovan and Tobias H{\"{o}}llerer.
\newblock {\em Modeling topic specific credibility on twitter}.
\newblock In 17th International Conference on Intelligent User Interfaces,
  {IUI} '12, Lisbon, Portugal, February 14-17, 2012, pages 179--188, 2012.

\bibitem[Karatzas~1991]{gbm}
Ioannis Karatzas and Steven~Eugene Shreve.
\newblock Brownian motion and stochastic calculus.
\newblock Graduate texts in mathematics. Springer-Verlag, New York, Berlin,
  Heidelberg, 1991.
\newblock Autres tirages corriges : 1996, 1997, 1999, 2000, 2005.

\bibitem[Kim~2006]{Kim:2006:AAR:1610075.1610135}
Soo{-}Min Kim, Patrick Pantel, Timothy Chklovski and Marco Pennacchiotti.
\newblock {\em Automatically Assessing Review Helpfulness}.
\newblock In {EMNLP} 2007, Proceedings of the 2006 Conference on Empirical
  Methods in Natural Language Processing, 22-23 July 2006, Sydney, Australia,
  pages 423--430, 2006.

\bibitem[Koller~2009]{KollerFriedman2009}
Daphne Koller and Nir Friedman.
\newblock Probabilistic graphical models - principles and techniques.
\newblock {MIT} Press, 2009.

\bibitem[Koren~2008]{korenKDD2008}
Yehuda Koren.
\newblock {\em Factorization meets the neighborhood: a multifaceted
  collaborative filtering model}.
\newblock In Proceedings of the 14th {ACM} {SIGKDD} International Conference on
  Knowledge Discovery and Data Mining, Las Vegas, Nevada, USA, August 24-27,
  2008, pages 426--434, 2008.

\bibitem[Koren~2010]{KorenKDD2010}
Yehuda Koren.
\newblock {\em Collaborative filtering with temporal dynamics}.
\newblock Commun. {ACM}, vol.~53, no.~4, pages 89--97, 2010.

\bibitem[Koren~2015]{koren2011advances}
Yehuda Koren and Robert~M. Bell.
\newblock {\em Advances in Collaborative Filtering}.
\newblock In Recommender Systems Handbook, pages 77--118. 2015.

\bibitem[Krallinger~2008]{Krallinger2008}
Martin Krallinger, Alfonso Valencia and Lynette Hirschman.
\newblock {\em Linking genes to literature: text mining, information
  extraction, and retrieval applications for biology}.
\newblock Genome Biology, vol.~9, no.~2, page~S8, 2008.

\bibitem[Krishnamurthy~2009]{Krishnamurthy2009}
Rajasekar Krishnamurthy, Yunyao Li, Sriram Raghavan, Frederick Reiss,
  Shivakumar Vaithyanathan and Huaiyu Zhu.
\newblock {\em Web Information Extraction}.
\newblock In Encyclopedia of Database Systems, pages 3473--3478. 2009.

\bibitem[Kumar~2016]{DBLP:conf/www/Kumar0L16}
Srijan Kumar, Robert West and Jure Leskovec.
\newblock {\em Disinformation on the Web: Impact, Characteristics, and
  Detection of Wikipedia Hoaxes}.
\newblock In Proceedings of the 25th International Conference on World Wide
  Web, {WWW} 2016, Montreal, Canada, April 11 - 15, 2016, pages 591--602, 2016.

\bibitem[Kwon~2013]{DBLP:conf/icdm/KwonCJCW13}
Sejeong Kwon, Meeyoung Cha, Kyomin Jung, Wei Chen and Yajun Wang.
\newblock {\em Prominent Features of Rumor Propagation in Online Social Media}.
\newblock In 2013 {IEEE} 13th International Conference on Data Mining, Dallas,
  TX, USA, December 7-10, 2013, pages 1103--1108, 2013.

\bibitem[Lakkaraju~2011]{lakkarajuSDM2011}
Himabindu Lakkaraju, Chiranjib Bhattacharyya, Indrajit Bhattacharya and Srujana
  Merugu.
\newblock {\em Exploiting Coherence for the Simultaneous Discovery of Latent
  Facets and associated Sentiments}.
\newblock In Proceedings of the Eleventh {SIAM} International Conference on
  Data Mining, {SDM} 2011, April 28-30, 2011, Mesa, Arizona, {USA}, pages
  498--509, 2011.

\bibitem[Lampe~2007]{lampe2007}
Cliff Lampe and R.~Kelly Garrett.
\newblock {\em It's All News to Me: The Effect of Instruments on Ratings
  Provision}.
\newblock In 40th Hawaii International International Conference on Systems
  Science {(HICSS-40} 2007), {CD-ROM} / Abstracts Proceedings, 3-6 January
  2007, Waikoloa, Big Island, HI, {USA}, page 180, 2007.

\bibitem[Lavergne~2008]{DBLP:conf/ecai/LavergneUY08}
Thomas Lavergne, Tanguy Urvoy and Fran{\c{c}}ois Yvon.
\newblock {\em Detecting Fake Content with Relative Entropy Scoring}.
\newblock In Proceedings of the ECAI'08 Workshop on Uncovering Plagiarism,
  Authorship and Social Software Misuse, Patras, Greece, July 22, 2008, 2008.

\bibitem[Le~2014]{doc2vec}
Quoc~V. Le and Tomas Mikolov.
\newblock {\em Distributed Representations of Sentences and Documents}.
\newblock In Proceedings of the 31th International Conference on Machine
  Learning, {ICML} 2014, Beijing, China, 21-26 June 2014, pages 1188--1196,
  2014.

\bibitem[Lewis~2010]{lewis2010}
Seth~C Lewis, Kelly Kaufhold and Dominic~L Lasorsa.
\newblock {\em Thinking about citizen journalism: The philosophical and
  practical challenges of user-generated content for community newspapers}.
\newblock Journalism Practice, vol.~4, no.~2, 2010.

\bibitem[Li~2011]{Li:2011:TVT:2004686.2005589}
Xian Li, Weiyi Meng and Clement~T. Yu.
\newblock {\em T-verifier: Verifying truthfulness of fact statements}.
\newblock In Proceedings of the 27th International Conference on Data
  Engineering, {ICDE} 2011, April 11-16, 2011, Hannover, Germany, pages 63--74,
  2011.

\bibitem[Li~2012]{DBLP:journals/pvldb/LiDLMS12}
Xian Li, Xin~Luna Dong, Kenneth Lyons, Weiyi Meng and Divesh Srivastava.
\newblock {\em Truth Finding on the Deep Web: Is the Problem Solved?}
\newblock {PVLDB}, vol.~6, no.~2, pages 97--108, 2012.

\bibitem[Li~2013]{Ott2013a}
Jiwei Li, Myle Ott and Claire Cardie.
\newblock {\em Identifying Manipulated Offerings on Review Portals}.
\newblock In Proceedings of the 2013 Conference on Empirical Methods in Natural
  Language Processing, {EMNLP} 2013, 18-21 October 2013, Grand Hyatt Seattle,
  Seattle, Washington, USA, {A} meeting of SIGDAT, a Special Interest Group of
  the {ACL}, pages 1933--1942, 2013.

\bibitem[Li~2014a]{Liu2014}
Huayi Li, Zhiyuan Chen, Bing Liu, Xiaokai Wei and Jidong Shao.
\newblock {\em Spotting Fake Reviews via Collective Positive-Unlabeled
  Learning}.
\newblock In 2014 {IEEE} International Conference on Data Mining, {ICDM} 2014,
  Shenzhen, China, December 14-17, 2014, pages 899--904, 2014.

\bibitem[Li~2014b]{Li2014}
Jiwei Li, Myle Ott, Claire Cardie and Eduard~H. Hovy.
\newblock {\em Towards a General Rule for Identifying Deceptive Opinion Spam}.
\newblock In Proceedings of the 52nd Annual Meeting of the Association for
  Computational Linguistics, {ACL} 2014, June 22-27, 2014, Baltimore, MD, USA,
  Volume 1: Long Papers, pages 1566--1576, 2014.

\bibitem[Li~2014c]{Li:2014:RCH:2588555.2610509}
Qi~Li, Yaliang Li, Jing Gao, Bo~Zhao, Wei Fan and Jiawei Han.
\newblock {\em Resolving conflicts in heterogeneous data by truth discovery and
  source reliability estimation}.
\newblock In International Conference on Management of Data, {SIGMOD} 2014,
  Snowbird, UT, USA, June 22-27, 2014, pages 1187--1198, 2014.

\bibitem[Li~2015a]{Li2015}
Huayi Li, Zhiyuan Chen, Arjun Mukherjee, Bing Liu and Jidong Shao.
\newblock {\em Analyzing and Detecting Opinion Spam on a Large-scale Dataset
  via Temporal and Spatial Patterns}.
\newblock In Proceedings of the Ninth International Conference on Web and
  Social Media, {ICWSM} 2015, University of Oxford, Oxford, UK, May 26-29,
  2015, pages 634--637, 2015.

\bibitem[Li~2015b]{YaliangLi:SIGKDD2015}
Yaliang Li, Jing Gao, Chuishi Meng, Qi~Li, Lu~Su, Bo~Zhao, Wei Fan and Jiawei
  Han.
\newblock {\em A Survey on Truth Discovery}.
\newblock {SIGKDD} Explorations, vol.~17, no.~2, pages 1--16, 2015.

\bibitem[Li~2015c]{Li:2015:DET:2783258.2783277}
Yaliang Li, Qi~Li, Jing Gao, Lu~Su, Bo~Zhao, Wei Fan and Jiawei Han.
\newblock {\em On the Discovery of Evolving Truth}.
\newblock In Proceedings of the 21th {ACM} {SIGKDD} International Conference on
  Knowledge Discovery and Data Mining, Sydney, NSW, Australia, August 10-13,
  2015, pages 675--684, 2015.

\bibitem[Lim~2010]{Liu2010}
Ee{-}Peng Lim, Viet{-}An Nguyen, Nitin Jindal, Bing Liu and Hady~Wirawan Lauw.
\newblock {\em Detecting product review spammers using rating behaviors}.
\newblock In Proceedings of the 19th {ACM} Conference on Information and
  Knowledge Management, {CIKM} 2010, Toronto, Ontario, Canada, October 26-30,
  2010, pages 939--948, 2010.

\bibitem[Lin~2008]{Lin2008}
Chih{-}Jen Lin, Ruby~C. Weng and S.~Sathiya Keerthi.
\newblock {\em Trust Region Newton Method for Logistic Regression}.
\newblock Journal of Machine Learning Research, vol.~9, pages 627--650, 2008.

\bibitem[Lin~2009]{linCIKM2009}
Chenghua Lin and Yulan He.
\newblock {\em Joint sentiment/topic model for sentiment analysis}.
\newblock In Proceedings of the 18th {ACM} Conference on Information and
  Knowledge Management, {CIKM} 2009, Hong Kong, China, November 2-6, 2009,
  pages 375--384, 2009.

\bibitem[Lin~2011]{Lin2011}
Chenghua Lin, Yulan He and Richard Everson.
\newblock {\em Sentence Subjectivity Detection with Weakly-Supervised
  Learning}.
\newblock In Fifth International Joint Conference on Natural Language
  Processing, {IJCNLP} 2011, Chiang Mai, Thailand, November 8-13, 2011, pages
  1153--1161, 2011.

\bibitem[Liu~2007]{liu-EtAl:2007:EMNLP-CoNLL2007}
Jingjing Liu, Yunbo Cao, Chin{-}Yew Lin, Yalou Huang and Ming Zhou.
\newblock {\em Low-Quality Product Review Detection in Opinion Summarization}.
\newblock In EMNLP-CoNLL 2007, Proceedings of the 2007 Joint Conference on
  Empirical Methods in Natural Language Processing and Computational Natural
  Language Learning, June 28-30, 2007, Prague, Czech Republic, pages 334--342,
  2007.

\bibitem[Liu~2008]{icdm2008}
Yang Liu, Xiangji Huang, Aijun An and Xiaohui Yu.
\newblock {\em Modeling and Predicting the Helpfulness of Online Reviews}.
\newblock In Proceedings of the 8th {IEEE} International Conference on Data
  Mining {(ICDM} 2008), December 15-19, 2008, Pisa, Italy, pages 443--452,
  2008.

\bibitem[Liu~2012]{Liu2012}
Bing Liu.
\newblock Sentiment analysis and opinion mining.
\newblock Synthesis Lectures on Human Language Technologies. Morgan {\&}
  Claypool Publishers, 2012.

\bibitem[Lu~2009]{DBLP:conf/www/LuZS09}
Yue Lu, ChengXiang Zhai and Neel Sundaresan.
\newblock {\em Rated aspect summarization of short comments}.
\newblock In Proceedings of the 18th International Conference on World Wide
  Web, {WWW} 2009, Madrid, Spain, April 20-24, 2009, pages 131--140, 2009.

\bibitem[Lu~2010]{Lu:2010:ESC:1772690.1772761}
Yue Lu, Panayiotis Tsaparas, Alexandros Ntoulas and Livia Polanyi.
\newblock {\em Exploiting social context for review quality prediction}.
\newblock In Proceedings of the 19th International Conference on World Wide
  Web, {WWW} 2010, Raleigh, North Carolina, USA, April 26-30, 2010, pages
  691--700, 2010.

\bibitem[Luca~2015]{Luca}
Michael Luca and Georgios Zervas.
\newblock {\em Fake It Till You Make It: Reputation, Competition, and Yelp
  Review Fraud}.
\newblock Technical report, Harvard Business School, 2015.

\bibitem[Lukasik~2016]{DBLP:conf/acl/LukasikSVBZC16}
Michal Lukasik, P.~K. Srijith, Duy Vu, Kalina Bontcheva, Arkaitz Zubiaga and
  Trevor Cohn.
\newblock {\em Hawkes Processes for Continuous Time Sequence Classification: an
  Application to Rumour Stance Classification in Twitter}.
\newblock In Proceedings of the 54th Annual Meeting of the Association for
  Computational Linguistics, {ACL} 2016, August 7-12, 2016, Berlin, Germany,
  Volume 2: Short Papers, 2016.

\bibitem[Ma~2011]{MaWSDM2011}
Hao Ma, Dengyong Zhou, Chao Liu, Michael~R. Lyu and Irwin King.
\newblock {\em Recommender systems with social regularization}.
\newblock In Proceedings of the Forth International Conference on Web Search
  and Web Data Mining, {WSDM} 2011, Hong Kong, China, February 9-12, 2011,
  pages 287--296, 2011.

\bibitem[Ma~2015]{Ma:2015:FFG:2783258.2783314}
Fenglong Ma, Yaliang Li, Qi~Li, Minghui Qiu, Jing Gao, Shi Zhi, Lu~Su, Bo~Zhao,
  Heng Ji and Jiawei Han.
\newblock {\em FaitCrowd: Fine Grained Truth Discovery for Crowdsourced Data
  Aggregation}.
\newblock In Proceedings of the 21th {ACM} {SIGKDD} International Conference on
  Knowledge Discovery and Data Mining, Sydney, NSW, Australia, August 10-13,
  2015, pages 745--754, 2015.

\bibitem[McAuley~2013a]{mcauleyrecsys2013}
Julian~J. McAuley and Jure Leskovec.
\newblock {\em Hidden factors and hidden topics: understanding rating
  dimensions with review text}.
\newblock In Seventh {ACM} Conference on Recommender Systems, RecSys '13, Hong
  Kong, China, October 12-16, 2013, pages 165--172, 2013.

\bibitem[McAuley~2013b]{mcauleyWWW2013}
Julian~John McAuley and Jure Leskovec.
\newblock {\em From amateurs to connoisseurs: modeling the evolution of user
  expertise through online reviews}.
\newblock In 22nd International World Wide Web Conference, {WWW} '13, Rio de
  Janeiro, Brazil, May 13-17, 2013, pages 897--908, 2013.

\bibitem[McCallum~2005]{mccullum2005}
Andrew McCallum, Kedar Bellare and Fernando C.~N. Pereira.
\newblock {\em A Conditional Random Field for Discriminatively-trained
  Finite-state String Edit Distance}.
\newblock In {UAI} '05, Proceedings of the 21st Conference in Uncertainty in
  Artificial Intelligence, Edinburgh, Scotland, July 26-29, 2005, pages
  388--395, 2005.

\bibitem[Mihalcea~2009]{Mihalcea2009}
Rada Mihalcea and Carlo Strapparava.
\newblock {\em The Lie Detector: Explorations in the Automatic Recognition of
  Deceptive Language}.
\newblock In {ACL} 2009, Proceedings of the 47th Annual Meeting of the
  Association for Computational Linguistics and the 4th International Joint
  Conference on Natural Language Processing of the AFNLP, 2-7 August 2009,
  Singapore, Short Papers, pages 309--312, 2009.

\bibitem[Miller~1995]{WordNet}
George~A. Miller.
\newblock {\em WordNet: A Lexical Database for English}.
\newblock Communications of the ACM, vol.~38, no.~11, pages 39--41, November
  1995.

\bibitem[Mimno~2008]{mimnoUAI2008}
David~M. Mimno and Andrew McCallum.
\newblock {\em Topic Models Conditioned on Arbitrary Features with
  Dirichlet-multinomial Regression}.
\newblock In {UAI} 2008, Proceedings of the 24th Conference in Uncertainty in
  Artificial Intelligence, Helsinki, Finland, July 9-12, 2008, pages 411--418,
  2008.

\bibitem[Mitchell~2016]{Fox:PewInternetAndAmericanLifeProject:2016}
Amy Mitchell, Jeffrey Gottfried, Michael Barthel and Elisa Shearer.
\newblock {\em Trust and Accuracy}.
\newblock Pew Internet and American Life Project, 2016.

\bibitem[Mudambi~2010]{Mudambi:2010:MHO:2017447.2017457}
Susan~M. Mudambi and David Schuff.
\newblock {\em What Makes a Helpful Online Review? {A} Study of Customer
  Reviews on Amazon.com}.
\newblock {MIS} Quarterly, vol.~34, no.~1, pages 185--200, 2010.

\bibitem[Mukherjee~2012]{Mukherjee2012}
Subhabrata Mukherjee and Pushpak Bhattacharyya.
\newblock {\em Sentiment Analysis in Twitter with Lightweight Discourse
  Analysis}.
\newblock In {COLING} 2012, 24th International Conference on Computational
  Linguistics, Proceedings of the Conference: Technical Papers, 8-15 December
  2012, Mumbai, India, pages 1847--1864, 2012.

\bibitem[Mukherjee~2013a]{Liu2013}
Arjun Mukherjee, Abhinav Kumar, Bing Liu, Junhui Wang, Meichun Hsu, Mal{\'{u}}
  Castellanos and Riddhiman Ghosh.
\newblock {\em Spotting opinion spammers using behavioral footprints}.
\newblock In The 19th {ACM} {SIGKDD} International Conference on Knowledge
  Discovery and Data Mining, {KDD} 2013, Chicago, IL, USA, August 11-14, 2013,
  pages 632--640, 2013.

\bibitem[Mukherjee~2013b]{Liu2013a}
Arjun Mukherjee, Vivek Venkataraman, Bing Liu and Natalie~S. Glance.
\newblock {\em What Yelp Fake Review Filter Might Be Doing?}
\newblock In Proceedings of the Seventh International Conference on Weblogs and
  Social Media, {ICWSM} 2013, Cambridge, Massachusetts, USA, July 8-11, 2013.,
  2013.

\bibitem[Mukherjee~2013c]{DBLP:conf/www/MukherjeeBJ13}
Subhabrata Mukherjee, Gaurab Basu and Sachindra Joshi.
\newblock {\em Incorporating author preference in sentiment rating prediction
  of reviews}.
\newblock In 22nd International World Wide Web Conference, {WWW} '13, Rio de
  Janeiro, Brazil, May 13-17, 2013, Companion Volume, pages 47--48, 2013.

\bibitem[Mukherjee~2014a]{mukherjee2014JAST}
Subhabrata Mukherjee, Gaurab Basu and Sachindra Joshi.
\newblock {\em Joint Author Sentiment Topic Model}.
\newblock In Proceedings of the 2014 {SIAM} International Conference on Data
  Mining, Philadelphia, Pennsylvania, USA, April 24-26, 2014, pages 370--378,
  2014.

\bibitem[Mukherjee~2014b]{mukherjee2014}
Subhabrata Mukherjee, Gerhard Weikum and Cristian Danescu{-}Niculescu{-}Mizil.
\newblock {\em People on drugs: credibility of user statements in health
  communities}.
\newblock In The 20th {ACM} {SIGKDD} International Conference on Knowledge
  Discovery and Data Mining, {KDD} '14, New York, NY, {USA} - August 24 - 27,
  2014, pages 65--74, 2014.

\bibitem[Mukherjee~2015a]{Subho:ICDM2015}
Subhabrata Mukherjee, Hemank Lamba and Gerhard Weikum.
\newblock {\em Experience-Aware Item Recommendation in Evolving Review
  Communities}.
\newblock In 2015 {IEEE} International Conference on Data Mining, {ICDM} 2015,
  Atlantic City, NJ, USA, November 14-17, 2015, pages 925--930, 2015.

\bibitem[Mukherjee~2015b]{SubhoCIKM2015}
Subhabrata Mukherjee and Gerhard Weikum.
\newblock {\em Leveraging Joint Interactions for Credibility Analysis in News
  Communities}.
\newblock In Proceedings of the 24th {ACM} International on Conference on
  Information and Knowledge Management, {CIKM} 2015, Melbourne, VIC, Australia,
  October 19 - 23, 2015, pages 353--362, 2015.

\bibitem[Mukherjee~2016a]{Subho:ECML2016}
Subhabrata Mukherjee, Sourav Dutta and Gerhard Weikum.
\newblock {\em Credible Review Detection with Limited Information Using
  Consistency Features}.
\newblock In Machine Learning and Knowledge Discovery in Databases - European
  Conference, {ECML} {PKDD} 2016, Riva del Garda, Italy, September 19-23, 2016,
  Proceedings, Part {II}, pages 195--213, 2016.

\bibitem[Mukherjee~2016b]{Subho:KDD2016}
Subhabrata Mukherjee, Stephan G{\"{u}}nnemann and Gerhard Weikum.
\newblock {\em Continuous Experience-aware Language Model}.
\newblock In Proceedings of the 22nd {ACM} {SIGKDD} International Conference on
  Knowledge Discovery and Data Mining, San Francisco, CA, USA, August 13-17,
  2016, pages 1075--1084, 2016.

\bibitem[Mukherjee~2017]{Subho:SDM2017}
Subhabrata Mukherjee, Kashyap Popat and Gerhard Weikum.
\newblock {\em Exploring Latent Semantic Factors to Find Useful Product
  Reviews}.
\newblock In Proceedings of the 2017 {SIAM} International Conference on Data
  Mining, Houston, Texas, USA, April 27-29, 2017, 2017.

\bibitem[Nakashole~2014]{DBLP:conf/acl/NakasholeM14}
Ndapandula Nakashole and Tom~M. Mitchell.
\newblock {\em Language-Aware Truth Assessment of Fact Candidates}.
\newblock In Proceedings of the 52nd Annual Meeting of the Association for
  Computational Linguistics, {ACL} 2014, June 22-27, 2014, Baltimore, MD, USA,
  Volume 1: Long Papers, pages 1009--1019, 2014.

\bibitem[Nber.org~]{nber}
Nber.org.
\newblock {\em Media Bias and Voting}.
\newblock {http://www.nber.org/digest/oct06/w12169.html}.
\newblock {Accessed: 2015-05-07}.

\bibitem[Nielsen~]{nielsen}
Corporations Nielsen.
\newblock {\em Global Online Shopping Report}.
\newblock
  \url{http://www.nielsen.com/us/en/insights/news/2010/global-online-shopping-report.html}.
\newblock [Online; accessed 10-Jun-2016].

\bibitem[Nytimes.com~]{boston}
Nytimes.com.
\newblock {\em Should Reddit Be Blamed for the Spreading of a Smear?}
\newblock {http://www.nytimes.com/2013/07/28/magazine/should-reddit-be-
  blamed-for-the-spreading-of-a-smear.html}.
\newblock {Accessed: 2015-05-07}.

\bibitem[O'Mahony~2009]{O'Mahony:2009:LRH:1639714.1639774}
Michael~P. O'Mahony and Barry Smyth.
\newblock {\em Learning to recommend helpful hotel reviews}.
\newblock In Proceedings of the 2009 {ACM} Conference on Recommender Systems,
  RecSys 2009, New York, NY, USA, October 23-25, 2009, pages 305--308, 2009.

\bibitem[Ott~2011]{Ott2011}
Myle Ott, Yejin Choi, Claire Cardie and Jeffrey~T. Hancock.
\newblock {\em Finding Deceptive Opinion Spam by Any Stretch of the
  Imagination}.
\newblock In The 49th Annual Meeting of the Association for Computational
  Linguistics: Human Language Technologies, Proceedings of the Conference,
  19-24 June, 2011, Portland, Oregon, {USA}, pages 309--319, 2011.

\bibitem[Ott~2013]{Ott2013}
Myle Ott, Claire Cardie and Jeffrey~T. Hancock.
\newblock {\em Negative Deceptive Opinion Spam}.
\newblock In Human Language Technologies: Conference of the North American
  Chapter of the Association of Computational Linguistics, Proceedings, June
  9-14, 2013, Westin Peachtree Plaza Hotel, Atlanta, Georgia, {USA}, pages
  497--501, 2013.

\bibitem[Pan~2004]{Pang2004}
{\em A Sentimental Education: Sentiment Analysis Using Subjectivity
  Summarization Based on Minimum Cuts}.
\newblock In Proceedings of the 42nd Annual Meeting of the Association for
  Computational Linguistics, 21-26 July, 2004, Barcelona, Spain., pages
  271--278, 2004.

\bibitem[Pang~2002]{pang2002}
Bo~Pang and Vaithyanathan~Shivakumar Lee~Lillian.
\newblock {\em Thumbs up?: sentiment classification using machine learning
  techniques}.
\newblock In Proceedings of the 2002 Conference on Empirical Methods in Natural
  Language Processing, EMNLP '02, 2002.

\bibitem[Pang~2007]{PangLee2007}
Bo~Pang and Lillian Lee.
\newblock {\em Opinion Mining and Sentiment Analysis}.
\newblock Foundations and Trends in Information Retrieval, vol.~2, no.~1-2,
  pages 1--135, 2007.

\bibitem[Pasternack~2010]{Pasternack:2010:KB:1873781.1873880}
Jeff Pasternack and Dan Roth.
\newblock {\em Knowing What to Believe (when you already know something)}.
\newblock In {COLING} 2010, 23rd International Conference on Computational
  Linguistics, Proceedings of the Conference, 23-27 August 2010, Beijing,
  China, pages 877--885, 2010.

\bibitem[Pasternack~2011]{DBLP:conf/ijcai/PasternackR11}
Jeff Pasternack and Dan Roth.
\newblock {\em Making Better Informed Trust Decisions with Generalized
  Fact-Finding}.
\newblock In {IJCAI} 2011, Proceedings of the 22nd International Joint
  Conference on Artificial Intelligence, Barcelona, Catalonia, Spain, July
  16-22, 2011, pages 2324--2329, 2011.

\bibitem[Pasternack~2013]{Pasternack:2013:LCA:2488388.2488476}
Jeff Pasternack and Dan Roth.
\newblock {\em Latent credibility analysis}.
\newblock In 22nd International World Wide Web Conference, {WWW} '13, Rio de
  Janeiro, Brazil, May 13-17, 2013, pages 1009--1020, 2013.

\bibitem[Paul~2013]{DBLP:conf/naacl/PaulD13}
Michael~J. Paul and Mark Dredze.
\newblock {\em Drug Extraction from the Web: Summarizing Drug Experiences with
  Multi-Dimensional Topic Models}.
\newblock In Human Language Technologies: Conference of the North American
  Chapter of the Association of Computational Linguistics, Proceedings, June
  9-14, 2013, Westin Peachtree Plaza Hotel, Atlanta, Georgia, {USA}, pages
  168--178, 2013.

\bibitem[Pennebaker~2001]{liwc}
J.W. Pennebaker, M.E. Francis and R.J. Booth.
\newblock Linguistic inquiry and word count: A computerized text analysis
  program.
\newblock Psychology Press, 2001.

\bibitem[Peterson~2003]{Peterson:JMedInternetRes:2003}
Geraldine Peterson, Parisa Aslani and A.~Kylie Williams.
\newblock {\em How do Consumers Search for and Appraise Information on
  Medicines on the Internet? A Qualitative Study Using Focus Groups}.
\newblock Journal of Medical Internet Research, vol.~5, no.~4, page e33, Dec
  2003.

\bibitem[Qazvinian~2011]{DBLP:conf/emnlp/QazvinianRRM11}
Vahed Qazvinian, Emily Rosengren, Dragomir~R. Radev and Qiaozhu Mei.
\newblock {\em Rumor has it: Identifying Misinformation in Microblogs}.
\newblock In Proceedings of the 2011 Conference on Empirical Methods in Natural
  Language Processing, {EMNLP} 2011, 27-31 July 2011, John McIntyre Conference
  Centre, Edinburgh, UK, {A} meeting of SIGDAT, a Special Interest Group of the
  {ACL}, pages 1589--1599, 2011.

\bibitem[Qin~2008]{qinNIPS2008}
Tao Qin, Tie{-}Yan Liu, Xu{-}Dong Zhang, De{-}Sheng Wang and Hang Li.
\newblock {\em Global Ranking Using Continuous Conditional Random Fields}.
\newblock In Advances in Neural Information Processing Systems 21, Proceedings
  of the Twenty-Second Annual Conference on Neural Information Processing
  Systems, Vancouver, British Columbia, Canada, December 8-11, 2008, pages
  1281--1288, 2008.

\bibitem[Radosavljevic~2010]{radosavljevicECAI2010}
Vladan Radosavljevic, Slobodan Vucetic and Zoran Obradovic.
\newblock {\em Continuous Conditional Random Fields for Regression in Remote
  Sensing}.
\newblock In {ECAI} 2010 - 19th European Conference on Artificial Intelligence,
  Lisbon, Portugal, August 16-20, 2010, Proceedings, pages 809--814, 2010.

\bibitem[Rahman~2015]{Rahman2015}
Mahmudur Rahman, Bogdan Carbunar, Jaime Ballesteros and Duen Horng~(Polo) Chau.
\newblock {\em To catch a fake: Curbing deceptive Yelp ratings and venues}.
\newblock Statistical Analysis and Data Mining, vol.~8, no.~3, pages 147--161,
  2015.

\bibitem[Ramage~2011]{ramageKDD2011}
Daniel Ramage, Christopher~D. Manning and Susan~T. Dumais.
\newblock {\em Partially labeled topic models for interpretable text mining}.
\newblock In Proceedings of the 17th {ACM} {SIGKDD} International Conference on
  Knowledge Discovery and Data Mining, San Diego, CA, USA, August 21-24, 2011,
  pages 457--465, 2011.

\bibitem[Recasens~2013]{recasens2013}
Marta Recasens, Cristian Danescu{-}Niculescu{-}Mizil and Dan Jurafsky.
\newblock {\em Linguistic Models for Analyzing and Detecting Biased Language}.
\newblock In Proceedings of the 51st Annual Meeting of the Association for
  Computational Linguistics, {ACL} 2013, 4-9 August 2013, Sofia, Bulgaria,
  Volume 1: Long Papers, pages 1650--1659, 2013.

\bibitem[Rosen{-}Zvi~2004a]{DBLP:conf/uai/Rosen-ZviGSS04}
Michal Rosen{-}Zvi, Thomas~L. Griffiths, Mark Steyvers and Padhraic Smyth.
\newblock {\em The Author-Topic Model for Authors and Documents}.
\newblock In {UAI} '04, Proceedings of the 20th Conference in Uncertainty in
  Artificial Intelligence, Banff, Canada, July 7-11, 2004, pages 487--494,
  2004.

\bibitem[Rosen{-}Zvi~2004b]{rosenzviUAI2004}
Michal Rosen{-}Zvi, Thomas~L. Griffiths, Mark Steyvers and Padhraic Smyth.
\newblock {\em The Author-Topic Model for Authors and Documents}.
\newblock In {UAI} '04, Proceedings of the 20th Conference in Uncertainty in
  Artificial Intelligence, Banff, Canada, July 7-11, 2004, pages 487--494,
  2004.

\bibitem[Sarawagi~2008]{Sarawagi2008}
Sunita Sarawagi.
\newblock {\em Information Extraction}.
\newblock Foundations and Trends in Databases, vol.~1, no.~3, pages 261--377,
  2008.

\bibitem[Shayne~2003]{wemedia}
Bowman Shayne and Willis Chris.
\newblock {\em We Media: How Audiences are Shaping the Future of News and
  Information}.
\newblock 2003.

\bibitem[Sloanreview.mit.edu~]{sloanreview}
Sloanreview.mit.edu.
\newblock {\em The Problem With Online Ratings}.
\newblock
  {http://sloanreview.mit.edu/article/the-problem-with-online-ratings-2/ }.
\newblock {Accessed: 2015-05-07}.

\bibitem[Snyder~2007]{DBLP:conf/naacl/SnyderB07}
Benjamin Snyder and Regina Barzilay.
\newblock {\em Multiple Aspect Ranking Using the Good Grief Algorithm}.
\newblock In Human Language Technology Conference of the North American Chapter
  of the Association of Computational Linguistics, Proceedings, April 22-27,
  2007, Rochester, New York, {USA}, pages 300--307, 2007.

\bibitem[Somasundaran~2009]{DBLP:conf/acl/SomasundaranW09}
Swapna Somasundaran and Janyce Wiebe.
\newblock {\em Recognizing Stances in Online Debates}.
\newblock In {ACL} 2009, Proceedings of the 47th Annual Meeting of the
  Association for Computational Linguistics and the 4th International Joint
  Conference on Natural Language Processing of the AFNLP, 2-7 August 2009,
  Singapore, pages 226--234, 2009.

\bibitem[Sridhar~]{sridhar:aclws14}
Dhanya Sridhar, Lise Getoor and Marilyn Walker.
\newblock {\em Collective Stance Classification of Posts in Online Debate
  Forums}.
\newblock In ACL Joint Workshop on Social Dynamics and Personal Attributes in
  Social Media 2014.

\bibitem[Strapparava~2004]{WNAffect}
Carlo Strapparava and Alessandro Valitutti.
\newblock {\em WordNet Affect: an Affective Extension of WordNet}.
\newblock In Proceedings of the Fourth International Conference on Language
  Resources and Evaluation, {LREC} 2004, May 26-28, 2004, Lisbon, Portugal,
  2004.

\bibitem[Stuart~2007]{allan2007}
Allan Stuart.
\newblock {\em Citizen Journalism and the Rise of `Mass Self-Communication':
  Reporting the London Bombings.}
\newblock Global Media, vol.~1, no.~1, 2007.

\bibitem[Suchanek~2007]{Yago}
Fabian~M. Suchanek, Gjergji Kasneci and Gerhard Weikum.
\newblock {\em Yago: a core of semantic knowledge}.
\newblock In Proceedings of the 16th International Conference on World Wide
  Web, {WWW} 2007, Banff, Alberta, Canada, May 8-12, 2007, pages 697--706,
  2007.

\bibitem[Suchanek~2013]{SuchanekWeikum2013}
Fabian~M. Suchanek and Gerhard Weikum.
\newblock {\em Knowledge harvesting from text and Web sources}.
\newblock In 29th {IEEE} International Conference on Data Engineering, {ICDE}
  2013, Brisbane, Australia, April 8-12, 2013, pages 1250--1253, 2013.

\bibitem[Sun~2013]{Sun2013}
Huan Sun, Alex Morales and Xifeng Yan.
\newblock {\em Synthetic review spamming and defense}.
\newblock In The 19th {ACM} {SIGKDD} International Conference on Knowledge
  Discovery and Data Mining, {KDD} 2013, Chicago, IL, USA, August 11-14, 2013,
  pages 1088--1096, 2013.

\bibitem[Sutton~2012]{Sutton2012}
Charles~A. Sutton and Andrew McCallum.
\newblock {\em An Introduction to Conditional Random Fields}.
\newblock Foundations and Trends in Machine Learning, vol.~4, no.~4, pages
  267--373, 2012.

\bibitem[Tang~2013]{Tang:2013:CRH:2507157.2507183}
Jiliang Tang, Huiji Gao, Xia Hu and Huan Liu.
\newblock {\em Context-aware review helpfulness rating prediction}.
\newblock In Seventh {ACM} Conference on Recommender Systems, RecSys '13, Hong
  Kong, China, October 12-16, 2013, pages 1--8, 2013.

\bibitem[Titov~2008]{DBLP:conf/acl/TitovM08}
Ivan Titov and Ryan~T. McDonald.
\newblock {\em A Joint Model of Text and Aspect Ratings for Sentiment
  Summarization}.
\newblock In {ACL} 2008, Proceedings of the 46th Annual Meeting of the
  Association for Computational Linguistics, June 15-20, 2008, Columbus, Ohio,
  {USA}, pages 308--316, 2008.

\bibitem[Turney~2002]{Turney2002}
Peter~D. Turney.
\newblock {\em Thumbs Up or Thumbs Down?: Semantic Orientation Applied to
  Unsupervised Classification of Reviews}.
\newblock In Proceedings of the 40th Annual Meeting on Association for
  Computational Linguistics, ACL '02, pages 417--424, Stroudsburg, PA, USA,
  2002. Association for Computational Linguistics.

\bibitem[Vydiswaran~2011a]{DBLP:conf/kdd/VydiswaranZR11}
V.~G.~Vinod Vydiswaran, ChengXiang Zhai and Dan Roth.
\newblock {\em Content-driven trust propagation framework}.
\newblock In Proceedings of the 17th {ACM} {SIGKDD} International Conference on
  Knowledge Discovery and Data Mining, San Diego, CA, USA, August 21-24, 2011,
  pages 974--982, 2011.

\bibitem[Vydiswaran~2011b]{Vydiswaran:2011:GID:2023582.2023589}
V.G.~Vinod Vydiswaran, ChengXiang Zhai and Dan Roth.
\newblock {\em Gauging the Internet Doctor: Ranking Medical Claims Based on
  Community Knowledge}.
\newblock In Proceedings of the 2011 Workshop on Data Mining for Medicine and
  Healthcare, DMMH '11, pages 42--51, New York, NY, USA, 2011. ACM.

\bibitem[Vydiswaran~2012]{VZRPCIKM12}
V.~G.~Vinod Vydiswaran, ChengXiang Zhai, Dan Roth and Peter Pirolli.
\newblock {\em BiasTrust: teaching biased users about controversial topics}.
\newblock In 21st {ACM} International Conference on Information and Knowledge
  Management, CIKM'12, Maui, HI, USA, October 29 - November 02, 2012, pages
  1905--1909, 2012.

\bibitem[Walker~2012]{DBLP:conf/naacl/WalkerAAG12}
Marilyn~A. Walker, Pranav Anand, Rob Abbott and Ricky Grant.
\newblock {\em Stance Classification using Dialogic Properties of Persuasion}.
\newblock In Human Language Technologies: Conference of the North American
  Chapter of the Association of Computational Linguistics, Proceedings, June
  3-8, 2012, Montr{\'{e}}al, Canada, pages 592--596, 2012.

\bibitem[Wallach~2009]{wallach}
Hanna~M. Wallach, Iain Murray, Ruslan Salakhutdinov and David~M. Mimno.
\newblock {\em Evaluation methods for topic models}.
\newblock In Proceedings of the 26th Annual International Conference on Machine
  Learning, {ICML} 2009, Montreal, Quebec, Canada, June 14-18, 2009, pages
  1105--1112, 2009.

\bibitem[Wang~2006]{Wang2006}
Xuerui Wang and Andrew McCallum.
\newblock {\em Topics over time: a non-Markov continuous-time model of topical
  trends}.
\newblock In Proceedings of the Twelfth {ACM} {SIGKDD} International Conference
  on Knowledge Discovery and Data Mining, Philadelphia, PA, USA, August 20-23,
  2006, pages 424--433, 2006.

\bibitem[Wang~2010]{DBLP:conf/kdd/WangLZ10}
Hongning Wang, Yue Lu and Chengxiang Zhai.
\newblock {\em Latent aspect rating analysis on review text data: a rating
  regression approach}.
\newblock In Proceedings of the 16th {ACM} {SIGKDD} International Conference on
  Knowledge Discovery and Data Mining, Washington, DC, USA, July 25-28, 2010,
  pages 783--792, 2010.

\bibitem[Wang~2011a]{Liu2011}
Guan Wang, Sihong Xie, Bing Liu and Philip~S. Yu.
\newblock {\em Review Graph Based Online Store Review Spammer Detection}.
\newblock In 11th {IEEE} International Conference on Data Mining, {ICDM} 2011,
  Vancouver, BC, Canada, December 11-14, 2011, pages 1242--1247, 2011.

\bibitem[Wang~2011b]{wang2011}
Hongning Wang, Yue Lu and ChengXiang Zhai.
\newblock {\em Latent aspect rating analysis without aspect keyword
  supervision}.
\newblock In Proceedings of the 17th {ACM} {SIGKDD} International Conference on
  Knowledge Discovery and Data Mining, San Diego, CA, USA, August 21-24, 2011,
  pages 618--626, 2011.

\bibitem[Wang~2012]{BleiCTM}
Chong Wang, David~M. Blei and David Heckerman.
\newblock {\em Continuous Time Dynamic Topic Models}.
\newblock CoRR, vol.~abs/1206.3298, 2012.

\bibitem[West~2014]{West-etal:2014}
Robert West, Hristo~S. Paskov, Jure Leskovec and Christopher Potts.
\newblock {\em Exploiting Social Network Structure for Person-to-Person
  Sentiment Analysis}.
\newblock {TACL}, vol.~2, pages 297--310, 2014.

\bibitem[Westnet~2009]{Westney1986}
P.~Westnet.
\newblock {\em HOW TO BE MORE OR LESS CERTAIN IN ENGLISH: SCALARITY IN
  EPISTEMIC MODALITY}.
\newblock {IRAL - International Review of Applied Linguistics in Language
  Teaching}, vol.~24, no.~1-4, pages 311--336, 2009.

\bibitem[White~2014a]{White2014}
R~W White, R~Harpaz, N~H Shah, W~DuMouchel and E~Horvitz.
\newblock {\em Toward Enhanced Pharmacovigilance Using Patient-Generated Data
  on the Internet}.
\newblock Clinical Pharmacology \& Therapeutics, vol.~96, no.~2, pages
  239--246, 2014.

\bibitem[White~2014b]{white2014health}
Ryen~W. White and Eric Horvitz.
\newblock {\em From health search to healthcare: explorations of intention and
  utilization via query logs and user surveys}.
\newblock Journal of the American Medical Informatics Association, vol.~21,
  no.~1, pages 49--55, 2014.

\bibitem[Wiebe~2005]{Wiebe2005}
Janyce Wiebe and Ellen Riloff.
\newblock {\em Creating Subjective and Objective Sentence Classifiers from
  Unannotated Texts}.
\newblock In Computational Linguistics and Intelligent Text Processing, 6th
  International Conference, CICLing 2005, Mexico City, Mexico, February 13-19,
  2005, Proceedings, pages 486--497, 2005.

\bibitem[Wiebe~2011]{Wiebe2011}
Janyce Wiebe and Ellen Riloff.
\newblock {\em Finding Mutual Benefit between Subjectivity Analysis and
  Information Extraction}.
\newblock {IEEE} Transactions on Affective Computing, vol.~2, no.~4, pages
  175--191, 2011.

\bibitem[Wolf~2004]{Wolf2004}
Florian Wolf, Edward Gibson and Timothy Desmet.
\newblock {\em Discourse coherence and pronoun resolution}.
\newblock Language and Cognitive Processes, vol.~19, no.~6, 2004.

\bibitem[Xiang~2010]{XiangKDD2010}
Liang Xiang, Quan Yuan, Shiwan Zhao, Li~Chen, Xiatian Zhang, Qing Yang and
  Jimeng Sun.
\newblock {\em Temporal recommendation on graphs via long- and short-term
  preference fusion}.
\newblock In Proceedings of the 16th {ACM} {SIGKDD} International Conference on
  Knowledge Discovery and Data Mining, Washington, DC, USA, July 25-28, 2010,
  pages 723--732, 2010.

\bibitem[Xiong~2010]{xiongSDM2010}
Liang Xiong, Xi~Chen, Tzu{-}Kuo Huang, Jeff~G. Schneider and Jaime~G.
  Carbonell.
\newblock {\em Temporal Collaborative Filtering with Bayesian Probabilistic
  Tensor Factorization}.
\newblock In Proceedings of the {SIAM} International Conference on Data Mining,
  {SDM} 2010, April 29 - May 1, 2010, Columbus, Ohio, {USA}, pages 211--222,
  2010.

\bibitem[Xu~2012a]{DBLP:conf/coling/XuZ12}
Qiongkai Xu and Hai Zhao.
\newblock {\em Using Deep Linguistic Features for Finding Deceptive Opinion
  Spam}.
\newblock In {COLING} 2012, 24th International Conference on Computational
  Linguistics, Proceedings of the Conference: Posters, 8-15 December 2012,
  Mumbai, India, pages 1341--1350, 2012.

\bibitem[Xu~2012b]{DBLP:journals/jamia/XuHTC12}
Yan Xu, Kai Hong, Junichi Tsujii and Eric~I{-}Chao Chang.
\newblock {\em Feature engineering combined with machine learning and
  rule-based methods for structured information extraction from narrative
  clinical discharge summaries}.
\newblock {JAMIA}, vol.~19, no.~5, pages 824--832, 2012.

\bibitem[Yang~2012]{Yang:2012:ADR:2350190.2350203}
Fan Yang, Yang Liu, Xiaohui Yu and Min Yang.
\newblock {\em Automatic Detection of Rumor on Sina Weibo}.
\newblock In Proceedings of the ACM SIGKDD Workshop on Mining Data Semantics,
  MDS '12, pages 13:1--13:7, New York, NY, USA, 2012. ACM.

\bibitem[Yin~2008]{Yin:2008:TDM:1399100.1399392}
Xiaoxin Yin, Jiawei Han and Philip~S. Yu.
\newblock {\em Truth Discovery with Multiple Conflicting Information Providers
  on the Web}.
\newblock IEEE Transactions on Knowledge and Data Engineering, vol.~20, no.~6,
  pages 796--808, June 2008.

\bibitem[Yoo~2009]{Yoo2009}
Kyung~Hyan Yoo and Ulrike Gretzel.
\newblock {\em Comparison of Deceptive and Truthful Travel Reviews}.
\newblock In Information and Communication Technologies in Tourism, {ENTER}
  2009, Proceedings of the International Conference in Amsterdam, The
  Netherlands, 2009, pages 37--47, 2009.

\bibitem[Yu~2003]{Yu2003}
Hong Yu and Vasileios Hatzivassiloglou.
\newblock {\em Towards Answering Opinion Questions: Separating Facts from
  Opinions and Identifying the Polarity of Opinion Sentences}.
\newblock In Proceedings of the 2003 Conference on Empirical Methods in Natural
  Language Processing, EMNLP '03, pages 129--136, Stroudsburg, PA, USA, 2003.
  Association for Computational Linguistics.

\bibitem[Yu~2011]{DBLP:conf/acl/YuZWC11}
Jianxing Yu, Zheng{-}Jun Zha, Meng Wang and Tat{-}Seng Chua.
\newblock {\em Aspect Ranking: Identifying Important Product Aspects from
  Online Consumer Reviews}.
\newblock In The 49th Annual Meeting of the Association for Computational
  Linguistics: Human Language Technologies, Proceedings of the Conference,
  19-24 June, 2011, Portland, Oregon, {USA}, pages 1496--1505, 2011.

\bibitem[Zhao~2012a]{DBLP:journals/qdb/ZhaoRGH12}
Bo~Zhao and Jiawei Han.
\newblock {\em A Probabilistic Model for Estimating Real-valued Truth from
  Conflicting Sources}.
\newblock {QDB}, 2012.

\bibitem[Zhao~2012b]{Zhao:2012:BAD:2168651.2168656}
Bo~Zhao, Benjamin I.~P. Rubinstein, Jim Gemmell and Jiawei Han.
\newblock {\em A Bayesian Approach to Discovering Truth from Conflicting
  Sources for Data Integration}.
\newblock Proceedings of VLDB Endowment, vol.~5, no.~6, pages 550--561, 2012.

\bibitem[Zhao~2012c]{DBLP:journals/pvldb/ZhaoRGH12}
Bo~Zhao, Benjamin I.~P. Rubinstein, Jim Gemmell and Jiawei Han.
\newblock {\em A Bayesian Approach to Discovering Truth from Conflicting
  Sources for Data Integration}.
\newblock {PVLDB}, vol.~5, no.~6, pages 550--561, 2012.

\bibitem[Zhi~2015]{Zhi:2015:MTE:2783258.2783339}
Shi Zhi, Bo~Zhao, Wenzhu Tong, Jing Gao, Dian Yu, Heng Ji and Jiawei Han.
\newblock {\em Modeling Truth Existence in Truth Discovery}.
\newblock In Proceedings of the 21th {ACM} {SIGKDD} International Conference on
  Knowledge Discovery and Data Mining, Sydney, NSW, Australia, August 10-13,
  2015, pages 1543--1552, 2015.

\bibitem[Zhu~2003]{Zhu2005}
Xiaojin Zhu, Zoubin Ghahramani and John~D. Lafferty.
\newblock {\em Semi-Supervised Learning Using Gaussian Fields and Harmonic
  Functions}.
\newblock In Machine Learning, Proceedings of the Twentieth International
  Conference {(ICML} 2003), August 21-24, 2003, Washington, DC, {USA}, pages
  912--919, 2003.

\end{thebibliography}


\begin{thebibliography}{xxxx}

\end{thebibliography}
